\pdfoutput=1
\documentclass[a4paper,10pt]{book}

\usepackage{amsmath,amsthm,amssymb}
\usepackage{setspace}
\usepackage{amssymb,amsmath,amsthm,graphicx,epsfig,latexsym}
\usepackage[pdftex]{hyperref}

\newtheorem{theorem}{Theorem}

\newtheorem{lemma}{Lemma}
\newtheorem{corollary}{Corollary}
\newtheorem{claim}{Claim}
\newtheorem{definition}{Definition}

\theoremstyle{remark}
\newtheorem{example}{Example}[section]

\newcommand\beq{\begin{equation}}
\newcommand\eeq{\end{equation}}
\newcommand\bea{\begin{eqnarray}}
\newcommand\eea {\end{eqnarray}}
\newcommand\ba{\begin{array}}
\newcommand\ea {\end{array}}

\def\la{{\langle}}

\def\a{{\alpha}}
\def\b{{\beta}}
\def\g{{\gamma}}

\def\e{{\epsilon}}

\def\P{\mathbb{P}}      % Probability Measure
\def\CP{\mathbb{CP}}    % Cumulative Probability
\def\O{\Omega}
\def\la{\Lambda}
\def\ZZ{\mathbb{Z}}
\def\QQ{\mathbb{Q}}
\def\R{\mathbb{R}}
\font\german=eufm10 at 10pt
\def\Buchstabe#1{{\hbox{\german #1}}}

\def\11{B_{11}}
\def\7{B_{7}}

\def\go{\gamma_{11}}
\def\gs{\gamma_{17}}
\def\gos{\gamma_{117}}
\def\pl{\phi^{\oplus}} % linear dual
\def\D{\overline{D}}
\def\pe{\phi_\epsilon}      % approximate co-event
\def\EA{\Buchstabe{A}}      % event algebra
\def\Z2{\mathbb{Z}_2}
\def\p{\phi}            % co-event
\def\CE{{\EA^*}}        % space of co-events
\def\HL{(\Lambda,\EA_\Lambda,\mu_\Lambda)}     % A partitioned histories theory
\def\H{(\Omega,\EA,\mu)}     % A histories theory
\def\S{{\cal{S}}(\Omega,\EA,\mu)} % A co-event scheme
\def\L{{\cal{L}}(\Omega,\EA,\mu)} % Linear co-events
\def\LU{{\cal{L}}_U(\Omega,\EA,\mu)} % Unital Linear co-events
\def\C{{\cal{C}}(\Omega,\EA,\mu)} % Classical co-events
\def\Q{{\cal{Q}}(\Omega,\EA,\mu)} % Quadratic co-events
\def\QU{{\cal{Q}}_U(\Omega,\EA,\mu)} % Unital Quadratic co-events
\def\Pr{{\cal{P}}(\Omega,\EA,\mu)} % The ideal of preclusive co-events
\def\I{{\cal{I}}(\Omega,\EA,\mu)} % Ideal co-events
\def\M{{\cal{M}}(\Omega,\EA,\mu)} % Multiplicative co-events
\def\ML{{\cal{M}}(\Lambda,\EA_\Lambda,\mu_\Lambda)} % Multiplicative co-events
\def\Homp{Hom^{\oplus}(\EA,\Z2)} % The space of linear co-events

\newcommand{\ket}[1]{|#1\rangle}
\newcommand{\bra}[1]{\langle#1|}
\newcommand{\braket}[2]{\langle#1|#2\rangle}
\newcommand{\m}[1]{\overline{#1}}
\newcommand{\Pn}[1]{{{\cal{P}}_{#1}}(\Omega,\EA,\mu)}

\begin{document}

%\singlespacing
\pagestyle{empty}
\begin{titlepage}

\newcommand*{\titleTMB}{\begingroup
%\drop=0.1\textheight
\centering
\settowidth{\unitlength}{\LARGE QUANTUM MEASURE THEORY:}
\vspace*{\baselineskip}
%{\large\scshape Yousef Ghazi-Tabatabai}\\[\baselineskip]
\rule{\unitlength}{1.6pt}\vspace*{-\baselineskip}\vspace*{2pt}
\rule{\unitlength}{0.4pt}\\[\baselineskip]
{\LARGE QUANTUM MEASURE THEORY:}\\[\baselineskip]
{\LARGE A NEW INTERPRETATION}\\[\baselineskip]
%{\itshape Yousef Ghazi-Tabatabai \\ Imperial College}\\[0.2\baselineskip]
\rule{\unitlength}{0.4pt}\vspace*{-\baselineskip}\vspace{3.2pt}
\rule{\unitlength}{1.6pt}\\[\baselineskip]\vspace{3.2pt}
{\Large\itshape Yousef Ghazi-Tabatabai \\ \vspace{3.2pt} Imperial College}\par
%{\large\scshape $e$}\par
\vfill
{\large\scshape A Thesis Submitted for the Degree of \\ Doctor of Philosophy of the University of London \\ and the Diploma of Imperial College}\\[\baselineskip]
%{\large\scshape the publisher}\\[\baselineskip]
{\small\scshape January 2009}\par
%\vspace*{\drop}
\endgroup}

\titleTMB

%\maketitle

%\pagenumbering{roman}
\end{titlepage} %\setcounter{page}{1}
%\doublespacing 

\chapter*{Declaration}
\pagenumbering{roman}
\pagestyle{plain}
\setcounter{page}{1}

Unless specifically mentioned otherwise the work presented in this thesis is my own.
\\
\\
\\
\\
\\
\\
\\
\\
\\
\\
\begin{center}
    Yousef Ghazi-Tabatabai \\
January 2009
\end{center}

\chapter*{Abstract}

Quantum measure theory can be introduced as a histories based reformulation (and generalisation) of Copenhagen quantum mechanics in the image of classical stochastic theories. These classical models lend themselves to a simple interpretation in which a single history (a single element of the sample space) is deemed to be `real'; we require only that this real history should not be ruled out by the dynamics, the axioms of which ensure that not all histories are precluded. However, applying this interpretation naively to quantum measure theory we can find experimentally realisable systems (notably the Peres-Kochen-Specker system) in which every history is ruled out by the dynamics, challenging us to formulate a deeper realist framework.

Our first response is to hold on to our existing interpretative framework and attempt a revision of the dynamics that would reduce quantum measure theory to a classical dynamics. We explore this approach by examining the histories formulation of a stochastic-collapse model on a simple (discrete) null-lattice, concluding that the drawbacks of this approach outweigh the benefits.

Our second response is to abandon our classically inspired interpretation in favour of Sorkin's `co-events', a more general ontology that still allows for strict realism. In this case the `potentially real' objects of the theory (the `beables' in Bell's language) are not individual histories but truth valuation maps, or co-events. We develop \& evaluate various co-event schemes that have been suggested to date, finally adopting the multiplicative scheme; the current working model of co-event theory and a promising interpretation of quantum measure theory, though still a work in progress. We conclude by exploring the expression of the dynamics \& predictions in this new framework.

\chapter*{Acknowledgements}

Firstly I would like to thank Fay Dowker for her support, encouragement \& supervision of this thesis, and the two papers we have written together.

Thanks are certainly due to Rafael Sorkin, whose creative ideas I have spent most of this thesis developing. I am indebted to Rafael for his many insightful, profound \& encouraging comments, for his time given in proof reading drafts of parts of this thesis, and for the many conversations on the research presented here.

Thanks are also due to Petros Wallden, my collaborator for the paper on which the final chapter of this thesis is based, and who has always been a source of practical and perceptive analysis.

Thanks also to David Rideout, Surya Sumati and Wajid Mannan for the many useful discussions regarding the research presented in this thesis. Further thanks to Anna Gustavsson, Alexander Haupt and Wajid Mannan for proof reading drafts of parts of this thesis. This research was funded by a PPARC/STFC studentship. 

\tableofcontents

\chapter*{Opening Comments}
\addcontentsline{toc}{chapter}{Opening Comments}
\pagenumbering{arabic}
\setcounter{page}{1}
\pagestyle{myheadings}
\markboth{OPENING COMMENTS \hfill}{OPENING COMMENTS \hfill}

New analysis of a series of now famous experiments early in the $20$th century led to the falsification and abandonment of previous physical theories and the adoption of General Relativity and Quantum Mechanics, which became the two pillars of $20$th century physics. While both theories proved remarkably successful in their predictive ability, Quantum Mechanics remained open to interpretative questions leading eventually to the adoption of the Copenhagen Interpretation (usually associated with Bohr \& Heisenberg), though there were notable dissenters (in particular Einstein \& Schrodinger) and the search for a `realist' and `observer independent' interpretation continued. Indeed, today the interpretation of quantum mechanics has become a field in its own right, and is given additional impetus by the growing realisation that an observer independent understanding of quantum mechanics may well be a prerequisite for the construction of a viable theory of Quantum Gravity.

The Histories Approach begun by Dirac \cite{Dirac:1933} and Feynman \cite{Feynman:1948} suggests a new, spacetime, way of looking at quantum mechanics in which the (spacetime) histories are fundamental rather than the state vector. This approach presents a new framework for quantum mechanics in which a more observer independent and fully spacetime view of reality can be pursued, and thus has been suggested as the `correct' framework for the development of quantum gravity \cite{Dowker:private}.

However the question of the interpretation remains; we seek one that is both realist and observer independent. It makes sense at this point to expand on what we mean by this, though of course the discussion of the meaning of such terms is a field in itself and we are here concerned only with a brief clarification that will assist our understanding of what follows.

Interpretation is part of what distinguishes a physical theory from a mere model. In the case of a model we care only for the useful predicative power and would not be dismayed at the model's falsification outside the `region' in which we intend to apply it. In the case of a theory we expect some of the `working parts' to be imbued with `meaning', and if falsified we would reject the theory or at least downgrade it to an `emergent theory' or model. Thus though Newtonian Dynamics has been falsified as a theory it remains as a useful model. It is precisely this `meaning' that constitutes the theory's interpretation, thus the interpretation of a theory assumes all of the predictions thereof and the search for an interpretation differs markedly from the search for a deeper theory. In particular, in choosing to look for an interpretation of quantum mechanics we implicitly assume its predictive content and reject any interpretation that leads to a measurable contradiction therewith.

If a theory has a realist interpretation we expect this meaning to be concerned with ontology; we would require that some of the mathematical objects in the theory be regarded as representing `reality'. Further if the interpretation is observer independent we require that the theory and its meaning, its ontology in the case of a realist interpretation, could be described without reference to, and in the absence of, any observer. Indeed, we would assume that such an observer would itself be governed and described by our theory.

To date, the most developed interpretation of the histories approach is the consistent histories interpretation (section \ref{sec:the consistent histories interpretation}), which, though it may be argued to be observer independent, is not unambiguously realist. In this thesis we will introduce a new interpretation that is both realist and observer independent.

Chapter $1$ is introductory; we embark upon a brief review of the histories interpretation of quantum mechanics, which will be the framework for all that follows. We then sharpen our focus, reformulating the histories approach as ``quantum measure theory'', which will be the language we will use in the remainder of this thesis. We present quantum measure theory as an attempt to recast the histories approach in the image of classical stochastic mechanics, and introduce the `naive interpretation', that `one history is real'.

Chapter $2$ examines a gedankenexperimental realisation of the Peres-Kochen-Specker (PKS) system in terms of a series of Stern-Gerlach beam splitters and recombiners, leading to the conclusion that the naive interpretation must be abandoned. This chapter is based largely on \cite{Dowker:2007zz}.

In chapter $3$ we look into a histories formulation of a stochastic-collapse model, seeking to describe the system with a purely classical mechanics and thus avoiding the problems encountered in chapter $2$. We explore the implications of this approach and conclude that, in the author's view, its drawbacks outweigh its achievements. We draw the implication that an interpretation of the quantum dynamics can not be avoided. This chapter is based largely on \cite{Dowker:2007ma}.

In chapter $4$ we introduce the co-event interpretation, outlining its motivation and philosophy before describing the various co-event schemes that have been considered to date. We finish by discussing the interpretation of co-events themselves. This chapter draws from \cite{CoEventSchemes}.

In chapter $5$ we delve into the linear scheme, historically the first scheme to be suggested and one of the most studied. We explore the scheme's basic properties, examine its application to several simple systems and prove that it can succeed where the naive interpretation failed in describing the PKS system. However, we then find fault with the scheme, and reject it on grounds of inconsistency and contradiction with the predictions of quantum mechanics. This chapter draws from \cite{Sorkin:2006wq,Sorkin:2007,CoEventSchemes}.

In chapter $6$ we turn to higher order polynomial schemes, which were historically introduced as generalisations of the linear scheme after its failure and are structured so as to avoid the problems of the linear scheme. Unfortunately we find that generalisations of these problems can be found that are fatal for these schemes. This chapter is drawn largely from \cite{CoEventSchemes}.

In chapter $7$ we finally come to the multiplicative scheme, the most successful scheme to date and the current working model of the co-event interpretation. We explore the basic properties of the scheme, examine its application to simple systems and prove that it can successfully describe the PKS system, a description which we use to gain a deeper insight into the ``anhomomorphic'' nature of this scheme. We then address the issue of consistency and ``emergent classical logic''. Parts of this chapter are drawn from \cite{CoEventSchemes,Sorkin:2007,Dowker:2007zz}.

In chapter $8$ we ask if we can rephrase the dynamics of a histories theory in terms of the associated co-events, which would complete the shift from histories theories to co-event theories. This chapter is based largely upon \cite{ApproximatePreclusion}.

\part{Quantum Measure Theory}
\pagestyle{headings}
%\singlespacing
\chapter[Introducing The Histories Approach]{Introducing The Histories Approach}
%\doublespacing

\section{Decoherent Histories}\label{sec:decoherent histories}

As we will be mainly interested in interpretations of quantum measure theory (which we will define in section \ref{sec:quantum measure theory}), we will assume the reader has some degree of familiarity with the histories formulation of quantum mechanics \cite{Griffiths:1984rx,Omnes:1988ek,Gell:1990,Hartle:1992as}, on which quantum measure theory \cite{Sorkin:1994dt} is founded. We will however begin by reviewing some of the core results of the histories formulation, in particular those that will be essential to our later results. For more detail the reader is referred to Hartle \cite{Hartle:1992as}, Gell-Mann and Hartle \cite{Gell:1990,GellMann:1992kh}, Griffiths \cite{Griffiths:1984rx,Griffiths:1993,Griffiths:1996,Griffiths:1998}, Omnes \cite{Omnes:1988ek,Omnes:1988em,Omnes:1988ej,Omnes:1988fv,Omnes:1992ag,Omnes:1996yi} and Halliwell \cite{Halliwell:1994,Halliwell:2003}.

The histories approach is a whole attitude to quantum mechanics in which the (spacetime) histories are fundamental rather than the state vector. Developed initially by Dirac \& Feynman as the sum over histories approach \cite{Dirac:1933,Feynman:1948,Feynman:1965}, it aimed to formulate a new framework for quantum mechanics in which a more observer independent and fully spacetime view of reality can be pursued. Indeed Feynman's original paper was entitled ``Space-time approach to non-relativistic quantum mechanics'' \cite{Feynman:1948} and in his Nobel Lecture he described how ``[he] was becoming used to a physical point of view different from the more customary point of view. [...] The behaviour of nature is determined by saying her whole spacetime path has a certain character.'' \cite{Feynman:1965}

The histories approach can be defined axiomatically without reference to `state-vector' or `Copenhagen' quantum mechanics \cite{Hartle:1992as}, and indeed in the following chapters we will regard the axiomatic structure of a histories theory (section \ref{sec:the axiomatic approach}) as fundamental. However there is a significant overlap between the two formulations, which can be used to describe the same systems, with recent results suggesting that the histories formalism is the more general \cite{Dowker:inpreparation}. Because the `state-vector' formulation of quantum mechanics historically preceded the histories approach, and because it is more frequently used, we shall begin with a brief description of the histories formalism as it applies to non-relativistic quantum mechanics.

\subsection{From States to Histories}\label{sec:histories}

The histories approach is a reformulation (and potential generalisation) of quantum mechanics based on spacetime histories rather than the state vector. Following \cite{Hartle:1991bb,Isham:1994uv} closely we can assume `standard' non-relativistic quantum mechanics and describe the histories approach from this vantage point, and indeed such a description preceded the axiomatic formulation of the histories approach \cite{Hartle:1992as} which we will introduce in section \ref{sec:the axiomatic approach}.

We begin by assuming a quantum system, described at each time $t\in \R$ by a state $\ket{\psi_t}$ in a Hilbert space ${\cal{H}}_t$ with Hamiltonian $H$. We require all of the ${\cal{H}}_t$ to be isomorphic via canonical isomorphisms (so we know how to uniquely identify the bases of each Hilbert space), thus we can regard $\ket{\psi_t}$ as a vector evolving in a single Hilbert space, the state space, ${\cal{H}}$. Given two times $t_1<t_2$ we can construct the evolution operator $\ket{\psi_{t_2}}=U_{t_1}^{t_2}\ket{\psi_{t_1}}$, these evolution operators will obey the consistency condition $U_{t_1}^{t_3}=U_{t_2}^{t_3}U_{t_1}^{t_2}$ for $t_1\leq t_2\leq t_3$. We typically constrain the evolution operators to be unitary, setting $U_{t_1}^{t_2}=e^{-iH(t_2-t_1)/\hbar}$.

The description of a quantum system thus requires the specification of a Hilbert space ${\cal{H}}$ and a Hamiltonian $H$, to which we shall add an initial time $t_0$ and associated initial state $\ket{\psi_{t_0}}$. Because we shall be primarily concerned with finite systems we will depart from the literature at this point to specify a set of `(physically) interesting' times $T\subset\R$, with the intention of considering as relevant only those states $\ket{\psi_t}$ where $t\in T$, and of making reference these states and no other $\ket{\psi_t}$'s in any description of the system. We shall require $T$ to be bounded below and to contain its greatest lower bound, which we shall refer to as the \emph{initial time} $t_0$; we call the associated state $\ket{\psi_{t_0}}$ the \emph{initial state}. If $T$ is bounded above we require it to contain its least upper bound, which we call the \emph{final time}, associated with the \emph{final state}. We refer to the quadruple $({\cal{H}},H,\ket{\psi_{t_0}},T)$ as a \emph{Hilbert space theory}. Unless specifically noted otherwise we shall assume $T$ to be finite.

Given a Hilbert space theory $({\cal{H}},H,\ket{\psi_{t_0}},T)$ a measurement at time $t$ corresponds to a projection onto the subspace of $H$ corresponding to the outcome of the measurement. Thus we can associate (possible) measurement outcomes with projectors, and reversing this we can think of every projector as associated with a theoretically possible measurement outcome, or event (`propositions' in \cite{Isham:1994uv}), even though the measurement in question may not be experimentally realisable in practise.

We will be concerned with time ordered sequences of such projections, $\{P_{t_i}(\g_i)\}_{i=1}^n$, onto the associated outcomes $\g=\{\g_i\}_{i=1}^{n}$, where $t_i\in T$ and $t_i<t_{i+1}$.  Then we can define $T_\g=\{t_i\}_{i=1}^n$ and call $T_\g\subset T$ the \emph{indexing set} or \emph{temporal support} of the sequence \cite{Isham:1994uv}. Such a sequence of projections is called a \emph{homogenous history}, or sometimes simply \emph{coarse grained history} \cite{Griffiths:1984rx,Omnes:1988ek,Gell:1990,Isham:1994uv,Hartle:1991bb}. We will follow \cite{Isham:1994uv} by more formally defining a homogenous history, in the context of non-relativistic quantum mechanics, as a `proposition valued function'\footnote{though note that \cite{Isham:1994uv} assumes $T=\R$}:

\begin{definition}\label{def:history}
Given a Hilbert space theory $({\cal{H}},H,\ket{\psi_{t_0}},T)$ a \textbf{homogenous history} $\g$ with temporal support $T_\g\subset T$ is a map:
\beq\label{eq:history}
\g :T\rightarrow {\cal{P}}({\cal{H}}),
\eeq
where ${\cal{P}}({\cal{H}})$ is the space of projectors onto ${\cal{H}}$, and
\beq
\g(t) = \left\{\ba{cc} P_{t}(\g_{t}) & t\in T_\g \\ I & t\not\in T_\g, \ea\right.
\eeq
where $P_{t_i}(\g_{t_i})$ projects onto the subspace associated with the (theoretically possible measurement) outcome $\g_{t_i}$, and $I$ is the identity operator (we can assume $P_{t_0}(\g_{t_0})$ is always the identity). If $T_\g= T$ and every projector $P_{t_i}(\g_{t_i})$ is one dimensional for $i>0$ then we say that $\g$ is a \textbf{fine grained history} \cite{Hartle:1992as}.
\end{definition}

We will only need to use Hilbert space theories in which $T$ is finite, however we note that when $T$ is continuous a further, continuity, condition is typically imposed on the homogenous histories. Intuitively this condition associates the fine grained histories with continuous spacetime paths; formally this has been achieved by regarding homogenous histories as tensor products (rather than simply sequences) of projectors, and using this structure to define \emph{continuous tensor products}. For further detail the reader is referred to \cite{Isham:1998ct}.

Now two sequences of projections may differ in their temporal supports or their measurement outcomes $\g_i$. Notice that even within a single sequence these outcomes need not conform to a single observable, $\g_1$ may relate to a position measurement while $\g_2$ relates to momentum. We will find it simpler to restrict to a given `measurement basis', choosing a basis for ${\cal{H}}$ and requiring all our projections to map onto rays or subspaces defined by that basis. Thus, unless specifically mentioned otherwise, we will always assume the position basis (and thus assume `spacetime histories'), following Feynman's \cite{Feynman:1965} and Hartle's \cite{Hartle:1991bb} argument that the position basis is physically fundamental and hence the `correct' choice of a `physical' basis.

Some histories are refinements of others; given a history $\g$ we can make it more detailed by taking more (or more precise) measurements. Such a refinement is called a \textbf{fine graining} \cite{Griffiths:1984rx,Omnes:1988ek,Gell:1990}:

\begin{definition}\label{def:coarse graining}
Given a Hilbert space theory $({\cal{H}},H,\ket{\psi_{t_0}},T)$ and two homogenous histories $\g,~\tilde{\g}$, we say that $\g$ is a \textit{fine graining} of $\tilde{\g}$ (alternatively $\tilde{\g}$ is a \textit{coarse graining} of $\g$) if
\begin{enumerate}
 \item $T_{\tilde{\g}}\subset T_\g$
 \item The codomain of $P_{t}(\g_{t})$ is a subspace of the codomain of $P_{t}(\tilde{\g}_{t})$ for all $t\in T_{\tilde{\g}}$.
\end{enumerate}
\end{definition}

It is easy to see that all homogenous histories are coarse grainings of fine grained histories, and conversely that all fine grained histories are fine grainings of homogenous histories. Thus if we use $\O_{({\cal{H}},H,\ket{\psi_{t_0}},T)}$, which we abbreviate to $\O$, to denote the space of fine grained histories associated with a Hilbert space theory $({\cal{H}},H,\ket{\psi_{t_0}},T)$, we can associate every homogenous history with the set of fine grained histories that are fine grainings of $\O$. In this way we see that the space of homogenous histories is a subset of $P\O$, the power set\footnote{the \textit{power set} $PX$ or $2^X$ of a set $X$ is defined to be the set of subsets of $X$. If $X$ is a set then $PX$ is always a set in Zermelo set theory, as is every subset of $PX$} of $\O$. One of the critical insights of the histories approach lies in generalising this set to all of $P\O$ \cite{Hartle:1991bb,Isham:1994uv}. We now call \textit{any} partition of $\O$ a coarse graining thereof. A coarse graining whose elements are given by definition \ref{def:coarse graining} will be called a \textit{homogeneous coarse graining}, whereas one whose elements are not given by definition \ref{def:coarse graining} will be called an \textit{inhomogeneous coarse graining}. We wish to the retain the flexibility to restrict to a proper subset of $P\O$, thus we will use the space $\EA\subset P\O$ of what Hartle calls \textit{allowed coarse grainings} \cite{Hartle:1992as}, though unless noted otherwise we assume that $\EA=P\O$.

The elements of $\EA$ are often referred to as \emph{coarse grained histories} \cite{Hartle:1992as}, however to avoid confusion we will refer to them as \emph{events}, and unless specifically noted otherwise we will reserve the term \emph{history} to mean the fine grained histories (the elements of $\O$). Note that under this terminology a homogenous history is in fact an event, and indeed we can even think of a fine grained history $\g$ as the event $\{\g\}\in \EA$ (assuming it is one of the `allowed' coarse grainings). We will tend to use lowercase Greek letters to denote (fine grained) histories, and uppercase Latin letters to denote events, though there will be exceptions.

\subsection{The Sum Over Histories Approach}\label{sec:the sum over histories approach}

The notion of quantum histories was first introduced by Dirac and Feynman \cite{Dirac:1933,Feynman:1948}, who began the reformulation of the dynamics in terms of these histories. In the previous section we assumed an initial state $\ket{\psi_{t_0}}$ at the initial time $t_0$, discussed the the possibility of a sequence of measurements $\{P_{t_i}(A_i)\}_{i=1}^n$ resulting in outcomes $A=\{A_i\}_{i=1}^{n}$, and defined the conditions under which this sequence could be called a `homogenous history'.

We now ask what the probability of the sequence of outcomes $A=\{A_i\}_{i=1}^{n}$ might be. We can answer this by computing the state that would be the result of the measurements $\{P_{t_i}(A_i)\}_{i=1}^n$ resulting in outcomes $A=\{A_i\}_{i=1}^{n}$, and following the practise of standard quantum mechanics we achieve this by evolving the initial state with our evolution operator until the time of the first measurement, applying the relevant projector, normalising, then evolving again and so on. Thus with each homogenous history $A$ we can, assuming a finite temporal support, associate the \emph{class operator}:
\beq\label{eq:class operator}
C_A = P_{t_n}(A_{t_n})U_{t_n}^{t_{n-1}}P_{t_{n-1}}(A_{t_{n-1}})\ldots U_{t_0}^{t_1},
\eeq
and a by simple inductive argument the final state
\beq
\ket{\psi_{t_n}|_A}= \frac{C_A\ket{\psi_{t_0}}}{\bra{\psi_{t_0}}C_A^\dagger C_A\ket{\psi_{t_0}}}.
\eeq
The denominator is a normalisation factor, so this suggests the importance of the term
\beq\label{eq:probability amplitude}
\ket{\psi_{t_n}^A}= C_A\ket{\psi_{t_0}},
\eeq
which called the \textit{vector valued amplitude} (the \textit{probability amplitude} in Feynman's terminology \cite{Feynman:1948} or the \textit{branch} according to Hartle \cite{Hartle:1992as}). We refer the reader to \cite{Isham:1998ct,Feynman:1948} for the generalisation of these concepts to the infinite (including the continuous) case. Then the probability of the outcomes $A=\{A_i\}_{i=1}^{n}$ occurring upon measurement is simply
\beq\label{eq:probability in standard quantum mechanics}
\P(A) = \braket{\psi_{t_n}^A}{\psi_{t_n}^A}.
\eeq
Note that this would still hold were $A$ a general sequence of projections rather than a homogenous history. Now one of the key insights of the path integral framework is that the amplitude of a homogenous history $A$ is equal to the sum of the amplitudes of the fine grained histories $\g$ that are fine grainings of $A$. This allows us to view vector valued amplitudes as a sum over (the class operator corresponding to) a set of (fine grained) histories acting on the initial state; in the infinite case this is the intuitive thinking behind the `\textit{path integral}' \cite{Dirac:1933,Feynman:1948}. The core of the sum over histories approach can be stated thus \cite{Feynman:1948,Hartle:1992as}:
\begin{lemma}\label{lemma:sum over histories}
\beq\label{eq:sum over histories}
\ket{\psi_{t_n}^A}=\displaystyle{\sum_{\begin{array}{c} \g \text{\tiny{ fine grained history}} \\ \g \text{\tiny{ fine graining of A}}\end{array}}} \ket{\psi^\g_{t_n}}.
\eeq
\end{lemma}
Recalling that we can associate each homogenous history $A$ with the subset of $\O$ consisting of the fine grained histories that are fine grainings of $A$ we can rewrite equation \ref{eq:sum over histories} as:
\beq\label{eq:sum over histories class operators}
C_A=\sum_{\g\in A}C_\g.
\eeq
We can use this to generalise our construction; if $A$ is now taken to be a general event in $\EA$ we can use equation \ref{eq:sum over histories class operators} to define the associated class operator $C_A$, allowing us to construct an amplitude for (and assign a probability to) every event $A\in\EA$ \cite{Hartle:1992as}. Note crucially that if $A,B\in\EA$ are disjoint we have
\bea
C_{A\sqcup B} &=& \sum_{\g\in A\sqcup B}C_\g, \nonumber \\
&=& (\sum_{\g\in A}+\sum_{\g\in A})C_\g, \nonumber \\
&=& C_A + C_B,
\eea
with the immediate consequence that
\beq
\ket{\psi^{A\sqcup B}_{t_n}} = \ket{\psi^{A}_{t_n}} +\ket{\psi^{B}_{t_n}}.
\eeq

\subsection{The Decoherence Functional}\label{sec:the decoherence functional}

The sum over histories formalism outlined above is still very much tied to the Hilbert space framework, though the thinking is now organised around the histories. Later work has taken this program further, altogether eliminating the need for the Hilbert space in the formulation of the dynamics. This is achieved by noticing that the predictive power of a Hilbert space theory lies in the probability assertions that are norms of the states. Applying this to the sum over histories formalism, we see that the norms of the vector valued amplitudes of coarse grained histories are nothing but sums of the inner products of the `constituent' fine grained histories. This leads us to claim that the information content of a Hilbert space theory (or it sum over histories reworking) lies entirely in the inner products of the vector valued amplitudes, in particular those associated with the fine grained histories. This assertion has recently been validated \cite{Dowker:inpreparation}, as we shall see below, putting the histories approach on a sure footing.

Now since each amplitude corresponds to an event we can combine all of these inner products into one object, a map from pairs of events to the complex numbers. This is the decoherence functional \cite{Hartle:1992as,Isham:1994uv,Isham:1994zz}:
\bea
D:\EA\times\EA&\rightarrow&\mathbb{C} \nonumber \\
D(A,B) &=& \braket{\psi^A_{t_n}}{\psi^B_{t_n}}. \label{eq:decoherence functional on amplitudes}
\eea
This automatically obeys the following conditions:
\begin{enumerate}
\item \textit{Hermiticity}
\bea
D(A,B) &=& \braket{\psi^A_{t_n}}{\psi^B_{t_n}},\nonumber\\
&=& \braket{\psi^B_{t_n}}{\psi^A_{t_n}}^\dagger, \nonumber \\
&=& D(B,A)^\dagger. \label{eq:decoherence functional on amplitudes hermiticity}
\eea
\item \textit{Linearity}\footnote{This condition is called \textit{superposition} by Hartle \cite{Hartle:1992as}}
\bea
D(A\sqcup B,C)&=& \braket{\psi^{A\sqcup B}_{t_n}}{\psi^C_{t_n}}, \nonumber \\
&=&(\bra{\psi^A_{t_n}}+\bra{\psi^B_{t_n}})\ket{\psi^C_{t_n}},\nonumber\\
&=&\braket{\psi^A_{t_n}}{\psi^C_{t_n}}+\braket{\psi^B_{t_n}}{\psi^C_{t_n}},\nonumber\\
&=&D(A,C)+D(B,C).\label{eq:decoherence functional on amplitudes linearity}
\eea
\item \textit{Positivity}
\bea
D(A,A) &=& \braket{\psi^A_{t_n}}{\psi^A_{t_n}}, \nonumber \\
&=& \P(A), \nonumber \\
&\geq& 0. \label{eq:decoherence functional on amplitudes positivity}
\eea
\end{enumerate}

Using hermiticity and linearity we can derive the value of $D$ on any pair of events from its value on pairs of (fine grained) histories, and so can consider $D$ as a complex valued hermitian form over the fine grained histories,
$$D:\O\times\O\rightarrow\mathbb{C}.$$
We can represent this with a matrix, which by abuse of notation we will also call the decoherence functional. Notice that this matrix will always have non-negative eigenvalues, because of the Cauchy-Schwarz inequality and because the diagonal terms will always be non-negative (by positivity).

Coarse grainings are easily expressed in this formalism. If $\la=\{A_i\}_{i=1}^m$ is a partition of $\O$ such that $A_i\in\EA_\O$ then we can generate a subset $\EA_\la$ of $\EA_\O$ by taking the transitive closure of $\la$ using the operations $\cap$ \& $\cup$. Notice that $\EA_\la$ is also a subset of $P\la$, and that $\EA_\O=P\O\Rightarrow \EA_\la=P\la$. Then a decoherence functional $D_\O$ on $\EA_\O$ implies a `coarse grained' decoherence functional $D_\la$ on $\EA_{\la}$ by restriction:
\bea
D_\la : \EA_{\la}\times \EA_{\la} &\rightarrow& \mathbb{C}, \nonumber \\
D_\la(A,B)&=& D_\O(A,B). \label{eq:coarse grained decoherence functional}
\eea

\subsection{Decoherence and Emergent Classicality}\label{sec:decoherence and emergent classicality}

We can model a classical system using Hilbert space quantum mechanics by setting all the interference terms to zero,
\beq\label{eq:decoherence with states}
i\neq j\Rightarrow\braket{\psi_i}{\psi_j}=0.
\eeq
Translating this into decoherence functional language we see that all the `off-diagonal' terms of the associated matrix are zero. In a general system it may be that a particular coarse graining results in a diagonal decoherence functional while the decoherence functional corresponding to the fine grained histories does not posses this property. A partition resulting in a diagonal decoherence functional matrix is said to be \textit{decoherent} or \textit{consistent} \cite{Hartle:1992as}. Notice that every coarse graining of a decoherent set will itself be decoherent, however there may also be `incompatible' decoherent sets that are not coarse grainings of one another

The concept of \textit{emergent classicality} is often used to explain the discrepancy between the observed classical macro-world and the hypothesised quantum micro-world. Emergent classicality occurs when a system whose microstates obey quantum dynamics behaves classically at the macro-level. The consistent histories framework lends itself naturally to such analysis, a system in which the fine grained histories are not decoherent may yet contain decoherent coarse grainings. In fact, given a decoherent set of histories, it is possible to interpret the diagonal elements of the decoherence functional as probabilities, as we have seen above (equation \ref{eq:decoherence functional on amplitudes positivity}).

To this end, the concept of \textit{weak decoherence} \cite{Diosi:2003tv} has also been put forward, requiring only that the non-diagonal terms are imaginary. Although his does allow the diagonal terms of a given decoherence functional to be interpreted as probabilities Diosi \cite{Diosi:2003tv} has pointed out potentially fatal problems for this condition, namely its dynamical instability and its failure to be consistent under the composition of systems. We will always use the `full' decoherence condition defined in equation \ref{eq:decoherence with states}.

Because of its relation to emergent classicality, decoherence has been suggested (for example in \cite{Hartle:1991bb}) as an observer independent alternative (strictly a generalisation) of the notion of measurement in Copenhagen quantum mechanics, allowing us to assign probabilities to elements of a decoherent partition without reference to an external observer or a measurement. The histories approach has subsequently been described as a more appropriate formalism for the study of closed system, in particular quantum gravity and quantum cosmology \cite{Hartle:1992as} in which the inherently spacetime nature of the approach is a further strength.

\subsection{The Axiomatic Approach}\label{sec:the axiomatic approach}

To complete our transition from a quantum theory formulated in terms of Hilbert space states to one whose basic objects are histories we will follow Hartle \cite{Hartle:1992as} in defining a histories theory from scratch without reference to a Hilbert space theory. The crucial step lies in changing the way we think about histories, moving from time ordered sequences of projectors to thinking of a history as a `irreducible theoretical entity in its own right' \cite{Isham:1994uv}. To this end we define a \textit{histories theory} in more general terms, as a triple $(\O,\EA,D)$ \cite{Hartle:1992as}, the constituents of which we can define in terms of abstract `histories' with no reference to a Hilbert space theory.

We start with a set $\O$ that we will use as our set of fine grained histories, and the related set of `allowed' subsets of $\O$ (generalising Hartle's `allowed coarse grainings \cite{Hartle:1992as}) which we call \textit{events} (or \textit{propositions} \cite{Isham:1994uv,Griffiths:1998}) $\EA\subset P\O$. When $\O$ is finite we will take $\EA=P\O$, but in the infinite case we may want to restrict it to a proper subset. More technically $P\O$ forms a boolean algebra and we require $\EA$ to be a subalgebra, a requirement that we will discuss in greater depth below. However it should be noted that, as previously mentioned, the specification of $\EA$ is yet to be developed in the infinite case. We are now in a position to define a decoherence functional abstractly \cite{Isham:1994uv}:
\begin{definition}\label{def:decoherence functional}
A decoherence functional on a sample space $\O$ with associated event algebra $\EA$ is a map $D:\EA\times\EA\rightarrow\mathbb{C}$ obeying:
\begin{enumerate}
\item \textit{Hermiticity}
\beq\label{eq:decoherence functional hermiticity} D(A,B)=D(B,A)^\dagger,\eeq
\item \textit{Linearity}
\beq\label{eq:decoherence functional linearity} D(A\sqcup B,C)=D(A,C) + D(B,C),\eeq
\item \textit{Positivity}
\beq\label{eq:decoherence fucntional positivity} D(A,A)\geq 0. \eeq
\end{enumerate}
\end{definition}
As before, using hermiticity and linearity we can derive the value of $D$ on any pair of events from its value on pairs of (fine grained) histories, and so can consider $D$ as a complex valued hermitian form over the fine grained histories,
$$D:\O\times\O\rightarrow\mathbb{C}$$
We can represent this with a matrix, which by abuse of notation we will also call the decoherence functional. However our observation in section \ref{sec:the decoherence functional} that the eigenvalues of the decoherence functional matrix are non-negative no longer holds in general. This leads us to define:
\begin{definition}\label{def:strong positivity}
Let $D$ be a decoherence functional on a sample space $\O$ with an associated event algebra $\EA$. We say that $D$ obeys \textbf{strong positivity} and is \textbf{strongly positive} if for any finite collection of (not necessarily disjoint) events $A_1,A_2, \ldots, A_n\in\EA$  the $n\times n$ matrix $M_{ij}=  D(X_i,X_j)$ is positive semi-definite.
\end{definition}
When $\O$ is finite strong positivity implies that the matrix of the decoherence functional is itself positive semi-definite, so that its eigenvalues are non-negative. Conversely it an be shown that for finite $\O$ every decoherence functional with a positive semi-definite matrix obeys strong positivity, therefore whenever $\O$ is finite a decoherence functional obeys strong positivity if and only if its eigenvalues are non-negative.

It can be shown that every histories theory derived from a Hilbert space theory possesses a decoherence functional obeying strong positivity \cite{Dowker:inpreparation}, thus such theories are a proper subset of the set of histories theories. For an example of a system that can be described by a histories theory but not by a Hilbert space theory see \cite{Popescu:1994,Khalfin:1985,Barnett:2007}.

The positivity condition can be seen as a generalisation of strong positivity; it is required if we are to interpret the diagonal terms as probabilities upon decoherence. To this end, we will also require \textit{unitality} \cite{Isham:1994uv}, though this condition is not always imposed on decoherence functionals:
\beq\label{eq:decoherence functional unitality} D(\O,\O)=1 \eeq

A (finite) partition $\Lambda+\{A_i\}_{i=1}^n$ of $\O$ where $A_i\ni\EA$ will generate subalgebra $\EA_{\Lambda}$ of $\EA$, and we can define $D_{\Lambda}$ as the restriction of $D$ to $\EA_{\Lambda}$ as before. A partition $\la=\{A_i\}$ is said to be \textit{decoherent} or \textit{consistent} if $i\neq j\Rightarrow D(A_i,A_j)=0$. In this case we can define:
\bea
\P:\EA_{\Lambda}&\rightarrow& \R, \nonumber \\
\P(A_i) &=& D(A_i,A_i), \label{eq:decoherence and probability}
\eea
and $\P$ will obey the axioms of a classical probability measure. Further, when the decoherence functional has been derived from a Hilbert space formulation of quantum mechanics equations \ref{eq:probability in standard quantum mechanics},\ref{eq:decoherence functional on amplitudes positivity} show that the valuation of $\P$ on the sets $A_i\in\Lambda$ agrees with the probabilities assigned to these events by the Hilbert space theory from which the decoherence functional was defined.

As noted above this approach is more general than the Hilbert space state formulation of quantum mechanics, for as we have seen above in finite dimensions every decoherence functional derived from the Hilbert space formalism possess the property of strong positivity in addition to the defining properties of a decoherence functional. Thus there are systems we can describe in terms of the histories approach that we could not formulate in terms of Hilbert spaces whereas as we have seen above every Hilbert space system admits a histories description. Further, it has recently been shown \cite{Dowker:inpreparation} that given a histories theory with a decoherence functional obeying strong positivity it is possible to reconstruct the Hilbert space, demonstrating that all the information present in the Hilbert space formulation is present in the histories description, which is thus placed on a firm footing.

\subsection{The Consistent Histories Interpretation}\label{sec:the consistent histories interpretation}

We have seen above that the elements of $P\O$ (and thus $\EA$) can be interpreted as \textit{events} or \textit{propositions} \cite{Isham:1994uv,Griffiths:1998}. If our histories theory has been derived from Hilbert space quantum mechanics every (fine grained) history will be associated with a set of (potential) measurement outcomes, thus the related event would be the realisation of such outcomes and we have the related proposition that such an event occurred. Further, any event can be thought of as a set of (fine grained) histories. This set can be constructed from its constituent elements using the set operations (ie $A=\bigcup_{\g\in A} \g$), which can be thought of as operations of Boolean logic. Thus since we can interpret the (fine grained) histories as propositions we can think of any event as a proposition constructed from its constituent histories.

Every coarse graining of a decoherent partition will itself be decoherent, and we will consider a partition $\{A_i\}$ of $\O$ to be `less than' a partition $\{B_j\}$ of $\O$ if $\{B_j\}$ can be considered as a partition of $\{A_i\}$. Alternatively some authors say that the \textit{framework} $\{A_i\}$ is a \textit{consistent refinement} of the framework $\{B_j\}$ \cite{Dowker:1994dd,Griffiths:1996}. A consistent set is said to be \textit{maximally refined} if it has no consistent refinement \cite{Dowker:1994dd}. We say that $\{A_i\}$ and $\{B_j\}$ are \textit{incompatible} (or \textit{incomparable}) if neither is less than or equal to the other \cite{Griffiths:1996}, otherwise they are \textit{compatible} (or \textit{comparable}). Now equation \ref{eq:decoherence and probability} shows how we can interpret the propositions contained in a decoherent partition $\{A_i\}$ using the usual classical tools, and if two partitions are compatible we know how to treat one as a classical subsystem of the other. Indeed, if the system is classical the set $\O$ of fine grained histories itself will be a decoherent partition, and will be a refinement of every other decoherent partition, thus every two incompatible decoherent partitions will share a decoherent refinement. However, this is not the case for a general quantum system, in which there is in general no unique maximally refined consistent partition.

As pointed out in \cite{Dowker:1994dd}, this poses an interpretational problem. Given a classical system we know how to assign truth valuations to all propositions simultaneously, however the quantum case is more difficult, presenting us with a plurality of incompatible classical coarse grainings which all have equal status as regards the dynamics. Every `classical proposition' (an event that is a member of a consistent set\footnote{We call such propositions suggestively, histories theories derived by coarse graining to consistent sets are dynamically classical, and we hope to be able to treat them as classical theories, `ignoring' the non-classical fine graining}) is `\textit{conditional}' \cite{Dowker:1994dd} on the (maximally refined) consistent set in which it participates. Classical statements made in one consistent set may preclude classical statements made in other consistent sets, the translation of the uncertainty principle into the histories formalism. Further, the dynamics affords us no means by which to `choose' a privileged consistent set, and so we are not in a position to make `unconditional' classical assertions \cite{Dowker:1994dd}. The incompatibility of certain measurements may be acceptable in the standard formulation of quantum mechanics, in which we assume an external observer who will `choose' the measurement basis, however this state of affairs is more problematic in histories theories for they attempt a description of a closed system (as we would expect in quantum gravity) in which no external observer can be assumed. The theory has thus been criticised as `incomplete' \cite{Dowker:1994dd}.

To date the most developed approach to this problem has been the \textit{consistent histories interpretation}, which embraces the existence of incompatible classical interpretations of a given system, affording equal status to each one and adjusting the rules of inference to prevent statements in one `framework' contradicting statements from other, incompatible, frameworks \cite{Griffiths:1996,Griffiths:1998}. Essentially, two questions can only be simultaneously answered if they both participate in a common consistent set: ``\ldots neither ``contrary'' nor ``contradictory'' can be defined as logical relationships unless both of the properties (or histories) being compared are found in the same consistent family. Thus the formalism never allows an inference to either ``contrary'' or contradictory'' pairs of propositions\ldots'' \cite{Griffiths:1998}. The recent work of D\"{o}ring and Isham \cite{Doring:2007ib,Doring:2007ic,Doring:2007id,Doring:2007ie,Doring:2008gv} has developed these ideas further and placed them on a more rigorous basis, although at the time of writing their results pertain only to single time quantum mechanics and are not yet able to treat histories theories. Note that this interpretation takes the various classical interpretations of the system, corresponding to the various maximally refined consistent sets, as fundamental, and thus has nothing to say about propositions concerning the non-classical `micro-realm' lying `beneath' the level of decoherent partitions, these propositions being `disallowed' under the new laws of inference.

An alternate approach to the problem of non-unique maximally refined consistent sets is to abandon our reliance on consistent sets as the pillar of our interpretative framework, seeking instead a `realist' account of the micro-realm that reduces to a classical interpretation on physically observable consistent sets. The recent \textit{co-event} approach \cite{Sorkin:2006wq,Sorkin:2007,Dowker:2007zz}, based upon quantum measure theory \cite{Sorkin:1994dt,Sorkin:1995nj}, is an attempt at such an interpretation, and its development will be the major theme of this work.

\section{Quantum Measure Theory}\label{sec:quantum measure theory}

Quantum measure theory \cite{Sorkin:1994dt,Sorkin:1995nj} builds on the histories approach to quantum mechanics, aiming to reformulate it in the image of classical stochastic mechanics as opposed to the classical hamiltonian mechanics that underlies the mathematics and the thinking behind the Hilbert space formulation of quantum mechanics.

A classical stochastic theory can be phrased in terms of a triple $(\O,\EA,\P)$ of a sample space $\O$, an `event algebra' $\EA$ of measurable subsets of the sample space and a probability measure $\P$ encoding the dynamics. The coincidence of this notation and that used above is not accidental. Quantum measure theory is essentially a histories theory of quantum mechanics, and in a sense it is a measure theory approach to the histories approach. This in itself is not new, as we have seen histories theories have already been defined in terms of a triple $(\O,\EA,D)$. Though quantum measure theory perhaps puts more emphasis on this `measure theory structure' than previous accounts, its real innovation lies in the recasting of the dynamics, represented in decoherent histories by the decoherence functional, as a \textit{quantum measure} $\mu$ that seeks to resemble the `classical' probability measure $\P$ as closely as possible. To this end, we begin with a review of the classical theory.

\subsection{Stochastic Theories}\label{sec:stochastic mechanics}

As mentioned above a classical stochastic theory can be phrased in terms of a triple $(\O,\EA,\P)$ of a sample space $\O$, an `event algebra' $\EA$ of measurable subsets of the sample space and a probability measure $\P$ encoding the dynamics.

The sample space is usually identified with the space of `beables' \cite{Bell:1987iii} or `possible realities' of the theory. The event algebra is a subset, and indeed a subalgebra, of $P\O$, every element of which is thought of as a question or event (see above). $P\O$ has a natural boolean structure, indeed power sets are the canonical examples of boolean algebras, and we usually require the event algebra to be a $\sigma$-algebra of $P\O$.

We can think of $P\O$ (and boolean algebras in general) as an algebra over $\Z2$. Power sets come equipped with the standard set operations, $\cup,\cap,\neg$, and Boolean algebras have analogous operations. Thus we can define $\forall~A,~B\in P\O$:
\bea
A+B &:=& A\triangle B, \nonumber \\
&=& (A\cup B)\setminus (A\cap B), \nonumber \\
&=& (A\cup B)\cap(\neg (A\cap B)), \label{eq:set addition} \\
AB &:=& A\cap B. \label{eq:set mutiplication}
\eea
Then it is easy to see that
\bea
AB &=& BA, \nonumber \\
\emptyset + A &=& A, \nonumber \\
A\emptyset &=& \emptyset, \nonumber \\
\O + A &=& \O, \nonumber \\
A\O &=& A, \nonumber \\
AA &=& A, \nonumber \\
A + A &=& \emptyset,
\eea
so $P\O$ forms an algebra over $\Z2$ under these operations, with $\emptyset$ as the zero element and $\O$ as a unit. Using this formulation, we can constrain the event algebra $\EA$ by requiring it to be a subalgebra of $P\O$. When $\O$ is finite we take all of $P\O$ as the event algebra.

The dynamics are encoded in the \textit{probability measure}, a real valued function on $\EA$ obeying:
\begin{enumerate}
\item \textit{Positivity} \beq\label{eq:probability measure positivity} \P(A)\geq 0.\eeq
\item \textit{Kolmogorov Sum Rule} \beq\label{eq:kolmogorov sum rule} \P(A\sqcup B)=\P(A)+\P(B).\eeq
\item \textit{Unitality} \beq\label{eq:probability measure unitality} \P(\O)=1.\eeq
\end{enumerate}
Where $A,~B\in\EA$.

The usual interpretation of a stochastic theory asserts the reality of one member of the sample space, one history in our terms. Events are true if and only if they contain the real history. However, the real history is not uniquely determined by the dynamics. In fact, the only statement we can make with certainty is that the real history is not a member of a null set (a set of measure zero). We call this concept \textit{preclusion}, and it will be a central theme of our attempt to generalise this interpretation to accommodate quantum mechanics. More formally, preclusion states that if the real history is $r\in\O$, then $\P(A)=0\Rightarrow r\not\in A$, where $A\in\EA$.

The centrality of preclusion to our arguments promotes the study of null sets and their structure. Under a classical measure, we note that all \textit{negligible sets}, which are measurable subsets of null sets, are themselves null. Further, as a direct consequence of the Kolmogorov sum rule the disjoint union of two null sets is also null.

\subsection{The Quantum Measure}\label{sec:the quantum measure}

We aim to take as much of this structure as is possible across to the study of quantum mechanics, reformulating it as a `quantum measure theory' based upon the histories approach outlined above.

The sample space and event algebra can be transferred across relatively unaltered. In our histories based approach the sample space will simply be the space $\O$ of (fine grained) histories that we used above. Similarly the event algebra $\EA$ will once more be a subalgebra of $P\O$ and in the finite case we will always use $\EA=P\O$. In the classical case, when $\O$ is infinite we define $\EA$ to be the set of measurable subsets of $\O$, however in the absence of a fully developed theory of quantum measures and integration we will in what follows specify $\EA$ in each infinite system we consider. This lack of a full quantum measure theory for infinite sample spaces is an impediment to the interpretation and application thereof, as we shall see in later chapters. For a more detailed discussion of the implications for our own work see section \ref{sec:infinite sample spaces}.

It is in the measure that the differences between classical and quantum mechanics become apparent. In the above (for example see equation \ref{eq:decoherence functional on amplitudes}) we have already expressed the dynamics of quantum mechanics in terms of histories and events, we thus need only to rephrase the decoherence functional as a generalised \textit{quantum measure}. We will want this quantum measure to behave classically when the system, or a coarse graining thereof, is classical. By `behaving classically' on a coarse graining we mean that the restriction of the quantum measure to the corresponding subalgebra of $\EA$ (unless noted otherwise we will henceforth assume that every partition of a sample space $\O$ consists of events $A_i\in\EA$ so that $\EA_\la$ is a subalgebra of $\EA$) should to obey the axioms defining a classical measure, and should yield the same probabilities as the corresponding Hilbert space theory. But in our histories formulation a `classical subsystem' means a decoherent one, and we have already seen (equation \ref{eq:decoherence and probability}) that if $\Lambda=\{A_i\}$ is a decoherent partition then the relevant probabilities are given by the diagonal terms, $\P(A_i)=D(A_i,A_i)$. This suggests \cite{Sorkin:1994dt}:
\begin{definition}\label{def:quantum measure}
Given a sample space $\O$, event algebra $\EA$ and associated decoherence functional $D$ the quantum measure of the system is defined by:
\bea \mu:\EA &\rightarrow& \R, \nonumber \\
\mu(A)&:=& D(A,A).\label{eq:quantum measure definition}  \eea
\end{definition}
Given a coarse graining (a partition) $\Lambda$ of $\O$ and the associated subalgebra $\EA_{\Lambda}$ of $\EA$ we can define the implied measure $\mu_{\Lambda}$ as a restriction of $\mu$ to $\EA_{\Lambda}$: $\mu_{\Lambda}(A)=\mu(A)$ for $A\in\Lambda$. Thus if $\Lambda$ is a decoherent partition, equation \ref{eq:decoherence and probability} implies that $\mu_{\Lambda}$ obeys the axioms of a probability measure. Further, if the underlying decoherence functional was derived from a Hilbert space theory the valuations of $\mu_{\Lambda}$ on $\EA_{\Lambda}$ will indeed be the probabilities predicted by theory. We say that $\mu$ \textit{behaves classically} on the partition $\Lambda$.

We can't always assume decoherence, and in the general case the properties of the decoherence functional, from definition \ref{def:decoherence functional}, imply \cite{Sorkin:1994dt,Salgado:1999pu}:
\begin{enumerate}
\item \textit{Positivity} \beq\label{eq:quantum measure positivity} \mu(A)\geq 0. \eeq
\item \textit{Sum Rule} \beq\label{eq:quantum measure sum rule} \mu(A\sqcup B\sqcup C) = \mu(A\sqcup B) + \mu(B\sqcup C) + \mu(A\sqcup C) -\mu(A) - \mu(B) -\mu(C). \eeq
\item \textit{Unitality} \beq\label{eq:quantum measure unitality} \mu(\O)=1. \eeq
\end{enumerate}
Where $A,~B,~C\in\EA$. We say that $\mu$ obeys strong positivity if the related decoherence functional possess this quality. Note that $\mu$ is equivalent to the real part of the decoherence functional, which we can reconstruct using
\beq\nonumber
\mu(A\sqcup B) = \mu(A) + \mu(B) + 2Re(D(A,B)),
\eeq
which yields \cite{Sorkin:1994dt}:
\beq
D(A,B) = \mu(A+B) + \mu(A) + \mu(B) - 2\mu(A + AB) - 2\mu(B + AB).
\eeq
Thus we could define the quantum measure axiomatically using the equations above and construct a decoherence functional from it. We have taken the decoherence functional as the more fundamental object because it contains more information (its imaginary part). This extra information does not contribute to measurable quantities (the probabilities), however it is used when coupling several systems. This does mean that for a closed system (as we assume the `universe' to be) the decoherence functional and quantum measure will have equivalent information content. In what follows when specifying a quantum measure we will always assume a decoherence functional in the background from which it is defined.

Now $\mu$ differs from a classical measure in that it does not obey the Kolmogorov sum rule, equation \ref{eq:kolmogorov sum rule}, due to the appearance of non-zero off-diagonal `interference' terms in the decoherence functional obstructing decoherence. However this interference is not unconstrained, and instead of the Kolmogorov sum rule we have the weaker equation \ref{eq:quantum measure sum rule}, where we can think of $\mu(A\sqcup B)-\mu(A)-\mu(B)$ as the interference term, or obstruction to classicality. This can be placed in a hierarchy of interference terms, and thus measure theories, defined in terms of a general positive real valued function on the event algebra measure $\mu:\EA\rightarrow \R$ \cite{Sorkin:1994dt,Salgado:1999pu}:
\bea
I_1(A) &:=& \mu(A), \nonumber \\
I_2(A_1,A_2) &:=& \mu(A_1\sqcup A_2) - \mu(A_1) - \mu(A_2), \nonumber \\
I_3(A_1,A_2,A_3) &:=& \mu(A_1\sqcup A_2 \sqcup A_3) - \mu(A_1\sqcup A_2) - \mu(A_2\sqcup A_3) - \mu(A_1\sqcup A_3) \nonumber \\
&& + \mu(A_1) + \mu(A_2) + \mu(A_3), \nonumber \\
&\vdots& \nonumber \\
I_{n}(A_{1},A_{2},\ldots ,A_{n+1}) &=& \mu(A_{1}\sqcup A_{2} \sqcup
\ldots A_{n}) - \sum \mu((n)sets) \nonumber \\
&& + \ldots \pm \sum_{j=1}^{n}\mu(A_{j}), \label{eq:order n interference term}
\eea
and so on, where $A,~B~C$ are disjoint elements of the event algebra. We are now in a position to define the basic object of the histories approach:

\begin{definition}\label{def:histories theory}
A \textbf{(level k) histories theory}, sometimes called a \textbf{(level k) measure theory}, is a triple $(\O,\EA,\mu)$ in which the measure satisfies the \textit{sum rule} $I_{k+1}=0$.
\end{definition}

It is known that $I_{k+1}=0$ implies that all higher sum rules are automatically satisfied, namely $I_{n+k}=0$ for all $n\geq 1$ \cite{Sorkin:1994dt,Salgado:1999pu}.

A level $1$ theory is thus one in which the measure satisfies the Kolmogorov sum rule, whereas a level $2$ theory is one in which the Kolmogorov sum rule may be violated but $I_3$ is nevertheless always zero. Any unitary quantum theory can be cast in the form of a generalised measure theory and its measure will satisfy $I_3=0$. We therefore refer to level $2$ theories as \textit{quantum measure theories}.

For definiteness in what follows we will, unless explicitly stated otherwise, assume that all histories theories are level $2$; in other words that they are quantum measure theories. However most of our definitions and results will hold in the general case.

\subsection{Null Sets}\label{sec:null sets}

Null sets will be of great importance to us in the development of a new interpretation of quantum measure theory. Given a histories theory $\H$ we define a null set to be an event of measure zero, $A\in\EA$ such that $\mu(A)=0$. We can further define a negligible set to be an event that is a subset of a null set.

The structure of null sets is markedly different under a quantum measure as compared with a classical one, in particular because of (destructive) interference negligible sets are no longer null in general. However if the measure obeys strong positivity then the disjoint union of two null sets is still null.

Unfortunately we lack general results about the structure of null sets, which is a hinderance to the development of our interpretations of quantum measure theory. There is recent research activity in this area, for example work on the implications of the existence of `antichains' of null sets \cite{Surya:2008}.

\subsection{The Naive Interpretation}\label{sec:the naive interpretation}

In classical mechanics the usual interpretation states that given a theory $(\O,\EA,\P)$, one history $r\in\O$ is real and that the real history is not an element of a null set, $\P(A)=0\Rightarrow r\not\in A$, the concept of preclusion. Events are \textit{true} if and only if they contain the real history (otherwise they are \textit{false}). Following our program of `stochasticising' quantum mechanics, much of what follows will be dedicated to constructing a generalisation of this classical interpretation in a manner compatible with quantum mechanics.

As a start, wielding Occam's razor, the most obvious generalisation of the classical interpretation is to make no change to it, asserting that given a theory $\H$ one history $r\in\O$ is real, that events $A\in\EA$ are \textit{true} if and only if they contain the real history, $r\in A$, (otherwise they are false) and that this history is not an element of any null set, $\mu(A)=0\Rightarrow r\not\in A$. We will call this the \textit{naive interpretation}.

\section{A Simple Example: The Double Slit System}\label{sec:double slit}

We end with a simple example that serves to illustrate the reformulation of Hilbert space quantum mechanics into the histories formalism. Consider an idealised double slit system, a particle is fired from an emitter and can pass through slits $A$ or $B$ before ending on the final screen, either at the detector $D$ or elsewhere $\overline{D}$ (not $D$). This can be represented as a trivial unitary system in which the evolution operator is the identity. We denote the initial state $\ket{\psi}$, and the projector corresponding to finding the particle at slit $A$ upon measurement by $P_A$, with the associated state $\ket{A}$ such that $P_A=\ket{A}\bra{A}$. We define the projectors $P_B,P_D$ and their associated states $\ket{B},\ket{D}$ in a similar fashion; note that the vectors corresponding to the two slits are mutually orthogonal. Finally, we can define the projector corresponding to $\overline{D}$ as $P_{\overline{D}} = \mathbb{I}-P_D$.

Formally, to define a Hilbert space theory we must specify each member of $({\cal{H}},H,\ket{\psi},T)$. We set the Hilbert itself to be the space of the vectors corresponding to our two slits, ${\cal{H}} = span(\ket{A},\ket{B})$. Our Hamiltonian $H$ is the null operator, so our evolution operators are the identity, and our temporal support is $T = \{0,1,2\}$, consisting of an initial time, an intermediate time at which our projectors $P_A,P_B$ act, and a final time at which our projectors $P_D,P_{\overline{D}}$ act. Note that this is an idealised and minimal treatment of the double slit system, a more complete account might involve an infinite dimensional Hilbert space.

Because our Hilbert space is the span of the projector states, we must specify the intimal state $\ket{\psi}$ and the detector state $\ket{D}$  in terms of $\ket{A},\ket{B}$. We constrain the initial state to treat the two slits symmetrically, so we set $\ket{\psi}=\frac{1}{\sqrt{2}}(\ket{A}+\ket{B})$. Further, we can place our detector in a dark fringe, a point of destructive interference so that $\ket{D} = \frac{1}{\sqrt{2}}(\ket{A}-\ket{B})$, and thus $\bra{D}(P_A+P_B)\ket{\psi}=0$. The reader is referred to appendix \ref{sec:many slit} for a discussion of the gedankenexperimental realisation of this system.

How do we analyse this in a histories formalism? Using the `minimal' Hilbert space theory $({\cal{H}},H,\ket{\psi},T)$ outlined above our histories are $\O=\{P_AP_D,P_BP_D,P_AP_{\overline{D}},P_BP_{\overline{D}}\}$. We will use the notation $AD = P_AP_D$ and so on (so that $AD$ represents the path of the particle passing through slit $A$ and ending at the detector $D$) to give us $\O=\{AD,BD,A\overline{D},B\overline{D}\}$. Since $\O$ is finite, we take the entirety of $P\O$ as our space of propositions, or event algebra, $\EA = P\O$.

If we had instead used a more complete, infinite dimensional, Hilbert space theory to describe the double slit system each history would (still) be a spacetime path for the particle; so in terms of this `preferred basis' the sample space would be the set of all the spacetime histories. However each path could pass through $A$ or $B$ then end at $D$ or $\overline{D}$, so although there are many paths passing through each slit we would only be interested in the coarse graining $\O=\{AD,BD,A\overline{D},B\overline{D}\}$, justifying our use of the minimal Hilbert space theory. Such a coarse graining is standard, in fact every sample space we consider is in fact a coarse graining of the full sample space of the universe.

We can associate a vector valued amplitude to each history in the usual manner, for example $\ket{AD} = P_D P_A \ket{\psi}$. Then it is easy to compute the decoherence functional for this system, for example:
\bea
{\textbf{D}}(AD,BD) &=& \bra{\psi}P_A P_D P_D P_B\ket{\psi},\nonumber \\
&=& 1/2\braket{A}{D}\braket{D}{B},\nonumber \\
&=& -1/4.
\eea
Computing the other entries in a similar fashion, in the `basis' $\{AD,BD,A\overline{D},B\overline{D}\}$ the decoherence functional can be represented by the matrix:
\beq
{\textbf{D}}=\begin{pmatrix}
1/4 & -1/4 & 0 & 0 \\
-1/4 & 1/4 & 0 & 0 \\
0 & 0 & 1/4 & 1/4 \\
0 & 0 & 1/4 & 1/4
\end{pmatrix}.
\eeq
So that, for example, ${\textbf{D}}(B\overline{D},B\overline{D})=1/4$. Notice that the histories ending at $D$ have no interference with histories ending at $\overline{D}$. This is a general feature of systems arising from a Hilbert space formulation and is due to the presence of a \textit{final time projector}. In this case, the class operators for histories ending at $D$ will always end in $P_D$ whereas the class operators corresponding to histories ending at $\overline{D}$ will always end in $P_{\overline{D}}$. As these two projectors are orthogonal, the inner product of the two vector valued amplitudes will always be zero. In general, if the decoherence functional is derived from a Hilbert space, it will split up into blocks as above, with no interference between fine grained histories from separate blocks.

This system admits two natural coarse grainings, we can partition the sample space according to the final time projector, or we can partition the sample space based on the two slits. In the first case we set $D=\{AD,BD\},~\overline{D}=\{A\overline{D},B\overline{D}\}$ to get the decoherence functional:
\beq
{\textbf{D}}=\begin{pmatrix}
0 & 0 \\
0 & 1
\end{pmatrix}.
\eeq
Notice that the set $D$ is of measure zero, this is the manifestation of the `preclusion' of the dark fringe in the histories formulation. In the second case we set $A=\{AD,A\overline{D}\},~B=\{BD,B\overline{D}\}$ to get the coarse grained decoherence functional:
\beq
{\textbf{D}}=\begin{pmatrix}
1/2 & 0 \\
0 & 1/2
\end{pmatrix}.
\eeq
Note that both of these coarse grainings are decoherent, although the only shared fine graining (the only partition of $\O$ which is a fine graining of both these partitions) is the full set of fine grained histories, $\O$ itself, which is not decoherent. This is an example of the non-uniqueness problem faced by the consistent histories interpretation.

%Introduction to histories
%\singlespacing
\chapter[The Kochen-Specker Theorem]{The Kochen Specker Theorem and the Failure of the Naive Interpretation}\label{chapter:kochen-specker}
%\doublespacing

\section{The Kochen-Specker Theorem}\label{sec:PKS}

The Kochen-Specker (KS) theorem \cite{Kochen:1967} is often cited as a key obstacle to a realist interpretation of quantum mechanics, and as such it is an issue that any aspiring interpretation must be able to address. In this chapter, following \cite{Dowker:2007zz} closely, we will focus on the Peres proof of the Kochen-Specker theorem \cite{Peres:1991}, recasting it into the histories formalism, noting the implications for the naive interpretation then finally discussing a concrete realisation of the system as a Peres-Kochen-Specker gedankenexperiment in terms of the spacetime paths of a spin-1 particle passing through a sequence of beam splitters and recombiners. The Peres-Kochen-Specker system furnishes us with a natural testing ground for our new interpretations; we will return to this system more than once in the following chapters, regarding a successful treatment of the Peres-Kochen-Specker system as an essential requirement of our new interpretations of quantum measure theory.

\subsection{The Peres Formulation}

\begin{figure}
\begin{center}
\includegraphics[width=0.3\textwidth]{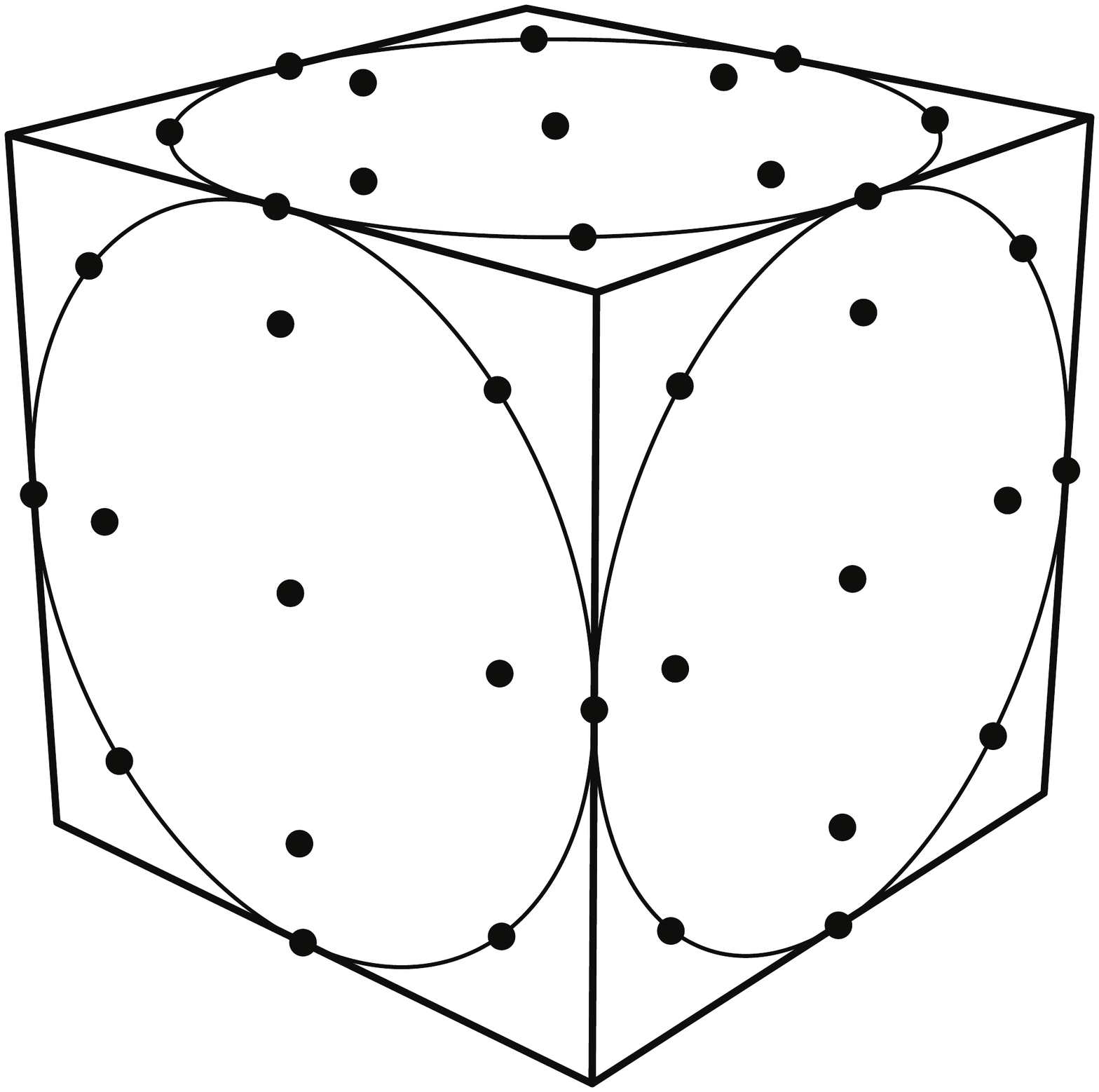}
\caption{\small{The $\pm 33$ directions are defined by the
lines joining the center of the cube to the
$\pm 6$ mid--points of the edges and the
$\pm3$ sets of 9 points of the $3 \times 3$ square arrays
shown inscribed in the incircles of its faces}} \label{Peres Set}
\end{center}
\end{figure}

Peres defines 33 rays in $\mathbb{R}^{3}$ from which 16 orthogonal bases can be formed. Peres defines the rays as being the 33 for which the squares of the direction cosines are one of the combinations:
\begin{equation*} 0+0+1=0+1/2+1/2=0+1/3+2/3=1/4+1/4+1/2 \,. \end{equation*}
Figure \ref{Peres Set} (taken with permission from \cite{Conway:2006}) illustrates the points where the rays intersect a unit cube centered on the origin.

To describe the rays, Peres employs a shorthand notation that we will find useful, writing $\m{x} = -x$ we define
\begin{equation*}
\m{x}yz = \alpha\left(\begin{array}{c}
-x\\y\\z\end{array}\right)\in\mathbb{R}^{3}
\end{equation*}
where $\m{xyz} = xyz$ as we are dealing with rays rather than vectors. We will call this set of rays the Peres Set, $PS$, and sometimes it will be convenient to refer to them as labelled from 1 to 33: $PS = \{u_i| i = 1,2,\dots 33\}$ (in some fixed but arbitrary order).

In what follows it will be important to understand the structure \& the symmetries of $PS$. Examining the magnitude of the angle between each ray and its nearest neighbour, we find that the rays can be divided into four types:

\begin{table}[ht]
\begin{center}
\begin{tabular}{l}
Type I: $001,010,100$
\\Type II: $011,01\m{1},101,10\m{1},110,1\m{1}0$
\\Type III: $012,0\m{1}2,021,02\m{1},102,\m{1}02,201,20\m{1},120,\m{1}20,210,2\m{1}0$
\\Type
IV: $112,\m{1}12,1\m{1}2,\m{1}\m{1}2,121,12\m{1},\m{1}21,\m{1}2\m{1},211,21\m{1},2\m{1}1,2\m{1}\m{1}$
\end{tabular}
\label{table:Ray Types}
\caption{The Peres rays divide into 4 Types.}
\end{center}
\end{table}

With reference to figure \ref{Peres Set}, Type I corresponds to the midpoints of faces of the cube, Type II to the midpoints of edges, Type III to the remaining points in the interior of the incircles of the faces and Type IV to the remaining points on the incircles of the faces. It can be seen that each symmetry of the projective cube induces a permutation of $PS$. The symmetry group of the projective cube, $H$, is of order 24 and is generated by rotations by $\pi/2$ around co-ordinate axes and reflections in co-ordinate planes. Since the ray types are defined in terms of angles, the induced permutations on the elements of $PS$ will preserve the type of each ray, {\it i.e.} the  permutation reduces to permutations on the four subsets of same-type rays. By inspection, $H$ is transitive on each type: this means that for any two rays $u$, $v$ of the same type $\exists$ $g\in H$ such that $v=g(u)$.

In the Peres version of the Kochen-Specker theorem, this set of 33 rays is mapped in the obvious way into a set of rays in a 3 dimensional Hilbert space, where two rays are orthogonal in the Peres set if and only if they are mapped to orthogonal rays in Hilbert space. We will therefore talk about the Peres rays both as rays in real 3-D space and in Hilbert space. To each orthogonal basis of rays, $\{u_{i},u_{j},u_{k}\}$, in Hilbert space can be associated an observable with three distinct outcomes, each outcome corresponding to one of the rays in the basis (the eigenvector of the outcome/eigenvalue). If that observable is measured, one of the three outcomes will be obtained and the quantum state collapses onto exactly one of the three basis rays, which corresponds to the outcome. If it is assumed that the result of the measurement of the observable exists within the system before or independent of the measurement being taken -- a non-contextual hidden variable -- then we would conclude that one of the basis rays corresponds to the actual value of the observable being measured, and is labelled ``true'', and the other two (using classical logic) are labelled ``false''. We follow Peres in considering the ``true'' ray to be coloured green whereas the other two are coloured red.

Assuming that an experimenter can freely choose to measure any of the observables associated to the 16 orthogonal bases in $PS$, if we assume that the result of the measurement actually done is encoded in the system beforehand, then the results of all potential measurements must be encoded in the system. Thus, all of the rays in $PS$ will be coloured. We call a map
$$\gamma:PS\rightarrow \{green,red\}$$
a colouring. We call $\gamma$ a {\it consistent} colouring if it colours exactly one ray green out of every basis in $PS$ (we assume henceforth that ``basis'' implies ``orthogonal basis'') and does not colour any pair of orthogonal rays both green. Note there are some orthogonal pairs in $PS$ that
are not contained in a basis in $PS$.

\begin{table}[ht]
\begin{center}
\begin{tabular}{ccc}
\hline
Basis & Basis Rays & Other Orthogonal Rays \\
\hline
$B_1$ & $\textbf{001}~100~010$ & $110~1\m{1}0$ \\
$B_2$ & $\textbf{101}~\m{1}01~\textit{010}$ & \\
$B_3$ & $\textbf{011}~0\m{1}1~\textit{100}$ & \\
$B_4$ & $\textbf{1$\m{\textbf{1}}$2}~\m{1}12~\textit{110}$ & $20\m{1}~021$ \\
$B_5$ & $\textbf{102}~\textit{20$\m{\textit{1}}$}~\textit{010}$ & $2\m{11}$ \\
$B_6$ & $\textbf{211}~\textit{0$\m{\textit{1}}$1}~\textit{2$\m{\textit{11}}$}$
& $\m{1}02$\\
$B_7$ & $\textbf{201}~\textit{010}~\textit{$\m{\textit{1}}$02}$ & $\m{11}2$\\
$B_8$ & $\textbf{112}~\textit{1$\m{\textit{1}}$0}~\textit{$\m{\textit{11}}$2}$
& $02\m{1}$\\
$B_9$ & $\textbf{012}~\textit{100}~\textit{02$\m{\textit{1}}$}$
& $\m{1}2\m{1}$\\
$B_{10}$ & $\textbf{121}~\textit{$\m{\textit{1}}$01}~\textit{$\m{\textit{1}}
$2$\m{\textit{1}}$}$ & $0\m{1}2$\\
$B_{11}$ & $\textit{100}~\textit{021}~\textit{0$\m{\textit{1}}$2}$ &
\end{tabular}
\caption{Peres' Proof of the Kochen-Specker Theorem}\label{table:Peres}
\end{center}
\end{table}

\begin{theorem}\label{thm:PKS}
 {\textbf{Peres-Kochen-Specker} \cite{Peres:1991}} \\There is no consistent colouring of $PS$
\end{theorem}
\begin{proof}
The proof is based on table \ref{table:Peres}. We first consider a colouring, $\g_P$ of the four bases, $B_1,~B_2,~B_3,~B_4$, as defined in table \ref{table:Peres}, which colours green the first ray in each of these bases (highlighted in bold in the table)\footnote{$\g_P$ colours green the first ray in each basis shown in table \ref{table:Peres} except for $B_{11}$.}. We then work down the table basis by basis starting at $B_5$ to try to extend $\gamma_P$ to a consistent colouring of $PS$. For each basis $B_i$ in turn (for $5\leq i\leq 10$) we find that two of the basis elements (italicised) have already been coloured red, and so in each case the choice of basis ray to colour green is forced by consistency. This continues until we reach basis $B_{11}$, which by then has all three basis rays coloured red, meaning that $\gamma_P$ cannot be extended to a consistent colouring of the whole Peres Set.

We now use symmetry considerations to extend this argument to rule out any consistent colouring of the Peres Set. First note that any symmetry $g\in H$ of the projective cube induces an action on the colourings of $PS$. We will also denote this action by $g\in H$, so that $g\g(u)=\g(g(u))$ for $u\in PS$. We start by assuming the existence of a consistent colouring $\g_C$ on $PS$, we then `step down' through table \ref{table:Peres} to find a $g\in H$ such that $\g_P=g\g_C$.

Consider $B_1$, since $\g_P$ is consistent there exists some $v_1\in B_1$ such that $\g_C(v)=green$ if $v=v_1$ and $\g_C(v)=red$ if $v\in B_1$ but $v\neq v_1$. The let $\tilde{g}_1$ be a rotation of $2\pi/3$ around the ray $111$, cyclically permuting the elements of $B_1$ such that $\tilde{g}_1(001)=010,~\tilde{g}_1(010)=100,~\tilde{g}_1(100)=001$. Then either $\tilde{g}_1(001)=v_1$, $\tilde{g}_1\tilde{g}(001)=\tilde{g}_1^2(001)=v_1$ or $001=\tilde{g}_1\tilde{g}_1\tilde{g}_1(001)=\tilde{g}_1^3(001)=v_1$. In any case, let $g_1$ be the rotation $\tilde{g}_i^n$ that maps $001$ to $v_1$; then $g_1\g_C(B_1)=\g_P(B_1)$.

We next consider $g_1(B_2)$. As in table \ref{table:Peres}, $g_1(010)$ is orthogonal to $v_1=g_1(001)$, so the consistency of $\g_C$ implies that exactly one of $\{g_1(101),g_1(\m{1}01)\}$ must be coloured green by $\g_C$; denote this ray $v_2$. Now let $\tilde{g}_2\in H$ be reflection in the plane orthogonal to $100$, so that $\tilde{g}_2(101)=\m{1}01$ and $\tilde{g}_2^2$ is the identity. Then as above either $\tilde{g}_2(101)=v_2$ or $\tilde{g}_2^2(101)=v_2$, and we define $g_2$ to be the $\tilde{g}_2^n$ which maps $101$ to $v_2$. Then noticing that $B_1$ is invariant under $g_2$ we see that $g_2g_1\g_C(B_i)=\g_P(B_i)$ for $i=1,2$.

Carrying on in this fashion we can find $g_3$ and $g_4$ such that $g_4g_3g_2g_1\g_C(B_i)=\g_P(B_i)$ for $i=1,\ldots 4$. Now as we saw in table \ref{table:Peres} the colouring of the bases $B_1$ to $B_4$ uniquely determines the colouring of the bases $B_i$ for $i>4$, similarly if we continue to `step down' the table to seek symmetries $g_i$, we find that $g_i$ is the identity whenever $i>4$. Thus setting $g=g_4g_3g_2g_1$ we find that $g\g_C(B_i)=\g_P(B_i)$ for $i=1,\ldots 10$. However, as in table \ref{table:Peres}, once we have coloured every ray in bases $g(B_1)$ to $g(B_10)$, and used the assumed consistency of $\g_C$ to colour red any ray orthogonal to the green rays in bases $g(B_1)$ to $g(B_{10})$, we realise that we have inadvertently already coloured $g(B_{11})$ inconsistently. Therefore there is no consistent colouring of the Peres Set.
\end{proof}

\subsection{Events and Null Sets}\label{sec:PKS null sets}

Let us now restate the theorem in terms of event algebras and co-events. We take as our sample space the set $\O$ of all (green/red) colourings, $\gamma$ of the Peres set $PS$. The event algebra is the Boolean algebra of subsets of $\Omega$ as before. Given a subset $S$ of the Peres Set we write:
\begin{align}
R_S =& \{\gamma\in\Omega |~\gamma(u_i)= red ~\forall u_i\in S\} \\
G_S =& \{\gamma\in\Omega |~\gamma(u_i)= green ~\forall u_i\in S\}
\end{align}

Now the Peres-Kochen-Specker result depends crucially upon disallowing, or precluding, non-consistent colourings. Thus in measure theory language, we wish to consider the sets corresponding to these failures in consistency to have measure zero (though we will not explicitly construct a measure until section \ref{sec:Stern-Gerlach}). These inconsistencies arise in two ways, an orthogonal pair of rays being coloured green or an orthogonal basis of rays being coloured red, thus we treat the following two types of set as if they are null:

\begin{align}\label{eq:kssets}
R_B && B~\text{an orthogonal basis} \\
G_P && P~\text{an orthogonal pair}
\end{align}

We will call these the Peres-Kochen-Specker (PKS) events, or PKS sets. Note that the Peres Set includes orthogonal pairs that are not subsets of orthogonal bases.

Now (as already implicitly assumed in the proof of Theorem \ref{thm:PKS}) every symmetry $g\in H$ of the projective cube induces an action on $\Omega$ and thus on $\EA$, via its action on $PS$. We will also denote this action by $g\in H$, so that $g\cdot\gamma(u_i)=\gamma(g\cdot u_i)$ and $g(A) = \{g(\gamma) |\gamma\in A\}$. Crucially, note that the PKS sets are permuted by the symmetries in $H$, via $g(G_P)=G_{g(P)}$ and $g(R_B)=R_{g(B)}$.

\subsection{The Failure of the Naive Interpretation}\label{sec:failure of the naive interpretation}

Recall that the naive interpretation (section \ref{sec:the naive interpretation}) ascribes ontology to exactly one, `real', history, which can not be an element of a null set. Events are interpreted as `true' if and only if they contain the real history. Though this is, perhaps, the simplest generalisation of the standard interpretation of stochastic mechanics to quantum measure theory, we find that as a direct consequence of theorem \ref{thm:PKS} it can not successfully explain the Peres-Kochen-Specker system:

\begin{lemma}\label{lemma:failure of the naive interpretation}
Let $\mu$ be a measure on the space $\Omega$ of colourings of $PS$ that is zero valued on the PKS sets. Then under the naive interpretation no element of the sample space can be real.
\end{lemma}
\begin{proof}
By Theorem \ref{thm:PKS}, every element of $\Omega$ is inconsistent on at least one basis or pair and therefore lies in at least one PKS set. Since every PKS set is null, every element of the sample space is an element of a null set and thus no element of the sample space can be `real'.
\end{proof}

This is fatal for the naive interpretation, for an event is true if and only if it contains the real history, the lack of which will therefore render every event false. Thus any physical system possessing a measure containing the null PKS sets required for theorem \ref{thm:PKS} would be problematic, even fatal, for the naive interpretation. We will use the remainder of this chapter to construct such a system.

\section{Expressing the Peres-Kochen-Specker setup in terms of Spacetime
Paths}\label{sec:Stern-Gerlach}

Thus far our analysis has been completely abstract and we now seek to embed our mathematical Peres-Kochen-Specker system into a more physical, even perhaps experimentally realisable, framework. To this end we  imagine a sequence of Stern-Gerlach apparatus to translate each colouring in $\Omega$ into
a spacetime history. We will construct a quantum measure on the corresponding event algebra in which the translations of the Kochen-Specker sets are null, though further `accidental' null sets are introduced as will be discussed.

\subsection{The Stern-Gerlach Apparatus}

The Stern-Gerlach apparatus allows spin to be expressed in terms of paths in spacetime. In the original experiment (1922) a beam of spin $1/2$ silver atoms was sent through an inhomogeneous magnetic field, splitting the beam into two components according to the spin of the particles.

\begin{figure}
\begin{center}
\includegraphics[width=0.45\textwidth,angle=-90]{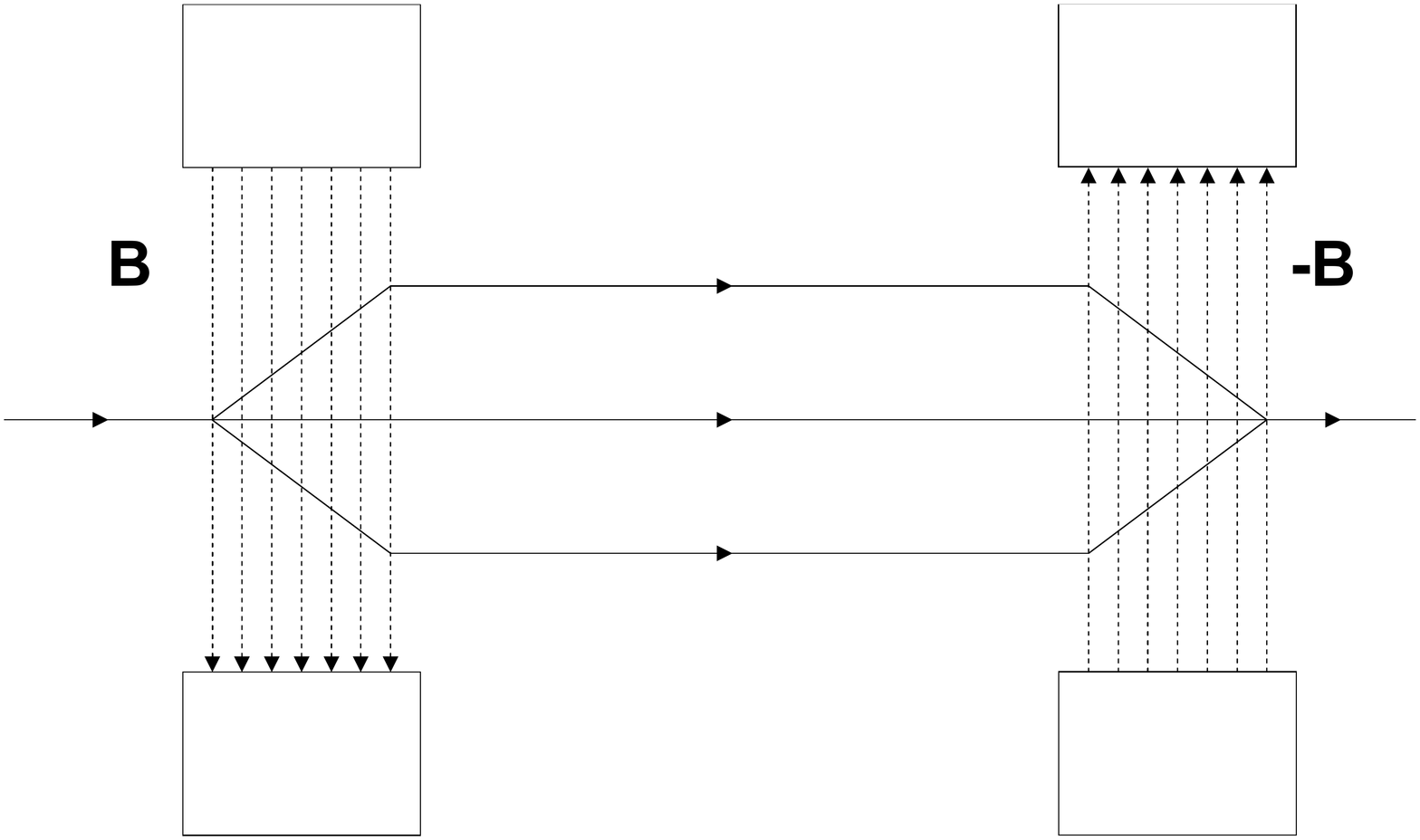}
\caption{Stern-Gerlach Apparatus} \label{Single Stern-Gerlach}
\end{center}
\end{figure}

In our gedankenversion of the Stern-Gerlach  experiment we imagine spin 1 particles sent though a parallel, static, inhomogeneous magnetic field $\textbf{B}$. By parallel we mean that vectors $\textbf{B}(x,t)$ are parallel in the vicinity of the particle beam. By static we mean that $\textbf{B}(x,t)$ is independent of t. We shall also assume that the gradient of the $\textbf{B}$ field is constant close to the beam. Now a particle with spin $\textbf{S}$ will have an effective magnetic moment in the $-\textbf{S}$ direction. Hence the particle will experience a force:
\begin{equation*}
{\bf F} \propto \nabla(\textbf{S}\cdot \textbf{B})
\end{equation*}
The force is proportional to the component of the spin in the direction of the magnetic field and so the beam will split into three branches, corresponding to the three possible spin states. So for a single particle a measurement ascertaining which branch holds the particle constitutes a measurement of the component of the spin in the direction of the magnetic field. Note that we can thus measure the spin in any direction other than along the path of the beam. However, if such a measurement is not taken, the three branches can be coherently recombined by application of the field $-\textbf{B}$ (see figure \ref{Single Stern-Gerlach}).

The spin Hilbert space $\cal{H}$ for a spin 1 particle is isomorphic to $\mathbb{C}^{3}$. Let $S_{x}, ~S_{y},~S{z}$ denote the observables corresponding to spin measurements in the x, y and z directions. Then we can choose a basis of  $\cal{H}$, $\{ \ket{0, z}, \ket{+1, z}, \ket{-1, z}\}$, in which  $S_{z}$ is diagonal. $S_{x}, ~S_{y},~S_{z}$ are then represented by the standard $3\times 3$ spin matrices. A measurement of spin in the $\textbf{B}$ direction corresponds to a basis in $\cal{H}$ consisting of the eigenvectors of ${\textbf{S.B}} =\text{B}_{x}S_{x}+\text{B}_{y}S_{y}+\text{B}_{z}S_{z}$.
The eigenvectors of spin in the $x$ and $y$ directions are
\begin{align}\label{eq:eigenvectors}
\ket{0,x} = &\frac{1}{\sqrt{2}}(\ket{+1,z} - \ket{-1,z})\, ,\\
\ket{+1,x} = &\frac{1}{{2}}(\ket{+1,z} + \sqrt{2} \ket{0,z} + \ket{-1,z})\, ,\\
\ket{-1,x} = &\frac{1}{{2}}(\ket{+1,z} - \sqrt{2} \ket{0,z} + \ket{-1,z})\, ,\\
\ket{0,y} = &\frac{1}{\sqrt{2}}(\ket{+1,z} + \ket{-1,z})\, ,\\
\ket{+1,y} = &\frac{1}{{2}}(\ket{+1,z} + i\sqrt{2} \ket{0,z} - \ket{-1,z})\, ,\\
\ket{-1,y} = &\frac{1}{{2}}(\ket{+1,z} - i\sqrt{2} \ket{0,z} - \ket{-1,z})\, .\label{eq:eigenvectors end}
\end{align}

\subsection{Using  Stern-Gerlach Apparatus to realise the Peres-Kochen-Specker setup}

Instead of spin we will be interested in spin squared, {\it i.e.} $S_i^2$ in the $i$ direction (no sum on $i$). There are now two possible outcomes of a
measurement, $S_i^{2}=0$ (corresponding to $S_i=0$) and $S_i^{2}=1$ (corresponding to $S_i=+1$ or $S_i=-1$). This can be mirrored within the Stern-Gerlach framework by lumping the two outer ($S_i=+1$ and $S_i=-1$) beams together and labelling them together as ``the red beam'' and labelling the middle beam as ``the green beam''. This can be done ``mentally'' by simply ignoring the fine grained detail of which of the outer beams the particle is in, or ``physically'' by coherently recombining the two outer beams into a single beam using a reversed Stern-Gerlach apparatus (whilst keeping the middle beam separated by diverting it out of the way).  Since in Anhomomorphic Logic we may not be able to reason (about fine and coarse graining for example) using classical rules, it will be clearer to assume that we have set up a physical recombiner so there really is a single ``red beam'' corresponding to $S_i^2 = 1$. Let us further imagine appending the exact reverse of this apparatus at the end which will coherently recombine the red and the green beams into a single beam again. We call this whole apparatus a ``spin-squared beam splitter and recombiner (bsr) in the $i$ direction.''

For a spin-1 particle, $S_i^2$ and $S_j^2$ commute if the $i$ and $j$ directions are orthogonal, {\textit {i.e.}} the squared $3\times 3$ spin representation matrices $\sigma_{x}^{2}, ~\sigma_{y}^{2},~\sigma_{z}^{2}$ commute. In the standard Copenhagen interpretation three (consecutive or simultaneous) spin-squared measurements in mutually orthogonal directions will necessarily result in one outcome of $0$ and two outcomes of $1$.

This can be seen directly in terms of projectors onto the relevant eigenspaces in the Hilbert space. Let $P_i^0 = \ket{0,i}\bra{0,i}$ and $P_i^1 = {\bf 1} - P_i^0$, be projectors onto the spin-squared (in direction $i$) eigenspaces corresponding to eigenvalues 0 and 1 respectively. Then, looking at equations \ref{eq:eigenvectors}-\ref{eq:eigenvectors end}, we have $P_i^0 P_j^0 = 0$ for any orthogonal pair of directions $i$ and $j$, and for a basis $u_i,u_j,u_k$ we have $P_i^a P_j^b P_k^c = 0$ unless exactly one of $a$, $b$ or $c$ equals 0.

To translate this into spacetime paths, we imagine a spin-1 particle (in any spin state) passing through a sequence of three spin-squared bsr's, one in each of the three orthogonal directions (no one of which coincides with the direction of motion of the particle). The classical realist picture is that the particle must pass through the green beam in exactly one bsr and through the red beam in the other two.

We want to translate colourings of the entire Peres set $\{u_i\}$, not just one basis, into spacetime paths. We imagine a sequence of 33 spin-squared bsr's in the directions $\{u_i\}$, so the first bsr will be in direction $u_{1}$, the second in $u_2$, {\textit etc.}. The particle trajectories form the space $\Omega$ in this setup and each one follows either the red or green beam through each bsr in turn and so every colouring can be realised by an element of $\Omega$. Strictly, there are many particle trajectories in each beam, with slight variations in positions, but we will ignore this finer grained detail and assume that $\Omega$ consists of the $2^{33}$ trajectories distinguished only by which beam is passed through in each bsr. This space is then in one-to-one correspondence with the space of colourings of the Peres Set and we identify the two in the obvious way, noting that the particle paths contain additional information, namely the choice of {\textit {order}} of the Peres rays in the experimental set up, not present in the original space of colourings.

Let the initial spin state of the particle be $\ket{\psi}$. Then a decoherence functional, and hence a quantum measure, on $\Omega$ can be defined as follows. Let $\gamma$ be an element of $\Omega$, so $\gamma(u_i)$ is a colour for each Peres direction $u_i$. As we saw above, colour ``green'' is identified with spin-squared value zero and  colour ``red'' is identified with spin-squared value one. We can construct a ``path state'' $\ket{\gamma}$ via
\begin{equation*}
\ket{\gamma} = P_{33}^{\gamma(u_{33})} \dots
P_2^{\gamma(u_2)} P_1^{\gamma(u_1)} \ket{\psi}
\end{equation*}
where $P_i^{green} \equiv P_i^{0}$ and $P_i^{red}\equiv P_i^{1}$ are the projection operators defined previously. For each event $A\in \EA$ we can define an ``event state''
\begin{equation*}
\ket{A} = \sum_{\gamma \in A} \ket{\gamma}
\end{equation*}
and
the decoherence functional is then defined
by
\begin{equation*}
D(A,B) = \braket{A}{B}
 \,.
\end{equation*}

If the spin state of the particle is mixed the decoherence functional is a convex combination of such terms.

\begin{claim}
The quantum measure on $\Omega$ defined as above values the PKS sets and their disjoint unions zero.
\end{claim}

\begin{proof}
The PKS sets are of the form $R_B$ or $G_P$ where $B$ is a basis and $P$ an orthogonal pair. We first consider PKS sets of the form $R_{\{u_i,u_j,u_k\}}$ for a basis $\{u_i,u_j,u_k\}$. Then the ``event state'' $\ket{R_{\{u_i,u_j,u_k\}}}$ is given by a sum over ``path states'' $\ket{\gamma}$, one for each $\gamma$ in $R_{\{u_i,u_j,u_k\}}$. This sum involves a sum over the colouring of all the Peres rays which are not $u_i$, $u_j$ or $u_k$ and so the projection operators for all these rays sum to the identity and leave:
\begin{equation*}
\ket{R_{\{u_i,u_j,u_k\}}} = P_i^{0}P_j^{0}P_k^{0}\ket{\psi}
\end{equation*}
which equals zero for any state $\ket{\psi}$ because the product of those projectors is zero. Similarly, PKS set corresponding to pairs, $G_{\{u_i,u_j\}}$ where $u_i,u_j$ are orthogonal, correspond to event states:
\begin{equation*}
\ket{G_{\{u_i,u_j\}}} = P_i^{1}P_j^{1}\ket{\psi}
\end{equation*}
which as before equals zero for any state $\ket{\psi}$ because the product of those projectors is zero.

An event which is a disjoint union of PKS events has a corresponding event state which is a sum of terms, one for each PKS event in the union, each of which is zero.

Hence the result.
\end{proof}

For a given initial state and a choice of ordering for the bsr's we therefore have an explicit realisation of a quantum measure in which the PKS sets are null. The statement of the Peres-Kochen-Specker theorem in the context of this gedankenexperiment is that every trajectory that a spin 1 particle can take through the apparatus is in one of the PKS precluded sets. The naive interpretation would thus imply that the particle can take no path through such an apparatus, or in other words that `there is no reality' \cite{Isham:private}.
%KS in QMT
%\singlespacing
\chapter{Stochastic Collapse}\label{chapter:stochastic collapse}
%\doublespacing

\section{Opening Comments}

One natural response to the failure of the naive interpretation is to replace quantum mechanics with a classical stochastic theory which would enable us to apply the better understood interpretation of classical mechanics. The well studied field of stochastic collapse models \cite{Bassi:2003gd} viewed through the prism of the Bell ontology \cite{Bell:1987i} achieves this by introducing new classical variables coupled to our existing quantum dynamics, then ascribing ontology to the classical variables, relegating the quantum dynamics to the status of a calculational tool. In this chapter, following \cite{Dowker:2007ma}, we will explore such models, placing them within the histories framework which furnishes a natural setting for these theories in which the classical and quantum components of the dynamics can be treated on the same footing. We then explore the coupling between the classical and quantum dynamics before returning to the interpretational implications and evaluating the ability of stochastic collapse under the Bell ontology to provide a satisfactory interpretation for quantum mechanics.

\section{Introduction}

Models of ``spontaneous localisation'' or ``dynamical wavefunction collapse'' are observer independent alternatives to standard Copenhagen quantum theory (see \cite{Bassi:2003gd} for a review). These models have a generic structure: there is a quantum state $\Psi$ which undergoes a stochastic
evolution in Hilbert space and there is a ``classical'' (c-number) entity -- call it $\alpha$ -- with a stochastic evolution in spacetime. The stochastic dynamics for the two entities -- $\Psi$ and $\alpha$ -- are coupled together. The stochastic dynamics in Hilbert space tends to drive $\Psi$ into an eigenstate of an operator $\hat{\alpha}$ that corresponds to $\alpha$. And the probability distribution for the realised values of $\alpha$ depends on $\Psi$ so that the history of $\alpha$ follows, noisily, the expectation value of $\hat\alpha$ in $\Psi$.

That collapse models have both quantum and classical aspects has been pointed out before, notably by Di\'osi. The nature of this interaction between the classical and quantum parts of these models is, however, somewhat obscured by the profound difference in the nature of their descriptions: the classical variable traces out a history in spacetime and the quantum state traces out its evolution in Hilbert space.

In order to illuminate the nature of the quantum-classical coupling within collapse models we will, in the case of a concrete and specific example, recast the formalism into the histories framework (sections \ref{sec:decoherent histories},\ref{sec:quantum measure theory}) which allows us to treat the classical and quantum systems on an equal footing, both being described by spacetime histories.

The model we will focus on is a discrete, finite, 1+1 dimensional lattice field theory. This is a useful model because it is completely finite (so long as we restrict ourselves to questions involving finite times) and thus expressions can be written down exactly and also because there is a well-defined background with non-trivial causal structure, so that questions of causality can be explored.

We will show that the model contains both ``classical'' and ``quantum'' histories, and demonstrate the nature of their interaction. We will show that one  choice of ontology for collapse models, the Bell ontology \cite{Bell:1987i}, corresponds to coarse graining over the quantum histories. We will also show how the well-known relationship between collapse models and open quantum systems coupled to an environment reveals itself in this histories framework.

\section{The lattice field model}

We review the lattice field model \cite{Dowker:2002wm,
Dowker:2004zn}
whose structure we will investigate.  The model is based on a
unitary QFT on a 1+1 null lattice \cite{Destri:1987ze}, which becomes a collapse
model on the introduction of local ``hits'' driving
the state into field eigenstates.

\begin{figure}[thb]
\begin{center}
\includegraphics[width=0.5\textwidth,angle=-90]{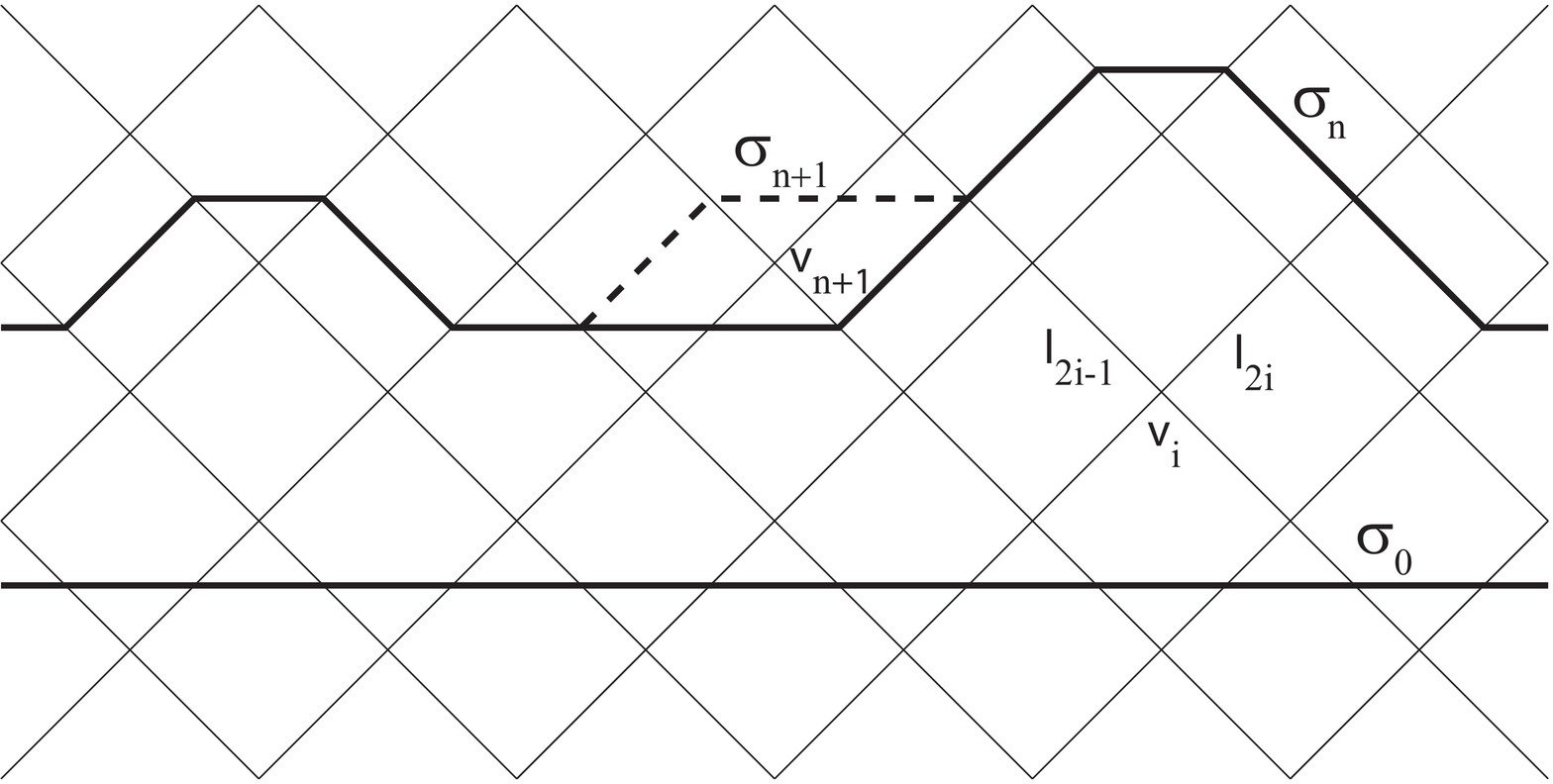}
\caption{The light cone lattice. $\sigma_0$ is the initial
surface and $\sigma_n$ is a generic spacelike surface. The surface
$\sigma_{n+1}$ is shown after the vertex $v_{n+1}$ is evolved over.
A vertex $v_i$ is shown with its two outgoing links: $l_{2i-1}$ to the
left and $l_{2i}$ to the right.}
\label{fig:lattice}
\end{center}
\end{figure}

The spacetime lattice is a lightcone discretisation of a cylinder, $N$ vertices wide and periodic in space. It extends to the infinite future, and the links between the lattice vertices are left or right going null rays. Figure \ref{fig:lattice} shows a part of such a spacetime lattice, identifying the leftmost vertices with the rightmost vertices we see that $N=6$. A spacelike surface $\sigma$ is maximal set of mutually spacelike links, and consists of $N$ leftgoing links and $N$ rightgoing links cut by the surface; an example of a spatial surface is shown in figure \ref{fig:lattice}. We assume an initial spacelike surface $\sigma_0$.

An assignment of labels, $v_1, v_2, v_3, \dots$,
to the vertices to the future of $\sigma_0$
 is called ``natural'' if $i< j$ whenever the vertex labelled $v_i$ is to the
causal past of the vertex labelled $v_j$. A natural labelling is
equivalent to a linear
extension of the (partial) causal order of the vertices. A natural
labelling, $v_1, v_2, \dots$ is also equivalent to a sequence of
spatial  surfaces, $\sigma_1, \sigma_2, \dots$ where the surface
$\sigma_n$ is defined such that between it and $\sigma_0$, lie
exactly the vertices $v_1, \dots v_n$. One can think of the
natural labelling as giving an ``evolution'' rule for the
spacelike surfaces: at time step $n$ the surface
creeps forward by one ``elementary motion'' across vertex $v_n$.
For the purpose of this paper, it is convenient to
consider a fixed natural labelling.
Nothing will depend on the natural labelling chosen, all mathematical
quantities will be independent of the choice.

The local field variables $\Phi$ live on the links. These field
variables take only two values $\{0,1\}$, so that on each link
there is a qbit Hilbert space spanned by the two field eigenstates
$\{\ket{0}, \ket{1}\}$.
As the field variables live on the links, it is convenient to
have a labelling of the links. We choose a labelling $l_a$,
$a = 1,2,\dots$, such that $l_{2i-1}$ and $l_{2i}$ are the
left-going and right-going outgoing links, respectively, from vertex $v_i$ (see figure \ref{fig:lattice}).
So, as vertex label $i$ increases from 1 to $n$, the link
label $a$ runs from $1$ to $2n$. We denote the qbit Hilbert space related to link $l_a$ by $H_{l_a}$.

The initial state $|\psi_0\rangle$ on surface $\sigma_0$ is
an element of the $2^{2N}$ dimensional Hilbert space $H_{\sigma_0}$
which is a tensor
product of the $2N$ 2-dimensional
Hilbert spaces on each link cut by $\sigma_0$, $H_{\sigma_0}=\displaystyle{\bigotimes_{l_a\in\sigma_0}}H_{l_a}$.
Similarly there is a $2^{2N}$ dimensional Hilbert space for each spacelike
surface $\sigma_i$ and they are isomorphic via the
isomorphisms, tied to the lattice, which map each link's
qbit Hilbert space onto the Hilbert spaces for the links
vertically above it on the lattice. In this way we can identify the Hilbert spaces $H_{\sigma_i}$ ($=\displaystyle{\bigotimes_{l_a\in\sigma_i}}H_{l_a}$) on each surface and describe the time evolution with a state evolving in a single Hilbert space $H_q$ ($\simeq H_{\sigma_i}$) of the system.

\subsection{The unitary theory}

In the standard unitary version of this local field theory, there
is a local unitary evolution operator, $R_i$, for each
$v_i$, which acts unitarily on the 4-dimensional factor of the Hilbert space
associated to the two ingoing and two outgoing links for $v_i$,
and acts as the identity operator on all other factors.
The state vector is evolved from $\sigma_{i-1}$
to $\sigma_i$ by applying $R_i$ \cite{Destri:1987ze}.

So in figure \ref{fig:lattice} we see that the surface $\sigma_n$
evolves `over' vertex $v_{n+1}$ to give us surface $\sigma_{n+1}$.
Now if $l_j,l_k$ are the two links going `into' vertex $v_{n+1}$,
and $l_{2(n+1)-1},l_{2(n+1)}$ the two outgoing links, the operator
$R_{n+1}$ maps $H_{l_j}\otimes H_{l_k}$ to
$H_{l_{2(n+1)-1}}\otimes H_{l_{2(n+1)}}$.
Further, for the links in the intersection of
$\sigma_n$ and $\sigma_{n+1}$, $R_{n+1}$ acts as the identity.
Since the surfaces $\sigma_n, \sigma_{n+1}$ only differ on the
links $l_j,l_k,l_{2(n+1)-1},l_{2(n+1)}$, we can put this
together to get $R_{n+1}:H_{\sigma_n}\rightarrow H_{\sigma_{n+1}}$.

Since we have identified the Hilbert spaces $H_{\sigma_i}$, we
regard $R_{n+1}$ as evolving a state in the `system Hilbert space'
$H_q$, so we write:
\begin{eqnarray}
\ket{\psi_{n+1}}&=&R_{n+1}\ket{\psi_n}\nonumber \\
&=& R_{n+1} R_n \ldots R_1 \ket{\psi_0}\,. \label{eqn:quantum evolution}
\end{eqnarray}
We define the unitary time evolution operator, $U(n)$, by
\begin{equation}\label{eq:Uoperator}
U(n) \equiv R_n\; R_{n-1} \ldots R_1\;.
\end{equation}

To cast the theory into a quantum measure theory framework,
we need to identify the space, $\Omega_q$ of histories, an event
algebra, $\EA_q$, of suitable subsets of $\Omega_q$ and the
decoherence functional, $D_q(\,\cdot\,\,,\,\cdot\,)$.

In the lattice field theory the set of histories, $\Omega_q$, is the set
of all field configurations on the lattice to the future of $\sigma_0$.
A field configuration, $\Phi$, is an assignment of $0$ or $1$ to every link,
in other words $\Phi$ is a function from the infinite set of links,
$\{l_a: a = 1,2,\dots\}$,
to $\mathbb{Z}_2$.

The  events that we want to consider are those which refer to properties
of the histories which are bounded in time. In other words
for $A \subset \Omega_q$ to be an event there must exist an integer $m$
such that to determine whether or not a field
configuration, $\Phi$ is in
$A$ it is only necessary to
know the values of $\Phi$ between $\sigma_0$ and $\sigma_m$.
For example, the subset
\begin{equation*}
E_k =\{\Phi \in \Omega_q: \Phi(l_{2k}) = 1\}
\end{equation*}
is an event for any fixed $k$. But the subset
\begin{equation*}
E = \{\Phi \in \Omega_q: \exists k \ \text{s.t.} \ \Phi(l_{2k}) = 1 \}
\end{equation*}
is {\it not} an event  (at least not for the purposes of the current paper).

We want to consider all events that are bounded in time. To this
end, for each positive integer $n$ we define $\Omega_q^n$ to be the
set of field
configurations, $\Phi^n$, on the
first $2n$ links, $l_1,\dots l_{2n}$,
outgoing from the first $n$ vertices $v_1,\dots v_n$. (Recall that
we have chosen an arbitrary, but fixed, natural labelling of the
vertices which gives unambiguous meaning to ``the first $2n$ links''.)
We define the cylinder set $Cyl(\Phi^n)$
to be the set of all elements of $\Omega_q$ which coincide with
$\Phi^n$ on $l_1,\dots l_{2n}$:
\begin{equation*}
Cyl(\Phi^n) \equiv \{\Phi \in \Omega_q | \Phi = \Phi^n\ {\text{when restricted
to the first}} \ 2n \ {\text{links}} \}\;.
\end{equation*}

Each cylinder set, $Cyl(\Phi^n)$
is an event that is bounded in time: it is the
event ``the first $2n$ values of the field agree with
$\Phi^n$.''
The event algebra, $\EA_q$, then, is the (unital)
ring of sets generated, under finite union and intersection,
by all the cylinder sets, $Cyl(\Phi^n)$, for all $n$ and all
$\Phi^n \in \Omega^n_q$.

Two cylinder sets have nonempty intersection if and only if one
contains the other and the complement of a cylinder set (that
for $\Phi^n$, say) is
a disjoint union of finitely many cylinder sets (those
for all the
configurations on $l_1 \dots l_{2n}$ that are {\textit{not}} $\Phi^n$).
Thus, all elements of $\EA_q$ are finite, disjoint unions of cylinder sets.
Given an event, $A \in \EA_q$,
there is indeed an integer, $m$, such that to determine whether or not a field
configuration, $\Phi$ is in $A$ it is only necessary to
know the values of $\Phi$ between $\sigma_0$ and $\sigma_m$.
We will refer to the minimum such $m$ as the {\it {time extent}}
of $A$.
The time extent of the cylinder set $Cyl(\Phi^n)$ is clearly
$n$ and the time extent of an event $A$ is no greater than the
maximum of the time extents of the cylinder sets whose union $A$ is.

Consider the example given previously, $E_k$. We can see that
this is the union of all the cylinder sets for the $\Phi^k$ such that
$\Phi^k(l_{2k}) = 1$:
\begin{equation}
E_k =
\mathop{\bigcup_{\Phi^k\ {\text s.t.}} }_{\Phi^k(l_{2k}) =1}
Cyl(\Phi^k)\;.
\end{equation}
The time extent of event $E_k$ is $k$.

A cylinder set is an event which corresponds to the
history of the field up to a finite time.
For each cylinder set, $Cyl(\Phi^n)$,
the {\textit {class}} operator, $C(\Phi^n)$ \cite{Hartle:1992as},
for that finite history is given by
\begin{equation}\label{eq:classoperator}
C(\Phi^n) \equiv P^H_{2n}({\Phi^n_{2n}})\;
P^H_{2n-1}({\Phi^n_{2n-1}}) \dots P^H_2({\Phi^n_{2}})\;
P^H_1({\Phi^n_{1}}) \; .
\end{equation}
$P^H_a({\Phi^n_a})$ is the projection operator onto the eigenspace
corresponding to the value, $\Phi_a^n = 0$ or $1$, of $\Phi^n$  at link $l_a$,
in the Heisenberg Picture:
\begin{equation}
P^H_a(\Phi^n_a) = U([(a+1)/2])^\dagger\; P_a(\Phi^n_a)\; U([(a+1)/2])
\end{equation}
where $P_a(\Phi^n_a)$ is the Schr\"odinger Picture projector,
$U(k)$ is the unitary time evolution operator (\ref{eq:Uoperator})
and
$[\cdot]$ denotes integer part. The Schr\"odinger picture projector is
\begin{equation}
P_a(\Phi^n_a) =
 \ket{\Phi^n_a} \bra{\Phi^n_a} \; ,
\end{equation}
acting on the factor of $H_q$ associated with $l_a$
(tensored with the identity operator on the other factors).

Expressed in the Schr\"odinger Picture the class operator is
\begin{align}\label{eq:classoperatoragain}
C(\Phi^n) &=  U(n) P_{2n}({\Phi^n_{2n}})\;
P_{2n-1}({\Phi^n_{2n-1}})\; R_n \dots\nonumber\\
{}& \quad \quad \dots P_4({\Phi^n_{4}})\;
P_3({\Phi^n_{3}}) \; R_2 \;P_2({\Phi^n_{2}})\;
P_1({\Phi^n_{1}}) \;R_1\;,
\end{align}
which might be summarised by the
slogan ``evolve, project, evolve, project...''

We define a useful vector valued amplitude for the finite history
$\Phi^n$ by applying its class operator to the initial state,
\begin{equation}\label{eq:unitaryamp}
\ket{\Phi^n} \equiv C(\Phi^n) \ket{\psi_0}\; .
\end{equation}
This vector is sometimes referred to in the literature as a ``branch''
\cite{Hartle:1992as}.

The decoherence functional, $D_q$, is
defined on cylinder sets by the standard expression \cite{Hartle:1992as}
\beq\label{eq:Dquantum}
D_q(Cyl(\Phi^n)\,, Cyl(\m{\Phi}{}^m))
\equiv \braket{\Phi^n}{{}\,\m{\Phi}{}^m}\,.
\eeq
The decoherence functional is defined on the whole event
algebra, $\EA_q$, by additivity since all events are finite disjoint
unions of cylinder sets. Although we have used the natural labelling
 that we chose for the vertices at the beginning, the decoherence
functional thus constructed is independent of the chosen order and
depends only on the vertices' causal order because the projectors
and unitary evolution operators for spacelike separated vertices and
links commute \cite{Dowker:2002wm}.

Note that the properties of the projectors ensure that the formula
(\ref{eq:Dquantum}) for the decoherence functional is consistent with
the condition of additivity when one cylinder set is a disjoint
union of other cylinder sets. For example, $Cyl(\Phi^n)$ is a disjoint
union of all events $Cyl(\Phi^{n+1})$  such that $\Phi^{n+1}$ agrees
with $\Phi^n$ on the first $2n$ links and the decoherence functional
of $Cyl(\Phi^n)$ (with any other event $B$) is indeed given as a sum:
\begin{equation}
D_q(Cyl(\Phi^n)\,,B) =
\mathop{\sum_{\Phi^{n+1}\ {\text s.t.}} }_{\Phi^{n+1}|_n = \Phi^n}
D_q(Cyl(\Phi^{n+1})\,, B)\, ,
\end{equation}
where the sum is over all four field configurations on the
 first $2(n+1)$ links which
agree with $\Phi^n$ on the first $2n$ links.

If the initial
state is a mixed state then the decoherence functional is
a convex combination of pure state decoherence functionals.

This decoherence functional gives a level $2$\ measure, $\mu_q$,
on $\EA_q$ (see section \ref{sec:the quantum measure}).

\subsection{The collapse model with the Bell ontology}

The above unitary quantum field theory inspired a collapse model
field theory \cite{Dowker:2002wm} which, with the Bell ontology, can be
understood as a level 1 (classical) measure theory in the
Sorkin hierarchy (see section \ref{sec:the quantum measure}) as follows.

The space, $\Omega_c$ of all possible histories/formal trajectories is
an identical copy of that for the quantum field theory, namely
the set of all field configurations on the semi-infinite lattice
to the future of $\sigma_0$.
We will refer to field configurations in
$\Omega_c$ as $\alpha$ in order to distinguish
them from the elements of $\Omega_q$ which we refer to
(as above) as $\Phi$. The event algebra $\EA_c$
consists of finite unions of cylinder sets of elements of $\Omega_c$
and so is isomorphic to $\EA_q$.

The dynamics of the collapse model
is given by a classical (level 1) measure.
Since a level 1 measure is also
level 2 -- each level of the hierarchy includes the
levels below it -- a classical measure can also be given in terms of
a decoherence functional
and in this case the decoherence
functional, $D_c$ is given as follows.

Let $\alpha^n$ be a field configuration on the first $2n$ links.
Define a vector valued amplitude $\ket{\alpha^n}\in H_q$ for each cylinder
set $Cyl(\alpha^n)$:
\begin{equation}
\ket{\alpha^n} \equiv J_{2n}({\alpha^n_{2n}})\;
J_{2n-1}({\alpha^n_{2n-1}})\; R_{n} \dots R_2\; J_2({\alpha^n_{2}})\;
J_1({\alpha^n_{1}})\; R_1 \ket{\psi_0}\, ,
\end{equation}
where $\ket{\psi_0}$ is the initial state on $\sigma_0$ and
$J_a({\alpha^n_a})$ is the Kraus operator implementing a
``partial collapse'' onto the eigenspace corresponding to the value
of $\alpha^n$  at link $l_a$. More precisely,
\begin{align}
J_a(0) &=\frac{1}{\sqrt{1+X^2}}(\ket{0}\bra{0} + X \ket{1}\bra{1}) \label{eqn:J(0)}
\\J_a(1) &=\frac{1}{\sqrt{1+X^2}}\left(X \ket{0} \bra{0} + \ket{1}\bra{1}\right)\, \label{eqn:J(1)}
\end{align}
(where $0\le X\le 1$) acting on the factor of
$H_q$ associated with link $l_a$
(tensored with the identity operator for the other factors).

Then the decoherence functional, $D_c$ is
defined on cylinder sets by
\begin{equation}\label{eq:Dclassical}
D_c(Cyl(\alpha^n)\,, Cyl(\m{\alpha}{}^n)) \equiv
\braket{\alpha^n}{{}\,\m{\alpha}{}^n}\delta_{\alpha^n\, \m{\alpha}{}^n}\, ,
\end{equation}
where $\delta_{\alpha^n\, \m{\alpha}{}^n}$ is a Kronecker delta which is
1 if the two field configurations are identical on all $2n$ links
and zero otherwise.

 The decoherence functional is then extended to the whole
event algebra, $\EA_q$ by additivity since all events are finite
disjoint unions of cylinder sets. In particular, if $m>n$, the
cylinder set $Cyl(\Phi^n)$ with time extent $n$ is a disjoint union
of cylinder sets with time extent $m$, and so it suffices to define $D_q$
as above for cylinder sets of the same time extent:
$D_c(Cyl(\alpha^n)\,, Cyl(\m{\alpha}{}^m))$ is given by additivity.

Again, the decoherence functional thus constructed is independent of
the chosen natural labelling and depends only on the vertices' causal order
because of spacelike commutativity of the evolution operators and
Kraus operators.

$D_c$ is well-defined, in particular the additivity condition is
consistent with the definition (\ref{eq:Dclassical}). For example,
consider
\begin{equation*}
D_c(Cyl(\alpha^n)\,,Cyl(\alpha^n))\,.
\end{equation*}
The event $Cyl(\alpha^n)$ is a disjoint union of all events
$Cyl(\alpha^{n+1})$ for which $\alpha^{n+1}$ agrees with $\alpha^n$
on the first $2n$ links and indeed we have:
\begin{equation}
D_c(Cyl(\alpha^n)\,, Cyl(\alpha^n)) = \mathop{\sum_{\alpha^{n+1}\
{\text s.t.}}}_{\alpha^{n+1}|_n = \alpha^n}\;
\mathop{\sum_{\m{\alpha}{}^{n+1}\ {\text s.t.}}}_{\m{\alpha}{}^{n+1}|_n
=
 \alpha^n}
D_c(Cyl(\alpha^{n+1})\,, Cyl(\m{\alpha}{}^{n+1}))\, .
\end{equation}
In verifying this, the crucial property is that
of the Kraus operators: $J_0^2 + J_1^2 = 1$ and the fact that distinct histories
have no interference, as expressed by the Kronecker delta. Note that
without the Kronecker delta, equation (\ref{eq:Dclassical}) would not
be a consistent definition of a decoherence functional satisfying
additivity.

This decoherence functional is level 1 (classical): it satisfies
\begin{equation}
D_c(Y\,, Z) = D_c(Y\cap Z\,, Y\cap Z)
\end{equation}
and this implies the Kolmogorov sum rule is satisfied by the
measure $\mu_c$ defined by $\mu_c(Y) \equiv D_c(Y\,,Y)$.
Being a level 1 measure, $\mu_c$
has a
familiar interpretation as a probability measure. Indeed the
measure $\mu_c$ defined on the cylinder sets is enough, via the
standard methods of measure theory, to define a unique probability
measure on the whole sigma algebra generated by the cylinder sets.
There is, as yet, no  analogous result for a quantal measure such
as $\mu_q$. Moreover,
there is, as yet, no consensus on how to {\textit {interpret}} a quantum
measure theory; a subject we will return to in later chapters.

\subsection{Quantum and Classical}

In every collapse model there is a coupling
between classical stochastic variables and a quantum state. How is
this classical-quantum coupling manifested in the generalised
measure theory form of the lattice collapse model just given? We now
show that there is indeed a quantum measure lurking within
and we will expose the nature of the interaction of the quantal variables
with the classical
 variables.

Consider a space of histories $\Omega_{qc}$ which is a direct product of
the two spaces introduced above, $\Omega_{qc}= \Omega_q \times \Omega_c$,
so that elements of $\Omega_{qc}$ are pairs of lattice field configurations,
$(\Phi, \alpha)$. We will refer to the elements of
$\Omega_q$ as {\textit{quantum histories/fields}}
and those of $\Omega_c$ as {\textit {classical histories/fields}}.
The event algebra $\EA_{qc}$ is the ring of sets
generated by the cylinder sets, $Cyl(\Phi^n, \alpha^n)$, where the
cylinder set contains all pairs $(\Phi, \alpha)$ such that $\Phi$
coincides with $\Phi^n$
 and $\alpha$ coincides with $\alpha^n$ on the first $2n$ links.

We now construct a decoherence functional on $\EA_{qc}$ by taking the
unitary decoherence functional, $D_q$ on $\EA_q$, defined above and
``tying'' the classical histories to the quantum histories by
suppressing the decoherence functional by an amount that depends on
how much the classical and quantum field configurations differ. The
more they differ, the greater the suppression. In detail, define
$D_{qc}$ on $\EA_{qc}$ by first defining it on the
cylinder sets:
\begin{align}\label{eq:Dcoupled}
D_{qc}(Cyl(\Phi^n,\alpha^n)\,,\, &Cyl(\m{\Phi}{}^n,\m{\alpha}{}^n)) \equiv
\nonumber \\
{}&D_q(Cyl(\Phi^{n})\,, Cyl( \m{\Phi}{}^{n}))\; \frac{ X^{d(\Phi^n,\alpha^n)+
d(\m{\Phi}{}^n,\m{\alpha}{}^n)} } { (1+X^2)^{2n}} \;
\delta_{\alpha^n\,\m{\alpha}{}^n}
\end{align}
where $0\le X \le 1$ and $d(\Phi^n,\alpha^n)$ is equal to the number of links
on which $\Phi^n$ and $\alpha^n$ differ. As usual it suffices to
define $D_{qc}$ for arguments which have the same time extent, $n$,
because a cylinder set with time extent $m < n$ is a finite disjoint
union of cylinder sets with time extent $n$. $D_{qc}$ is extended to the
full event algebra by additivity.

Checking that the definition (\ref{eq:Dcoupled}) of $D_{qc}$
on the cylinder sets is consistent with the
property of additivity follows the same steps
as for $D_c$ and $D_q$. $D_{qc}$ is
level 2 in the Sorkin hierarchy, although it is clearly
classical on $\Omega_c$.

We now prove some lemmas regarding $D_{qc}$ which
lay bare the structure of our collapse model
of a lattice field in histories
form.

\begin{lemma}
Let $(\Omega_q, \EA_q, D_q)$,
 $(\Omega_c, \EA_c, D_c)$ and  $(\Omega_{qc}, \EA_{qc}, D_{qc})$
be defined as above for the lattice field theory.
Then the decoherence functional for the collapse model, $D_c$
is equal to $D_{qc}$ coarse grained over $\Omega_q$:
\begin{equation}
D_c(A\,,\m{A}) = D_{qc}(\Omega_q\times A\,, \Omega_q\times \m{A}) \ \ \forall
A, \m{A} \in \EA_c\;.
\end{equation}
\end{lemma}
\begin{proof}

It suffices to prove that
\begin{equation}
D_c(Cyl(\alpha^n)\,,Cyl(\m{\alpha}{}^n)) =
\sum_{\Phi^n, \m{\Phi}{}^n}
D_{qc}(Cyl(\Phi^n,\alpha^n)\,,Cyl(\m{\Phi}{}^n,\m{\alpha}{}^n))\;,
\end{equation}
where the double sum is over all field configurations,
$\Phi^n$ and $\m{\Phi}{}^n$,
on the first $2n$ links.
The result follows by additivity because
\begin{equation}
\bigcup_{\Phi^n}Cyl(\Phi^n,\alpha^n) = \Omega_q \times Cyl(\alpha^n)\;.
\end{equation}

Recall
the definition of $D_c$,
\begin{equation*}
D_c(Cyl(\alpha^n)\,, Cyl(\m{\alpha}{}^n)) =
\braket{\alpha^n}{{}\,\m{\alpha}{}^n}\delta_{\alpha^n\, \m{\alpha}{}^n}\, ,
\end{equation*}
where
\begin{equation*}
\ket{\alpha^n} =  J_{2n}({\alpha^n_{2n}})\,
J_{2n-1}({\alpha^n_{2n-1}}) \, R_n \dots R_2 \,J_2({\alpha^n_{2}})\,
J_1({\alpha^n_{1}}) \,R_1 \ket{\psi_0}\, .
\end{equation*}
 Each jump operator $J_a(\alpha^n_a)$
is a linear combination of the two projection
operators $P_a(1)= \ket{1}\bra{1}$ and $P_a(0)= \ket{0}\bra{0}$
on link $l_a$ (see equations \ref{eqn:J(0)} and \ref{eqn:J(1)}). Substituting in this linear combination of projectors for each
 $J_a(\alpha^n_a)$ and expanding out,
 we see that the ket becomes a sum of
$2^{2n}$ terms, one for each possible field configuration
-- call it $\Phi^n$ -- on the $2n$ links. Each such term
is precisely the vector valued amplitude
$\ket{\Phi^n}$ (\ref{eq:unitaryamp}) and
each term is weighted by a factor
\begin{equation*}
\frac{X^{d(\alpha^n, \Phi^n)}}{(1+X^2)^n}
\end{equation*}
from which the result follows.
\end{proof}

The next lemma shows that
if we coarse grain $D_{qc}$ over the classical histories
instead, we find a
quantum theory exhibiting the
symptoms of environmental decoherence.
\begin{lemma}\label{lemma:decoh}
Define a decoherence functional $\widetilde{D}_q$ on $\Omega_q$
by
\begin{equation}
\widetilde{D}_q(F\,, \m{F})  \equiv
D_{qc}(F\times \Omega_c\,, \m{F}\times \Omega_c)\ \ \forall F, \m{F} \in \EA_q \;.
\end{equation}
Then
\begin{equation}
\widetilde{D}_q(Cyl(\Phi^n)\,, Cyl(\m{\Phi}{}^n))=
\left(\frac{2X}{1+X^2}\right)^{d(\Phi^n, \m{\Phi}{}^n)}
D_q(Cyl(\Phi^n)\,, Cyl(\m{\Phi}{}^n))\;.
\end{equation}
\end{lemma}

We leave the proof to the appendix. Note that the factor suppresses off-diagonal terms in the decoherence functional and so looks as if it is the result of environmental decoherence.

\subsection{Equivalence to a model with environment}

The system described by decoherence functional $D_{qc}$ on
the joint space $\Omega_{qc}$ was not derived from any
physical consideration but simply invented as a way to unravel
the decoherence functional of the collapse model. However,
once obtained, the urge to coarse
grain $D_{qc}$ over the classical histories
is irresistible and
the ``approximately diagonal'' form of the resulting
decoherence functional, $\widetilde{D}_q$ on $\Omega_q$ suggests it
can be interpreted as having arisen from coupling to
an environment.

Indeed, the mathematics of collapse models and of
open quantum systems that result from coarse graining
over an ignored environment are known to be closely related
and so it is of no surprise to discover that our
current model can be understood in this way.
Indeed, the classical histories in the collapse model
can simply be reinterpreted as
histories of an environment consisting of variables, one
per link, that interact impulsively with the field there, and then
have no further dynamics.

Let the quantum lattice field,
$\Phi$, interact with
a collection of environment variables, one for each link, taking
values $0$ or $1$.
The space of histories for the whole system
is $\Omega_{qe} \equiv \Omega_q \times \Omega_e$,
where the space of environment
histories, $\Omega_e$, is yet another copy of the same space of
 $\{0,1\}$-field configurations on the semi-infinite lattice. We denote
an element of $\Omega_e$ by $E$, an environment configuration on the
first $2n$ links by $E^n$, the corresponding cylinder set
by $Cyl(E^n)$, and the value of the environment
variable on link $a$ by $E^n_a$.

In the standard state vector language, the Hilbert
space of the whole system of field, $\Phi$,
and the environment is  $H_{qe} \equiv H_q \otimes H_e$
where the environment
Hilbert space, $H_e$, is an infinite tensor product of qubit Hilbert spaces,
$H_{e_a}$, $a = 1,2,\dots$,
one for each link $l_a$ on the lattice to the future of $\sigma_0$.

\begin{lemma}
There is a unitary dynamics of this system such that the unitary
decoherence functional which encodes it, $D_{qe}$,
is equal to $D_{qc}$ if the
environment histories are identified with the classical
histories.
\end{lemma}

\begin{proof}
The proof is by construction of such a dynamics. We add,
to the unitary dynamics of the field $\Phi$,
a one-time interaction
between $\Phi$ and the environment variable on each link
which establishes a partial correlation
between them. Since each environment state lives on
exactly one link, it interacts only once and is
then fixed, which means that
the decoherence functional is diagonal on
the environment histories.

We begin with the space of histories $\Omega_{qe} = \Omega_q \times \Omega_e$
and the Hilbert space $H_{qe} =  H_q \otimes H_e$
where $H_e = \otimes_{a=1}^\infty H_{e_a}$ and each $H_{e_a}$ is a
qubit space.

 The initial state is a
tensor product:
\begin{equation}\label{eq:Psi0}
\ket{\Psi_0} = \ket{\psi_0}_q \otimes_{a =1}^\infty
\ket{0}_{e_a}
\end{equation}
where $ \ket{\psi_0}_q \in H_q$ is the same initial state
for the field $\Phi$ as we had before.

After each elementary unitary evolution $R_i$ is
applied over vertex $i$, two unitary ``partial measurement'' operators
$U_{2i-1}$ and $U_{2i}$ -- to be defined --
 are applied to the Hilbert spaces
associated with
the outgoing links $l_{2i-1}$ and $l_{2i}$, respectively.

Consider a single link, $l_a$. The factor of the total Hilbert space
associated with $l_a$ is
the four-dimensional tensor product of the qubit space, $H_{q_a}$,
of the $\Phi$ states on $l_a$ and the qubit space $H_{e_a}$.
In the field representation, the basis of
this link Hilbert space is $\{\ket{0}_{q_a}\ket{0}_{e_a},
\ket{1}_{q_a}\ket{0}_{e_a},\ket{0}_{q_a}\ket{1}_{e_a},\ket{1}_{q_a}\ket{1}_{e_a}\}$.

The unitary partial measurement operator $U_a$ is defined by
its action on this basis:
\begin{align}
U_a\ket{0}_q\ket{0}_e &= \frac{1}{\sqrt{1+X^2}}\; \ket{0}_q \big(\ket{0}_e + X \ket{1}_e\big)\nonumber\\
U_a\ket{1}_q\ket{0}_e &= \frac{1}{\sqrt{1+X^2}}\; \ket{1}_q \big(X\ket{0}_e +  \ket{1}_e\big)\nonumber\\
U_a\ket{0}_q\ket{1}_e &= \frac{1}{\sqrt{1+X^2}}\; \ket{0}_q \big(X\ket{0}_e - \ket{1}_e\big)\nonumber\\
U_a\ket{1}_q\ket{1}_e &= \frac{1}{\sqrt{1+X^2}}\; \ket{1}_q \big(\ket{0}_e - X \ket{1}_e\big)\label{eq:defU}\;,
\end{align}
where $0\le X\le 1$ and
we have suppressed the $a$ label on all the kets.
$U_a$ acts as the identity on all other factors in
the tensor product Hilbert space for the system.

The action of $U_a$ is to leave $\Phi$ eigenstates alone and put the
initial $\ket{0}_e$ environment state into a superposition of $\ket{0}_e$
and $\ket{1}_e$, so that the environment eigenstate that is correlated
with the $\Phi$ eigenstate is relatively enhanced by a factor $X^{-1}$.

For each cylinder set $Cyl(\Phi^n, E^n)$
we define a vector
valued amplitude, $\ket{\Phi^n, E^n}_{qe} \in H_{qe}$
by evolving the state over each vertex, applying the unitary partial
measurements on the outgoing links and projecting onto the
values of $\Phi^N$ and $E^N$ on the links:
\begin{align}
\ket{\,\Phi^n, E^n}_{qe} \equiv \,&
Q_{2n}(E^n_{2n})\, P_{2n}(\Phi^n_{2n})
\, Q_{2n-1}(E^n_{2n-1})\, P_{2n-1}(\Phi^n_{2n-1})\nonumber\\
{}&\ \ \ \ U_{2n}\, U_{2n-1}\, R_n \dots \nonumber\\
{}&\ \ \ \ \ \ \ \ \dots Q_2(E^n_2)\, P_2(\Phi^n_{2})\,
Q_1(E^n_1) \,P_1(\Phi^n_{1})\nonumber\\
{}&\ \ \ \ \ \ \ \ \ \ \ \ \ \ \ \ \ \ U_2\, U_1 \,R_1 \ket{\Psi_0}\, ,
\end{align}
where $\ket{\Psi_0}$ is defined in (\ref{eq:Psi0}),
$P_a({\Phi^n_a})$ is the projection operator onto the eigenspace
corresponding to the value of $\Phi^n$  at link $l_a$
and $Q_a(E^n_a)$ is the projection operator onto the
eigenspace corresponding to the value of $E^n$ at link
$l_a$. $P_a({\Phi^n_a})$ is only non-trivial on the
factor in $H_q$ associated with link $l_a$ and
$Q_a(E^n_a)$ is only non-trivial on the factor in $H_e$
associated with link $l_a$. As a consequence,
the $P$ projectors and $Q$ projectors
commute.

The initial state is a product, each $U_a$ leaves $\Phi$-eigenstates
alone and the $Q$ projectors act only on the environment states.
We claim that therefore $\ket{\Phi^n, E^n}_{qe}$ is a product,
\begin{equation}
\ket{\Phi^n, E^n}_{qe} = \ket{\Phi^n}_q \ket{E^n}_e\,,
\end{equation}
where $\ket{\Phi^n}_q \in H_q$ is the vector valued amplitude (\ref{eq:unitaryamp})
for the plain vanilla unitary field theory
and
\begin{equation}\label{eq:ketE}
\ket{E^n}_{e} = \frac{X^{d(\Phi^n, E^n)}}{(1+X^2)^n}\;
\ket{E^n_1}_{e_1} \ket{E^n_2}_{e_2}\dots \ket{E^n_{2n}}_{e_{2n}}
\end{equation}
where we have left off the factors of $\ket{0}$ for all the
infinitely many links
to the future of $\sigma_n$, which play no role.

The proof of this claim is given in the appendix.

The decoherence functional, $D_{qe}$, for the total system
is given by
\begin{align}
D_{qe}(Cyl(\Phi^n,E^n)\,,Cyl(\m{\Phi}{}^n,\m{E}{}^n)) &\equiv
\braket{\Phi^n,E^n}{\m{\Phi}{}^n,\m{E}{}^n} \\
{}& = \braket{\Phi^n}{\m{\Phi}{}^n}_q \braket{E^n}{\m{E}{}^n}_e\,.
\end{align}
Using (\ref{eq:ketE}), we see that the decoherence functional is
zero unless $E^n = \m{E}{}^n$ and we have
\begin{align}
D_{qe}(Cyl(\Phi^n,E^n)\,,\,&Cyl(\m{\Phi}{}^n,\m{E}{}^n)) =& \nonumber\\
& D_q(Cyl(\Phi^{n})\,, Cyl(\m{\Phi}{}^{n}))\; \frac{X^{d(\Phi^n,E^n)+
d(\m{\Phi}{}^n,\m{E}{}^n)}}{(1+X^2)^{2n}} \;\delta_{E^n\,\m{E}{}^n}\;.
\end{align}

As usual, we only need to define the decoherence functional
for cylinder sets of equal time extent. We see that this is
equal to $D_{qc}$, the decoherence functional of the
collapse model (\ref{eq:Dcoupled}).
\end{proof}

The model is technically unitary and so falls into
the category of ordinary quantum theory, but the classicality of the
environment variables is achieved by the device of postulating an infinite
environment and one-time interactions.

\section{Discussion}
None of the physics we have presented is new.
We have merely provided a novel perspective
on a known model that arises when spacetime
and histories are given a central role.
Di\'osi stressed that both classical variables and quantum state are
present in a collapse model and advocates ascribing reality to them both
\cite{Diosi:2004}.
We have replaced the formalism of
quantum state with quantum histories and
by placing quantum and classical variables on the same
footing in spacetime
we can see more clearly the character of the interaction
between them.

We claim that the structure outlined above for the collapse model
for a lattice field theory, is generic to collapse models.
There is always, more or less hidden in the model,
a  space of histories which is a product of a
space of quantum histories and a space of classical histories,
with a decoherence functional on it.
For example, in the case of the GRW model \cite{Ghirardi:1986mt}
the classical histories
are countable subsets of Galilean spacetime, to the future of some
initial surface, $t=0$. The elements of such a countable subset are
the ``collapse centres'' $(x_i, t_i), i = 1,2, \dots$.
The probability distribution on these
classical histories is given by a classical
decoherence functional $D_c$, which is, essentially, set out in
\cite{Kent:1998bc}.
In order to follow the steps taken in this paper of
unraveling $D_c$ into $D_{qc}$, the positive operators, Gaussians,
that correspond to the classical events are
expressed as
integrals of projection operators and the evolution between
collapses expressed using the Dirac-Feynman propagator as
a sum over the histories.
The quantum histories, then, are precisely the
histories summed over in the Dirac-Feynman path integral.

The continuum limit of the GRW model is the
continuous spontaneous localisation model for a single particle
\cite{Diosi:1988a,Diosi:1988b} and this too can be cast into the generic
form as can be seen
from the formulation of the model in terms of
a ``restricted propagator'' as described in
references \cite{Mensky:1979, Mensky:1994, Diosi:1995}.
Although the analysis in these references uses phase space path
integrals, if it is the position operator whose eigenstates are collapsed onto, as is the case for the continuum limit of GRW, the path
integrals can be transformed into configuration space path integrals.
In this case, the quantum histories are again the continuous paths
that contribute to the Dirac-Feynman sum-over-histories,
but the classical histories are very noisy, and not continuous
paths at all.

Note that in the lattice field theory the spaces of classical and quantum
histories in this case are isomorphic, whereas in the GRW model and its
continuum limit the quantum and classical histories are very different.
In all cases, however, it is the quantum histories that bear all the consequence
of dynamical law encoded in a local spacetime
action, whereas the classical histories
are simply dragged along by being tied to the quantum histories.

This state of affairs is illuminated further by considering
coupling together two separate collapse models X and Y. Each model will
contain both quantum and classical histories and the
coupling between X and Y will be achieved by
an appropriate term in the action involving the quantum histories alone.
It is the quantum histories of X which directly touch the quantum histories of
Y. The classical variables of X only react to the classical
variables of Y because they are restricted to be close to the
quantum variables which interact with the quantum variables
of Y to which the classical variables of Y must, in their turn, be close.

Two important reasons for pursuing collapse models with the Bell ontology are that the models are already in spacetime form and the stochasticity involved is completely classical so all the familiar machinery of stochastic processes can be brought to bear: the stochasticity of collapse models causes no more interpretational difficulty than does the randomness of Brownian motion. The theory concerns the classical variables only and the quantum histories are relegated to some sort of auxiliary, hidden status, despite the fact that the dynamics of the model is most easily described in terms of these quantum histories. In order to pursue this direction, therefore, one must pay the price of ignoring the quantum histories as far as the ontology is
concerned: ``Pay no attention to that man behind the curtain'' \cite{Oz:1939}.

On the other hand, if the quantum histories are kept in the theory
to be treated on the same footing, a priori, as the
classical histories, then
the question of the physical meaning of the
quantum measure on them has to be wrestled with:
what {\it is} the ontology in a quantum measure theory?
But if {\textit {this}} thorny problem is to be tackled, then
one might start by trying to address it in the
case of unitary quantum mechanics in the
first instance. It may be that an interpretation of the
quantum measure can be discovered that,
 by itself, provides
a solution to the interpretational problems of quantum mechanics,
while yet maintaining unitary dynamics
and without need of new quantum-classical couplings.

%Stochastic Collapse
\part{Co-Event Schemes for Quantum Measure Theory}
%\singlespacing
\chapter{Introducing Co-Events}\label{chapter:introducing coevents}
%\doublespacing

\section{Summary}

The `co-event interpretation' proposed by Sorkin \cite{Sorkin:2006wq,Sorkin:2007} is a new approach to interpreting the histories formulation that can engage with the whole event algebra rather than a decoherent subalgebra, and thus meaningfully describe the `micro-level'. Sorkin's `co-events' are based on truth valuations, thinking of reality in terms of yes/no answers to all questions about the system, or alternatively as a set of `true propositions'.

Starting with a classical histories theory $(\O,\EA,\P)$ (definition \ref{def:histories theory}), and recalling that the event algebra can be seen as the space of (meaningful) propositions concerning the system (section \ref{sec:the axiomatic approach}), we proceed by adding two further objects to the theory, a space of ``answers'' (or ``truth values'') which classically is $\Z2 = \{0,1\}= \{false, true\} = \{no, yes\}$, and the space of allowed ``answering maps'' $\phi: \EA \rightarrow \Z2$, which in the classical theory is the space of non-zero homomorphisms. As we shall see, the fact that the answering maps are homomorphisms reflects the use of Boolean (ordinary) logic to reason about the system. These answering maps, or {\it co-events}, are the fundamental elements of Sorkin's new interpretation; indeed they will be identified as the `beables' of a histories theory, as we shall see in some detail below.

To generalise this structure to an arbitrary histories theory $\H$, and in particular to the quantum case, Sorkin proposes to keep both the ``event algebra'' $\EA$ and the truth values $\Z2$ as they are but relax the condition that the answering map/co-event $\phi$ must be a homomorphism. Thus the general framework can be described as one that embraces ``Anhomomorphic Logic''.

However, relaxing a condition leads immediately to the question of the alternative; if we are no longer demanding that our co-events be homomorphisms then what condition are we requiring of them? There have, to date, been several proposals, or {\it schemes}, meeting various degrees of success, from the original linear scheme \cite{Sorkin:2006wq} to the current favoured candidate the multiplicative scheme \cite{Sorkin:2007,Dowker:2007zz,Furey:2007}. Most of these proposals have encountered problems, some fatal, leading to a number of them (including the original linear scheme) being abandoned. However both the multiplicative and the ideal based approach have `survived' all the tests they have been put through (so far), and the multiplicative scheme in particular is regarded as the `success story' of the co-event interpretation.

These schemes have so far been defined for histories theories possessing finite sample spaces, both for simplicity and because quantum measure theory has not yet been fully developed in the infinite case, as we touched on in section \ref{sec:quantum measure theory}. A beginning has been made, however, in the endeavor to generalise the multiplicative scheme to the infinite case, and we will return to this issue in section \ref{sec:infinite sample spaces}. In the interim, unless explicitly stated otherwise, we will in the remainder of this thesis assume that all sample spaces (and thus all event algebras) are finite, and that $\EA=P\O$.

In this chapter we will lay out the framework and concepts that we will be using in the remaining chapters. We warm up by applying the co-event approach to classical stochastic theories, before formally defining a framework for the application of these ideas to quantum (or higher order) histories theories. We then review the major proposals for a replacement of the classical co-events before setting out the criteria by which the success of these schemes should be judged. We finish by stepping back to consider the `interpretation of the interpretation', discussing how we should understand co-events.

\section{Co-Events in Classical Stochastic Theories}\label{sec:coevents in classical stochastic mechanics}

As we have previously discussed (section \ref{sec:the naive interpretation}), the standard interpretation of a classical stochastic theory distinguishes exactly one of the fine grained histories (say $r\in\O$) as `the real history', which `actually occurs'. Unlike a deterministic theory, the stochastic dynamics (even including the initial condition) does not identify the real history, though it does rule out (or preclude) histories that are elements of null sets. Thus the set of fine grained histories that are not elements of any null set constitutes the space of \textit{potential realities}. We will refer to the `real' history that `actually occurs' as the \emph{actually real history} to avoid confusion with the \emph{potentially real histories}, which are the elements of the space of potential realities. We require that the actually real history is one of the potentially real histories.

As we have previously noted (section \ref{sec:the naive interpretation}), in classical theories events are \textit{true} if they contain the actually real history $r$, and \textit{false} otherwise. Thus for every potential reality $\g\in\O$ we have a truth valuation map $\g^*$:
\bea
\g^*:\EA&\longrightarrow& \mathbb{Z}_{2}, \nonumber \\
\g^*(A) &=& \left\{\begin{array}{cc} 1  & \g\in A \\ 0 &
\g\not\in A. \end{array}\right. \label{eq:r* action}
\eea
We intuitively think of $\g^*$ as the \textit{dual} of the history $\g$, or the event $\{\g\}$, from whence we derive the term \textit{co-event}.

The co-event approach, as proposed by Sorkin \cite{Sorkin:2006wq,Sorkin:2007}, rests on a subtle shift in the way we think about this interpretation. If we know which history $\g$ is actually real, $\g^*$ gives us the truth or falsity for each event. Conversely, if we know the truth or falsity of every event we would be able to reconstruct $\g^*$, and hence identify $\g$. The knowledge of the truth or falsity of all events is thus equivalent to the knowledge of reality, of which history is actually real; in other words our knowledge of the truth valuation map is equivalent to our knowledge of the ontology.

Sorkin proposes that we treat the truth valuation map as fundamental, with the intention of generalising it when we come to quantum mechanics. There is more than one way to understand this `shift in thinking'; while all of our results will hold for every interpretation of Sorkin's proposal, for definiteness we will adopt an interpretation that identifies the ontology with the truth valuation map itself, returning to discuss the various alternatives in greater depth in section \ref{sec:the interpretation of the interpretation}. Thus instead of thinking of $\g$ as the actually real history, we will think of $\g^*$ as \textit{the actually real co-event}. Our space of potential realities will now be the set of co-events, rather than histories, that are not ruled out by the dynamics. Since a history that is not ruled out by the dynamics is not an element of any null set, the corresponding concept for co-events is given by \textit{preclusion}; a co-event $\g^*$ is preclusive if it obeys the following condition:
\beq
\mathbb{P}(A) = 0 \Rightarrow \g^*(A) = 0,
\eeq
and the set of preclusive co-events is our `new' space of potential realities. We can call the preclusive co-events the \emph{potentially real co-events}. Notice that $\g^*$ is preclusive if and only if $\g$ is not contained in any null set, so that $\g^*$ is a potentially real co-event if and only if $\g$ is a potentially real history.

This shift in thinking is non-trivial because these co-events have a natural algebraic structure, which can be used to generalise them. $\Z2$ is a field, and hence a ring, under the standard multiplication and addition operators inherited from $\mathbb{Z}$. As discussed in section \ref{sec:stochastic mechanics}, as a subset of $P\O$ the event algebra $\EA$ also admits a ring or algebraic structure (in fact it is a boolean algebra), with symmetric difference used as addition (equation \ref{eq:set addition}) and intersection used as multiplication (equation \ref{eq:set mutiplication}). It is easy to see that $\g^*$ is a homomorphism between these two rings because it obeys linearity and multiplicativity:
\begin{eqnarray}
\g^*(A+B) &=& \g^*(A) + \g^*(B), \ \
(linearity)\nonumber \\
\g^*(AB) &=& \g^*(A)\g^*(B). \ \ (multiplicativity) \nonumber
\end{eqnarray}
Where on the left hand side addition and multiplication of events is in $\EA$, and on the right hand side addition and multiplication are in $\Z2$. In fact, it can be shown that the $\g^*$'s are {\it all} the non-zero homomorphisms. We state the following lemma without proof:

\begin{lemma}\label{lemma:classical coevents are homomorphisms}
Given a finite sample space $\Omega$ and the associated event algebra $\EA=P\O$
$$\{\g^* | \g\in\Omega\}=Hom(\EA,\mathbb{Z}_{2}),$$
where $Hom(\EA,\Z2)$ is the space of ring (or algebra) homomorphisms from $\EA$ to $\Z2$, excluding the zero map.
\flushright{$\square$}
\end{lemma}

We can illustrate these ideas using the example of a simple coin.

\begin{example}\label{example:coin for classical dynamics coevents}
Consider a single toss of a coin as described by classical stochastic dynamics. We assume that we will have either a `heads' (`h') or a `tails' (`t') outcome. Then our sample space is given by
\beq
\O=\{h,t\}.
\eeq
Our event algebra is simply the power set of $\O$,
\beq
\EA = P\O = \{\emptyset,\{h\},\{t\},\{h,t\}=\O\},
\eeq
and our classical measure is $\P$:
\bea
\P(\emptyset) &=& 0, \nonumber \\
\P(\{h\}) &=& p, \nonumber \\
\P(\{t\}) &=& 1-p, \nonumber \\
\P(\{h,t\}) &=& 1.
\eea
We assume that $p\not\in\{0,1\}$ so that neither $\{h\}$ nor $\{t\}$ is null. Then using lemma \ref{lemma:classical coevents are homomorphisms} we have:
\beq
Hom(\EA,\mathbb{Z}_{2}) = \{h^*,t^*\}.
\eeq
Both of these co-events can be better understood by consideration of their action on $\EA$. Recall that a co-event $\g^*$ maps to unity events that would be true if that co-event were the actual reality. Notice that the internal logic of both $h^*$ and $t^*$ is Boolean:
$$
\begin{array}{c|cccc}
 & \emptyset & \{h\} & \{t\} & \{h,t\}=\O \\
\hline
h^* & 0 & 1 & 0 & 1 \\
t^* & 0 & 0 & 1 & 1
\end{array}
$$
Under the naive interpretation our space of potential realities would be $\O = \{h,t\}$, and if $h$ were the actual reality the events $\{h\}$ and $\{h,t\}$ would be true. Under the co-event interpretation our space of potential realities would be $Hom(\EA,\mathbb{Z}_{2}) = \{h^*,t^*\}$, and if $h^*$ were the actual reality the events $\{h\}$ and $\{h,t\}$ would, again, be true.

If we now set $p=1$ then $\{t\}$ will be null. Then under the naive interpretation the history $t$ would be `disallowed' by the dynamics, and the space of potential realities would be $\{h\}$. Similarly, because $t^*(\{t\})=1$ the co-event $t^*$ is no longer preclusive, hence the space of potential realities under the co-event interpretation is $\{h^*\}$.
\flushright{$\square$}
\end{example}

\section{Co-Events for Quantum Mechanics}

So how should we adapt this to quantum mechanics, or to a more general histories theory? The most naive approach would be to apply the co-event structure of classical stochastic theories (section \ref{sec:coevents in classical stochastic mechanics}) unaltered to quantum mechanics, using $\mu(A)=0\Rightarrow \g^*(A)=0$ as the preclusion rule. This is in fact equivalent to the naive interpretation (``exactly one history in $\O$ is real'', see section \ref{sec:the naive interpretation}) we examined earlier, and thus breaks down when applied to systems based on the Kochen-Specker theorem \cite{Kochen:1967,Dowker:2007zz} (chapter \ref{chapter:kochen-specker}).

It therefore becomes necessary to generalise the co-events used in stochastic theories if this framework is to be applicable to quantum mechanics. Several attempts have been made at such a generalisation, with varying degrees of success, leading up to the `multiplicative scheme' which is the working co-event scheme at the time of writing. We will review some of these schemes later in the chapter (section \ref{sec:coevent schemes}), but first we need to define what we mean by a co-event and a co-event scheme in the context of a general histories theory.

\begin{definition}
Given a histories theory $(\Omega,\EA,\mu)$, a co-event is a map
$$\p:\EA\rightarrow\mathbb{Z}_{2},$$
satisfying
$$\p(\emptyset)=0.$$
A co-event $\p$ is preclusive if it satisfies
$$\mu(A) = 0 \Rightarrow \p(A) = 0.$$
We call the set of all co-events $\EA^*$, and the set of all preclusive co-events $\Pr$.
\end{definition}

Note that this definition would still make sense if the sample space were infinite. Under this more general definition, co-events are no longer required to be homomorphisms, or to be `dual' to single histories as they were in classical stochastic theories. However, these concepts remain key to our notions of `classicality'.

\begin{definition}\label{def:classicality for coevents}
Let $\H$ be a histories theory with a finite sample space and let $\la$ be a subalgebra of $\EA$. We say that a co-event $\p\in\EA^*$ is \textbf{classical on $\la$} if the restriction of $\p$ to $\la$ is a homomorphism. We say that $\p$ is classical if it is classical on $\EA$.
\end{definition}

We can use `classical' (Boolean) logic to reason about a subsystem precisely when the actually real co-event is classical on the corresponding subalgebra of $\EA$. When the actually real co-event is classical we can use Boolean logic to reason about the whole system. Lemma \ref{lemma:classical coevents are homomorphisms} then informs us that the classical co-events are precisely the co-events $\g^*$ defined in the previous section for $\g\in\O$. The use of the term \emph{classical co-event} to describe the $\g^*$ is consistent with Sorkin's terminology \cite{Sorkin:2007}, further note that these are precisely the co-events that are the potential realities in classical stochastic mechanics (section \ref{sec:coevents in classical stochastic mechanics}). Classical co-events are sometimes referred to as the \emph{duals of the singleton sets}\footnote{A singleton set is a set of cardinality one} ($\g^*$ is referred to as the \emph{dual} of $\g$), or as \emph{atomic co-events} (see below).

As discussed in section \ref{sec:stochastic mechanics}, $\EA$ is an algebra over $\Z2$; we can give $\EA^*$ a similar algebraic structure. We state the following lemma without proof:

\begin{lemma}
Let $\H$ be a histories theory with a finite sample space. Then the following binary operations on $\EA^*$ are well defined:
\bea
(\p + \psi)(A) &=& \p(A) + \psi(A), ~(addition) \nonumber\\
(\p\psi)(A) &=& \p(A)\psi(A),~(multiplication) \nonumber
\eea
for all $A\in\EA$ and for all $\p,\psi\in\EA^*$. Under these operations $\EA^*$ is both a ring and an algebra over $\Z2$. Considering $\EA^*$ as a ring, $\Pr$ is a two-sided ideal in $\EA^*$.
\flushright{$\square$}
\end{lemma}

The singleton sets $\{\g\}$ have a privileged status as \textit{atoms} of $\EA$\footnote{An atom $A$ of an algebra $\EA$ is an element $A\in\EA$ such that $AB\in\{\emptyset,A\}$ for all $B\in\EA$}. A similar algebraic status in $\EA^*$ is afforded to classical co-events, justifying the use of the term `atomic co-events' to describe them. As noted in \cite{Sorkin:2007} the classical co-events generate the whole ring $\EA^*$. We state the following lemma without proof:

\begin{lemma}\label{lemma:classical coevents generate all coevents}
Given a histories theory $\H$, where $\O$ is finite, $\EA^*$ (considered as a ring or an algebra) is generated by the classical co-events. In particular, every co-event $\p\in\EA^*$ can be expressed as a polynomial in the classical co-events.
\flushright{$\square$}
\end{lemma}

We can illustrate these concepts and the nature of our generalisation by returning to our example of a simple coin.

\begin{example}\label{example:coin for general coevents}
Consider the system described in example \ref{example:coin for classical dynamics coevents}. Using lemma \ref{lemma:classical coevents generate all coevents} we see that the space of co-events is
\beq
\EA^* = \{\textbf{0},h^*,t^*,h^*+t^*,h^*t^*,h^*t^*+h^*,h^*t^*+t^*,h^*t^*+h^*+t^*\},
\eeq
where $\textbf{0}$ is the zero map. As before we can calculate the truth valuation table using each of the co-events in $\EA^*$:
$$
\begin{array}{c|cccc}
 & \emptyset & \{h\} & \{t\} & \{h,t\}=\O \\
\hline
\textbf{0} & 0 & 0 & 0 & 0 \\
h^* & 0 & 1 & 0 & 1 \\
t^* & 0 & 0 & 1 & 1 \\
h^* + t^* & 0 & 1 & 1 & 0 \\
h^*t^* & 0 & 0 & 0 & 1 \\
h^*t^* + h^* & 0 & 1 & 0 & 0 \\
h^*t^* + t^* & 0 & 0 & 1 & 0 \\
h^*t^* + h^* + t^* & 0 & 1 & 1 & 1 \\
\end{array}
$$
Recall that by construction all co-events are zero valued on the empty set and that every combination of truth values on the events in $\EA$ is represented by one of the co-events, which are all expressed as polynomials in the classical co-events. Notice that only the co-events $h^*,t^*$  are classical and thus possess a Boolean `internal logic'. For example, if $h^*t^*$ were the actual reality both the events $\{h\}$ and $\{t\}$ would be false whilst the event $\{h,t\}$ would be true. We can rephrase the truth table of $h^*t^*$ in `question/answer' format:
\\
\\
\begin{tabular}{c|c}
\bf{Question} & \bf{Answer} \\
\hline
Does the coin show heads? & no \\
Does the coin show tails? & no \\
Does the coin show one of heads or tails? & yes \\
\end{tabular}
\\
\\
As a second example, $h^*t^* + h^* + t^*$ also yields `non-classical' results:
\\
\\
\begin{tabular}{c|c}
\bf{Question} & \bf{Answer} \\
\hline
Does the coin show heads? & yes \\
Does the coin show tails? & yes \\
Does the coin show one of heads or tails? & yes \\
\end{tabular}
\\
\\
As before, we can set $p=1$ so that $\{t\}$ becomes null. Then the co-events $t^*,h^* + t^* ,h^*t^* + t^*$ and $h^*t^* + h^* + t^*$ are no longer preclusive since they all map $\{t\}$ to unity. We then have:
\beq
{\cal{P}}(\O,\EA,\P) = \{\textbf{0},h^*,h^*t^*,h^*t^* + h^*\}.
\eeq
However, we have not yet made clear what the space of potential realities should be, and this will be discussed in the next section.
\flushright{$\square$}
\end{example}

\section{The Various Co-Event Schemes}\label{sec:coevent schemes}

Following the discussion in section \ref{sec:coevents in classical stochastic mechanics} we require that any `potentially real' co-event should be preclusive so as to respect the dynamical laws. This might suggest that given a histories theory $\H$ we should take $\Pr$ as the space of potential realities in place of $\O$. However, as we can see from our coin tossing example (example \ref{example:coin for general coevents}), $\Pr$ is `too large' in practice; in particular when the measure $\mu$ is classical $\Pr$ is strictly larger than the space of homomorphisms. This would raise the possibility of `non-classical behaviour' of dynamically classical systems; we may be faced with a situation in which non-Boolean logic is applied to observable, `macroscopic', outcomes.

We therefore want to pick a subset of $\cal{P}$ to represent the potentially real co-events in a manner that ensures Boolean logic is applied to observable outcomes but nevertheless retains the flexibility required to describe quantum systems. There are several \emph{schemes} determining how this choice should be made.

\begin{definition}
A co-event scheme ${\cal{S}}$ is a map from the space of histories theories with finite sample spaces, taking as its argument a triple $\H$ where $\O$ is finite and returning (mapping it to) a subset of $\Pr$,
\beq\nonumber
{\cal{S}}:\H\mapsto\S\subset\Pr.
\eeq
\end{definition}

Note that it is \emph{not} the intention to apply one co-event scheme to one histories theory and a second scheme to another histories theory. We envision that there is \emph{one} correct co-event scheme that should be applied to every system. However during the (ongoing) development of the co-event interpretation various proposals have been put forward as to what the `correct' scheme should be, therefore it is helpful to maintain the framework of multiple co-event schemes. If, say, ${\cal{S}}_c$ were the `correct' scheme, then the set ${\cal{S}}_c\H$ would be our space of potential realities for the histories theory $\H$.

A co-event scheme can be thought of as a methodology for restricting from $\Pr$ to a subset thereof in some systematic fashion. In our treatment of classical stochastic mechanics we restricted the space of possible co-events by imposing `rules', namely the preservation of the linear and multiplicative structure that is implicit under an algebra homomorphism, and this proved sufficient to define the possible co-events. Most of the proposed schemes for quantum mechanics restrict $\Pr$ by first imposing a rule then imposing a `{\it primitivity condition}' that picks the `finest grained' preclusive co-events obeying our rule. Primitivity is sometimes referred to as \textit{minimality} when the concept of the fine graining of co-events can be defined in terms of a partial order. Recalling that every co-event can be expressed as a polynomial in the classical co-events (lemma \ref{lemma:classical coevents generate all coevents}), we note that these `rules' can often be expressed in algebraic terms as conditions on the polynomials in $\Pr$.

We are now in a position to introduce some proposed co-event schemes for quantum mechanics, and will explore most of them in greater depth in subsequent chapters. Where relevant, we will for each scheme detail the defining rule, its algebraic formulation and the appropriate primitivity condition.

\subsection{Classical Co-Events}\label{sec:classical coevents}

We begin by revisiting the co-events that were appropriate for classical stochastic mechanics.
\begin{description}
  \item[Rule:]
   We require our co-events to be homomorphisms. Starting with a histories theory $(\Omega,\EA,\mu)$, where $\mu$ is not necessarily classical, for $\phi\in \Pr$ and $A,~B\in\EA$ we impose:
    \begin{eqnarray}
    \phi(A+ B) &=& \phi(A) + \phi(B), \ \ (linearity)\label{eq:linearity} \\
    \phi(AB) &=& \phi(A)\phi(B). \ \
    (multiplicativity)\label{eq:multiplicativity}
    \end{eqnarray}
  Further we rule out the zero map (as we will in every scheme\footnote{We could rule out the zero map in our definition of $\cal{P}$, however this would mean that $\cal{P}$ would lose its ideal structure. Apart from its aesthetic value, this algebraic structure will be relevant to one of our co-event schemes (see section \ref{sec:ideal coevents})}).

  \item[Algebraic Formulation:]
  Lemma \ref{lemma:classical coevents are homomorphisms} forces $\p=\g^*$ for some $\g\in\O$, conversely every $\g^*$ obeys equations \ref{eq:linearity}, \ref{eq:multiplicativity}. Thus a co-event obeys the rule of the classical scheme (and is a homomorphism) if and only if it is atomic.

  \item[Primitivity:]
  Not required; intuitively we could say that `maximal fine graining' is built into the rule.
\end{description}

We denote the subset of $\Pr$ obeying equations \ref{eq:linearity} \& \ref{eq:multiplicativity} by $\C$. It is clear from the arguments of section \ref{sec:coevents in classical stochastic mechanics} that the adoption of the classical co-event scheme is equivalent to the naive interpretation of quantum measure theory. Thus recalling the failure of the naive interpretation in explaining the Peres-Kochen-Specker system in chapter \ref{chapter:kochen-specker}, we can reword lemma \ref{lemma:failure of the naive interpretation} in terms of classical co-events.

\begin{lemma}\label{lemma:failure of classical scheme}
Let $\mu$ be a measure on the space $\O$ of colourings of $PS$ that is zero valued on the PKS sets. Then $\C=\emptyset$
\end{lemma}
\begin{proof}
By Theorem \ref{thm:PKS}, every element of $\Omega$ is inconsistent on at least one basis or pair and therefore lies in at least one PKS set. Consider  $\gamma^*$, a classical co-event. $\gamma$ lies in at least one PKS set and therefore $\gamma^*$ is not preclusive, therefore $\C$ is empty.
\end{proof}

Although this means that the classical scheme cannot explain all quantum systems it is nevertheless a useful concept to keep in mind, for we will require the other co-event schemes to produce classical co-events given a classical measure.

\subsection{Linear Co-Events}\label{sec:linear coevents}

The linear scheme was historically the first attempt to generalise from classical co-events \cite{Sorkin:2006wq}. The idea is to relax the rules governing classical co-events, dropping multiplicativity (\ref{eq:multiplicativity}) whilst retaining linearity (\ref{eq:linearity}). We will discuss the linear scheme in greater depth in chapter \ref{chapter:linear scheme}, exploring and justifying some of the statements we make below.

\begin{description}
  \item[Rule:]
  \begin{equation}
  \phi(A+ B) = \phi(A) + \phi(B), \ \ (linearity)\label{liearity1}
  \end{equation}
  As before we rule out the zero map. Co-events obeying the linear rule are referred to as \emph{linear co-events}.

  \item[Algebraic Formulation:]
  Linear co-events are linear polynomials in the classical co-events $\g^*$, and a general linear co-event is of the form:
  \begin{equation}
  \p = \sum_{\gamma_a\in\Omega}n_a\gamma_a^*,
  \end{equation}
  for $n_a\in\mathbb{Z}_2$.

  \item[Primitivity:]
  Our rule leaves us with `too many' co-events and we impose a primitivity condition to pick out the `fundamental' or `primitive' ones. Given a linear co-event $\phi$, we say it is `{\it dominated}' by the linear co-event $\psi$ if:
  \begin{equation}\label{eq:linear domination}
  \psi(\{\gamma\})=1\Rightarrow\phi(\{\gamma\})=1 ~ \forall~\g\in\O.
  \end{equation}
  A preclusive linear co-event is primitive if it is not dominated by any other preclusive linear co-event.
\end{description}

We denote the set of primitive preclusive linear co-events $\L$.

\subsection{Quadratic Co-events}\label{sec:quadratic coevents}

We will discuss the quadratic scheme in greater depth in chapter \ref{chapter:polynomial}, exploring and justifying some of the statements we make below.

\begin{description}
  \item[Rule:]
  The quadratic scheme \cite{Sorkin:private,CoEventSchemes} generalises the linear scheme by altering the rule from linearity to a quadratic law in line with the properties of the quantum measure (equation \ref{eq:quantum measure sum rule}).
  More specifically, for $\p\in \Pr$ and $A,~B,~C\in\EA$ the quadratic rule is:
  \begin{equation}\label{eq:quadratic rule}
  \phi(A+ B+ C) = \phi(A+ B) + \phi(B+ C) +
  \phi(A+ C) + \phi(A) + \phi(B) + \phi(C).
  \end{equation}
  As before we rule out the zero map. Notice that we use addition where in the sum rule of the quantum measure (equation \ref{eq:quantum measure sum rule}) we use subtraction, this is because $+1$ and $-1$ are identified in $\Z2$.

  \item[Algebraic Formulation:]
  Quadratic co-events are algebraically quadratic or second order polynomials in the classical co-events:
  \begin{eqnarray}
  \p &=& \sum_{\substack{\gamma_a,\gamma_b\in\Omega \\ a\neq b}}n_{ab}\gamma_a^*\gamma_b^* + \sum_{\gamma_c\in\Omega}n_c\gamma_c^*, \nonumber\\
  &=& \sum_{\gamma_a,\gamma_b\in\Omega}n_{ab}\gamma_a^*\gamma_b^*,
  \end{eqnarray}
  where $n_{ab},n_c\in\Z2$.

  \item[Primitivity:]
  The quadratic scheme uses a primitivity condition based upon that of the linear scheme, though in this case the definition is more involved. First, every primitive linear co-event is admitted as a primitive quadratic co-event. Then if $\p$ is a quadratic, non-linear co-event it is dominated by another quadratic non-linear co-event $\psi$ if both the following conditions are met:
  \bea
  \psi(\{\g\})=1 &\Rightarrow& \p(\{\g\})=1, \nonumber \\
  \psi(\{\g,\overline{\g}\})+\psi(\{\g\})+\psi(\{\overline{\g}\})=1 &\Rightarrow& \p(\{\g,\overline{\g}\})+\p(\{\g\})+\p(\{\overline{\g}\})=1, \nonumber \\
  && \label{eq:quadratic domination by quadratic coevents}
  \eea
  for all $\g,\overline{\g}\in\O$. $\p$ is dominated by a linear co-event $\psi$ if either both conditions in equation \ref{eq:quadratic domination by quadratic coevents} are met or if
  \beq\label{eq:quadratic domination by linear coevents}
  \psi(A)=1 \Leftarrow \p(A)=1,
  \eeq
  for all $A\in\EA$. A preclusive, non-linear, quadratic co-event is primitive if it is not dominated by a preclusive quadratic co-event, linear or non-linear.
\end{description}

We denote the set of primitive preclusive quadratic co-events by $\Q$.

\subsection{General n-Polynomial Co-Events}\label{sec:order n coevents}

We will discuss the general $n$-polynomial scheme \cite{Sorkin:private,CoEventSchemes} in greater depth in chapter \ref{chapter:polynomial}, exploring and justifying some of the statements we make below.

\begin{description}

  \item[Rule:]
  The linear and quadratic schemes could be defined in terms of an `addition rule', and we can generalise the $n$-polynomial schemes along these lines. The linear scheme employs the same `additive' rule as the classical scheme, whereas the quadratic scheme generalises this in line with the sum rule obeyed by the quantum measure. By considering linear co-events as $1$-polynomial co-events, and quadratic co-events as order $2$-polynomial co-events, it is possible to generalise to $n$-polynomial co-events, following the sum rules obeyed by the hierarchy measures introduced in equation \ref{eq:order n interference term} to yield co-events that are cubic, quadratic, quintic and so forth. The sum rule for an $n$-polynomial co-event is:
  \beq\label{eq:order n additive rule}
  \p(\sum_{i=1}^{n+1} A_{i})=\sum_{m=1}^n \sum_{i_j<i_{j+1}} \p(\sum_{j=1}^m A_{i_j}).
  \eeq
  In other words:
  \bea
  \p(A_1 + A_2 + \ldots + A_{n+1}) &=& \p(A_2 + A_3 + \ldots + A_{n+1}) + \p(A_1 + A_3 + \ldots + A_{n+1}) \nonumber \\ && + \ldots + \p(A_1+A_2+\ldots + A_n) + \nonumber \\
  &&  \vdots \nonumber \\
  && + \p(A_1+A_2) + \p(A_1+A_3) + \ldots + \p(A_n+ A_{n+1}) \nonumber \\
  && + \p(A_1)+ \p(A_2) + \ldots + \p(A_{n+1}). \nonumber
  \eea

  \item[Algebraic Formulation:]
  Every co-event can be expressed as a polynomial in the classical co-events (lemma \ref{lemma:classical coevents generate all coevents}), and the linear and quadratic schemes are examples of $n$-polynomial co-events, schemes in which the rule simply restricts the degree of the polynomial to some integer $n$. Given a histories theory $\H$ it seems natural to generalise to the $n$th order scheme, $\Pn{n}$, in which co-events are restricted to be polynomials of degree at most $n$ in the classical co-events, with the zero map ruled out as before. Thus $\Pn{1}=\L$ and $\Pn{2}=\Q$.

  \item[Primitivity:]
  We can generalise the notion of primitivity used for the quadratic scheme, defining the $n$th order scheme iteratively. We must decide which of the co-events with polynomials of degree $n$ or less are primitive. If the degree $deg(\p)$ of the polynomial of a co-event $\p$ is less than $n$, we say that it is primitive if and only if the co-event is primitive in the lower order schemes, so that

  $$deg(\p)<n \Rightarrow  [\p\in\Pn{n-1} \Leftrightarrow \p\in \Pn{n}].$$

  The problem is then reduced to deciding which of the degree $n$ co-events are primitive. To this end we define

  \bea
  S^{\p}_1(\g) &=& \p(\{\g\}),  \nonumber \\
  S^{\p}_2(\g_1,\g_2) &=& \p(\{\g_1,\g_2\})+\p(\{\g_1\})+\p(\{\g_2\}),  \nonumber \\
  S^{\p}_3(\g_1,\g_2,\g_3) &=& \p(\{\g_1,\g_2,\g_3\})+\p(\{\g_1,\g_2\})+\p(\{\g_2,\g_3\})+\p(\{\g_1,\g_3\}) \nonumber \\
  && +\p(\{\g_1\})+\p(\{\g_2\})+\p(\{\g_3\}),  \nonumber \\
  &\vdots& \nonumber \\
  S^{\p}_n(\g_1,\ldots,\g_n) &=& \p(\{\g_1,\ldots,\g_n\})+\sum_{i_1,\ldots,i_{n-1}}\p(\{\g_{i_1},\ldots,\g_{i_{n-1}}\}) \nonumber \\
  && + \sum_{i_1,\ldots,i_{n-2}}\p(\{\g_{i_1},\ldots,\g_{i_{n-2}}\}) + \ldots + \sum_i \p(\{\g_i\}). \label{eq:order n primitivity definitions}
  \eea

  We can now say that a degree $n$ co-event $\p$ is dominated by another co-event $\psi$ of degree $n$ or less if
  \beq\label{eq:order n additive domination}
  S^{\psi}_m(\{\g_1,\ldots,\g_m\})=1 \Rightarrow S^{\p}_m(\{\g_1,\ldots,\g_m\})=1 \forall~m\leq n,\g_i\in\O,
  \eeq
  or,
  \beq\label{eq:order n multiplicative domination}
  \psi(A)=1 \Leftarrow \p(A)=1 ~\forall~A\in\EA.
  \eeq
  A preclusive degree $n$ co-event is primitive if it is not dominated by any other preclusive co-event. Thus $\Pn{n}$ is the space of primitive degree $n$ co-events union $\Pn{n-1}$.

\end{description}

\subsection{Multiplicative Co-Events}\label{sec:multiplicative scheme}

At the time of writing, the multiplicative scheme \cite{Sorkin:2007,Dowker:2007zz,CoEventSchemes} is the working model for the co-event interpretation. This scheme takes a different approach to the `additive' polynomial schemes we have discussed so far, which are based upon modified sum rules. Going back to the classical scheme, we generalise by abandoning the linear sum rule entirely (\ref{eq:linearity}), instead retaining the multiplicative rule (\ref{eq:multiplicativity}), in contrast to the additive approach. We will discuss the multiplicative scheme in greater depth in chapter \ref{chapter:the multiplicative scheme}, exploring and justifying some of the statements we make below.

\begin{description}
  \item[Rule:]
  Thus the `rule' we impose on $\Pr$ is:
  \begin{equation}
  \phi(AB) = \phi(A)\phi(B) \ \ (multiplicativity)
  \end{equation}
  Again we rule out the zero map. It is easy to see that the support of $\phi$ (the inverse image of $1$ in $\EA$) forms a filter in $\EA$, an issue we shall explore in greater depth in chapter \ref{chapter:the multiplicative scheme}.

  \item[Algebraic Formulation:]
  Algebraically the multiplicative co-events are monomials:
  \beq
  \phi = \prod_{\gamma_a\in A}\gamma_a^*,
  \eeq
  where $A\in\EA$. In this case we refer to $\phi$ as $A^*$, which can be seen as generalising our existing notation since we have:
  \begin{eqnarray}
  \{\gamma\}^* &=& \gamma^*, \nonumber \\
  A^*(B) &=& \left\{\begin{array}{cc} 1  & A\subset B \\ 0 &
  A\not\subset B. \end{array}\right. \label{eq:multiplicative coevent base}
  \end{eqnarray}
  A multiplicative co-event that can be written in the form $\p=A^*$ for some $A\in\EA$ is said to be \textit{dual} to $A$.

  Comparing equations \ref{eq:multiplicative coevent base} \& \ref{eq:r* action} we see that multiplicative co-events generalise classical co-events by a `smearing out' or an `ontological coarse graining'. Our ontology constitutes a plurality of histories rather than a single history, and only the properties common to all of these histories are true (see section \ref{sec:the interpretation of the interpretation}).

  \item[Primitivity:]
  In this case we use a different notion of primitivity, saying that a multiplicative co-event $\phi$ is dominated by the multiplicative co-event $\psi$ if:
  \begin{equation}
  \psi(A)=1\Leftarrow\phi(A)=1.\label{multiplicative domination}
  \end{equation}
  A preclusive multiplicative co-event is primitive if it is not dominated by any other preclusive multiplicative co-event.
\end{description}
We denote the set of primitive preclusive multiplicative co-events ${\cal{M}}(\Omega,\EA,\mu)$.

\subsection{Ideal Based Co-Events}\label{sec:ideal coevents}

The ideal based scheme is one of the two schemes (the other being the multiplicative scheme) that have `passed' all the tests they have been subjected to and thus remain as `viable' candidates for defining the co-event interpretation. However the multiplicative scheme is generally used as the working model for various reasons. Firstly the ideal based scheme is computationally intractable, it is difficult to calculate the ideal based co-events for all but the smallest of sample spaces. Further, the multiplicative scheme boasts several advantages; it has a simple interpretation (section \ref{sec:the interpretation of the interpretation}), we have more positive results concerning its consistency (section \ref{sec:classical partitions}) and we have a better idea of how we might generalise it to histories theories with infinite sample spaces (section \ref{sec:infinite multiplicative scheme}). For these reasons little work has been done on the ideal based scheme. We define it here for completeness and because it is one of the two remaining `viable' schemes, however we will not explore it further.

As previously mentioned the ideal based scheme stands out in that it does not use a rule, but imposes its primitivity condition directly onto $\Pr$. Consequently the primitivity condition is more involved than those used by the multiplicative or polynomial schemes. Essentially the idea is to take the ring structure of $\EA^*$ seriously by insisting that the potentially physical co-events be fundamental with respect to this structure. This is implemented by requiring the primitive co-events to be a generating set for the ideal $\Pr$. The issue is complicated by the non-uniqueness of such a generating set, so an algorithm is used to pick out the primitive co-events.

We start by `grading' or ordering $\Pr$, which we can think of as a graded ideal. This is because each of its elements can be thought of as a polynomial over $\Z2$ in the classical co-events, and we can grade each polynomial by its degree. Within each tier, we can grade the ideal further, since every polynomial is a (unique) sum of monomials we can define the `order' of a polynomial by the number of its constituent monomial terms. For example, the co-event $\g_1^*+\g_2^*+\g_1^*\g_3^*$ is a sum of the three monomials $\g_1^*,~\g_2^*,~\g_1^*\g_3^*$ and thus has order three, and degree two. For any subset $A$ of $\Pr$, we will use $[A]^m_n$ to denote the degree $m$, order $n$  elements of $A$.

The algorithm then passes through $\Pr$ tier by tier, picking out co-events. We begin by `selecting' all degree $1$, order $1$ elements of $\Pr$, call this set ${\cal{I}}_1^1$. We then factor out this set, and pass onto the degree $1$ order $2$ elements and so on until we have exhausted the degree $1$ elements and can proceed onto the degree two co-events, and so on. Using the notation $P^i_j$ to refer to the degree $i$ order $j$ elements of a set $P$ of polynomials we have:
\bea
{\cal{I}}_1^1 &=& [\Pr]_1^1, \nonumber \\
{\cal{I}}_2^1 &=& [\Pr / {\cal{I}}_1^1]_2^1, \nonumber \\
&\vdots& \nonumber \\
{\cal{I}}_i^1 &=& [\Pr / {\cal{I}}_{i-1}^1]_i^1, \nonumber \\
&\vdots& \nonumber \\
{\cal{I}}_1^2 &=& [\Pr / {\cal{I}}_m^1]_1^2, \nonumber \\
&\vdots& \nonumber \\
\I &=& {\cal{I}}_1^N,\label{eq:ideal scheme algorithm}
\eea
where $m$ is the maximum order of polynomials of degree $1$, and $N$ is the cardinality of the sample space. Note that though $\I$ is a generating set, it is not necessarily a minimal generating set since there may be elements of the same order and degree that are not independent.

\section{Evaluating the Co-Event Schemes}\label{sec:the criteria}

Given the proliferation of co-event schemes it becomes important to establish a clear set of criteria by which the success of these schemes can be judged. It is, of course, easy to propose arbitrary schemes; the challenge lies in finding meaningful and useful schemes that do not contradict known results. We will at this point put forward criteria that have been used to evaluate the success of the co-event schemes to date. These criteria should not be regarded as a comprehensive list, indeed there may be specific factors particular to some of our schemes that make them more or less appealing. Rather these points should be seen as a set of basic requirements; a failure to satisfactorily treat any of these points would effectively rule out a co-event scheme as an aspiring interpretation of quantum measure theory.

\begin{itemize}
    \item \textbf{Existence:} Given a physically realisable histories theory, our co-event scheme must be non-empty, otherwise we have no description for the system.
    \item \textbf{Compatibility} with Copenhagen quantum mechanics: For every experimentally measurable outcome that is not of probability 1 or 0 our co-event scheme should contain a co-event that maps the subspace corresponding to that outcome to 1, and a co-event that maps it to 0.
    \item \textbf{Emergent Classicality:} We require that both the following conditions are met:
                                          \begin{itemize}
                                            \item \textbf{Weak Emergent Classicality:} Our co-event scheme should agree with the classical co-event scheme whenever the measure is classical.
                                            \item \textbf{Strong Emergent Classicality:} Our co-event scheme should agree with the classical co-event scheme on partitions that we deem to be `classical'. As research into strong emergent classicality is still a work in progress the term `classical partition' is at present somewhat ill-defined, however it should at the very least include `observable coarse grainings', so that we only observe classical logic (see section \ref{sec:classical partitions} for further discussion).
                                          \end{itemize}
    \item \textbf{Consistency:} We demand consistency under coarse graining so we can intelligently talk about sub-systems (and do not have to consider the measure of the whole universe every time we wish to describe a system).
\end{itemize}

In this language we would say that the classical co-event scheme (aka the naive interpretation) fails on the existence criterion, as demonstrated by the Peres-Kochen-Specker system (lemma \ref{lemma:failure of classical scheme}). Because of this, in what follows we will use the PKS system as a testing ground for our co-event schemes, satisfying ourselves that they can succeed where the classical scheme failed.

\section{The Interpretation of the Interpretation}\label{sec:the interpretation of the interpretation}

Before we embark upon the exploration of each co-event scheme we will pause for a moment to reflect on the `meaning' of the co-event approach itself. From the outset there have been two interpretations, one emphasising the events and the other emphasising the dual co-events.

Consider the classical scheme with $\g^*$ as the `correct' co-event, by which we mean `the co-event that describes what actually happens' \cite{Sorkin:2007}. In the first interpretation we say that the history $\g$ is actually real, as we would in the naive interpretation, and that the space of potential realities is $\O$ (more accurately the subset thereof corresponding to histories that are not an element of any null set). In the second interpretation we say that the co-event $\g^*$ itself is actually real, while the space of potential realities is given by $\C$. Generalising to an arbitrary scheme $\cal{S}$, the second interpretation would consider a single co-event $\p\in\S$ as actually real, while the space of potential realities is given by $\S$. Returning to the first interpretation we would, loosely, think of reality as a single `quantum history' that obeys a non-standard logic described by the correct co-event $\p$. To quote Sorkin \cite{Sorkin:private} (who on this occasion used the term `trajectory' or `gamma' to refer to mean a history, and `Omega' to mean the sample space) these two interpretations can be summarised as follows:
\begin{enumerate}
  \item Trajectories are real, more precisely reality is described by a single``quantal trajectory'' gamma.  An answer to a question gives information on gamma.  The complete set of answers -- a co-event -- fully characterizes gamma.  However, the answers obey a nonclassical logic. (gamma can possess ``contradictory attributes'')
  \item The coevent itself is the reality, taking over the role that an element of Omega plays in classical logic/physics.
\end{enumerate}

There are two key points to consider in understanding these interpretations; firstly the question of which mathematical object in the theory represents reality, and secondly the nature of the logic we use in describing this reality. As mentioned above the second interpretation identifies $\S$ as the space of potential realities, and the correct co-event $\p\in\S$ as the actual reality. The logic we use to reason about the potential realities is Boolean, a feature we shall revisit shortly. In the first interpretation the space of potential realities is $\O$, and we can say that the actual reality is a single history in $\O$. However the logic we use to reason about the potential realities is in general non-Boolean, and is described by the correct co-event $\p$. Thus we might for example assert that none of the individual histories $\g\in\O$ is the actually real history, yet still claim that the actually real history is an element of $\O$; thus circumventing theorem \ref{thm:PKS} and in a way regaining the naive interpretation by abandoning standard logic. Note that our space of potential realities is now all of $\O$ rather than the subset thereof corresponding to the non-precluded histories.

In both interpretations we use anhomomorphic logic in our reasoning concerning propositions in the event algebra. In contrast, both interpretations use `ordinary' boolean logic in reasoning concerning the co-events themselves. This `two-tiered' use of logic and meta-logic is common to theories of quantum logic \cite{Birkhoff:1936,Mackey:1957,Chiara:2001}, or the more recent work of D\"{o}ring and Isham \cite{Doring:2007ib,Doring:2007ic,Doring:2007id,Doring:2007ie,Doring:2008gv}. In fact, the mathematical treatment of logic tends to employ `ordinary language' as the meta-language in which the logical system under consideration can be defined and discussed \cite{Machover:1996}; we typically assume that this language corresponds to a Boolean meta-logic.

In co-event theory the use of a Boolean meta-logic is explicit, for our analysis of co-events utilises not only `ordinary language' but also `standard' mathematics, defined through Zermelo set theory \cite{Machover:1996} and its underlying Boolean logic. Indeed a theory, such as the first interpretation, that fundamentally rejects Boolean logic would lead to a reformulation of mathematics. Further, unless the theory claimed that its own proposals, rather than Boolean logic, actually correspond to our `ordinary' meta-logic, the thought processes leading up to the adoption of this theory would be called into question as well. For these reasons, the author is inclined to the use of Boolean logic in the fundamental description of reality, and so prefers the second interpretation to the first.

While the second interpretation also applies anhomomorphic logic to the event algebra, we are no longer regarding histories as the potential realities, or beables, of our theory. In a sense we could think of $\S$ as the `true sample space' of our theory, and thus while we use standard (Boolean) logic to describe co-events we encounter `strange' or anhomomorphic answers when dealing with $\EA$ because we are `asking the wrong questions'. Because the histories sample space $\O$ is not the `true' sample space, its elements are not potential realities but theoretical constructs, and questions concerning them may not be well-posed statements concerning the true potential realities of our theory (the co-events). Thus there is no reason to expect `meaningful' answers to questions that are not well-posed.

However, in criticism of the second interpretation, if we were to a priori believe that the counterintuitive nature of quantum measure theory was the result of our use of the `wrong' sample space, leading to our propositions not being well posed, we would not necessarily look to find the `true' sample space in sets of anhomomorphic truth valuation maps.

The multiplicative scheme raises a third interpretation

\begin{enumerate}
\setcounter{enumi}{2}
\item If $A^*$ is the `correct' co-event then $A\in\EA$ is the actual reality
\end{enumerate}

This is well defined for as we shall see in section \ref{sec:mult filters} every multiplicative co-event can be expressed as $A^*$ for some $A\in\EA$. In this way our `actual reality' is no longer a single history, but a plurality of histories, and only properties common to all of these histories are true. This `ontological coarse graining' fits well with the spirit of the histories approach, in which coarse graining has a central role. As with the second interpretation above, we can claim that the `true' sample space, which should correspond to the space of `potential realities', is not $\O$ but the subset of $\EA$ corresponding to the multiplicative co-events. As before we use homomorphic logic in our reasoning concerning the potential realities, and `blame' the appearance of anhomomorphic logic on the `mistake' made in believing the `true' sample space to be $\O$.

All three interpretations are `mathematically equivalent', and all of our results will apply equally to all of them. However the interpretation colours our understanding and evaluation of these results, and under different interpretations different directions of research may seem more or less important. In what follows we will tend to sideline the first interpretation, which we shall refer to as the `anhomomorphic logic interpretation', in favor of the second, `real co-events' and third `ontological coarse graining' interpretations. For co-event schemes other than the multiplicative scheme we will use the second interpretation, and for the multiplication scheme will consider both the second and third interpretations. This shift in thinking from the sample space to our new beables suggests that we should be able to reformulate the entirety of quantum measure theory in terms of these beables, or potential realities, and in particular that we should be able to express the dynamics in terms of the co-events. We will return to this subject in chapter \ref{chapter:dynamics}.

%Introducing Co-Events
%\singlespacing
\chapter{The Linear Scheme}\label{chapter:linear scheme}
%\doublespacing

\section{Basic Properties}

To recap on section \ref{sec:linear coevents} the linear scheme generalises from the classical scheme by dropping the multiplicativity rule (equation \ref{eq:multiplicativity}) while keeping linearity (equation \ref{eq:linearity}):
$$\p(A+B)=\p(A)+\p(B).$$
A linear co-event $\phi$ is dominated by a linear co-event $\psi$ if:
$$\psi(\{\gamma\})=1\Rightarrow\phi(\{\gamma\})=1.$$
A preclusive linear co-event is primitive if it is not dominated by any other preclusive linear co-event. The set of primitive, preclusive, linear co-events is called ${\cal{L}}(\Omega,\EA,\mu)$.

The linear scheme was historically the first attempt to generalise from classical co-events \cite{Sorkin:2006wq}. At the time it was felt that one of the two conditions (`linearity' \& `multiplicativity', see equations \ref{eq:linearity} \& \ref{eq:multiplicativity}) defining classical co-events as homomorphisms should be dropped. Multiplicativity was seen as a more `classical' condition, in which every superset of a `true' event (an event mapped to one by the co-event representing the system) would have to be true, whereas linearity seemed to capture more of the essence of quantum interference, allowing a `true' event to be a subset of a `false' one \cite{Sorkin:private}.

The linear scheme is not well defined in the infinite case, principally because our notion of primitivity would not make sense were the sample space infinite. To this end, unless specifically noted otherwise, in this chapter we will assume that all the co-events we are using are linear and that all sample spaces (and thus all event algebras) are finite.

In this chapter we will `get our hands dirty' with the technicalities of the linear scheme; we explore the scheme's basic properties, examine its application to several simple systems and prove that it can succeed where the naive interpretation failed in describing the PKS system. However, we then find fault with the scheme, and reject it on grounds of inconsistency and contradiction with the predictions of quantum mechanics. This chapter draws from \cite{Sorkin:2006wq,Sorkin:2007,CoEventSchemes}.

\subsection{The Linear Dual}

Given a histories theory $\H$ it is useful to talk about all the linear co-events in $\EA^*$ without restricting ourselves to $\L$. To this end Sorkin has introduced the notation:

\begin{definition}
Let $\H$ be a histories theory with a finite sample space. Then we define:
$$\Homp = \{\p\in\EA^*|\p(A+B)=\p(A)+\p(B)\}~\forall A,B\in\EA.$$
\end{definition}

In this section we will introduce a `duality' between the event algebra $\EA$ and the space of linear co-events $\Homp$.

\begin{definition}\label{def:linear dual one}
Let $\H$ be a histories theory with a finite sample space. Then for every $\p\in\Homp$ we define the \textbf{linear dual} $\pl$ as follows:
\bea
\pl &=& \{\g\in\O | \p(\{\g\}) = 1\}, \nonumber \\
&=& \sum_{\p(\{\g\})=1}\{\g\} . \nonumber
\eea
We can use this to define a map
\bea
\oplus : \Homp &\rightarrow& \EA, \nonumber \\
\oplus : \p &\mapsto& \pl, \nonumber
\eea
which we will also call the linear dual.
\end{definition}
It is easy to see that $\Homp$ is a commutative group under addition, similarly since $\EA$ is a ring it is also a commutative group under addition. We can show that the linear dual is in fact an isomorphism between these two groups.

\begin{lemma}
Let $\H$ be a histories theory with a finite sample space. Then considering $\EA$ and $\Homp$ as groups under addition the map
$$\oplus : \Homp \rightarrow \EA,$$
is an isomorphism.
\end{lemma}
\begin{proof}
We need to show that $\oplus$ is injective, surjective and homomorphic.
\begin{description}
  \item[Injective:]
  We can describe the action of any $\p\in\Homp$ on $\EA$ in terms of its linear dual:
  \bea
  \p(A) &=& \p(\sum_{\g\in A}\{\g\}), \nonumber \\
  &=& \sum_{\g\in A} \p(\{\g\}), \nonumber \\
  &=& |(\pl\cap A)|~mod~2, \label{eq:linear coevent evaluation}
  \eea
  where $|A|$ denotes the cardinality of the set $A$. Thus $\psi^{\oplus}=\pl\Leftrightarrow \psi=\p$, and the linear dual is injective.

  \item[Surjective:]
  We prove surjectivity by constructing the inverse map. First define:
  \bea
  \oplus_*:\EA&\rightarrow&\CE, \nonumber \\
  \oplus_*(A) &=& \sum_{\g\in\O} \g^*(A)\g^*. \label{eq:linear dual inverse definition}
  \eea
  We can compute the action of $\oplus_*(A)$ on $\EA$:
  \bea
  [\oplus_*(A)](B) &=& \sum_{\g\in\O} \g^*(A)\g^*(B), \nonumber \\
  &=& \sum_{\g\in A\cap B}, \nonumber \\
  &=& |A\cap B|~mod~2. \nonumber
  \eea
  Contrasting this with equation \ref{eq:linear coevent evaluation} we see that
  $$\oplus_*(\pl) = \p.$$
  Thus $\oplus_*$ restricted to the range of $\oplus$ is the inverse map $\oplus^{-1}$. Further, we can show that $\oplus_*(A)$ is always linear
  \bea
  [\oplus_*(A)](B)+ [\oplus_*(A)](C) &=& |A\cap B|~mod~2 + |A\cap C|~mod~2, \nonumber \\
  &=& |A\cap (B\cup C)|~mod~2 + |A\cap (B\cap C)|~mod~2, \nonumber \\
  &=& |A\cap (B + C)|~mod~2 + (2*|A\cap (B\cap C)|)~mod~2, \nonumber \\
  &=& [\oplus_*(A)](B+C),
  \eea
  so the range of $\oplus_*$ is a subset of $\Homp$. But then if $\p\in\Homp$, $\p\not\in range(\oplus_*)$ then $\pl\in\EA$ and $\oplus_*(\pl)=\p$, contradicting $\p\not\in range(\oplus_*)$. Therefore $\oplus_*$ maps onto $\Homp$.

  Similarly,
  \bea
  [\oplus_*(A)](\{\overline{\g}\}) &=& \sum_{\g\in\O}\g^*(A)\g^*(\overline{\g}), \nonumber \\
  &=& \left\{\ba{cc} 1 & \overline{\g}\in A \\ 0 & \overline{\g} \not\in A \ea\right. \nonumber
  \eea
  thus
  $$[\oplus_*(A)]^{\oplus} = A.$$
  So if we now take $A\in\EA$, $A\not\in range(\oplus)$ we will find a contradiction as before, thus $range(\oplus)=\EA$, and $\oplus$ is surjective with inverse $\oplus_*$.

  \item[Homomorphic:]
  \bea
  [\oplus_*(A)+\oplus_*(B)](C) &=& [\oplus_*(A)](C)+[\oplus_*(B)](C), \nonumber \\
  &=& |A\cap C|~mod~2 + |B\cap C|~mod~2, \nonumber \\
  &=& |(A+B)\cap C|~mod~2 + (2*|A\cap (B\cap C)|)~mod~2, \nonumber \\
  &=& [\oplus_*(A+B)](C).
  \eea

\end{description}

\end{proof}

We have an immediate corollary.

\begin{corollary}\label{corollary:linear co-events are linear polynomials}
Let $\H$ be a histories theory with a finite sample space. Then every $\p\in\Homp$ can be expressed as a linear polynomial in the classical co-events:
$$\p = \sum_{\g\in\pl}\g^*.$$
\end{corollary}
\begin{proof}
Let $\p\in\Homp$. Then $\pl\in\EA$ and
$$\pl = \sum_{\g\in\pl}\{\g\}.$$
Then acting on this with $\oplus_*$ and noting that $\oplus_*(\{\g\})=\g^*$ we get
$$\p = \sum_{\g\in\pl}\g^*.$$
\end{proof}

As might be expected, we can express the properties of a linear co-event $\p$ in terms of its linear dual. For example it can be verified that:

\begin{lemma}\label{lemma:linear dual properties}
Let $\H$ be a histories theory with a finite sample space and let $\p,\psi\in\Homp$. Then if $Z\in\EA$ is null we have:
\begin{enumerate}
  \item $\p$ is dominated by $\psi$ if and only if $\psi^{\oplus}\subset\pl$.
  \item If $\p$ is preclusive then $\pl\cap Z$ has even cardinality.
\end{enumerate}
\flushright{$\square$}
\end{lemma}

To make the `linear duality' between $\EA$ and $\Homp$ more clear, we note that $\EA$ can be thought of as a space of maps on $\Homp$, defining
\beq
A[\p] = \p(A),
\eeq
where $A\in\EA$ and $\p\in\Homp$, we see that
\bea
A[\p+\psi] &=& (\p+\psi)(A), \nonumber \\
&=& \p(A) + \psi(A), \nonumber \\
&=& A[\p] + A[\psi],
\eea
for all $A\in\EA$, $\p,\psi\in\Homp$, thus the action of $\EA$ on $\Homp$ is linear. Further, this action respects $\EA$'s additive structure. For $A,B\in\EA$, $\p\in\Homp$ we have
\bea
(A+B)[\p] &=& \p(A+B), \nonumber \\
&=& \p(A) + \p(B), \nonumber \\
&=& A[\p] + B[\p].
\eea
In fact, it can be shown that:
\begin{lemma}
Let $\H$ be a histories theory with a finite sample space. Then
$$\EA = Hom^{\oplus}(\Homp,\Z2).$$
\flushright{$\square$}
\end{lemma}
Because of the symmetry of this structure, we will henceforth denote the `inverse' dual $\oplus_*$ by the same notation we use for $\oplus$. Essentially we extend the domain of the linear dual to make it an involution.
\begin{definition}\label{def:linear dual}
Let $\H$ be a histories they with a finite sample space. Then for $A\in\EA$ we define
\beq
A^\oplus = \oplus_*(A).
\eeq
This allows us to extend the domain of the linear dual:
\bea
\oplus: \EA\cup\Homp &\rightarrow&  \EA\cup\Homp, \nonumber \\
\oplus(\Upsilon) &=& \Upsilon^{\oplus},
\eea
where $\Upsilon\in\EA\cup\Homp$.
\end{definition}
Note that this is consistent with our previous definition of $\oplus$, we have simply extended the domain of the map. Because $\oplus$ is an involution, we now have $(\pl)^{\oplus}=\p$ and $(A^{\oplus})^{\oplus}=A$.

\subsection{Primitivity \& Weak Emergent Classicality}

In this section we will outline the thought process that led to the adoption of the primitivity condition used in the linear scheme (see section \ref{sec:linear coevents}), before demonstrating that this condition does indeed cause the linear scheme to obey weak emergent classicality.

We generalise from $\C$ to $\L$, dropping the multiplicativity condition, in the hope that our new scheme will be able to effectively treat systems, such as the $PKS$ set-up, for which the classical structure appears too restrictive. However by simply relaxing the `rules' we get `too many' co-events, specifically given a histories theory $\H$ with a classical measure $\mu$, $\C$ is a proper subset of $Hom^{\oplus}$ raising the possibility of the `non-classical' behaviour of classical systems, contrary to observation. We must therefore impose some other restriction on $Hom^{\oplus}$ that will force the linear co-events to be classical when the measure is classical. Thus the observation that classical co-events $\g^*$ are in some way `simpler' than non-classical linear co-events such as $\g_1^*+\g_2^*$ leads us to the notion of primitivity.

To define primitivity we need to be more precise about what we mean by `simpler'. The singleton sets $\{\g\}$, which are the duals of the classical co-events, are the `finest grained' elements of $\EA$ in that they are atoms of the algebra, or equivalently they are minimal in the partial order defined by inclusion. However, algebraically atoms are defined using multiplication, which is not defined in $\Homp$. Thus it is not in general possible to identify the classical co-events using only the internal algebraic structure of $\Homp$. However using the linear dual we can utilise the multiplicative (equivalently the set inclusion) structure of $\EA$, which allows us to define a partial order (based on inclusion) in which the duals of the classical co-events are minimal. Then $\g_1^*$ is simpler than $\g_1^*+\g_2^*$ in that the linear dual of the later contains the linear dual of the former, leading us to exclude from $\L$ those co-events whose linear duals contain the linear duals of other preclusive co-events. It is easy to see that this is equivalent to our definition of dominance and primitivity described by equation \ref{eq:linear domination}.

The following shows that our definition of primitivity is indeed successful in ensuring that the linear scheme reduces to classical co-events in the case of a classical measure, in accordance with our weak emergent classicality criterion (section \ref{sec:the criteria}).  The result is an immediate corollary of the following lemma.
\begin{lemma}\label{lemma:linear scheme classical null sets}
If $\H$ is a histories theory with finite sample space and a measure $\mu$ such that every negligible set\footnote{Recall that a negligible set is a subset of a null set} is null, then $\p\in\L$ is a homomorphism.
\end{lemma}
\begin{proof}
Let $\p\in\L$ and let $A\in\EA$ be null. Assuming $\pl \cap A \neq \emptyset$, there exists some $\gamma\in\pl\cap A$ and by the hypothesis $\mu(\{\gamma\})=0$. However $\phi(\{\gamma\}) = |\{\gamma\}|~mod~2 = 1$, which contradicts the preclusivity of $\phi$. Thus $\mu(A)=0 \Rightarrow\pl \cap A = \emptyset$. Thus no $\g\in\pl$ is an element of any null set, thus $\{\gamma \}^\oplus$ is preclusive and dominates $\phi$. Since $\phi$ is by assumption primitive, we must have: $\phi = \{\gamma\}^\oplus = \gamma^*$.
\end{proof}

\begin{corollary}\label{corollary:linear_scheme_emergent_classicality}
Let $\H$ be a histories theory in which $\O$ is finite. Then:
$$\mu(A\sqcup B)=\mu(A)+\mu(B)~\forall ~A,B\subset\Omega~\Rightarrow ~\L=\C.$$
\end{corollary}
\begin{proof}
Follows trivially from lemma \ref{lemma:linear scheme classical null sets}.
\end{proof}

\subsection{Unitality}\label{sec:linear unitality}

Unitality is a critical concept for the linear scheme, and in fact for all the $n$-polynomial schemes. Intuitively a \textit{unital co-event} is one that answers the question `does anything happen?' or `does anything exist?' in the affirmative. More technically,

\begin{definition}\label{def:unitality}
A co-event $\p$ related to a histories theory $\H$ is unital if it obeys
$$\p(\O)=1.$$
We write $\LU$ to denote the co-events that are primitive among the set of preclusive, linear, unital co-events
\end{definition}

Thus the adoption of $\LU$ is tantamount to introducing unitality as an additional `rule' in the linear scheme. The following lemma, due to Rafael Sorkin \cite{Sorkin:private}, classifies the measures that admit unital co-events,

\begin{lemma}\label{lemma:linear null sets prevent unitality}
Let $\H$ be a histories theory with a finite sample space, then $\LU=\emptyset$ if and only if there exist null sets $A_i\in\EA$ such that $\sum_I A_i =\O$.
\end{lemma}
\begin{proof}
First assume there exist null sets $A_i\in\EA$ such that $\sum_I A_i =\O$. Then preclusivity means that $\p(A_i)=0$ for all $i$ and linearity means that $\p(\O)=\p(\sum_I A_i)=\sum_I\p(A_i)=0$.

To prove the converse we assume that $\not\exists$ null sets $A_i\in\EA$ such that $\sum_I A_i =\O$. Since $\EA$ is an algebra over $\Z2$ it can be thought of as a vector space. Thus if we denote the collection of null sets $N_i$ we can pass to a linearly independent subset $N_{i_j}$ and extend this to a basis $N_{i_1},\ldots,N_{i_n},B_1,\ldots,B_m$. By assumption $\sum_j N_{i_j} \neq \O$, thus $m\geq 1$. Now we can view a linear co-event $\p$ as a linear functional on the vector space $\EA$, and uniquely define it by its action on any basis of $\EA$. Now preclusivity forces $\p(N_{i_j})=0$, so if we set $\p(B_i)=\delta_{1i}$ and extend linearly $\p$ will be a preclusive linear co-event such that $\p(\O)=1$.
\end{proof}

Unitality may be desirable simply on `philosophical' grounds, though its importance becomes apparent in the consideration of sub-systems where non-unital co-events fail the criteria we set out in \ref{sec:the criteria}. It is not so much the lack of unitality itself that is the issue, rather the very same structures of null sets that force non-unitality also cause problems for consistency. We can illustrate the problem with the following simple example,

\begin{example}\label{example:linear unitality problem}
Let $\Omega$ be a tensor product sample space $\Omega=\Omega_{1}\times\Omega_{2}$, with the associated measure $\mu=\mu_{1}\mu_{2}$. Now let $\{A_{i}\}$ be null subsets of $\Omega_{1}$ (under $\mu_{1}$), such that $\sum_{i}A_{i}=\Omega_{1}$. Then if $\phi$ is a preclusive linear co-event and $B$ is any subset of $\O_2$ we have:
\begin{eqnarray}
\mu(A_i\times B) &=& \mu_{1}(A)\times\mu_{2}(B), \nonumber \\
\Rightarrow \mu(A_{i}\times B) &=& 0\times\mu_{2}(B), \nonumber \\
&=& 0, \nonumber \\
\Rightarrow \phi(  A_{i}\times B) &=& 0, \nonumber \\
\therefore \phi(\Omega_{1}\times B) &=& \phi(\sum_{i}A_{i}\times B), \nonumber \\
&=& \sum_{i}\phi(A_{i}\times B), \nonumber \\
&=& 0. \nonumber
\end{eqnarray}
\flushright{$\square$}
\end{example}

Essentially this means we would have to consider the full fine grained measure of the whole universe before deriving any co-event, in contrast with our consistency criteria (section \ref{sec:the criteria}), a subject we shall return to in section \ref{sec:linear failure}. Because of this we will henceforth focus on $\LU$ rather than $\L$.

\section{Simple Examples}\label{sec:linear double and triple slit}

We will now work through three simple examples, familiarising ourselves with the workings of the linear scheme in this way.

\subsection{The Linear Coin}\label{sec:linear coin}

Consider the simple system outlined in examples \ref{example:coin for classical dynamics coevents} \& \ref{example:coin for general coevents}. We considered a single toss of a coin as described by classical stochastic dynamics, assuming that we would have either a `heads' (`h') or a `tails' (`t') outcome. Our sample space is given by:
\beq
\O=\{h,t\}.
\eeq
Our event algebra is simply the power set of $\O$,
\beq
\EA = P\O = \{\emptyset,\{h\},\{t\},\{h,t\}=\O\},
\eeq
and our classical measure is $\P$:
\bea
\P(\emptyset) &=& 0, \nonumber \\
\P(\{h\}) &=& p, \nonumber \\
\P(\{t\}) &=& 1-p, \nonumber \\
\P(\{h,t\}) &=& 1.
\eea
So that the space of co-events is
\beq
\EA^* = \{\textbf{0},h^*,t^*,h^*+t^*,h^*t^*,h^*t^*+h^*,h^*t^*+t^*,h^*t^*+h^*+t^*\},
\eeq
where $\textbf{0}$ is the zero map. It is easy to see that the linear co-events are:
$$h^*,t^*,h^*+t^*.$$
We assume $p\not\in\{0,1\}$ so all three linear co-events are preclusive. Then to determine primitivity we can examine the action of these co-events on the singleton sets:
$$
\begin{array}{c|cc}
  & \{h\} & \{t\}\\
\hline
h^* & 1 & 0 \\
t^* & 0 & 1 \\
h^* + t^* & 1 & 1 \\
\end{array}
$$
Thus $h^*$ and $t^*$ are both primitive, and both dominate $h^*+t^*$ which is therefore not primitive. Thus we have:
\bea
\L &=& \{h^*,t^*\}, \nonumber \\
&=& \C,
\eea
as we would expect from corollary \ref{corollary:linear_scheme_emergent_classicality}.

\subsection{The Double Slit System}\label{sec:linear double slit}

Recall the histories description of this system we outlined in section \ref{sec:double slit}. Our fine grained histories are spacetime paths that can pass through one of slits $A$ and $B$ before ending either at the detector $D$ or elsewhere $\overline{D}$. Our sample space is $\O=\{AD,BD,A\overline{D},B\overline{D},\}$, and our event algebra is $\EA=P\O$. We have two natural decoherent partitions, $A,~B$ where $A=\{AD,A\overline{D}\},~B=\{BD,B\overline{D}\}$ and $D,~\overline{D}$ where  $D=\{AD,BD\},~\overline{D}=\{A\overline{D},B\overline{D}\}$. The decoherence functional is
\beq
{\textbf{D}}=\begin{pmatrix}
1/4 & -1/4 & 0 & 0 \\
-1/4 & 1/4 & 0 & 0 \\
0 & 0 & 1/4 & 1/4 \\
0 & 0 & 1/4 & 1/4
\end{pmatrix},
\eeq
in the `basis' $\{AD,BD,A\overline{D},B\overline{D}\}$ so that, for example, ${\textbf{D}}(B\overline{D},B\overline{D})=1/4$. The only null set in $\EA$ is $D$.

Now the histories $A\overline{D},B\overline{D}$ do not participate in any null set, thus the classical co-events $A\overline{D}^*,B\overline{D}^*$ are preclusive. Since classical co-events are always linear and can not be dominated by any other co-event, they are members of $\L$ whenever they are preclusive. The histories $AD,BD$ are elements of the null set $D$, so the corresponding classical co-events are not preclusive. The linear co-event $D^{\oplus}=AD^*+BD^*$ is linear and preclusive, however it is not unital. Any linear co-event whose dual contains either $A\overline{D}$ or $B\overline{D}$ will be dominated by one of the two corresponding classical co-events, and thus will not be primitive. This exhausts $\EA$, so we can conclude that:
\bea
\L &=& \{ A\overline{D}^{\oplus},B\overline{D}^{\oplus},D^{\oplus}\}, \nonumber \\
\LU &=& \{ A\overline{D}^{\oplus},B\overline{D}^{\oplus}\}, \nonumber \\
&=& \C.
\eea
Thus in the linear scheme the double slit system is classical, a surprising result for one of the seminal experiments of quantum mechanics and a system that is often regarded as epitomising `quantum behaviour'.

\subsection{The Triple Slit System}\label{sec:triple slit}

The (idealised) triple slit system is a straightforward generalisation of the double slit, simply substituting three slits $A,B,C$ for the two slits $A,B$ we had previously. As before we denote the initial state $\ket{\psi}$, and the projector corresponding to finding the particle at slit $A$ upon measurement by $P_A$, with the associated state $\ket{A}$ such that $P_A=\ket{A}\bra{A}$. We define the projectors $P_B,P_C,P_D$ and their associated states $\ket{B},\ket{C},\ket{D}$ in a similar fashion; note that the vectors corresponding to the three slits are mutually orthogonal. Finally we can define the projector corresponding to $\overline{D}$ as $P_{\overline{D}} = \mathbb{I}-P_D$.

As in the double slit system we define our minimal Hilbert space theory $({\cal{H}},H,\ket{\psi},T)$ by setting ${\cal{H}} = span(\ket{A},\ket{B},\ket{C})$, setting $H$ to be the null operator, so our evolution operators are the identity, and setting our temporal support to be $T = \{0,1,2\}$, consisting of an initial time, an intermediate time at which our projectors $P_A,P_B,P_C$ act, and a final time at which our projectors $P_D,P_{\overline{D}}$ act.

Since our Hilbert space is the span of the detector states, we must define our initial state $\ket{\psi}$ and the detector state $\ket{D}$ in terms of the $\ket{a_i}$. As discussed in appendix \ref{appendix:many slit}, we can construct a gedankenexperiment in which $\ket{\psi}=\frac{1}{\sqrt{3}}(\ket{A}+\ket{B}+\ket{C})$ and $\ket{D} = \frac{1}{\sqrt{3}}(\ket{A}-\ket{B}+\ket{C})$. Note that this gives us $\bra{D}(P_A+P_B)\ket{\psi}=\bra{D}(P_B+P_C)\ket{\psi}=0$.

Our sample space is now $\O=\{AD, BD, CD, A\overline{D}, B\overline{D}, C\overline{D}\}$ and our event algebra $\EA=P\O$ as before. We can define the events $A,B,C,D,\overline{D}$ in the obvious manner, $A=\{AD,A\overline{D}\}$ etc. Our decoherence functional is
\beq
{\textbf{D}}=\frac{1}{9}\begin{pmatrix}
1 & -1 & 1 & 0 & 0 & 0 \\
-1 & 1 & -1 & 0 & 0 & 0 \\
1 & -1 & 1 & 0 & 0 & 0 \\
0 & 0 & 0 & 2 & 1 & -1 \\
0 & 0 & 0 & 1 & 2 & 1 \\
0 & 0 & 0 & -1 & 1 & 2
\end{pmatrix}.
\eeq
As with the double slit system, we have two natural decoherent partitions, $\{D,\overline{D}\}$ with decoherence functional
\beq
{\textbf{D}}=\frac{1}{9}\begin{pmatrix}
1 & 0 \\
0 & 8
\end{pmatrix},
\eeq
where ${\textbf{D}}(\overline{D},\overline{D})=8/9$, and the partition according to the slit, $\{A,B,C\}$, which has decoherence functional
\beq
{\textbf{D}}=\frac{1}{3} \begin{pmatrix}
1 & 0 & 0 \\
0 & 1 & 0 \\
0 & 0 & 1
\end{pmatrix}.
\eeq
Thus unlike the double slit system neither of the `final states' is precluded, however this system does contain two null sets, $\{AD,BD\}$ and $\{BD,CD\}$. As before the fine grained histories $A\overline{D}, B\overline{D}, C\overline{D}$ are not elements of any null set so their duals $A\overline{D}^*, B\overline{D}^*, C\overline{D}^*$ are classical co-events, which also means these histories will not participate in any other primitive co-events. As to the histories ending at the detector, $D$, $AD$ and $BD$ are elements of the null set $\{AD,BD\}$ while $CD$ is an element of the null set $\{BD,CD\}$, thus none of the corresponding classical co-events are preclusive. The co-event $\{AD,BD\}^{\oplus}$ maps $\{BD,CD\}$ to one and thus is not preclusive; similarly $\{BD,CD\}^{\oplus}$ maps $\{AD,BD\}$ to one and $\{AD,CD\}^{\oplus}$ maps both $\{AD,BD\}$ and $\{BD,CD\}$ to one. However, unlike the double slit system there is a preclusive co-event made up from the histories ending at $D$, corresponding to the non-preclusion of the set $D$ in the triple slit as opposed to the double slit system. The whole set $D^{\oplus}=AD^*+BD^*+CD^*$ is preclusive, and furthermore is unital. We have now exhausted $\Homp$, so there can be no further primitive co-events. To summarise:
\bea
\LU = \L &=& \{A\overline{D}^{\oplus}, B\overline{D}^{\oplus}, C\overline{D}^{\oplus}, D^{\oplus}\} \nonumber \\
&\neq& \C.
\eea
So unlike the double slit system the triple slit displays genuinely `quantum' or `anhomomorphic' behaviour, because of $D^{\oplus}$. To see what this means we examine the valuation of $D^{\oplus}$ on the events in the subalgebra of $\EA$ generated by $A,B,C$.
\bea
D^{\oplus}(A) &=& 1, \nonumber \\
D^{\oplus}(B) &=& 1, \nonumber \\
D^{\oplus}(C) &=& 1, \nonumber \\
D^{\oplus}(\{A,B\}) &=& 0, \nonumber \\
D^{\oplus}(\{B,C\}) &=& 0, \nonumber \\
D^{\oplus}(\{C,A\}) &=& 0, \nonumber \\
D^{\oplus}(\{A,B,C\}) &=& 1. \nonumber
\eea
Assuming that $D^{\oplus}$ is the `actual reality' we can rephrase this in `questions \& answers' format, exposing the `anhomomorphic' nature of this theory.
\\
\\
\begin{tabular}{c|c}
\bf{Question} & \bf{Answer} \\
\hline
Does the particle pass through slit $A$? & yes \\
Does the particle pass through slit $B$? & yes \\
Does the particle pass through slit $C$? & yes \\
Does the particle pass through one of slits $A$ or $B$? & no \\
Does the particle pass through one of slits $B$ or $C$? & no \\
Does the particle pass through one of slits $C$ or $A$? & no \\
Does the particle pass through one of slits $A$, $B$ or $C$? & yes \\
\end{tabular}
\\
\\
\\
Note that applied to the observable partition $\{D,\overline{D}\}$ all the above co-events restrict to classical ones, in particular $D^{\oplus}$ simply becomes $D^*$, thus we do not observe anhomomorphic behaviour.

\section{The Kochen-Specker-Peres System}\label{sec:linear PKS}

Recall the histories formulation of the Peres-Kochen-Specker system as outlined in chapter \ref{chapter:kochen-specker}. As discussed in section \ref{sec:the criteria}, it is this system that demonstrates the failure of the classical co-event scheme and so it is natural to use this system as a testing ground for our alternative schemes. In particular, the classical scheme fails on the existence criterion (lemma \ref{lemma:failure of classical scheme}) so we are especially keen to establish the existence of (unital) linear co-events describing the $PKS$ system. Further, the thinking behind the Kochen-Specker theorem centers around picking bases corresponding to potential measurements and asking which of the basis rays would be labelled `true' (or 'green' in our language) by such a measurement. In this light we also seek, for each ray $u_i$ in each basis $u_i,u_j,u_k$ in the Peres Set, a preclusive unital linear co-event that maps to one the event $G_{u_i}$ (which we will abbreviate to $G_i$) that our chosen ray is green and the events $R_{u_j},R_{u_k}$ (again abbreviated to $R_j,R_k$) that the other two rays are coloured red. This goes some way toward showing that the linear scheme's treatment of the $PKS$ system meets the compatibility criterion, however an actual measurement would alter the measure and thus may also alter the nature of the primitive co-events describing it.

To rule out a description of the $PKS$ system by classical co-events we used a general measure (on the set of colourings of the Peres set) whose null sets included the $PKS$ sets (see lemma \ref{lemma:failure of classical scheme} and section \ref{sec:failure of the naive interpretation}). However the derivation of co-events requires an explicit knowledge of the measure, and in particular the full list of null sets. As discussed in section \ref{sec:Stern-Gerlach} the very size of the system (there are $2^{33}$ fine grained histories) makes it unwieldy. Though we are able to construct measures, realisable as gedankenexperiments, that are zero valued on the PKS sets, we are at this point unable to fully characterise the null sets of these measures. The event algebra of the PKS system contains $2^{2^{33}}$ elements, and it has proved beyond our computational power to construct a full list of null sets \& prove that it is exhaustive. As a consequence we will simply assume that we can find a measure $\mu$ whose null sets are precisely the PKS sets and their disjoint unions\footnote{If the disjoint unions of our null sets were not null the measure could not obey strong positivity (see definition \ref{def:strong positivity} and section \ref{sec:null sets}), and thus our histories theory not arise from a Hilbert space theory (see section \ref{sec:the axiomatic approach}).}. The following lemma classifies these null sets,

\begin{lemma}\label{lemma:PKS unions}
The disjoint unions of the $PKS$ sets are of the form
$$R_B \sqcup G_P,$$
where $B$ is a basis and is an (orthogonal) pair such that $B\cap P\neq\emptyset$.
\end{lemma}
\begin{proof}
The question is which of the $PKS$ sets are disjoint. $PKS$ sets are either of the form $R_B$ where $B$ is a basis or $G_P$ where $P$ is an (orthogonal) pair. Now $R_{B_1}\cap R_{B_2}$ is never empty, for both $PKS$ sets contain $R_{PS}$, the singleton set containing the \textit{red map} which colours every ray red. Similarly $G_{P_1}\cap G_{P_2}\supset G_{PS}$. This leaves us with $R_B\cap G_P$; if $B\cap P=\emptyset$ this contains the colouring sending the rays in $B$ to red, the rays in $P$ to green and all other rays to red, and thus is non-empty. However if $B\cap P\neq\emptyset$ then any ray $u_i\in B\cap P$ will be coloured red by every element of $R_B$ and green by every element of $G_P$, therefore $R_B\cap G_P=\emptyset$ and $R_B\cup G_P = R_B\sqcup G_P$.
\end{proof}

In the remainder of this section we will prove the following theorem

\begin{theorem}\label{thm:linear PKS}
Let $\O$ be the space of (green/red) colourings of the Peres Set $PS$, with the associated event algebra $\EA=P\O$, and let $\mu$ be a measure on $\EA$ whose null sets are precisely the $PKS$ sets and their disjoint unions. Then using the abbreviated notation $G_i=G_{u_i}$ and $R_i=R_{u_i}$ we have:
\ \begin{enumerate}
    \item {$\LU \neq \emptyset$}.
    \item {For any basis $\{u_{1},u_{2},u_{3}\}$ of Peres Rays, and for any $1\leq i\leq 3$ there exists $\p\in\LU$ such that $\p(G_i) = 1$ and $\p(R_j) = 1$ for $i\neq j$}.
\end{enumerate}
\end{theorem}

We first introduce some notation. We say that a colouring $\g\in\O$  of $PS$ is consistent on a basis if it colours exactly one of the basis elements green. We say that $\g$ is consistent on an (orthogonal) pair if it does not colour both rays in the pair green. Then let $C \subset PS$, we say that $\g$ is \textit{consistent everywhere other than on $C$}, or \textit{has no  inconsistencies other than on $C$} if $\g$ is consistent on all bases and pairs not entirely contained in $C$. Let $s_{C}$ denote the set of all labellings that are consistent everywhere other than on (bases and orthogonal pairs in) $C$. Looking back at the proof of theorem \ref{thm:PKS}, notice that the `extension' of $\g_P$ (which we shall henceforth refer to simply as $\g_P$) is consistent everywhere other than on $B_{11}=\{100,021,0\m{1}2\}$ (see section \ref{sec:PKS}). Recall that we use $R_{C}$ to denote the set of all labellings which paint all the rays in $C$ red. Now let $r_{C}$ denote the set of all labellings with no inconsistencies other than on $C$, and with all the rays in $C$ labelled red: $r_C = s_C \cap R_C$. If $B$ is a basis, $r_B$ will be consistent everywhere other that on $B$, on which it is `red' valued, thus none of its elements can lie in any $PKS$ that does not contain $R_B$. Further, $r_B\subset R_B$, so using lemma \ref{lemma:PKS unions} we see that $r_B\cap Z$ is either empty or equal to $r_B$, for any null set $Z$. Notice that $\g_P\in r_{B_{11}}$.

We begin our construction with the colouring $\g_P$, which seems fitting given that it is the crux of theorem \ref{thm:PKS}, which the failure of the classical scheme is a corollary of. The most obvious co-event to construct from $\g_P$ is the classical $\g_P^{\oplus}=\g_P^*$, though this is not preclusive because it maps to one the $PKS$ set $R_{B_{11}}$. Perhaps the simplest generalisation of $\g^*$ is $r_{B_{11}}^{\oplus}$; to determine whether this co-event is preclusive it will be useful to better understand the structure of $r_{B_{11}}$.

\begin{lemma}\label{lemma:card rB equals 16}
$|r_{B_{11}}| = 16$\footnote{For a set $A$ we use $|A|$ to denote the cardinality of $A$.}.
\end{lemma}
\begin{proof}
Notice that $\g_P\in r_{B_{11}}$, so if $g\in H$ is a symmetry of $PS$ that leaves $B_{11}$ invariant then $g\g_P\in r_{B_{11}}$. Working through table \ref{table:Peres} and the proof of theorem \ref{thm:PKS}, it can be shown that every element of $r_{B_{11}}$ is of this form. Hence $|r_{B_{11}}| = |\{g\in H | g~stabilises~r_{B_{11}}\}=16$.
\end{proof}

Then recalling equation \ref{eq:linear coevent evaluation} the preclusivity and non-unitality of $r_{B_{11}}^{\oplus}$ is an immediate corollary,

\begin{corollary}
The co-event $r_{B_{11}}^\oplus$ is zero valued on the null sets and on $\O$.
\end{corollary}
\begin{proof}
We have seen above that for any null set $Z$ we have $r_{B_{11}}\cap Z $ is either $r_{B_{11}}$ or the empty set.  Using equation \ref{eq:linear coevent evaluation} this means either $r_{B_{11}}^\oplus (Z) = |r_{B_{11}}|~mod~2 = 0$ using lemma \ref{lemma:card rB equals 16}, or $r_{B_{11}}^\oplus (Z) = |\emptyset|~mod~2 = 0$. In a similar fashion we see that $r_{B_{11}}^{\oplus}(\O)=|r_{B_{11}}| ~ mod ~ 2 = 0$.
\end{proof}

Thus we have shown that $\L$ is not empty, but to prove the first part of theorem \ref{thm:linear PKS} we must construct a co-event that is unital. This requires more subtlety.

\begin{lemma}\label{lemma:phi exists}
$\LU\neq\emptyset$.
\end{lemma}
\begin{proof}
Consider the non-intersecting bases $\7$ \& $\11$ defined in table \ref{table:Peres}. $R_{\7}$ and $R_{\11}$ are $PKS$ sets, so by assumption $\mu(R_{11})=\mu(R_{\7})=0$. Further $\11\cup \7$ is not a basis and $R_{\11\cup\7}$ is neither a $PKS$ set nor an intersection thereof, so by assumption $\mu(R_{\11\cup \7})\neq 0$. Now consider $r_{\11\cup \7}\subset R_{\11\cup \7}$. We can choose distinct labellings $\go\in r_{\11}$,$\gs\in r_{\7}$ and $\gos\in r_{\11\cup \7}$\footnote{To prove this we would explicitly construct the labellings $\go,~\gs$ and $\gos$, a tedious calculation which we will suppress.}, with which we can define the co-event $\p^s$ that will be the backbone of our argument.
\beq
\p^s = \go^* + \gs^* + \gos^*.
\eeq

Then if $Z$ is a null set we have four possibilities

\beq\nonumber
\begin{array}{ccc}
1. & Z=G_P &  \\
2. & Z=R_B & B\not\in\{\11,\7\}  \\
3. & Z=R_B & B\in\{\11,\7\}     \\
4. & Z=N   &   N=R_B\sqcup G_P.
\end{array}
\eeq

In each case it is easy to see that $\p^s(Z)=0$:

\beq\nonumber
\begin{array}{ccc}
1. & \go,\gs,\gos\not\in Z & \Rightarrow \p^s(Z) = 0, \\
2. & \go,\gs,\gos\not\in Z & \Rightarrow \p^s(Z) = 0, \\
3. & if~B=\11~\go,\gos\in Z,\gs\not\in Z & \Rightarrow \p^s(Z)=0, \\
   & if~B=\7~\gs,\gos\in Z,\go\not\in Z & \Rightarrow \p^s(Z)=0, \\
4. & \p^s(Z) = \p^s(R_B)+\p^s(G_P) & \Rightarrow \p^s(Z)=0.
\end{array}
\eeq

Hence $\p^s$ is preclusive. As to unitality,
\bea
\p^s(\O) &=& |\p^{s\oplus}\cap\O|~mod~2, \nonumber \\
&=& |\p^{s\oplus}|~mod~2, \nonumber \\
&=& |\{\go,\gs,\gos\}|~mod~2, \nonumber \\
&=& 1.\nonumber
\eea
\end{proof}

This proves the first part of theorem \ref{thm:linear PKS}, we will build on these results to achieve the second part.

\begin{lemma}\label{lemma:rotating phi}
For any Peres ray $v$ there is a primitive, preclusive, unital co-event $\p$ and a labelling $\g$ such that:
\bea
\g(v) &=& green, \nonumber \\
\p(\{\g\}) &=& 1. \nonumber
\eea
\end{lemma}

\begin{proof}
We will prove the claim in several stages, essentially using the symmetries of Peres Set to `rotate' the co-event $\p^s$ defined above to apply to the Peres Ray $v$. First, recall that the symmetry group $H$ of the Peres Set corresponds to the rigid isometries of the projective cube (see section \ref{sec:PKS}), being generated by rotations by $\pi/2$ around co-ordinate axes and reflections in the co-ordinate planes. Further recall that the Peres Rays split into four types (see table \ref{table:Ray Types}). It is easy to see that $H$ acts transitively on each type, so that given two Peres Rays $u,v$ $\exists$ $g\in H$ such that $v=g(u)$ if and only if $u$ and $v$ are of the same type.

In section \ref{sec:PKS null sets} we defined the action of $H$ on $\O$ by

$$[g\g](v) = \g(g(v))$$

so that

$$\g(v) = [g\g](g^{-1}(v)),$$

and extended it to an action on the event algebra in the obvious fashion

$$gA = \{g\g|\g\in A\}.$$

It is easy to check that every $g\in H$ defines an automorphism\footnote{In other words it is injective and surjective.} on $\O$. Further, as mentioned in section \ref{sec:PKS null sets}, note crucially that the PKS sets are permuted by the symmetries in $H$, via $g(G_P)=G_{g(P)}$ and $g(R_B)=R_{g(B)}$.

We can use the above to define an action of $H$ on the `(linear) dual space' of linear co-events by requiring

$$[g\p](A) = \p(gA)~\forall A\in\EA,$$

which implies

$$g\p = (g^{-1}\pl)^{\oplus}.$$

We can show that this map preserves preclusivity, unitality and primitivity:

\begin{description}
  \item[Preclusivity:] As noted in section \ref{sec:PKS null sets} elements of $H$ permute the $PKS$ sets, thus if $Z\in\EA$ is null then so is $gZ$ for all $g\in H$. But then if $\p$ is a preclusive linear co-event, $g\in H$ and $Z\in\EA$ is null, then $g\p(Z) = \p(gZ) = 0$.
  \item[Unitality:] Because every $g\in H$ is an automorphism on the sample space, we have $g\O=\O$ for all $g\in H$, so if $\p$ is unital then $g\p(\O) = \p(g\O) = \p(\O) = 1$.
  \item[Primitivity:] Because every $g\in H$ is an automorphism on $\O$, for $A,B\in\EA$, $A\subset B\Rightarrow gA\subset gB$ and in particular $\pl\subset \psi^{\oplus}\Rightarrow g\pl \subset g\psi^{\oplus}$. Bearing in mind that $g$ preserves preclusivity and unitality, by lemma \ref{lemma:linear dual properties} this means that if $\psi$ dominates $\p$ then $g\psi$ dominates $g\p$, thus $g$ preserves primitivity.
\end{description}

Now for a general labelling $\g$ we can define $Green(\g)$ to be the set of rays coloured green by $\g$, notice that the `green set' $Green(\g)$ uniquely identifies $\g$. We can also identify a `red set' $Red(\g)$ associated with a labelling $\g$ in the obvious fashion (in fact $Green(\g)\sqcup Red(\g) = PS$).

Now consider our previous construction of a preclusive, unital, linear co-event $\p^s$. Recall that we chose $\go$ from $r_{\11}$, and we can fix $\go$ to be $\g_P$, the labelling used in the proof of theorem \ref{thm:PKS}. Then a quick look at table \ref{table:Peres} shows us that

$$Green(\go)=\{001,101,011,1\m{1}2,102,211,201,112,012,121\}.$$

Notice that $Green(\go)$ contains a ray of each type. So by the transitivity of $H$ acting on each type, given any ray $v$  there exists $g_v\in H$ such that $g_v(v)$ is labelled green by $\go$. Now $\go\in\p^{s\oplus}$, so $\p^s(\{\go\})=1$. Now $\p^s$ is preclusive, unital and primitive thus by the above so is $g_v^{-1}\p^s$. But, $g_v^{-1}\p^s(\{g_v\go\})=\p^s(\{\go\})=1$; so given any $v\in PS$, the labelling $g_v\go$ and the co-events $g_v^{-1}\p^s$ meet the conditions of the lemma.

\end{proof}

The next step is to find primitive, preclusive, unital co-events that map to one the events we are interested in.

\begin{lemma}\label{lemma:phi green and phi red}
Given any Peres Ray $u_i$ there exist primitive, preclusive, unital co-events $\p_i^{green}$, $\p_i^{red}$ that map $G_i$ and $R_i$ to $1$ respectively
\end{lemma}

\begin{proof}
Using our previous construction, we can pick $\gs$ and $\gos$ such that\footnote{To prove this we would explicitly construct the labellings $\gs$ and $\gos$, a tedious calculation which we will suppress}:

\bea
Green(\go)&=&
\{\overbrace{001}^{Type~I},\overbrace{101}^{Type~II},011,1\m{1}2,\overbrace{102}^{Type~III},\overbrace{211}^{Type~IV},201,112,012,121\}, \nonumber \\
Green(\gs)&=&
\{\overbrace{001}^{Type~I},011,\overbrace{101}^{Type~II},\m{1}12,012,121,021,112,\overbrace{102}^{Type~III},\overbrace{211}^{Type~IV}\},\nonumber \\
Green(\gos)&=&
\{\overbrace{001}^{Type~I},\overbrace{101}^{Type~II},011,1\m{1}2,\overbrace{102}^{Type~III},\overbrace{211}^{Type~IV},112,012,121\}.\nonumber
\eea

So that

$$Green(\go)\cap Green(\gs)\cap Green(\gos)\supset\{\overbrace{001}^{Type~I},\overbrace{101}^{Type~II},\overbrace{102}^{Type~III},\overbrace{211}^{Type~IV}\},$$

contains a ray of each type. Now since the symmetry group $H$ is transitive on each type of Peres Ray, given an arbitrary ray $u_i$ we can find $g_i\in H$ such that $g_i(u_i)\in Green(\go)\cap Green(\gs)\cap Green(\gos)$. But then following the argument we used in the proof of lemma \ref{lemma:rotating phi} all three labellings in $(g_i^{-1}\p^{s})^{\oplus}$ colour $u_i$ green, so $|(g_i^{-1}\p^s)^{\oplus}\cap G_i| = 3$ implying $[g_i^{-1}\p^{s}](G_i)=1$, and we can take $\p_i^{green} = g_i^{-1}\p^{s}$.

Similarly, $Red(\go)\cap Red(\gs)\cap Red(\gos)\supset\{100,110,120,\m{1}21\}$ contains a ray of each type, and by a similar argument we can find $\p_i^{red}$.
\end{proof}

We are now in a position to prove theorem \ref{thm:linear PKS}:

\begin{proof}{of theorem \ref{thm:linear PKS} \ }
We have proven part $1$ of the theorem in lemma \ref{lemma:phi exists}, and part 2 follows directly from lemma \ref{lemma:phi green and phi red}. Given a ray $u_i$ in a basis $\{u_i,u_j,u_k\}$ we have shown that there exists $\p^{green}_i$ mapping $G_i$ to $1$, however we want a co-event that goes further to also map $R_j$ and $R_k$ to $1$. It turns out that $\p^{green}_i$ meets these conditions. Note that by the construction of $r_B$ none of $\go,\gs$ or $\gos$ colour two orthogonal rays green, and since $H$ consists of rigid isometries this means that none of $g\go,g\gs,g\gos$ colour two orthogonal rays green for any $g\in H$. Therefore if we define $g_i$ as in the proof of lemma \ref{lemma:phi green and phi red}, the three labellings $g_i^{-1}\go,g_i^{-1}\gs$ and $g_i^{-1}\gos$ colour $u_i$ green and therefore must colour $u_j$ and $u_k$ red. These three labellings are the three elements of $\p^{green\oplus}_i$, which is therefore a subset of $R_j$ and of $R_k$ meaning that $|\p_i^{green\oplus}\cap R_j| = |\p^{green\oplus}_i\cap R_k| =3$, therefore $\p^{green}_i(R_j)=\p^{green}_i(R_k)=1$, thus $\p^{green}_i(R_j)$ satisfies the conditions of the theorem.
\end{proof}

\section{The Failure of the Linear Scheme}\label{sec:linear failure}

Further exploration of the linear scheme has uncovered several points at which it fails to meet the criteria we set out in section \ref{sec:the criteria}. We begin with two toy models that indicate the more serious problems to come. Firstly, we can find systems in which the linear scheme precludes events that are not precluded by the dynamics.

\begin{lemma}\label{lemma:linear compatibility failure}
We can find a histories theory $\H$ obeying strong positivity for which $\exists~A\in\EA$ such that $\mu(A)\neq 0$ yet $\p(A)=0$ for all $\p\in\L$.
\end{lemma}
\begin{proof}
We use a simple proof by construction. Consider the histories theory $\H$ with sample space $\O = \{a_{0},a_{1},a_{2},a_{3}\}$, event algebra $\EA=P\O$ and the associated strongly positive decoherence functional
$${\textbf{D}}=\left(\begin{array}{cccc}
   1&-1&-1& 2
\\-1& 1& 1&-2
\\-1& 1& 1&-2
\\ 2&-2&-2& 4
\end{array}\right)$$
The null sets are $\{a_{0},a_{1}\}, \{a_{0},a_{2}\}$ and $\{a_{1},a_{2},a_{3}\}$, as illustrated in figure \ref{fig:linear compatibility failure}. Now let $\p\in \L$ map $a_3$ to $1$, so that $a_3\in\pl$. Then since $\p$ is preclusive lemma \ref{lemma:linear dual properties} means that $\pl$ must have an intersection of even cardinality with the null set $\{a_{1},a_{2},a_{3}\}$, hence $\pl$ must include exactly one of $a_1$ or $a_2$. By the symmetry of the system, wlog assume $a_2\in \pl$. But then $\mu(\{a_{2},a_{0}\})=0\Rightarrow a_{0}\in\pl$, and then $\mu(\{a_{1},a_{0}\}=0 \Rightarrow a_{1}\in \pl$. But then $\{a_{1},a_{2},a_{3}\}\subset \pl$ which forces $\pl$ to have an odd intersection with $\{a_{1},a_{2},a_{3}\}$, a null set, contradicting our assumption that $\p$ is preclusive. Hence no $\p\in \L$ can map $a_{3}$ to 1. Note that $\L$ is non-empty, for example $\{a_{0},a_{1},a_{2}\}^\oplus\in\L$.

\begin{figure}[t]
\begin{center}
\includegraphics[width=50mm]{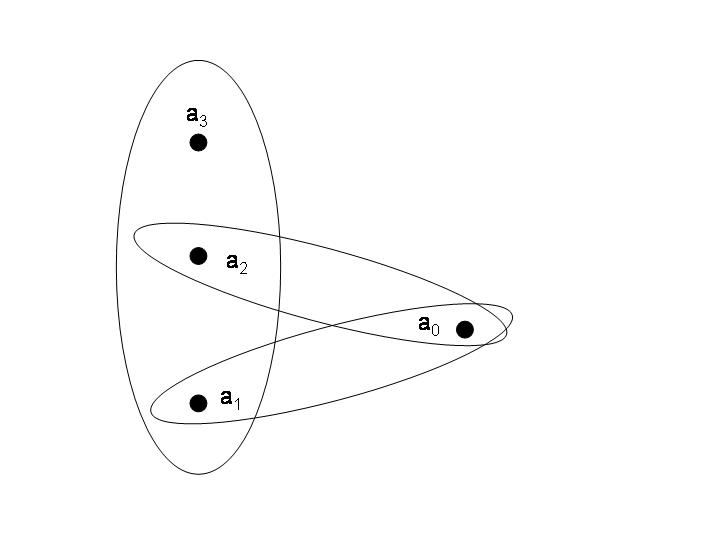}
\caption{Null Sets for lemma \ref{lemma:linear compatibility failure}} \label{fig:linear compatibility failure}
\end{center}
\end{figure}

\end{proof}

Note that this does not violate the compatibility criterion (section \ref{sec:the criteria}), which is defined in terms of experimentally measurable outcomes. However it does mean that the mathematical framework of the linear scheme does not rule out co-events which preclude events allowed by the dynamics, which is of concern.

\begin{figure}[t]
\begin{center}
\includegraphics[width=50mm]{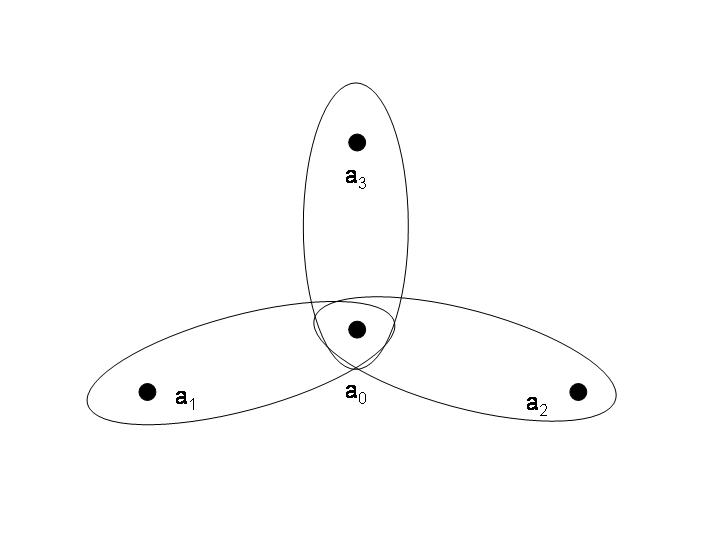}
\caption{Null Sets for lemma \ref{lemma:linear unitality counterexample}} \label{fig:linear unitality problem}
\end{center}
\end{figure}

Further, it turns out that there are measures that do not admit unital linear co-event, in other words:

\begin{lemma}\label{lemma:linear unitality counterexample}
There exists a histories theory $\H$ obeying strong positivity such that $\LU=\emptyset$.
\end{lemma}
\begin{proof}
We use a simple proof by construction. Consider the toy model histories theory with sample space $\O=\{a_0,a_1,a_2,a_3\}$, event algebra $\EA=P\O$ and the associated strongly positive decoherence functional
$${\textbf{D}}=\left(\begin{array}{cccc} 1&-1&-1&-1
\\-1&1&1&1
\\-1&1&1&1
\\-1&1&1&1\end{array}\right)$$
The null sets are $A_{i}=\{a_{0},a_{i}\}$ for $i=1,\ldots ,3$, as shown in figure \ref{fig:linear unitality problem}, and it is easy to see that $\sum_i A_i = \O$. Thus by lemma \ref{lemma:linear null sets prevent unitality} any $\phi(\O)=0~\forall\phi\in\L$, so that $\LU=\emptyset$.
\end{proof}

Both of these results are problematic for the linear scheme, and even on their own indicate that an alternate scheme should be pursued. However both of these lemmas use toy model measures, and, clinging to straws, one may argue that the linear scheme should remain in play until it fails on an experimentally realisable system. It is to this question that we now turn.

\subsection{The Four Slit System}\label{sec:four slit}

It turns out that we can realise the measure used in lemma \ref{lemma:linear unitality counterexample} as a part of the measure of an idealised four slit system, a generalisation of the double and triple slit systems we constructed in section \ref{sec:linear double and triple slit}. Using an argument similar to that employed in lemma \ref{lemma:linear unitality counterexample} we can deduce a lack of compatibility with experimentally testable predictions of quantum mechanics. The gedankenexperimental realisation of this system is discussed in appendix \ref{appendix:many slit}.

This time we have four slits, $A_0,A_1,A_2,A_3$, in place of the two or three slits used previously. We denote the initial state $\ket{\psi}$, and the projector corresponding to finding the particle at slit $A_i$ upon measurement by $P_i$, with the associated state $\ket{A_i}$ such that $P_i=\ket{A_i}\bra{A_i}$; note that the vectors corresponding to the four slits are mutually orthogonal. We define the projectors $P_D$ and the associated state $\ket{D}$ in a similar fashion. The projector corresponding to $\overline{D}$ is defined by $P_{\overline{D}} = \mathbb{I}-P_D$.

As before we define our minimal Hilbert space theory $({\cal{H}},H,\ket{\psi},T)$ by setting ${\cal{H}} = span(\ket{A_i})$, setting $H$ to be the null operator, so our evolution operators are the identity, and setting our temporal support to be $T = \{0,1,2\}$, consisting of an initial time, an intermediate time at which our projectors $P_i$ act, and a final time at which our projectors $P_D,P_{\overline{D}}$ act.

Since our Hilbert space is the span of the detector states, we must define our initial state $\ket{\psi}$ and the detector state $\ket{D}$ in terms of the $\ket{a_i}$. As discussed in appendix \ref{appendix:many slit}, we can construct a gedankenexperiment in which $\ket{\psi}=\frac{1}{2}(\ket{A_0}+\ket{A_1}+\ket{A_2}+\ket{A_3})$ and $\ket{D} = \frac{1}{2}(\ket{A_0}-\ket{A_1}-\ket{A_2}-\ket{A_3})$. Notice that this gives us $\bra{D}(P_0+P_i)\ket{\psi}=0$ for $1\leq i\leq 3$.

Denoting the path that passes through slit $A_i$ to reach $D$ ($\overline{D}$) by $D_i$ ($\overline{D}_i$), our sample space is now $\O=\{D_0, D_1,D_2,D_3, \overline{D_0}, \overline{D_1}, \overline{D_2},\overline{D_3}\}$ and our event algebra is $\EA=P\O$ as before. We can define the events $A_i,D,\overline{D}$ in the obvious manner, $A_i=\{D_i,\overline{D_i}\}_{i=0}^3$, $D=\{D_i\}_{i=0}^3$ etc. Our decoherence functional is
\beq
{\textbf{D}}=\frac{1}{16}\begin{pmatrix}
1 & -1 & -1 & -1 & 0 & 0 & 0 & 0 \\
-1 & 1 & 1 & 1 & 0 & 0 & 0 & 0 \\
-1 & 1 & 1 & 1 & 0 & 0 & 0 & 0 \\
-1 & 1 & 1 & 1 & 0 & 0 & 0 & 0 \\
0 & 0 & 0 & 0 & 3 & 1 & 1 & 1 \\
0 & 0 & 0 & 0 & 1 & 3 & -1 & -1 \\
0 & 0 & 0 & 0 & 1 & -1 & 3 & -1 \\
0 & 0 & 0 & 0 & 1 & -1 & -1 & 3 \\
\end{pmatrix}
\eeq
As with the double and triple slit systems, we have two natural decoherent partitions, $\{D,\overline{D}\}$ with decoherence functional
\beq
{\textbf{D}}=\frac{1}{4}\begin{pmatrix}
1 & 0 \\
0 & 3
\end{pmatrix}
\eeq
where ${\textbf{D}}(\overline{D},\overline{D})=8/9$, and the partition according to the slit, $\{A,B,C\}$, which has decoherence functional
\beq
{\textbf{D}}=\frac{1}{4} \begin{pmatrix}
1 & 0 & 0 & 0 \\
0 & 1 & 0 & 0\\
0 & 0 & 1 & 0\\
0 & 0 & 0 & 1\\
\end{pmatrix}
\eeq
Thus as with the triple slit system neither of the `final states' is precluded, and the system contains three null sets, $\{D_0,D_i\}$ for $1\leq i \leq 3$. As before the fine grained histories $\overline{D_i}$ are not elements of any null set so their duals $\overline{D_i}^*$ are classical co-events, which also means these histories will not participate in any other primitive co-events.

A comparison of the decoherence functional above with that used in lemma \ref{example:linear unitality problem} reveals that the `internal' dynamics of the histories ending at the detector $D$ is equivalent to that described in the lemma, and we can therefore apply the argument used in its proof. Let $\p\in\L$ be a co-event whose dual contains one of the $D_i$, then by the above $\pl$ does not contain any of the $\overline{D}_i$. Now first assume that $D_0\not\in\pl$, then it must contain $D_i$ for some $1\leq i \leq 3$. But $\{D_0,D_i\}$ is null, thus for $\p$ to be preclusive the intersection of $\pl$ and $\{D_0,D_i\}$ must be even. This means $D_i\in\pl\Rightarrow D_0\in\pl$ contradicting our assumption. Thus $D_0\in\pl$ for any $\p\in\L$ whose dual contains one of the $D_i$. But then the preclusivity of $\p$ means the intersection of $\pl$ with any of the null sets $\{D_0,D_i\}$ where $1\leq i \leq 3$ must be even, which means that $\pl$ must contain all of the $D_i$. Therefore the only co-event involving the histories ending at $D$ is $D^{\oplus}=D_0^*+D_1^*+D_2^*+D_3^*$. Note that because $|D|$ is even, $D^{\oplus}$ is not unital, and $D^{\oplus}(D)=0$. To summarise:
\bea
\LU &=& \{\overline{D}^*_0,\overline{D}^*_1,\overline{D}^*_2,\overline{D}^*_3\}, \nonumber \\
\L &=& \{\overline{D}^*_0,\overline{D}^*_1,\overline{D}^*_2,\overline{D}^*_3,D^{\oplus}\}. \nonumber
\eea
Crucially note that none of these co-events map $D$ to $1$. This means that according to the linear scheme there is no potential reality compatible with the dynamics in which the event $D$ occurs, or in other words the linear scheme predicts that $D$ will not occur. However, the partition $\{D,\overline{D}\}$ is classical, with a non-zero probability of $D$ occurring. More pertinently, the event $D$ is observable, corresponding to a detector reading, and the linear scheme's prediction that $D$ will never occur is a falsifiable claim at odds with the standard predictions of quantum mechanics; as such it is not an assertion that the author is happy to support. This finding constitutes a violation of our compatibility criterion, and has led to further research into linear scheme being abandoned.

%The Linear Scheme
%\singlespacing
\chapter{Polynomial Schemes}\label{chapter:polynomial}
%\doublespacing

In this chapter we examine the quadratic and general degree polynomial schemes. We examine the properties of each scheme and note that for any histories theory there exists an $n$ such that the $n$th order polynomial scheme contains unital co-events. However we then show that for any given $n$ there exists a histories theory in which there are no unital $n$-polynomial co-events, and conclude by using this construction to construct a Hilbert space theory which can be gedankenexperimentally realised (appendix \ref{appendix:many slit}), but for which the $n$-polynomial scheme makes predictions contrary to those of standard quantum mechanics. This chapter draws from \cite{CoEventSchemes}.

\section{The Quadratic Scheme}

Historically, the failure of the linear scheme led to a search for alternatives. It was felt that the linearity rule (equation \ref{eq:linearity}) was too strong, thus the quadratic scheme replaced linearity with a weaker `quadratic' sum rule which mirrored the sum rule obeyed by the quantum measure, whereas linearity seemed closer to the Kolmogorov sum rule (equation \ref{eq:kolmogorov sum rule}). To recap on section \ref{sec:quadratic coevents} the quadratic sum rule is:

$$\p(A+B+C)=\p(A+B)+\p(B+C)+\p(C+A)+\p(A)+\p(B)+\p(C).$$

The definition of primitivity for quadratic co-events is more complicated than the linear case, and will be justified in section \ref{sec:quadratic primitivity and emergent classicality}. First, every primitive linear co-event is admitted as a primitive quadratic co-event. Then if $\p$ is a quadratic, non-linear co-event it is dominated by a quadratic co-event $\psi$ if either both the following conditions are met:
\bea
\psi(\{\g\})=1 &\Rightarrow& \p(\{\g\})=1, \nonumber \\
\psi(\{\g,\overline{\g}\})+\psi(\{\g\})+\psi(\{\overline{\g}\})=1 &\Rightarrow& \p(\{\g,\overline{\g}\})+\p(\{\g\})+\p(\{\overline{\g}\})=1, \label{eq:quadratic domination by sum}
\eea
or if
\beq\label{eq:quadratic domination by product}
\psi(A)=1 \Leftarrow \p(A)=1,
\eeq
for all $A\in\EA$. A preclusive non-linear quadratic co-event is primitive if it is not dominated by a preclusive quadratic co-event, linear or non-linear. We denote the set of primitive quadratic co-events by $\Q$. As with the linear scheme, the quadratic scheme is only defined in the finite case, principally because our notion of primitivity would not make sense were the sample space infinite. This in this section, unless specifically noted otherwise, we will assume that all co-events are quadratic and all sample spaces finite.

Unitality (definition \ref{def:unitality}) plays an important role a role in the quadratic scheme as it did in the linear scheme, and for much the same reasons. As in the linear case, we denote the preclusive unital quadratic co-events that are not dominated by any other preclusive unital quadratic co-event by $\QU$. In this case, we would use $\LU$ in the definition of primitivity in place of $\L$.

\subsection{Primitivity \& Weak Emergent Classicality}\label{sec:quadratic primitivity and emergent classicality}

In this section we will outline the argument that led to the adoption of the primitivity condition used in the quadratic scheme (see section \ref{sec:quadratic coevents}), before demonstrating that this condition does indeed cause the quadratic scheme to obey weak emergent classicality.

The quadratic rule is a weakening of the linear rule, thus the linear scheme's problem of `too many' co-events is only exacerbated by the generalisation to the quadratic scheme. As in the linear case we seek a primitivity condition to reduce the number of co-events, and specifically to ensure weak emergent classicality.

In the linear scheme every co-event is a sum of degree one monomials, $\p=\sum_{\g\in\pl}\g^*$.We then say that the linear $\psi$ dominates the linear $\p$ if every summand in $\psi$ is a summand in $\p$, which means $\psi^{\oplus}\subset\p^{\oplus}$. In the quadratic scheme the possibilities are wider. We begin by accepting all primitive linear co-events as primitive quadratic co-events, and then must decide how to deal with the degree two polynomials, the quadratic, non-linear co-events. Looking forward to theorem \ref{thm:n-polynomial coevents are degree n polynomials} a quadratic co-event is a sum of order one and two monomials,
$$\p=\sum_{i\neq j}n^{\p}_{ij}\g_i^*\g_j^* + \sum_k n^{\p}_k \g_k^*,$$
where $n^{\p}_{ij},~n^{\p}_k\in\Z2$. We want a quadratic co-event $\p$ to be dominated by a quadratic co-event $\psi$ if all of the summands in $\psi$ are also summands of $\p$. In other words we want
\bea
n^{\psi}_{ij}=1 &\Rightarrow& n^{\p}_{ij}=1 \label{eq:degree two domination} \\
n^{\psi}_{k}=1 &\Rightarrow& n^{\p}_{k}=1. \label{eq:degree one domination}
\eea
But $\g_i^*\g_j^*(\{\g_m\}) =0$ if $i\neq j$, so that
$$\p(\{\g_m\})=n^{\p}_m,$$
meaning that equation \ref{eq:degree one domination} can be written
$$\psi(\{\g\})=1 \Rightarrow \p(\{\g\})=1,$$
Further,
\bea
\p(\{\g_m,\g_n\})+\p(\{\g_m\})+\p(\{\g_n\}) &=& n^{\p}_{mn}+n^{\p}_m+n^{\p}_n+n^{\p}_m+n^{\p}_n, \nonumber \\
&=& n^{\p}_{mn}. \nonumber
\eea
Thus equation \ref{eq:degree two domination} can be written
$$\psi(\{\g,\overline{\g}\})+\psi(\{\g\})+\psi(\{\overline{\g}\})=1 \Rightarrow \p(\{\g,\overline{\g}\})+\p(\{\g\})+\p(\{\overline{\g}\})=1,$$
explaining equation \ref{eq:quadratic domination by sum}. Finally, we also want co-events such as $\g_i^*\g_j^*$ to be dominated by their factors, in this case $\g_i^*$ and $\g_j^*$. But if $\p=\psi\chi$ then $\p(A)=\psi(A)\chi(A)$ so that $\p(A)=1\Rightarrow \psi(A)=\chi(A)=1~\forall A\in\EA$. This explains equation \ref{eq:quadratic domination by product}.

Notice that by accepting all primitive linear co-events as primitive quadratic co-events we have ensured $\L\subset\Q$ and $\LU\subset\QU$.

It is easy to show that this definition of primitivity does indeed lead to weak emergent classicality. As in the linear case the result is an immediate corollary of the following lemma:

\begin{lemma}\label{lemma:quadratic scheme classical null sets}
Let $\H$ be a histories theory with a finite sample space in which the measure $\mu$ is such that every subset of a null set is null, and the disjoint union of any two null sets is null. Then $\p\in\Q$ is a homomorphism.
\end{lemma}
\begin{proof}
Let $\p\in \Q$ be anhomomorphic. We have seen above that $\p$ is of the form
$$\p=\sum_{i\neq j}n^{\p}_{ij}\g_i^*\g_j^* + \sum_k n^{\p}_k \g_k^*.$$
Now assume $n^{\p}_k=1$ for some $k$. Then $\p(\{\g_k\})=1$ which implies that $\g_k$ is not an element of any null set, for otherwise by assumption $\{\g_k\}$ would be null. But then $\g_k^*$ is preclusive, and by assumption $\p$ is not a homomorphism, so $\p\neq\g_k^*$. Thus $\g_k^*$ dominates $\p$, contradicting our assumption that $\p\in Q$ and so is primitive. Therefore $n_k=0~\forall k$, and $\p$ is of the form
$$\p=\sum_{i\neq j}n^{\p}_{ij}\g_i^*\g_j^*.$$
Now assume that $n_{ij}=1$ for some $i,j$. $\p(\{\g_m,\g_n\})=n_{mn}$ so $\{\g_i,\g_j\}$ can not be null, and so can not be a subset of any null set. But then $\g_i^*\g_j^*$ is preclusive and so will dominate $\p$ unless
$$\p=\g_i^*\g_j^*.$$
Now $\g_i$ must be an element of a null set $N_i$, otherwise $\g_i^*$ will be preclusive and will dominate $\p$. Similarly $\g_j$ must be an element of a null set $N_j$, so $\{\g_i\}$ and $\{\g_j\}$ are subsets of null sets and therefore null, as is their disjoint union $\{\g_i,\g_j\}$. But then $\p$ is not preclusive, contrary to our assumptions. Therefore $\p$ must be a homomorphism.
\end{proof}

\begin{corollary}\label{corollary:quadratic_scheme_emergent_classicality}
Let $\H$ be a histories theory in which $\O$ is finite. Then:
$$\mu(A\sqcup B)=\mu(A)+\mu(B)~\forall ~A,B\subset\Omega~\Rightarrow ~\Q=\C.$$
\flushright{$\square$}
\end{corollary}

\subsection{Existence}

Existence is an immediate corollary of theorem \ref{thm:quadratic coevent existence} proved in appendix \ref{appendix:quadratic coevent existence}:

\begin{corollary}\label{corollary:quadratic existence}
Let $\H$ be a histories theory with finite $\O$. Then $\Q\neq\emptyset$.
\flushright{$\square$}
\end{corollary}

However, as we shall see in section \ref{sec:quadratic failure}, unitality plays a central role as it did in the linear scheme (section \ref{sec:linear unitality}). Thus our focus will be on $\QU$ rather than $\Q$.

\subsection{The Four Slit System}\label{sec:quadratic four slit}

Notice that the linear scheme is a special case of the quadratic scheme, and in particular $\LU\subset\QU$ so that all primitive preclusive unital linear co-events are automatically primitive preclusive unital quadratic co-events. Thus if there is a successful linear description of the system there will always be a successful quadratic description of the system, for example we are automatically assured that the quadratic scheme can describe the $PKS$ system. However the quadratic scheme is also able to treat systems that are problematic for the linear scheme, and as an example we shall examine the four slit system that led us to disregard the linear scheme.

In this section it will be convenient to let the indices $i,j$ run from $1$ to $3$ and the indices $\a,\b$ from $0$ to $3$. Then referring back to section \ref{sec:four slit} recall that our sample space is $\O=\{D_0,D_i, \overline{D_0},\overline{D_j}\}_{i,j=1}^3$, our event algebra $\EA=P\O$ and our null sets $\{D_0,D_i\}$ for $1\leq i\leq 3$. A general quadratic co-event for this system has the form:
\beq\label{eq:quadratic four slit general coevent}
\p = \sum_\a \overline{n}_\a \overline{D}_\a^* + \sum_\a n_\a D_\a^* + \sum_{\a\b} \overline{n}_{\a\b} \overline{D}_\a^*\overline{D}_\b^* + \sum_{\a\b} n_{\a\b} D_\a^* D_\b^* + \sum_{\a\b}\tilde{n}_{\a\b} \overline{D}_\a^* D_\b^*,
\eeq
where the $n$'s are elements of $\Z2$. Now starting with the co-events involving the `non-detector' histories $\overline{D}_\a$, it is easy to see that $\overline{D}_\a^*$, $\overline{D}_\a^*\overline{D}_\b^*$ and $\overline{D}_\a^* D_\b^*$ are preclusive, thus dominating any co-event containing them as a summand. Further, $\overline{D}_\a^*$ dominates the other two, do from equation \ref{eq:quadratic four slit general coevent} we see that $\overline{D}_\a^*\in\QU$, further no other co-event involving the non-detector histories can be primitive. Thus other than the $\overline{D}_\a^*$, $\p\in\Q$ is of the form
\beq\label{eq:quadratic four slit coevent with detector histories}
\p = \sum_\a n_\a D_\a^* + \sum_{\a\b} n_{\a\b} D_\a^* D_\b^*.
\eeq
We have already shown in section \ref{sec:four slit} that there are no unital linear co-events describing the remaining `detector histories' $D_\a^*$, the only preclusive linear co-event being the non-unital $D^{\oplus}$. However we can find preclusive unital degree two co-events.

The monomials $D_1^*D_2^*$, $D_2^*D_3^*$ and $D_3^*D_1^*$ are unital and preclusive, and are primitive because they could only be dominated by the singletons $D_i^*$, which in this case are not preclusive. This means the terms $D_i^*D_j^*$ can not be summands of any primitive co-event, so the most general primitive preclusive co-event is
$$\p=n_0D_0^*+\sum_i n_iD_i^* + \sum_j n_{0j} D_0^*D_j^*,$$
where the $n$'s are elements of $\Z2$. Now if we assume that $n_0=n_i=0$ we are left with $\sum_j n_{0j} D_0^*D_j^*$, and it is easy to see that $n_{0k}=1\Rightarrow \p(\{D_0,D_k\})=1$, so no such co-event can be preclusive. If we relax our assumption to $n_0=0$ we find that the co-events $D_0^*D_i^* + D_i^*$ are preclusive though not unital. They are also primitive in $\Q$, for in general the only candidates for dominating them are $D_0^*,D_i^*$ and $D_0^*D_i^*$, none of which are preclusive. However, it is not possible to extend any $D_0^*D_i^* + D_i^*$ to a primitive unital preclusive co-event $\p=D_0^*D_i^* + D_i^*+\psi$, for the preclusivity of $\p$ would imply the preclusivity of $\psi$ and the unitality of $\p$ would imply the unitality of $\psi$. But then $\psi$ would dominate $\p$, which thus could not be primitive. This exhausts our options for $n_0=0$.

Now assuming that $n_0=1$, note that $D_0^*(\{D_0,D_i\})=1$, so we need to add summands to set the valuation of our co-event on each null set to zero. For each null set $\{D_0,D_i\}$ we could add either $D_i^*$ or $D_0^*D_i^*$, which means that the most general primitive preclusive co-event assuming $n_0=1$ is
$$\p=D_0^* + \sum_i (n_i D_i^* + [1+n_i]D_0^*D_i^*).$$
These co-events are not unital, and following argument we used above can not be extended to a primitive preclusive unital co-event. This exhausts our options for $n_0=1$, and thus we have exhausted the list of primitive co-events. In summary:

\bea
\QU &=& \{\overline{D}_i^*,D_i^*D_j^*\}_{i,j=1}^3, \nonumber \\
\Q &=&  \QU\sqcup \{D_0^*D_i^* + D_i^*\}_{i,j=1}^3, \nonumber \\
&& \sqcup\{D_0^* + \sum_i (n_i D_i^* + [1+n_i]D_0^*D_i^*)|n_i\in\Z2\}. \nonumber
\eea

\subsection{The Failure of the Quadratic Scheme}\label{sec:quadratic failure}

Although the quadratic scheme can successfully describe the four slit system, which was the bane of the linear scheme, we can find generalisations that the quadratic scheme can not cope with. As with the linear scheme the problem is unitality, or rather the structures that prevent it. The next lemma identifies the `second order' null set structures that prevent quadratic co-events from being unital.

\begin{lemma}\label{lemma:quadratic problem structure}
Let $\H$ be a histories theory with a finite sample space. If $\exists$ $\{Z_i\}_{i=1}^m\subset\Omega$ such that:
\bea
\sum_i Z_i &=& \O, \nonumber \\
\mu(Z_i) &=& 0 ~\forall~i,\nonumber \\
\mu(Z_i+ Z_j)&=& 0~\forall~i,j. \nonumber
\eea
Then $\QU=\emptyset$.
\end{lemma}
\begin{proof}
Since $\O=\sum_i Z_i$, and noting that $\mu(\O)=1$ means that the above structure implies that $m>2$ (otherwise we would have $\mu(\O)=0$), we can use the quadratic sum rule (equation \ref{eq:quadratic rule}) to decompose the action of any quadratic co-event $\p$ into sums of its actions on the sets $Z_i$ and their pairwise sums:
$$\p(\O) = \sum_{i\neq j} n_{ij} \p(Z_i+Z_j) + \sum_k n_k \p(Z_k),$$
where $n_{ij},n_k\in\Z2$. Thus the assumed condition $\mu(Z_{i}) = \mu(Z_{i}+ Z_{j}) = 0$ implies that $\p(\O)=0$.
\end{proof}

As in the linear case (section \ref{sec:linear unitality}) unitality is important because the structures of null sets that rule out unital co-events also cause problems for consistency. More precisely, if a system that does not admit unital quadratic co-events is coupled to a second system with which it does not interact, we would not be able to coarse grain out the first system to focus on events in the second. This would mean that we could not make any statements without considering the measures of every subsystem that does not admit unital co-events; essentially we would have to know the fine grained measure of the whole universe to make any assertion.

\begin{example}\label{example:quadratic unitality problem}
Let $\Omega$ be a tensor product sample space: $\Omega=\Omega_{1}\times\Omega_{2}$, with the associated measure $\mu=\mu_{1}\mu_{2}$ (so that $\O_1$ and $\O_2$ are non-interacting). Now let $\{Z_{i}\}$ be a set of subsets of $\Omega_{1}$ satisfying:
\begin{eqnarray}
\sum_i Z_{i} &=& \Omega_{1}, \nonumber \\
\mu_{1}(Z_{i}) &=& 0, \nonumber \\
\mu_{1}(Z_{i}+ Z_{j}) &=& 0. \nonumber \\
\end{eqnarray}
Then if $\phi\in\Q$ and $B$ is any subset of $\O_2$ we have:
\begin{eqnarray}
\mu(Z_i\times B) &=& \mu_{1}(Z)\times\mu_{2}(B), \nonumber \\
\Rightarrow \mu(Z_{i}\times B) &=& 0, \nonumber \\
\Rightarrow \phi(  Z_{i}\times B) &=& 0, \nonumber \\
\text{Further,} &&\nonumber \\
\mu(Z_{i}+ Z_{j}\times B) &=& 0, \nonumber \\
\Rightarrow \phi(Z_{i}+ Z_{j}\times B) &=& 0, \nonumber \\
\therefore \phi(\Omega_{1}\times B) &=& \phi(\sum_{i}Z_{i}\times B), \nonumber \\
&=& N_{1}\sum_{i}\phi(Z_{i}\times B) + N_{2}\sum_{i\neq j}\phi(Z_{i}+ Z_{j}\times B), \nonumber \\
&=& 0. \nonumber
\end{eqnarray}
\flushright{$\square$}
\end{example}

As the next example shows, we can find `toy model' history theories that contain these `problem structures'.

\begin{example}\label{example:quadratic unitality counterexample}
We can construct a histories theory $\H$ with a strongly positive measure which admits $5$ measure zero sets $Z_i$ satisfying the requirements of Lemma \ref{lemma:quadratic problem structure}, hence this system will not admit unital quadratic co-events, and will cause problems for consistency. The sample space is to be a sum of null sets $Z_i$ whose pairwise sums $Z_i+Z_j$ are also null.
\bea
\sum_i Z_i &=& \O, \nonumber \\
\mu(Z_i) &=& 0, \nonumber \\
\mu(Z_i+Z_j) &=& 0. \nonumber
\eea
We populate $\O$ with sixteen elements, $\{a_i\}_{i=1}^5$, $\{a_{ijk}\}_{1\leq i <j<k\leq 5}$ and $a_{12345}$ as follows,
\bea
a_i \in Z_l &\Leftrightarrow& l=i, \nonumber \\
a_{ijk} \in Z_l &\Leftrightarrow& l\in\{i,j,k\}, \nonumber \\
a_{12345} \in Z_l &\Leftrightarrow& \forall l. \nonumber
\eea
Note that there are five elements $a_i$, which each lie in exactly one set $Z_i$, ten elements $a_{ijk}$, which each lie in exactly three sets $Z_l$ and one element $a_{12345}$ which lies in all five of the $Z_l$. Conversely we see that the null set $Z_i$ contains the element $a_i$, the six elements $a_{lmn}$ where $i\in\{l,m,n\}$ and the element $a_{12345}$. The null set $Z_i+Z_j$ contains $a_i,~a_j$ and the six elements $a_{lmn}$ where exactly one of $i,j$ is an element of $\{l,m,n\}$.

We now define a Hilbert space related to our system, using its structure to construct a strongly positive decoherence functional on $\EA=P\O$. We define ${\cal{H}}_\chi$ to the space of formal linear combinations of elements of $\O$ with real coefficients, essentially meaning that ${\cal{H}}_\chi$ is a $|\O|=16$ dimensional vector space with a canonical basis whose elements we can identify with the elements of $\O$:
\bea
a\in\O &\rightarrow& \ket{a}{\cal{H}}_\chi, \nonumber \\
\braket{a}{b} &=& \delta_{ab}~\forall a,b\in\O. \nonumber
\eea
Then we can define the vector (in ${\cal{H}}_\chi$):
\beq
\ket{\chi} = \sum_i N \ket{a_i} + \sum_{ijk} -\frac{N}{3} \ket{a_{ijk}} + N \ket{a_{ijkmn}},
\eeq
where $N=3/8$ is a normalisation factor. For computational ease we will enumerate the elements of $\O$, relabeling them $d_i$ as follows:
\beq
d_i = \left\{\ba{cc} a_i & 1\leq i \leq 5 \\ a_{p_iq_ir_i} & 5 < i \leq 15 \\ a_{12345} & i = 16, \ea\right.
\eeq
where the $a_{p_iq_ir_i}$  are all distinct. Then defining $\ket{d_i}$ in the obvious manner we have
\beq
\ket{\chi} = \sum_i c_i \ket{d_i},
\eeq
where
\beq
c_i = \left\{\ba{cc} N & 1\leq i \leq 5 \\ -N/3 & 5 < i \leq 15 \\ N & i = 16. \ea\right.
\eeq
We are now in a position to define our decoherence functional, which we construct from the operator $\ket{\chi}\bra{\chi}$:
\bea
{\textbf{D}_\chi}(d_i,d_j) &=& \bra{d_i}{\textbf{D}_\chi}\ket{d_j}, \nonumber \\
&=& \braket{d_i}{\chi}\braket{\chi}{d_j},\nonumber \\
&=& c_ic_j.
\eea
Because of its form, ${\textbf{D}_\chi}$ is strongly positive. It is easy to check that the associated measure is null on the sets $Z_i$, $Z_i+Z_j$. In blocks of  size $(5, 10, 1)\times (5,10,1)$ the matrix is:

$$
{\textbf{D}_\chi} =\frac{9}{64}
\left(\begin{array}{c|ccc}
 &5&10&1 \\
 \hline
 5&1&-1/3&1 \\
 10&-1/3&1/9&-1/3 \\
 1&1&-1/3&1
\end{array} \right)
$$

%\beq
%D_{\chi}=\frac{1}{9}
%\left(\begin{array}{cccccccccccccccc}
%9 & 9 & 9 & 9 & 9 & -3 & -3 & -3 & -3 & -3 & -3 & -3 & -3 & -3 & -3 & 9 \\
%9 & 9 & 9 & 9 & 9 & -3 & -3 & -3 & -3 & -3 & -3 & -3 & -3 & -3 & -3 & 9 \\
%9 & 9 & 9 & 9 & 9 & -3 & -3 & -3 & -3 & -3 & -3 & -3 & -3 & -3 & -3 & 9 \\
%9 & 9 & 9 & 9 & 9 & -3 & -3 & -3 & -3 & -3 & -3 & -3 & -3 & -3 & -3 & 9 \\
%9 & 9 & 9 & 9 & 9 & -3 & -3 & -3 & -3 & -3 & -3 & -3 & -3 & -3 & -3 & 9 \\
%-3 & -3 & -3 & -3 & -3 & 1 & 1 & 1 & 1 & 1 & 1 & 1 & 1 & 1 & 1 & -3 \\
%-3 & -3 & -3 & -3 & -3 & 1 & 1 & 1 & 1 & 1 & 1 & 1 & 1 & 1 & 1 & -3 \\
%-3 & -3 & -3 & -3 & -3 & 1 & 1 & 1 & 1 & 1 & 1 & 1 & 1 & 1 & 1 & -3 \\
%-3 & -3 & -3 & -3 & -3 & 1 & 1 & 1 & 1 & 1 & 1 & 1 & 1 & 1 & 1 & -3 \\
%-3 & -3 & -3 & -3 & -3 & 1 & 1 & 1 & 1 & 1 & 1 & 1 & 1 & 1 & 1 & -3 \\
%-3 & -3 & -3 & -3 & -3 & 1 & 1 & 1 & 1 & 1 & 1 & 1 & 1 & 1 & 1 & -3 \\
%-3 & -3 & -3 & -3 & -3 & 1 & 1 & 1 & 1 & 1 & 1 & 1 & 1 & 1 & 1 & -3 \\
%-3 & -3 & -3 & -3 & -3 & 1 & 1 & 1 & 1 & 1 & 1 & 1 & 1 & 1 & 1 & -3 \\
%-3 & -3 & -3 & -3 & -3 & 1 & 1 & 1 & 1 & 1 & 1 & 1 & 1 & 1 & 1 & -3 \\
%-3 & -3 & -3 & -3 & -3 & 1 & 1 & 1 & 1 & 1 & 1 & 1 & 1 & 1 & 1 & -3 \\
%9 & 9 & 9 & 9 & 9 & -3 & -3 & -3 & -3 & -3 & -3 & -3 & -3 & -3 & -3 & 9 \\
%\end{array}\right)
%\eeq

(each block has all entries equal)

Although we have involved a Hilbert space in the construction of our theory, we have not explicitly constructed a Hilbert space theory that gives rise to our histories theory. To do this, we would have to find a Hilbert space containing an initial state $\ket{\psi}$ and at least sixteen sequences of projectors, with a sequence corresponding to each of the elements $d_i\in\O$ acting on $\ket{\psi}$ to yield an analogue of $\ket{d_i}$. We construct such a Hilbert space theory in the next section.
\flushright{$\square$}
\end{example}

\subsection{The Sixteen Slit System}\label{sec:sixteen slit}

As in the linear case (section \ref{sec:four slit}) we can realise the measure outlined in example \ref{example:quadratic unitality counterexample} as part of the measure of an idealised sixteen slit system, a further generalisation of the double, triple and four slit systems we have encountered so far. This leads us, as it does in the linear case, to deduce a lack of compatibility with experimentally testable predictions of quantum mechanics.

We will need sixteen slits, $\{A_i\}_{i=1}^{16}$, to realise this system, corresponding to the dimensions of the matrix of the decoherence functional defined in example \ref{example:quadratic unitality counterexample} above. As before we denote the initial state $\ket{\psi}$ and the projector corresponding to finding the particle at slit $A_i$ upon measurement by $P_i$, with the associated state $\ket{A_i}$ such that $P_i=\ket{A_i}\bra{A_i}$; note that the vectors corresponding to the sixteen slits are mutually orthogonal. We define the projectors $P_D$ and the associated state $\ket{D}$ in a similar fashion. The projector corresponding to $\overline{D}$ is defined by $P_{\overline{D}} = \mathbb{I}-P_D$.

As before we define our minimal Hilbert space theory $({\cal{H}},H,\ket{\psi},T)$ by setting ${\cal{H}} = span(\ket{A_i})$, setting $H$ to be the null operator, so our evolution operators are the identity, and setting our temporal support to be $T = \{0,1,2\}$, consisting of an initial time, an intermediate time at which our projectors $P_i$ act, and a final time at which our projectors $P_D,P_{\overline{D}}$ act. By specifying the initial state $\ket{\psi}$ and the detector state $\ket{D}$ appropriately we are able to realise the decoherence functional constructed in example \ref{example:quadratic unitality counterexample} as the restriction to the histories ending at the detector of the decoherence functional of the histories theory corresponding to this Hilbert space theory.

We start with the sixteen real numbers $c_i$ used in example \ref{example:quadratic unitality counterexample}. In appendix \ref{appendix:many slit} we show that our Hilbert space theory is gedankenexperimentally realisable if we set:
\bea
\ket{\psi} &=& N_\psi \sum_i \sqrt{|c_i|} \ket{A_i} \nonumber \\
\ket{D}    &=& N_\psi    \sum_i sign(c_i)\sqrt{|c_i|} \ket{A_i}, \nonumber
\eea
where $sign(x) = 1$ when $x\geq 0$ and $-1$ otherwise. $N_\psi$ is a normalisation factor set so that $\braket{\psi}{\psi}=1$.

Denoting the path that passes through slit $A_i$ to reach $D$ ($\overline{D}$) by $D_i$ ($\overline{D}_i$), our sample space is now $\O=\{D_i\}_{i=1}^{16}\sqcup \{\overline{D}_i\}_{i=1}^{16}$ and our event algebra is $\EA=P\O$ as before. We can define the events $A_i,D,\overline{D}$ in the obvious manner, $A_i=\{D_i,\overline{D_i}\}_{i=1}^{16}$, $D=\{D_i\}_{i=1}^{16}$ etc. Our decoherence functional is, in block matrix form
$$
{\textbf{D}_\mu} = N\left(\begin{array}{c|c}
{\textbf{D}_\chi} & 0 \\
\hline
0 & E \end{array}\right)
$$
where $N$ is a normalisation factor, ${\textbf{D}_\chi}$ is the matrix we saw in example \ref{example:quadratic unitality counterexample} and $E$ is a $16\times 16$ matrix admitting no null sets. Thus the null sets of this decoherence functional will correspond to and will possess the same structure as the null sets of the decoherence functional ${\textbf{D}_\chi}$.
Now restricting to the paths ending at the detector we see that the null set structure of our decoherence functional ${\textbf{D}_\mu}$ is equivalent to that of the decoherence functional ${\textbf{D}_\chi}$; in fact ${\textbf{D}_\mu}(D_i,D_j)=N{\textbf{D}_\chi}(d_i,d_j)$. Thus the set $D\subset\O$ can be decomposed into a sum of null sets $Z_i$ whose pairwise sums are null, and the logic of lemma \ref{lemma:quadratic problem structure} implies that no unital (ie unit valued on $D$) quadratic co-event can be entirely constructed from the duals of the elements of $D$. Thus any quadratic co-event constructed entirely from the duals of the elements of $D$ will map $D$ to zero.  Now as in our previous many-slit gedankenexperiments the fine grained histories $\overline{D_i}$ are not elements of any null set so their duals $\overline{D_i}^*$ are classical co-events, which also means these histories will not participate in any other primitive co-events. Putting all this together means there can be no primitive quadratic co-event mapping $D$ to one. However $\{D,\overline{D}\}$ is a decoherent and measurable partition, the event $D$ is observable and does not have zero measure, thus putting the quadratic scheme in an experimentally testable contradiction with quantum mechanics.

\section{Higher Order Polynomial Schemes}\label{sec:higher order polynomial schemes}

\subsection{Basic Properties}

Following the failure of the linear and the quadratic schemes we generalise to an $n$th order polynomial scheme, as before assuming finite sample spaces throughout this section. We will refer to a co-event that obeys the $n$th order sum rule (equation \ref{eq:order n additive rule}) as an \emph{order $n$ co-event}, and will first justify our assertion in section \ref{sec:order n coevents} that order $n$ co-events are degree $n$ polynomials in the classical co-events. To prove the result we need two technical lemmas:

\begin{lemma}\label{lemma:polynomial coevent action decomosition}
Let $\H$ be a histories theory with a finite sample space. Then for every $A\in\EA$ and every $n<|\O|$ there exist events $A_i\in\EA$ such that
\bea
%A_i &\subset& A \nonumber \\
|A_i| &\leq& n, \nonumber \\
\p(A) &=& \sum_i \p(A_i), \label{eq:polynomial coevent action decomposition}
\eea
for all order $n$ co-events $\p\in\CE$.
\end{lemma}
\begin{proof}
We proceed by induction on the cardinality of $A$. If $|A|\leq n$ we are done, so assume that equation \ref{eq:polynomial coevent action decomposition} holds for all $A\in\EA$ such that $|A|< m$ for some integer $m>n$. Now consider the event

$$B = \{\g_1,\g_2,\ldots,\g_m\}.$$

We can partition $B$ into $n$ components $B_i$:

$$B_i = \left\{\ba{cc} \{\g_i\} & 1\leq i \leq n \\ \{\g_n,\ldots,\g_m\} & i=n+1.\ea\right.$$

Then the $B_i$ are disjoint and $B=\bigsqcup_{i=1}^{n+1} B_i$ so for any order $n$ co-event $\p$ we can use the order $n$ sum rule to yield
\bea
\p(B) = \p(\bigsqcup_{i=1}^{n+1} B_i) &=& \sum_j\p(\sum_{i\neq j} B_i) + \ldots \nonumber \\
&& + \sum_{j_1,\ldots,j_p}\p(\sum_{i\not\in\{j_1,\ldots,j_p\}} B_i) + \ldots \nonumber \\
&& + \sum_i \p(B_i), \nonumber \\
&=& \sum_k \p(C_k), \nonumber
\eea
where $C_1 = B_1\sqcup B_2 \sqcup\ldots\sqcup B_n$ and so on. Note that the $C_k$ are independent of the choice of $\p$, depending only on $\p$'s compliance with the order $n$ sum rule. But $|C_k|<m$ for all $k$, thus by the inductive assumption for each $C_k$ there exist events $A^k_l$ that obey equation \ref{eq:polynomial coevent action decomposition} for all order $n$ co-events. But then
\beq\nonumber
\p(B) = \sum_{k,l} \p(A^k_l),
\eeq
for all order $n$ co-events $\p$, hence the result.
\end{proof}

The proof of the following lemma is a technical \& uninformative argument which we will suppress:

\begin{lemma}\label{lemma:monomials obey n-sum rule}
Let $\H$ be a histories theory with a finite sample space and let $\p\in\CE$ be a non-zero monomial polynomial of degree less than or equal to $n$,
$$\p = \g_1^*\g_2^*\ldots\g_m^*~m\leq n.$$
Then $\p$ obeys the order $n$ sum rule.
\flushright{$\square$}
\end{lemma}

We are now in a position to prove the main result of this section:

\begin{theorem}\label{thm:n-polynomial coevents are degree n polynomials}
Let $\H$ be a histories theory with a finite sample space. Then $\p\in\CE$ obeys the order $n$ sum rule if and only if it is a polynomial of degree at most $n$ in the classical co-events.
\end{theorem}
\begin{proof}
First assume that $\p$ is a polynomial of degree at most $n$ in the classical co-events. It is easy to see that the sum of two co-events $\psi_1,\psi_2$ obeying the order $n$ sum rule itself obeys the order $n$ sum rule:
\bea
(\psi_1+\psi_2)(\sum_{i=1}^{n+1} A_{i})&=&\psi_1(\sum_{i=1}^{n+1} A_{i})+\psi_2(\sum_{i=1}^{n+1} A_{i}),\nonumber \\
&=&\sum_{m=1}^n \sum_{i_j<i_{j+1}} \psi_1(\sum_{j=1}^m A_{i_j}), \nonumber \\
&& + \sum_{m=1}^n \sum_{i_j<i_{j+1}} \psi_2(\sum_{j=1}^m A_{i_j}), \nonumber \\
&=& \sum_{m=1}^n \sum_{i_j<i_{j+1}} (\psi_1+\psi_2)(\sum_{j=1}^m A_{i_j}). \nonumber
\eea
Now since $\p$ is a polynomial of degree at most $n$ in the classical co-events, it is a sum of monomials of degree at most $n$. But then by lemma \ref{lemma:monomials obey n-sum rule} each of these monomials obeys the order $n$ sum rule, and by the above so does $\p$.

To prove the converse we now assume that $\p\in\CE$ obeys the $n$th order sum rule. By lemma \ref{lemma:classical coevents generate all coevents} $\p$ can be expressed as a polynomial in the classical co-events:
\beq\label{eq:polynomial coevent expression}
\p = \sum_{m=1}^{|\O|} \sum_I n_I^\p \g_{i_1}^*\ldots\g_{i_m}^*,
\eeq
for $n_I^\p \in\Z2$. Now given some $l\leq|\O|$ define the co-event $\p_l$ by `throwing away' the summands in equation \ref{eq:polynomial coevent expression} of degree greater than $l$,
\beq\label{eq:polynomial coevent expression cutoff}
\p_l = \sum_{m=1}^{l} \sum_I n_I^\p \g_{i_1}^*\ldots\g_{i_m}^*.
\eeq

Now noticing that a degree $l$ monomial co-event will always be zero valued on an event of cardinality less than $l$ we see that for $A\in\EA$ such that $|A|\leq l$ we have:
\bea
\p(A) &=& \sum_{m=1}^{|\O|}\sum_I(n_I^\p \g_{i_1}^*\ldots\g_{i_m}^*(A) ),\nonumber \\
&=& \sum_{m=1}^{l}\sum_I(n_I^\p \g_{i_1}^*\ldots\g_{i_m}^*(A) ),\nonumber \\
&=& \p_l(A). \label{eq:polynomial cutoff coevet equals coevent}
\eea
Now let $A$ be any element of $\EA$. By lemma \ref{lemma:polynomial coevent action decomosition} we can find events $A_i$ obeying $|A_i|<n$ such that
$$\psi(A) = \sum \psi(A_i),$$
for all $\psi\in\CE$ obeying the order $n$ sum rule. As this category includes both $\p$ (by assumption) and $\p_n$ (by the first part of the proof, because it is a polynomial of order at most $n$), using equation \ref{eq:polynomial cutoff coevet equals coevent} we have:
\bea
\p(A) &=& \sum_i \p(A_i), \nonumber \\
&=& \sum_i \p_n(A_i), \nonumber \\
&=& \p_n(A).
\eea
\end{proof}

We have an immediate corollary:

\begin{corollary}\label{corollary:general polynomial scheme Pn-1 subset of Pn}
Let $\H$ be a histories theory with a finite sample space, then if $\p\in\CE$ obeys the order $n$ sum rule then $\p$ obeys the order $n+k$ sum rule for any integer $k\geq 0$.
\end{corollary}

Put together with our definition of primitivity, corollary \ref{corollary:general polynomial scheme Pn-1 subset of Pn} ensures that $\Pn{n-1}\subset\Pn{n}$. Thus corollary \ref{corollary:quadratic existence} ensures existence for all $n$-polynomial schemes where $n\geq 2$. Further, using theorem \ref{thm:n-polynomial coevents are degree n polynomials} we can show that primitivity ensures weak emergent classicality.

\begin{lemma}\label{lemma:general_polynomial scheme_emergent_classicality}
Let $\H$ be a histories theory. Then:
$$\mu(A\sqcup B)=\mu(A)+\mu(B)~\forall ~A,B\subset\Omega~\Rightarrow \Pn{n}=\C.$$
\end{lemma}
\begin{proof}
Using theorem \ref{thm:n-polynomial coevents are degree n polynomials} we know that $\p\in\Pn{n}$ is a polynomial of degree at most $n$. Then we can generalise the proof of lemma \ref{lemma:quadratic scheme classical null sets} to show that $\p$ is a preclusive homomorphism and hence an element of $\C$. Finally we note that our definition of primitivity (along with corollary \ref{corollary:general polynomial scheme Pn-1 subset of Pn}) ensures that $\Pn{n-1}\subset\Pn{n}$, so $\C\subset\Pn{n}$. Hence the result.
\end{proof}

\subsection{The Failure of the Polynomial Schemes}

The higher order schemes share the weaknesses as well as the strengths of the quadratic scheme. Now, since $\mu(\Omega)\neq 0$ the scheme $\Pn{|\Omega|}$ always contains $\Omega^*:=\prod_{\gamma\in\Omega}\gamma^*$, which is trivially unital. However given any $n$ we can always find a histories theory containing an `$n$th order problem structure', in a simple generalisation of the problems with the linear and quadratic cases.

\begin{lemma}\label{lemma:general polynomial problem structure}
Let $\H$ be a histories theory with a finite sample space If there exists $\{Z_i\}_{i=1}^m\subset \O$ obeying the following two conditions:
\begin{enumerate}
\item The $Z_i$ sum to the sample space:
\beq\label{eq:Zi sum to Omega}
\sum_{i=1}^m Z_i = \O.
\eeq
\item For $1\leq p\leq n$, the sum of any $p$ of the $Z_i$ is of measure zero:
\beq\label{eq:Zi sum measure}
\mu(\sum_{\a=1}^p Z_{i_\a}) = 0~~~\forall 1\leq p\leq n,~\{i_\a\}_{\a=1}^p,
\eeq
\end{enumerate}
then there are no preclusive unital $n$-polynomial co-events describing this theory.
\end{lemma}
\begin{proof}
First note that $m>n$, otherwise our two conditions would mean that $\mu(\O)=0$. Then because of the first condition, equation \ref{eq:Zi sum to Omega}, we can use the generalised sum rule to decompose the action of any $n$-polynomial co-event $\p$ on the sample space into sums of its actions on the sets $Z_i$ and their sums:
$$\p(\O) = \sum_{p=1}^n(\sum_{i_\a~distinct} n_{i_1\ldots i_p} \p(\sum_{i_\a}Z_{i_\a})),$$
where $n_{i_1\ldots i_p}\in\Z2$. Thus the second condition, equation \ref{eq:Zi sum measure}, implies that $\p(\O)=0$.
\end{proof}

As before, it is not so much that lack of unitality itself that can cause problems, but that the structures of null sets assumed in the proof of lemma \ref{lemma:general polynomial problem structure} also cause problems for consistency.

\begin{example}\label{example:general polynomial scheme unitality problem structure}
Consider the $n$th order polynomial scheme. Let $\Omega$ be a tensor product sample space: $\Omega=\Omega_{1}\times\Omega_{2}$, with the associated measure $\mu=\mu_{1}\mu_{2}$ (so that $\O_1$ and $\O_2$ are non-interacting). Now let $\{Z_{i}\}_{i=1}^m$ be a set of subsets of $\Omega_{1}$ satisfying:
\begin{eqnarray}
\sum_i Z_{i} &=& \Omega_{1}, \nonumber \\
\mu(\sum_{\alpha=1}^p Z_{i_\alpha}) &=& 0. \nonumber
\end{eqnarray}
For $1\leq p \leq n$. Then if $\phi\in\Pn{n}$ and $B$ is any subset of $\O_2$ we have:
\begin{eqnarray}
\mu(\sum_{\alpha=1}^p Z_{i_\alpha}) &=& 0, \nonumber \\
\Rightarrow \phi(\sum_{\alpha=1}^p Z_{i_\alpha}) &=& 0, \nonumber \\
\therefore \phi(\Omega_{1}\times B) &=& \phi(\sum_{i}Z_{i}\times B), \nonumber \\
&=& \sum_j N_j \sum_{i_\alpha distinct} \phi(\sum_{\alpha=1}^j Z_{i_\alpha} \times B), \nonumber \\
&=& 0. \nonumber
\end{eqnarray}
Where $N_j\in\Z2$.
\flushright{$\square$}
\end{example}

As before this is fatal for consistency under coarse graining, meaning that we would have to consider the full fine grained measure of the whole universe before deriving a co-event to describe a small (semiclassical) subsystem such as a laboratory experiment. The next theorem shows that given any $n$, we can find a histories theory containing the `$n$th order problem structure', thus causal fatal problems for $\Pn{n}$'s consistency under coarse graining. This has led to the abandonment of the polynomial schemes.

\begin{theorem}\label{thm:general poynomial scheme unitality problem}
Given $n$ (representing the degree of our candidate polynomial co-event scheme) we can find a $m\in\mathbb{Z}$, and a histories theory $\H$ with a finite sample space and a strongly positive measure such that $\exists$ subsets $\{Z_{i}\}_{i=1}^m$ of $\O$ obeying the conditions of lemma \ref{lemma:general polynomial problem structure}, equations \ref{eq:Zi sum to Omega} \& \ref{eq:Zi sum measure}
\end{theorem}

\begin{proof}
We simply generalise the argument employed in example \ref{example:quadratic unitality counterexample} for the quadratic case.

Given $n$, fix $m$ for now. We proceed by generalising the argument we used for the quadratic scheme in section \ref{sec:quadratic failure}. Thus we start by constructing a sample space $\O$ with $m$ subsets $Z_i$ such that $\sum_{i=1}^m Z_i=\O$. We populate the space as follows, let:
\begin{eqnarray}
a_i^{(1)}\in Z_j &\Leftrightarrow& i=j, \nonumber \\
a_{i_1 i_2}^{(2)} \in Z_j &\Leftrightarrow& j\in\{i_1,i_2\}, \nonumber \\
a_{i_1 i_2 i_3}^{(3)} \in Z_j &\Leftrightarrow& j\in\{i_1,i_2,i_3\}, \nonumber \\
\vdots \nonumber \\
a_{1 \ldots m}^{(m)} \in Z_j && \forall j. \nonumber
\end{eqnarray}
We will use $a^{(q)}$ to refer to a general element $a^{(q)}_{i_1 \ldots i_q}$. Notice that there are $n(q) = {|\O|\choose q}$ distinct elements $a^{(q)}$ for each $q$.

Now $a^{(q)}\in\sum Z_i \Leftrightarrow q$ odd, so to enforce equation \ref{eq:Zi sum to Omega} we remove the even $a^{(q)}$'s:
\begin{equation}\nonumber
\Omega=\{a^{(q)}|q~odd\}.
\end{equation}
Now by symmetry the number of $a^{(q)}$'s in each term $\sum_{\alpha=1}^p Z_{i_\alpha}$ is independent of the actual choice of $Z_i$'s, we will call it $N_{pq}(n)$.

We now construct a Hilbert space from our system, using its structure to define a strongly positive decoherence functional on $\EA=P\O$. We define ${\cal{H}}_\chi$ to the space of formal linear combinations of elements of $\O$ with real coefficients, essentially meaning that ${\cal{H}}_\chi$ is a $|\O|$ dimensional vector space with a canonical basis whose elements we can identify with the elements of $\O$:
\bea
a\in\O &\rightarrow& \ket{a}\in {\cal{H}}_\chi, \nonumber \\
\braket{a}{b} &=& \delta_{ab}~\forall a,b\in\O. \nonumber
\eea
Then we can define the vector:
\beq
\ket{\chi} = \sum_{q odd} C_q (\sum_{i_1<i_2<\ldots<i_q} \ket{a^{(q)}_{i_1\ldots i_q}}) ,
\eeq
for some $C_q\in\R$. For computational ease it will be convenient to linearly order the $a^{(q)}$'s, relabeling them $\{d_i\}_{i=1}^{|\O|}$. We require that the $d_i$ ordering respect the (partial) ordering by the label $q$, so that if $d_i = a^{(q)}$ and $d_j = a^{(p)}$ then $q<p\Rightarrow i<j$:
\beq
d_i = \left\{\ba{cc} a^{(1)}_j & 0 < i \leq n(1) \\
a^{(2)}_{j_1j_2} & n(1)< i \leq n(1) + n(2) \\
\vdots \\
a^{(q)}_{j_1\ldots j_q} & \sum_{p=1}^{q-1} n(p) < i \leq \sum_{p=1}^{q} n(p) \\
\vdots \\
a_{1 \ldots m}^{(m)} & i= |\O|. \ea\right.
\eeq
The ordering induced by $d_i$ within each set of $a^{(q)}$'s is irrelevant to our purposes. Then following the above we can define $c_i\in\R$
\beq
c_i = C_q  ~~for~~ \sum_{p=1}^{q-1} n(p) < i \leq \sum_{p=1}^{q} n(p).
\eeq
%\beq
%c_i = \left\{\ba{cc}
%C_1 & 0 < i \leq n(1) \\
%C_2 & n(1) < i \leq n(1)+n(2) \\
%\vdots \\
%C_q  & \sum_{p=1}^{q-1} n(p) < i \leq \sum_{p=1}^{q} n(p) \\
%\vdots \\
%C_m & i= |\O|. \ea\right.
%\eeq
Then defining $\ket{d_i}$ in the obvious manner we have
\beq
\ket{\chi} = \sum_i c_i \ket{d_i}.
\eeq
We are now in a position to define our decoherence functional, which we construct from the operator $\ket{\chi}\bra{\chi}$:
\bea
{\textbf{D}_\chi}(d_i,d_j) &=& \braket{d_i}{\chi}\braket{\chi}{d_j}\nonumber \\
&=& c_ic_j.
\eea
Thus a specification of the $c_i$, or the underlying $C_q$, will determine the measure. Because of its form, ${\textbf{D}_\chi}$ is strongly positive. Now if we associate with every $A\in\EA$ the vector $\ket{A} = \sum_{a\in A}\ket{a}$, we see that
\bea
{\textbf{D}_\chi}(A,A) &=& \sum_{\{i,j|d_i,d_j\in A\}}c_ic_j, \nonumber \\
&=& (\sum_{d_i\in A} c_i )^2, \nonumber \\
&=& \braket{A}{\chi}\braket{\chi}{A} .\label{eq:polynomial thm proof action of D in H}
\eea
This using equation \ref{eq:polynomial thm proof action of D in H} we can impose the requirement that this measure conform to equation \ref{eq:Zi sum measure} as constraint equations:
\beq\label{eq:general polynomial scheme theorem quadratic constraint}
\mu(\sum_{\a=1}^p Z_{i_\alpha})=0 \Rightarrow \braket{\sum_{\alpha=1}^p Z_{i_\alpha}}{\chi}\braket{\chi}{\sum_{\alpha=1}^p Z_{i_\alpha}} = \delta_{mp}.
\eeq
Note that the $p=m$ equation ensures the required $\mu(\O)=1$ (from equation \ref{eq:decoherence functional unitality}). This yields quadratic constraints in the $c_i$, and thus the $C_q$. However we can rephrase it in terms of simpler, though stronger, linear constraints. We can impose:
\beq
\braket{\sum_{\alpha=1}^p Z_{i_\alpha}}{\chi} = 0 \Leftrightarrow p\neq m.
\eeq
It is easy to see that this implies the $p\neq m$ equations in \ref{eq:general polynomial scheme theorem quadratic constraint}, and thus equation \ref{eq:Zi sum measure}. This is equivalent to:
\beq\label{eq:general polynomial scheme theorem constraint}
\sum_{\substack{q=1\\q~odd}}^m N_{pq}C_q = 0 \Leftrightarrow p\neq m,
\eeq
to which we add the $p=m$ equation, which enforces the condition $\mu(\O)=1$ (from equation \ref{eq:decoherence functional unitality}):
\beq
\sum_{i=1}^{|\O|} c_i = \sum_{q odd} n(q) C_q = 1.
\eeq
Now this is simply a system of linear equations in the real variables $C_q$. We have $m/2$ unknowns if $m$ is even, and $(m+1)/2$ if $m$ is odd. However we have $n$ equations, one for each condition in equation \ref{eq:Zi sum measure}. Further we are given $n$, and can choose $m$ to be arbitrarily large. Thus we can always find such a system satisfying the constraints in equations \ref{eq:Zi sum to Omega} and \ref{eq:Zi sum measure}, and so for any $n$ we can find an $m$ such that we can construct a histories theory satisfying the requirements of the theorem.
\end{proof}

We have the immediate corollary:

\begin{corollary}
For all $n$ there exists a histories theory $\H$ with a finite sample space and a strongly positive measure that does not admit unital $n$-polynomial co-events.
\end{corollary}

\subsection{The Many Slit System}\label{sec:many slit}

As a final nail in the coffin of the polynomial schemes, we can generalise the gedanken experiments outlined in sections \ref{sec:four slit} \& \ref{sec:sixteen slit} that show the incompatibility of the linear \& quadratic schemes with quantum mechanics.

Starting with the general $n$-polynomial scheme theorem \ref{thm:general poynomial scheme unitality problem} gives us an integer $m$ and a histories theory $\HL$ with a strongly positive measure which contains events $\{Z_i\}$ such that
\begin{enumerate}
\item The $Z_i$ sum to the sample space:
\beq\label{eq:Zi sum to lambda}
\sum_{i=1}^m Z_i = \la.
\eeq
\item For $1\leq p\leq n$, the sum of any $p$ of the $Z_i$ is of measure zero:
\beq\label{eq:lambda Zi sum measure}
\mu(\sum_{\a=1}^p Z_{i_\a}) = 0 ~~~\forall 1\leq p \leq n,~\{i_\a\}_{\a=1}^p.
\eeq
\end{enumerate}
We can realise this as a part, though not a coarse graining of, an idealised many slit system. As with the four and sixteen slit systems (sections \ref{sec:four slit} \& \ref{sec:sixteen slit} respectively), we can find a many slit gedanken experiment represented by a histories theory $\H$ where $\la$ is a subset rather than a partition of $\O$, though we still get $\EA_\la\subset\EA$ and $\mu_\la = \mu|_{\EA_\la}$. In fact the restriction from $\EA$ to $\EA_\la$ will be achieved by `postconditioning' on the final state.

Given $\Pn{n}$, we will need $m$ slits $\{A_i\}_{i=1}^m$ to realise this system, corresponding to the number $m$ of null sets given by theorem \ref{thm:general poynomial scheme unitality problem}. As before we denote the initial state $\ket{\psi}$ and the projector corresponding to finding the particle at slit $A_i$ upon measurement by $P_i$, with the associated state $\ket{A_i}$ such that $P_i=\ket{A_i}\bra{A_i}$; note that the vectors corresponding to the slits are mutually orthogonal. We define the projectors $P_D$ and the associated state $\ket{D}$ in a similar fashion. The projector corresponding to $\overline{D}$ is defined by $P_{\overline{D}} = \mathbb{I}-P_D$.

As before we define our minimal Hilbert space theory $({\cal{H}},H,\ket{\psi},T)$ by setting ${\cal{H}} = span(\ket{A_i})$, setting $H$ to be the null operator, so our evolution operators are the identity, and setting our temporal support to be $T = \{0,1,2\}$, consisting of an initial time, an intermediate time at which our projectors $P_i$ act, and a final time at which our projectors $P_D,P_{\overline{D}}$ act. By specifying the initial state $\ket{\psi}$ and the detector state $\ket{D}$ appropriately we are able to realise ${\textbf{D}_\chi}$ as the restriction to the histories ending at the detector of the decoherence functional of the histories theory $\H$ corresponding to this Hilbert space theory. These histories will all have $P_D$ as their final time projector, and as mentioned above such a restriction can be referred to as a postconditioning on the final state.

We start with the real numbers $c_i$ found in theorem \ref{thm:general poynomial scheme unitality problem}. In appendix \ref{appendix:many slit} we show that our Hilbert space theory is gedankenexperimentally realisable if we set:
\bea
\ket{\psi} &=& N_\psi \sum_i \sqrt{|c_i|} \ket{A_i}, \nonumber \\
\ket{D}    &=& N_\psi    \sum_i sign(c_i)\sqrt{|c_i|} \ket{A_i}, \nonumber
\eea
where $sign(x) = 1$ when $x\geq 0$ and $-1$ otherwise. $N_\psi$ is a normalisation factor set so that $\braket{\psi}{\psi}=1$.

This construction is a generalisation of the double, triple, quadruple and sixteen slit systems defined in sections \ref{sec:double slit},\ref{sec:triple slit},\ref{sec:four slit} \& \ref{sec:sixteen slit}. Although we used $\ket{\psi}=\frac{1}{\sqrt{m}}\ket{A_i}$ in defining the double, triple and quadruple slit systems, in those cases $c_i=\pm 1$, so the resulting coefficients were equivalent to those given here for $n\in\{1,2,3,4\}$.

Denoting the path that passes through slit $A_i$ to reach $D$ ($\overline{D}$) by $D_i$ ($\overline{D}_i$), our sample space is now $\O=\{D_i\}_{i=1}^{m}\sqcup \{\overline{D}_i\}_{i=1}^{m}$ and our event algebra is $\EA=P\O$ as before. We can define the events $A_i,D,\overline{D}$ in the obvious manner, $A_i=\{D_i,\overline{D_i}\}_{i=1}^{m}$, $D=\{D_i\}_{i=1}^{m}$ and so forth. Note that this means $\EA_\la = PD$. We can therefore partition the sample space in terms of the `detector events', $\O=D\sqcup\overline{D}$, which are observable in our gedankenexperiment (appendix \ref{appendix:many slit}).

As suggested by the notation we can identify the histories $D_i\in\O$ with the histories $d_i\in\la$, identifying $\la$ itself with the subset $D$ in $\O$. We can do this because
\bea
\ket{D_i} &=& P_DP_i\ket{\psi}, \nonumber \\
&=& \ket{D}\braket{D}{A_i}\braket{A_i}{\psi}, \nonumber \\
&=& (\sqrt{c_i})^2 sign(c_i) N_\psi^2 \ket{D}, \nonumber \\
&=& N_\psi^2 c_i\ket{D}, \nonumber
\eea
which means that
\beq\label{eq:many slit decoherence functional}
{\textbf{D}_\mu}(D_i,D_j) = N_\psi^2 \braket{D}{D}c_ic_j,
\eeq
where ${\textbf{D}_\mu}$ is the decoherence functional associated with the measure $\mu$. Then setting $N=N_\psi^2 \braket{D}{D}$ we get:
\bea
{\textbf{D}_\mu}(D_i,D_j) &=& N c_ic_j, \nonumber \\
&=& N {\textbf{D}_\chi}(d_i,d_j). \label{eq:polynomial many slit measure restricted to detector events}
\eea
Now the null sets $Z_i$ we constructed in the proof of theorem \ref{thm:general poynomial scheme unitality problem} will correspond to events in $\EA_\la$, and thus $\EA$, which we shall also refer to by $Z_i$. Using equation \ref{eq:polynomial many slit measure restricted to detector events} we can see that these sets $Z_i$, as well as the relevant sums thereof (see proof of theorem \ref{thm:general poynomial scheme unitality problem}) will also be null in $\H$; thus $\EA_\la$, and thus $\EA$, contains null sets conforming to the `problem structure' described by equations \ref{eq:lambda Zi sum measure} \& \ref{eq:Zi sum to lambda}.

Since every co-event can be expressed as a polynomial in the classical co-events, we can decompose any $\p\in\Pn{n}$ as follows:
$$\p = \p_D + \p_{D\overline{D}} + \p_{\overline{D}},$$
where $\p_D$ is a sum of monomials $D_{i_1}^*\ldots D_{i_p}^*$ made up of products of classical co-events $D_{i_j}^*$ whose duals lie in $D$. $\p_{\overline{D}}$ is similarly built from classical co-events whose duals lie in $\overline{D}$ and $\p_{D\overline{D}}$ is a sum of monomials each of which contains at least one term whose dual lies in $D$ and at least one term whose dual lies in $\overline{D}$. Then
$$\p_{\overline{D}}(D)=\p_{D\overline{D}}(D)=0,$$
so that
$$\p(D)=\p_D(D).$$
But the co-events $\p_D$, or more accurately their restriction to $\EA_\la=PD$, are precisely the co-events we would find if we took $D$ to be our sample space and considered the histories theory $(D,\EA_\la,\mu_D)$, where $\mu_D = N_D\mu|_D$, with normalisation factor $N_D$. Then since the null sets $Z_i$ in $\EA_\la$ satisfy equations \ref{eq:lambda Zi sum measure} \& \ref{eq:Zi sum to lambda} we can apply lemma \ref{lemma:general polynomial problem structure} to show that the structure of our null sets forces
$$\p(D)=0,$$
$\forall \p\in\Pn{n}$. However the partition $\{D,\overline{D}\}$ is measurable, corresponding to wether or not we have a detector reading. Further, the measure of the event $D$ is non-zero, so standard quantum mechanics predicts that a detector reading occurs with a non-zero probability. So for every $n$ we can find an gedankenexperiment (outlined in appendix \ref{appendix:many slit}) in which the $n$-polynomial scheme makes a falsifiable claim at odds with the standard predictions of quantum mechanics; as such it is not an assertion that the author is happy to support. This finding constitutes a violation of our compatibility criterion, and has led to further research into polynomial schemes being abandoned.

%Polynomial Schemes
\part{The Multiplicative Scheme}
%\singlespacing
\chapter{The Multiplicative Scheme}\label{chapter:the multiplicative scheme}
%\doublespacing

In this chapter we finally come to the multiplicative scheme, the most successful scheme to date and the current working model of the co-event interpretation. We explore the basic properties of the scheme, examine its application to simple systems and prove that it can successfully describe the PKS system, a description which we use to gain a deeper insight into the ``anhomomorphic'' nature of this scheme. We then address the issue of consistency and ``emergent classical logic''. Parts of this chapter are drawn from \cite{CoEventSchemes,Sorkin:2007,Dowker:2007zz}.

\section{Basic Properties}

To recap on section \ref{sec:multiplicative scheme}, in contrast to the polynomial schemes we have studied so far the multiplicative scheme \cite{Sorkin:2007} generalises from the classical scheme by dropping the linearity rule (equation \ref{eq:linearity}) entirely while keeping multiplicativity (equation \ref{eq:multiplicativity}):
$$\p(AB) = \p(A)\p(B)~~\forall~A,B,C\in\EA.$$
A multiplicative co-event $\p$ is dominated by a multiplicative co-event $\psi$ if:
$$\p(A)=1 \Rightarrow \psi(A) =1~\forall~A\in\EA.$$
In other words if the support of $\p$ in $\EA$\footnote{The support of $\p$ is $\{A\in\EA|\p(A)=1\}$.} is a subset of the support of $\psi$. A preclusive multiplicative co-event $\p$ is primitive if it is not dominated by any other preclusive multiplicative co-event. Given a histories theory $\H$ the set of primitive preclusive multiplicative co-events is denoted $\M$.

In the multiplicative scheme we have in a sense `gone back to the drawing board' and pursued a course diametrically opposed to our original linear scheme, and the higher order polynomial generalisations thereof. Recall (chapter \ref{chapter:linear scheme}) that linearity was at first favoured for its perceived `quantum' feel, as opposed to multiplicativity which was seen as a more `classical' condition. However it turns out that the multiplicative scheme is more flexible and better suited to quantum mechanics than the schemes we have looked at so far. Further, although the multiplicative condition may in some ways be more `classical' than linearity, it does lead to genuinely (and perhaps uncomfortably) `anhomomorphic' descriptions of simple systems. Finally, the multiplicative scheme possesses a clearer and more cogent interpretation than the more general interpretations we would use for other co-event schemes, as discussed in section \ref{sec:the interpretation of the interpretation}. Unless explicitly stated otherwise, in this chapter we will assume that all co-events are multiplicative and all sample spaces finite.

\subsection{Filters \& Duals}\label{sec:mult filters}

It is natural to think of $P\O$ as a partially ordered set, ordered by inclusion, and $\EA$ inherits this order structure. This partial order proves to be a useful angle from which to explore multiplicative co-events, mainly because they have a natural order theoretic description. Recall the definition of a filter.
\begin{definition}\label{def:filters}
A non-empty proper subset $F$ of a partially ordered set $P$ is a \textit{filter} if
\begin{enumerate}
  \item {$\forall x,y\in F~\exists~z\in F$ such that $z\leq x,y$}.
  \item {$if x\in F$ and $y\in P$ with $x\leq y$ we have $y\in F$}.
\end{enumerate}
Given any $p\in P$ the filter $\{x\in P |p\leq x\}$ is called a \textit{principal filter}, which we say is \textit{generated} by the filter's  \textit{principle element} $p$. A filter is \textit{maximal} if it is not a proper subset of any other filter.
\end{definition}
Note that filters are upper sets closed under finite meets (which are intersections in our case). The empty set is excluded from every filter by the requirement that $F$ be a proper subset of $P$. A principal filter is the smallest filter containing its principle element, and if $P$ is a power set the principal element is the intersection of all of the elements of the principle filter. If the intersection of all the elements of a filter is empty, then the filter is said to be a \textit{free filter}.
\begin{lemma}\label{lemma:mult coevents are filters}
Let $\H$ be a histories theory with a finite sample space. If $\p\in\M$ then the set of events $\p^{-1}(1)$,
\begin{equation*}
\p^{-1}(1) = \{ A \in \EA | \p(A) = 1\}\, ,
\end{equation*}
is a filter.
\end{lemma}
\begin{proof}
First note that the zero map is excluded from $\M$ so that $\p^{-1}$ is non-empty. Then let $A \in \p^{-1}(1)$. If $A \subset B$ then $AB = A$ and therefore $1 = \p(A) = \p(AB) = \p(A)\p(B) = \p(B)$. Which means that $B \in \p^{-1}(1)$. Further, if $A,~B\in\p^{-1}(1)$ then $\p(AB)=\p(A)\p(B)=1$, so $A\cap B\in\p^{-1}(1)$, thus $\p^{-1}(1)$ is a filter.
\end{proof}

Notice that classical co-events correspond to maximal filters. It can be shown that every finite filter is a principal filter, so if $\O$ is finite then $\p^{-1}(1)$ is a principal filter. To see this first note that
because $\O$ and thus $\EA$ are finite, $\p^{-1}(1)$ must contain a minimal element (minimal under set inclusion) $\p^*\in\EA$. If $A\in\EA$ is also a minimal element of the filter then $\p^* \cap A$ is in the filter, is contained in both $\p^*$ and $A$ and therefore must be equal to both. Therefore $A=\p^*$ and $\p^*$ is the unique minimal element and the principal element of $\p^{-1}(1)$. Note that $\p^*$ is not empty as the empty set is explicitly excluded from all filters. In the terminology of reference \cite{Sorkin:2007}, $\p^*$ is called the support of the multiplicative
co-event $\p$, however we will use the word `support' to denote $\p^{-1}(1)$, and denote $\p^*$, when it exists, as the \textit{multiplicative dual} or simply the \textit{dual} of $\p$. The multiplicative dual is best defined in terms of events, as a map:
\bea
*:\ \  \EA &\rightarrow& \EA^*, \nonumber \\
*:\ \  A &\mapsto& A^*, \nonumber
\eea
where
\beq\label{eq:mult dual}
A^*(B) =  \ \left\{\begin{array}{cc} 1 & {\text{if} ~A \subset B} \\
                                      0 & {\text{otherwise.}}\end{array}\right.
\eeq
Note that, using this notation, $\{\gamma\}^*=\gamma^*$. We say that $A^*$ is the \textit{dual}, or \textit{multiplicative dual} of $A$, and a comparison with equation \ref{eq:r* action} shows that $*$ is a natural generalisation of our original shift from events to co-events in classical stochastic theories (see section \ref{sec:coevents in classical stochastic mechanics}).

The map $*$ is a bijection. Thus we can define the inverse map, which we shall also call $*$, taking $A^*$ back to $A$, so that $(A^*)^*=A$. It is easy to see that this use of $*$ coincides with the construction of $\p^*$ above, for a comparison of equation \ref{eq:mult dual} and definition \ref{def:filters} reveals that $A$ is the principal element of $A^*$.

Finally, notice that $[\g_1^*\g_2^*](A)=\g_1^*(A)\g_2^*(A)$ means that for all $A\in\EA$
\bea
[\g_1^*\g_2^*](A) = 1 &\Leftrightarrow& \g_1,\g_2\in A, \nonumber \\
&\Leftrightarrow& \{\g_1,\g_2\}\subset A, \nonumber \\
&\Leftrightarrow& \{\g_1,\g_2\}^*(A)=1, \nonumber \\
\therefore \g_1^*\g_2^* &=& \{\g_1,\g_2\}^*. \nonumber
\eea
We can generalise this to co-events of arbitrary degree
\beq\label{eqn:multiplicative dual product decomposition}
A^* = \prod_{\g\in A} \g^*.
\eeq

\subsection{Primitivity \& Weak Emergent Classicality}\label{sec:mult prim and emergent classicality}

As with previous schemes, by simply applying the multiplicativity rule we would end up with `too many' co-events, and in particular in the case of a classical measure we would have non-classical co-events. Indeed if there were no null sets the dual of every set would be a multiplicative co-event. Now as noted above the classical co-events correspond to maximal filters, leading to a concept of primitivity expressed in terms of the maximality of filers; the primitive preclusive co-events should be maximal among preclusive co-events. This means the support of a primitive preclusive co-event should not be a subset of the support of any other preclusive co-event, so that a preclusive co-event $\p$ is primitive if there does not exist a preclusive co-event $\psi$ such that
$$\p(A)=1\Rightarrow\psi(A)=1,$$
which is the definition of domination we gave above. We can also express this in terms of the multiplicative dual defined above; we can say that a preclusive co-event is primitive if there does not exist a preclusive co-event $\psi$ such that
$$\psi^* \subset \p^*.$$
It is easy to see that this is equivalent to the definition of primitivity given above. It is also straightforward to check that this condition does indeed yield weak emergent classicality.
\begin{lemma}\label{lemma:multiplicative emergent classicality}
Let $\H$ be a histories theory with a finite sample space and a classical measure, then $\M=\C$.
\end{lemma}
\begin{proof}
Let $\p\in \M$. The Kolmogorov sum rule implies that the union, $Z$, of null sets in $\EA$ is itself null, so the preclusivity of $\p$ implies $\p^*\not\subset Z$. Thus $\exists \g \in \p^*\setminus Z$, so $\g$ is not an element of any null set. This means $\g^*$ is preclusive, but then the primitivity of $\p$ means that $\p^* = \{\gamma\}$ so that $\p = \gamma^*$, a homomorphism, and $\p\in\C$. Further, any homomorphism is a multiplicative co-event, thus $\M=\C$.
\end{proof}

When the measure is not classical, $\M$ may include co-events that are not classical on all of $\EA$. So far, we have defined classical behaviour in terms of partitions, we can now (for multiplicative co-events) extend this definition to events themselves. Given a histories theory $\H$ and an event $A\in\EA$, the `smallest' (coarsest) partition of $\O$ containing $A$ is $\{A,\O+A\}$. This leads us to define:

\begin{definition}\label{def:mult coevent anhomomorphic behaviour}
Let $\H$ be a histories theory with a finite sample space. We say that a multiplicative co-event $\p$ is \textbf{homomorphic} on an event $A\in\EA$ if it is classical on the partition $\{A,\O+A\}$. Otherwise $\p$ is \textbf{anhomomorphic} on $A$.
\end{definition}

We can characterise this behaviour in terms of the dual,

\begin{lemma}\label{lemma:mult coevent anhomomorphic behaviour}
Let $\H$ be a histories theory with a finite sample space. A multiplicative co-event $\p$ is homomorphic on an event $A\in\EA$ if and only if either of the following conditions are met:
\begin{enumerate}
  \item $\p^* \subset A.$
  \item $\p^* \subset A + \O.$
\end{enumerate}
\end{lemma}
\begin{proof}
Since for any event $B$ we have seen that $(B^*)^*=B$, the result follows trivially from the definition of $\p^*$.
\end{proof}

\subsection{Existence \& Compatibility}\label{sec:mult existence}

%Existence is a direct corollary of the following lemma.

The existence of primitive preclusive multiplicative co-events for any histories theory is a simple corollary of the following lemma, which in itself goes some way in demonstrating compatibility.
\begin{lemma}\label{lemma:mult existence primitivity}
Let $\H$ be a histories theory with a finite sample space and let $A\in\EA$ be non-negligible\footnote{ie $A$ is not a subset of a null set}. Then $\exists \p\in\M$ such that $\p(A)=1$.
\end{lemma}
\begin{proof}
Assume that $\nexists \p\in\M$ such that $\p(A)=1$. Because $A$ is non-negligible we know that $A^*$ is preclusive. Then since $A^*(A)=1$, $A^*$ can not be primitive, so there exists some preclusive $\psi_1$ dominating $A^*$ and by the definition of domination $\psi_1(A)=1$. Then same argument shows that $\psi_1$ can not be primitive, so there exists a preclusive $\psi_2$ dominating $\psi_1$ with $\psi_2(A)=1$. Carrying on in this fashion we can find an infinite sequence of preclusive co-events $\{\psi_i\}_{i=0}^\infty$ with $\psi_0=A^*$, with $\psi_i$ dominating $\psi_{i-1}$ (and thus $\psi_{i-n}$) and with $\psi_i(A)=1$.

The sequence can not terminate, otherwise the final co-event would be primitive contradicting our assumption. Further, it is easy to see that domination induces a strict partial order on the preclusive multiplicative co-events so that the $\psi_i$ are all distinct. However since $\O$ is finite so is $\EA$ and thus $\EA^*$, thus the sequence must terminate, contradicting our assumption. Therefore $\exists$ a primitive preclusive $\p$ such that $\p(A)=1$, and so $\p\in\M$.
\end{proof}

Leading to
\begin{corollary}\label{corollary:mult existence}
For any histories theory $\H$ with a finite sample space, $\M$ is non-empty.
\end{corollary}
\begin{proof}
By construction $\mu(\O)=1$, thus $\O^*$ is preclusive and by lemma \ref{lemma:mult existence primitivity} $\exists~\p\in\M$ such that $\p(\O)=1$.
\end{proof}
Note that this also proves unitality, though this concept is perhaps not as crucial to the multiplicative scheme as it was to the polynomial schemes.

\section{Simple Examples}

We will now work through some of the simple examples we have introduced in previous chapters, familiarising ourselves with the multiplicative scheme in this way.

\subsection{The Double Slit System}

Recall the histories description of this system we outlined in section \ref{sec:double slit}. Our fine grained histories are spacetime paths that can pass through one of slits $A$ and $B$ before ending either at the detector $D$ or elsewhere $\overline{D}$. Our sample space is $\O=\{AD,BD,A\overline{D},B\overline{D},\}$, and our event algebra is simply $\EA=P\O$. We have two natural decoherent partitions, $\{A,~B\}$ where $A=\{AD,A\overline{D}\},~B=\{BD,B\overline{D}\}$ and $\{D,~\overline{D}\}$ where  $D=\{AD,BD\},~\overline{D}=\{A\overline{D},B\overline{D}\}$. The decoherence functional is
\beq
{\textbf{D}}=\begin{pmatrix}
1/4 & -1/4 & 0 & 0 \\
-1/4 & 1/4 & 0 & 0 \\
0 & 0 & 1/4 & 1/4 \\
0 & 0 & 1/4 & 1/4
\end{pmatrix}
\eeq
in the `basis' $\{AD,BD,A\overline{D},B\overline{D}\}$ so that, for example, ${\textbf{D}}(B\overline{D},B\overline{D})=1/4$. The only null set in $\EA$ is $D$.

The histories $A\overline{D},B\overline{D}$ do not participate in any null set, thus the classical co-events $A\overline{D}^*,B\overline{D}^*$ are preclusive. Since classical co-events are always multiplicative and can not be dominated by any other co-event, they are members of $\M$ whenever they are preclusive. The histories $AD,BD$ are elements of the null set $D$, so the corresponding classical co-events are not preclusive. Further, $D^*$ itself is not preclusive, and any co-event involving elements of $D$ and $\overline{D}$ will be dominated by the classical co-events corresponding to the elements of $\overline{D}$. This exhausts $\EA$, and we can conclude that:
\bea
\M &=& \{A\overline{D}^*,B\overline{D}^*\}, \nonumber \\
&=& \C. \label{eq:mult double slit}
\eea
As with the linear scheme, the multiplicative scheme treats the double slit system classically.

\subsection{The Triple Slit System}

Recall the histories description of this system we outlined in section \ref{sec:triple slit}. Our fine grained histories are spacetime paths that can pass through one of slits $A$, $B$ or $C$ before ending either at the detector $D$ or elsewhere $\overline{D}$. Our sample space is $\O=\{AD, BD, CD, A\overline{D}, B\overline{D}, C\overline{D}\}$, and since the system is finite our event algebra is simply $\EA=P\O$. We have two natural decoherent partitions, $\{A,~B,~C\}$ where $A=\{AD,A\overline{D}\}$ etc, and $\{D,~\overline{D}\}$ where  $D=\{AD,BD,CD\},~\overline{D}=\{A\overline{D},B\overline{D},C\overline{D}\}$. The decoherence functional is
\beq
D=\frac{1}{9}\begin{pmatrix}
1 & -1 & 1 & 0 & 0 & 0 \\
-1 & 1 & -1 & 0 & 0 & 0 \\
1 & -1 & 1 & 0 & 0 & 0 \\
0 & 0 & 0 & 2 & 1 & -1 \\
0 & 0 & 0 & 1 & 2 & 1 \\
0 & 0 & 0 & -1 & 1 & 2
\end{pmatrix}
\eeq
Unlike the double slit system neither of the `final states' is precluded, however this system does contain two null sets, $\{AD,BD\}$ and $\{BD,CD\}$. As before the fine grained histories $A\overline{D}, B\overline{D}, C\overline{D}$ are not elements of any null set so their duals $A\overline{D}^*, B\overline{D}^*, C\overline{D}^*$ are classical co-events, which also means these histories will not participate in any other primitive co-events. As to the histories ending at the detector, $D$, $AD$ and $BD$ are elements of the null set $\{AD,BD\}$ while $CD$ is an element of the null set $\{BD,CD\}$, thus none of the corresponding classical co-events are preclusive. Similarly the duals of the two null sets can not be preclusive, however $\{AD,CD\}^* = AD^*CD^*$ is preclusive and since it could only be dominated by $AD^*$ or $CD^*$, the fact that neither of these are preclusive makes $\{AD,CD\}^*$ primitive. $D^*$ is also preclusive, however it is dominated by $\{AD,CD\}^*$ and thus is not primitive. Further, the dual of any combination of histories in $D$ and $\overline{D}$ will be dominated by the classical co-event associated with the later. We have exhausted $\EA$, so there can be no further primitive co-events. In summary:
\beq\label{eq:mult triple slit}
\M = \{A\overline{D}^*, B\overline{D}^*, C\overline{D}^*,AD^*CD^*\}.
\eeq
As with the linear scheme, the multiplicative scheme's treatment of this system displays genuinely `quantum' or `anhomomorphic' behaviour, because of the presence of $AD^*CD^*$. To see what this means we examine the valuation of $AD^*CD^*$ on the events in the coarse graining $A,B,C$.
\bea
AD^*CD^*(A) &=& 0 \nonumber \\
AD^*CD^*(B) &=& 0 \nonumber \\
AD^*CD^*(C) &=& 0 \nonumber \\
AD^*CD^*(\{A,B\}) &=& 0 \nonumber \\
AD^*CD^*(\{B,C\}) &=& 0 \nonumber \\
AD^*CD^*(\{C,A\}) &=& 1 \nonumber \\
AD^*CD^*(\{A,B,C\}) &=& 1 \nonumber
\eea
Assuming that $AD^*CD^*$ is the actual reality we can rephrase this in `questions \& answers' format, exposing the `anhomomorphic' nature of this theory.
\\
\\
\begin{tabular}{c|c}
\bf{Question} & \bf{Answer} \\
\hline
Does the particle pass through slit $A$? & no \\
Does the particle pass through slit $B$? & no \\
Does the particle pass through slit $C$? & no \\
Does the particle pass through one of slits $A$ or $B$? & no \\
Does the particle pass through one of slits $B$ or $C$? & no \\
Does the particle pass through one of slits $C$ or $A$? & yes \\
Does the particle pass through one of slits $A$, $B$ or $C$? & yes \\
\end{tabular}
\\
\\
\\
Notice that, unlike the linear scheme, in the multiplicative scheme a primitive co-event may be zero valued on all singleton sets. This is in line with the interpretation of the multiplicative scheme as an `ontological coarse graining', the singleton sets can be thought of as corresponding to `meaningless', or `ill-posed', questions, and only the properties associated with \textit{all} the histories in the dual of the co-event are true (see section \ref{sec:the interpretation of the interpretation}).

Note that all the above co-events are classical on the observable partition $\{D,\overline{D}\}$, in particular $AD^*CD^*$ simply behaves like $D^*$, thus we do not \emph{observe} anhomomorphic behaviour.

\subsection{The Four Slit System}

Recall the histories description of this system we outlined in section \ref{sec:four slit}. Our fine grained histories are spacetime paths that can pass through one of the four slits $A_i$, $0\leq i\leq 3$, before ending either at the detector $D$ or elsewhere $\overline{D}$. Denoting the path that passes through slit $A_i$ to reach $D$ ($\overline{D}$) by $D_i$ ($\overline{D}_i$), our sample space is $\O=\{D_i\}_{i=0}^3\sqcup\{\overline{D}_i\}_{i=0}^3$, and our event algebra is simply $\EA=P\O$. We have two natural decoherent partitions, $\{A_i\}_{i=1}^3$ where $A_i=\{D_i,\overline{D}_i\}$ etc, and $\{D,~\overline{D}\}$ where $D=\{D_i\}_{i=0}^3,~\overline{D}=\{\overline{D}_i\}_{i=0}^3$. The decoherence functional is
\beq
D=\frac{1}{16}\begin{pmatrix}
1 & -1 & -1 & -1 & 0 & 0 & 0 & 0 \\
-1 & 1 & 1 & 1 & 0 & 0 & 0 & 0 \\
-1 & 1 & 1 & 1 & 0 & 0 & 0 & 0 \\
-1 & 1 & 1 & 1 & 0 & 0 & 0 & 0 \\
0 & 0 & 0 & 0 & 3 & 1 & 1 & 1 \\
0 & 0 & 0 & 0 & 1 & 3 & -1 & -1 \\
0 & 0 & 0 & 0 & 1 & -1 & 3 & -1 \\
0 & 0 & 0 & 0 & 1 & -1 & -1 & 3 \\
\end{pmatrix}
\eeq
The null sets are $\{D_0,D_i\}_{i=1}^3$. As usual, the histories $\overline{D}_i$ are not elements of any null set, so the classical co-events $\overline{D}_i^*$ are preclusive and therefore primitive. Each of the histories $D_i$ is contained in a null set. Moving up one `degree' the duals of the null sets can not be preclusive, however the sets $\{D_i,D_j\}$ for $1\leq i,j\leq 3$ are non-negligible thus their duals are preclusive. Since the $\{D_i,D_j\}^*$ could only be dominated by $D_i^*$ or $D_j^*$, neither of which are preclusive, it must be primitive. It is not possible to construct any other non-negligible set that does not contain the dual of one of our primitive co-events as a subset, therefore we have found all the primitive co-events:
\bea
\M &=& \{\overline{D}_1^*,\overline{D}_2^*,\overline{D}_3^*,D_1^*D_2^*,D_2^*D_3^*,D_3^*D_1^*\},\nonumber \\
&=& \QU. \label{eq:mult four slit}
\eea

\section{The Kochen-Specker-Peres System}

Recall the histories formulation of the Peres-Kochen-Specker system as outlined in chapter \ref{chapter:kochen-specker}. As discussed in section \ref{sec:the criteria}, it is this system that demonstrates the failure of the classical scheme and so it is natural to use it as a testing ground for our alternate schemes. Noting that the classical scheme fails on the existence criterion (lemma \ref{lemma:failure of classical scheme}), in section \ref{sec:linear PKS} we demonstrated the existence of (unital) linear co-events describing the system, and in theorem \ref{thm:linear PKS} showed that the linear scheme could provide a `satisfactory' treatment of the system. Because $\L\subset\Pn{n}$, this meant all the polynomial schemes were able to satisfactorily deal with the $PKS$ system. We now turn to the multiplicative scheme's treatment. As in section \ref{sec:linear PKS} we will assume a measure whose null sets are precisely the $PKS$ sets and their disjoint unions.

The existence of multiplicative co-events describing the PKS system is assured by corollary \ref{corollary:mult existence}. Further, the sets $G_i,~R_j$ are not subsets of any PKS set, and so are non-negligible, lemma \ref{lemma:mult existence primitivity} therefore ensures the existence of co-events $\p^g_i$ and $\p^r_j$ mapping $G_i$ and $R_j$ to $1$ respectively. Though this is weaker than our result in the linear case (theorem \ref{thm:linear PKS}), it is satisfactory bearing in mind that a actual measurement would alter the system, the measure and the co-events.

Although their existence is assured, it is nevertheless instructive to construct the co-events $\p^g_i$ and $\p^r_j$ and to probe their treatment of the complex $PKS$ system. As the system is finite, every multiplicative co-event corresponds to a principle filter and so possesses a dual, which can not be negligible. Now the Peres-Kochen-Specker theorem says that every $\g\in\O$ is an element of at least one $PKS$ set. Some elements of $\O$, however, lie in exactly one PKS set and so are good places to start building a primitive co-event. If we can find two elements $\g,\g' \in \O$ that each lie in exactly one PKS set, and if these two $PKS$ sets are neither equal nor disjoint, then the set $\{\g,\g'\}$ will be not contained in any PKS set or disjoint union of PKS sets and thus will be non-negligible.

Recall our analysis of the Peres Set in section \ref{sec:PKS}. Looking at the proof of theorem \ref{thm:PKS}, we see there is a unique way to extend $\g_P$ from the four bases $B_1$, $B_2$, $B_3$, $B_4$ to the whole of $PS$ so that it is consistent on all bases and pairs of rays except basis $B_{11}$. We will henceforth refer to this extension as the Peres colouring and use the same notation $\g_P$ for it. The Peres colouring lies in exactly one $PKS$ set, $R_{B_{11}}$. $\gamma_P$ is given in table \ref{table:mult}.

We can obtain another such colouring $\g_P'$ by acting on $\g_P$ by one of the symmetries of the projective cube: let us choose the reflection that exchanges the $x$ and $y$ axis. The colouring $\g_P'$ is then obtained from $\g_P$ by permuting the rays by the action of swapping the first and second labels (in the Peres notation we have used where a ray is labelled by the components, e.g. $001$). $\g_P'$ is also given in table \ref{table:mult}. By symmetry, $\g_P'$ is also contained in exactly one $PKS$ set, $R_{B_7}$.

The $PKS$ sets $R_{B_7}$ and $R_{B_{11}}$ are not disjoint because they share, for example, the colouring which is red on all rays.

Therefore, the multiplicative co-event $\p=\g_P^*\g_P'^* = \{\g_P, \g_P'\}^*$  will value all $PKS$ sets and disjoint unions of $PKS$ sets zero and it is primitive among multiplicative co-events that do so because neither of the atomic co-events are preclusive. Table \ref{table:mult} shows the valuation of $\g_P^*\g_P'^*$ on the events corresponding to either colouring on each ray. Note that among the rays $u_j$ for which $\p(G_j)=1$ there is a ray of each type (see section \ref{sec:PKS}), thus given any ray $u_i$ there exists a ray $u_j$ of the same type such that $\p(G_j)=1$. But then, as we saw in section \ref{sec:linear PKS}, there is a symmetry $g\in H$ of the Peres Set mapping $u_i$ to $u_j$, so that $\p(g(G_i))=1$. But then $g^{-1}(\p)=(g^{-1}(\g_P))^*(g^{-1}(\g_P'))^*$ will map $G_i$ to $1$, so we can set $\p^g_i=g^{-1}(\p)$. Similarly we can find a symmetry $h\in H$ such that $\p^r_i=h^{-1}(\p)$.

\begin{table}[ht]
\begin{center}
\begin{tabular}{|c|ccc|cc|}
\hline
 $u_{i}$& & $\g_P$& $\g_P'$& $\p(G_i)$& $\p(R_i)$ \\
\hline
$001$& & g& g &1 &0\\
$010$& & r& r &0 &1\\
$100$& & r& r &0 &1\\
$011$& & g& g &1 &0\\
$01\m{1}$& & r& r &0 &1\\
$101$& & g& g &1 &0\\
$10\m{1}$& & r& r &0 &1\\
$110$& & r& r &0 &1\\
$1\m{1}0$& & r& r &0 &1\\
$012$& & g& g &1 &0\\
$0\m{1}2$& & r& r &0 &1\\
$021$& & r& g &0 &0\\
$02\m{1}$& & r& r &0 &1\\
$102$& & g& g &1 &0\\
$\m{1}02$& & r& r &0 &1\\
$201$& & g& r &0 &0\\
$20\m{1}$& & r& r &0 &1\\
$120$& & r& r &0 &1\\
$\m{1}20$& & r& r &0 &1\\
$210$& & r& r &0 &1\\
$2\m{1}0$& & r& r &0 &1\\
$112$& & g& g &1 &1\\
$\m{1}12$& & r& g &0 &0\\
$1\m{1}2$& & g& r &0 &0\\
$\m{11}2$& & r& r &0 &1\\
$121$& & g& g &1 &0\\
$12\m{1}$& & r& r &0 &1\\
$\m{1}21$& & r& r &0 &1\\
$\m{1}2\m{1}$& & r& r &0 &1\\
$211$& & g& g &1 &0\\
$21\m{1}$& & r& r &0 &1\\
$2\m{1}1$& & r& r &0 &1\\
$2\m{11}$& & r& r &0 &1\\
\hline
\end{tabular}
\caption{A primitive unital multiplicative co-event sending the PKS sets to zero}\label{table:mult}
\end{center}
\end{table}

\subsection{Anhomomorphism Exposed}\label{sec:mult PKS logic}

Looking at table \ref{table:mult} we see that for most of the rays we have $\p(G_i) + \p(R_i) = 1$, which is the familiar situation in classical (homomorphic) logic, for example if it is false that $u_i$ is green then it is true that it is red. However, there are some rays, $021, 201, \m{1}12$ and $1\m{1}2$, for which both $\p(G_i)=0$ and $\p(R_i)=0$. Note that there are no rays for which $\phi(G_i)=1$ and $\phi(R_i)=1$. This is not a coincidence: $\p(X)\p(1+X) = \p(\emptyset) = 0$ ($\p(\emptyset) = 1$ would imply $\p = 1$ since every event contains the empty event) and so an event and its complement cannot both be valued true. An event and its complement {\it can} both be valued false as we see here, and indeed this is the epitome of the nonclassical logic that the multiplicative scheme tolerates.

This violation of classical rules of inference may at the first encounter seem unpalatable. However, it is the way in which a genuine coarse graining is expressed at the level of logical inference. As discussed in section \ref{sec:the interpretation of the interpretation}, one way to think of the multiplicative scheme is that reality corresponds to the dual of the `actually real' co-event and only {\it common} properties of all the formal trajectories in the dual are real properties: if the dual is contained in an event then that event happens. In this way, both an event and its complement -- {\it e.g.} ray $u_i$ is green and ray $u_i$ is red -- can be valued false if the support of the co-event intersects them both. As mentioned in section \ref{sec:the interpretation of the interpretation}, we may choose to interpret this anhomomorphic behaviour as arising from our misunderstanding of what the `true' event algebra is. Following this argument, the questions which receive anhomomorphic answers are `in reality' not well posed, because $\EA$ is not the `real' event algebra. Another interpretation, less palatable to the author, would be to accept the anhomomorphism as fundamental and to question the foundations of logic. In either case, we shall in this section review the questions on which $\p$ is not classical in an attempt to expose and illuminate the nature of the multiplicative scheme's anhomomorphism.

First we will introduce some notation. Given a basis $\{u_i,u_j,u_k\}$ we will want to talk about events such as `$u_i$ is green' or `$u_j$ is green and $u_k$ is red'. In our existing notation we would denote the first event $G_i$ and the second $G_j\cap R_k$. However we will now introduce a notation in which we will describe the first event as $\{g,-,-\}$ and the second as $\{-,g,r\}$.

Given a set of Peres Rays $S_I=\{u_i\}_{i\in I}$ corresponding to some indexing set $I$ we denote by the \emph{colouring} $\{a_i\}_{i\in I}$ the event $\bigcap_{i\in I} A_i$ where $a_i\in\{g,r,-\}$ and
$$A_i = \left\{\begin{array}{cc}
G_i & \text{if}~a_i=g \\
R_i & \text{if}~a_i=r \\
\O & \text{if}~a_i=-
\end{array}\right.$$

We now probe into the events on which $\p=\g_P^*\g_P'^*$ is anhomomorphic. By lemma \ref{lemma:mult coevent anhomomorphic behaviour} these are events $A\in\EA$ such that $A\neq\p^*\cap A\neq \emptyset$. Therefore $A$ must contain exactly one of the two histories $\g_P,~\g_P'$ in $\p^*$.We will focus on events of the form $\{a_i\}_{i\in I}$ as described above, corresponding to the Peres rays $\{u_i\}_{i\in I}$. Now every history $\g\in\O$ will correspond to a colouring of the rays $\{u_i\}_{i\in I}$, namely $\{\g(u_i)\}_{i\in I}$, and so to an event in $\EA$. Now if $\g_P(u_i)=\g_P'(u_i)$ for all $i\in I$, then either $\g_P(u_i)=g_P'(u_i)=a_i$ for all $i\in I$ or $\g_P(u_i)=g_P'(u_i)\neq a_i$ for all $i\in I$; thus either both $\g_P$ and $\g_P'$ are in $\{a_i\}_{i\in I}$ or neither $\g_P$ nor $\g_P'$ are in $\{a_i\}_{i\in I}$. In both cases $\p$ will be homomorphic on $\{a_i\}_{i\in I}$. Further, if $\g_P(u_j)\neq\g_P'(u_j)$ for some $j\in I$ then if $a_j=\g_P(u_j)$ we have $\g_P\in\{a_i\}_{i\in I}$ and $\g_P'\not\in\{a_i\}_{i\in I}$. Conversely if $a_j\neq\g_P(u_j)$ then $\g_P\not\in\{a_i\}_{i\in I}$ and $\g_P'\in\{a_i\}_{i\in I}$. In either case $\p$ will be anhomomorphic on $\{a_i\}_{i\in I}$. To conclude, $\p$ will be homomorphic on the event $\{a_i\}_{i\in I}$ corresponding to Peres rays $\{u_i\}_{i\in I}$ if and only if $\g_P$ and $\g_P'$ agree on every ray $u_i,~i\in I$.

We will explore some examples of such situations to further expose the anhomomorphic nature of $\p$. We restrict our attention to events defined by the colourings of individual rays, (orthogonal) pairs \& bases, which by the above must possess at least one element on which $\g_P$ and $\g_P'$ disagree to be relevant to this enquiry.

Now $\g_P$ and $\g_P'$ agree on all rays other than $D = \{021, ~201, ~1\m{1}2, ~\m{1}12\}$. On these rays we have:
\\
\\
\\
%\begin{table}[h]
\begin{tabular}{c|cccc}
 &$021$ & $201$ & $\m{1}12$ &$1\m{1}2$ \\
\hline
$\g_P$ & r & g & r & g \\
$\g_P'$ & g & r & g & r
\end{tabular}
%\end{table}
\\
\\
\\
\noindent We will be concerned with pairs and bases which have non-empty intersection with $D$. Such a pair could contain either one or two elements of $D$, and such a basis could contain one, two elements of $D$, since there are no bases entirely in $D$. In the interests of brevity we will only deal with one example from each type of pair and basis. We begin with the case of a pair, both rays of which are contained in $D$, our example being the pair $\{021,~1\m{1}2\}$. We can construct a truth table to understand the action of $\p$ on the events $\{a_i\}$ corresponding to this pair.
\\
\\
\\
%\begin{table}[h]
\begin{tabular}{cc|cc|c}
event & colouring & $\g_P^{*}$ & $\g_P'^{*}$ & $\p=\{\g_P,\g_P'\}^*$ \\
\hline
$G_{021}$ &                 g- & 0 & 1 & 0 \\
$R_{021}$ &                 r- & 1 & 0 & 0 \\
$G_{1\m{1}2}$ &             -g & 1 & 0 & 0 \\
$R_{1\m{1}2}$ &             -r & 0 & 1 & 0 \\
$G_{021}\cap G_{1\m{1}2}$ & gg & 0 & 0 & 0 \\
$R_{021}\cap G_{1\m{1}2}$ & rg & 1 & 0 & 0 \\
$G_{021}\cap R_{1\m{1}2}$ & gr & 0 & 1 & 0 \\
$R_{021}\cap R_{1\m{1}2}$ & rr & 0 & 0 & 0
\end{tabular}
%\end{table}
\\
\\
\\
\noindent Where, as we specified above, by the colouring $\{g-\}$ we mean the set of paths $\{\gamma | ~\gamma(021) ~= ~green\}$ and by $\{gr\}$ we mean the set of paths $\{\gamma | ~\gamma(021) ~=~green ~, ~\gamma(1\m{1}2) ~=~red \}$. We can present this in `question/answer format:
\\
\\
\\
\begin{tabular}{c|c}
\bf{Question} & \bf{Answer} \\
\hline
Is the ray $021$ green? & no \\
Is the ray $021$ red? & no \\
Is the ray $1\m{1}2$ green? & no \\
Is the ray $1\m{1}2$ red? & no \\
Are both rays $021$ and $1\m{1}2$ green? & no \\
Is the ray $021$ red and the ray $1\m{1}2$ green? & no \\
Is the ray $021$ green and the ray $1\m{1}2$ red? & no \\
Are both rays $021$ and $1\m{1}2$ red? & no \\
\end{tabular}
\\
\\
\\
There are no true statements for this pair.

Next we consider the pairs containing one ray in $D$ and one ray outside $D$, our example will be $\{021,~0\m{1}2\}$. We have:
\\
\\
\\
%\begin{table}[h]
\begin{tabular}{c|cc}
 &$021$ & $0\m{1}2$ \\
\hline
$\g_P$ & r & r \\
$\g_P'$ & g & r
\end{tabular}
%\end{table}
\\
\\
\\
\noindent So we get:
\\
\\
\\
%\begin{table}[h]
\begin{tabular}{cc|cc|c}
event & colouring & $\g_P^{*}$ & $\g_P'^{*}$ & $\p=\{\g_P,\g_P'\}^*$ \\
\hline
$G_{021}$ &                 g- & 0 & 1 & 0 \\
$R_{021}$ &                 r- & 1 & 0 & 0 \\
$G_{1\m{1}2}$ &             -g & 0 & 0 & 0 \\
$R_{1\m{1}2}$ &             -r & 1 & 1 & 1 \\
$G_{021}\cap G_{1\m{1}2}$ & gg & 0 & 0 & 0 \\
$R_{021}\cap G_{1\m{1}2}$ & rg & 0 & 0 & 0 \\
$G_{021}\cap R_{1\m{1}2}$ & gr & 0 & 1 & 0 \\
$R_{021}\cap R_{1\m{1}2}$ & rr & 1 & 0 & 0
\end{tabular}
%\end{table}
\\
\\
\\
\noindent As before can present this in `question/answer format:
\\
\\
\\
\begin{tabular}{c|c}
\bf{Question} & \bf{Answer} \\
\hline
Is the ray $021$ green? & no \\
Is the ray $021$ red? & no \\
Is the ray $1\m{1}2$ green? & no \\
Is the ray $1\m{1}2$ red? & yes \\
Are both rays $021$ and $1\m{1}2$ green? & no \\
Is the ray $021$ red and the ray $1\m{1}2$ green? & no \\
Is the ray $021$ green and the ray $1\m{1}2$ red? & no \\
Are both rays $021$ and $1\m{1}2$ red? & no \\
\end{tabular}
\\
\\
\\
\noindent So the only true statement is $-r$, or `$0\m{1}2$ is red'. Note that $0\m{1}2$ is the ray not in $D$.

Turning to the bases, first we examine the basis with two rays in $D$ and one ray outside $D$:
\\
\\
\\
%\begin{table}[h]
\begin{tabular}{c|ccc}
 &$1\m{1}2$ &$\m{1}12$ & $110$ \\
\hline
$\g_P$ & g & r & r  \\
$\g_P'$ & r & g & r
\end{tabular}
%\end{table}
\\
\\
\\
\noindent For brevity we will suppress the truth tables for the basis examples. It turns out that the only true statement is $\{--r\}$, or `$110$ is red'. Note that $110$ is the ray not in $D$. finally, considering the bases with one element in $D$ and two elements outside $D$, we take as our example the set $\{021,~0\m{1}2,~100\}$. We have:
\\
\\
\\
%\begin{table}[h]
\begin{tabular}{c|ccc}
 & $021$ & $0\m{1}2$ & $100$ \\
\hline
$\g_P$ & r & r & r  \\
$\g_P'$ & g & r & r
\end{tabular}
%\end{table}
\\
\\
\\
\noindent The true statements turn out to be $\{-rr\}$, $\{-r-\}$ and $\{--r\}$, in other words `$0\m{1}2$ and $100$ are red', `$0\m{1}2$ is red' and `$100$ is red'. Note again that $0\m{1}2$ and $100$ are the two rays not in $D$.

\section{Strong Emergent Classicality}

\subsection{Consistency \& Compatibility}

Anhomomorphism, as exposed in the last section, is a phenomenon that we have not observed. Thus consistent with the `quantum micro-world, classical macro-world' picture of emergent classicality we look for a mechanism by which `classical partitions' lead naturally to homomorphic outcomes. More specifically, we would like to be able to say that the co-events $\p\in\M$ are classical on the partition $\{A_i\}$ whenever it is a `classical partition' (whatever we mean by this).

A key question is of course what exactly we mean by a `classical partition'. The most obvious definition is that of `dynamical classicality', as used in the decoherent histories approach, that a decoherent partition should be considered classical since the measure restricts to a classical measure on the subalgebra generated by the partition. In fact, we could begin with the restriction of the measure  to a subalgebra, and calculate the co-events from there. When the subalgebra is generated by a decoherent partition emergent classicality (section \ref{sec:mult prim and emergent classicality}) will ensure that the co-events are all homomorphisms. Then an equivalent question is one of \textit{consistency}, given a subalgebra do the multiplicative co-events calculated from the full measure on the whole algebra restrict to the co-events calculated from the measure restricted to the subalgebra? This is of course a desirable property in and of itself, and is one of the criteria for a successful co-event scheme that we set out in section \ref{sec:the criteria}. Without this property we would have to consider the measure of the whole universe before making any statements about events we observe.

To proceed we will require a more technical definition of consistency, and fortunately we can provide two:

\begin{definition}
Let $\H$ be a histories theory and let the subalgebra $\EA_\Lambda$ correspond to the finite partition $\Lambda$ of $\O$. We say that a co-event scheme ${\cal{S}}$ is \textbf{weakly consistent under the coarse graining} $\Lambda$ if for every pair $A\in\EA_\Lambda$, $\p^{(\la)}\in {\cal{S}}(\Lambda,\EA_\Lambda,\mu_\Lambda)$ such that $\p^{(\la)}(A)=1$ there exists $\psi_A^{(\O)}\in \S$ such that $\psi_A^{(\O)}(A)=1$. We say that $\cal{S}$ is \textbf{weakly consistent} if it is weakly consistent for any histories theory and finite coarse graining.
\end{definition}

\noindent As before $\mu_\Lambda$ is the restriction of $\mu$ to the subalgebra $\EA_{\Lambda}$. The action of the co-events $\p\in {\cal{S}}(\Lambda,\EA_\Lambda,\mu_\Lambda)$ on the events in $\EA_\Lambda$ can thus be thought of as restrictions of the actions of co-events in ${\cal{S}}\H$. Given $A,B\in\EA_\Lambda$, $\p\in {\cal{S}}(\Lambda,\EA_\Lambda,\mu_\Lambda)$ such that $\p(A)=\p(B)=1$ weak consistency ensures that we can find $\psi_A,\psi_B\in {\cal{S}}\H$ such that $\psi_A(A)=\psi_B(B)=1$. However we have no guarantee that $\psi_A(B)=1$, and so we can not in general consider $\p$ as the restriction to $\EA_\Lambda$ of any co-event in ${\cal{S}}\H$. This is ensured by the stronger version of consistency.

\begin{definition}
Let $\H$ be a histories theory and let the subalgebra $\EA_\Lambda$ correspond to the finite partition $\Lambda$ of $\O$. We say that a co-event scheme ${\cal{S}}$ is \textbf{strongly consistent under the coarse graining} $\Lambda$ if for every $\p^{(\la)}\in {\cal{S}}(\Lambda,\EA_\Lambda,\mu_\Lambda)$ there exists $\psi^{(\O)}\in \S$ such that $\p^{(\la)}(A)=1\Rightarrow\psi^{(\O)}(A)=1$ for $A\in\EA_\Lambda$. We say that $\cal{S}$ is \textbf{strongly consistent} if it is strongly consistent for any histories theory and finite coarse graining.
\end{definition}

\noindent Weak consistency is trivially implied by strong consistency, which allows to think of $\p\in {\cal{S}}(\Lambda,\EA_\Lambda,\mu_\Lambda)$ as a restriction of $\psi\in {\cal{S}}\H$.

It can be shown that the ideal based scheme is weakly consistent \cite{Sorkin:private}, though it fails the test of strong consistency. The multiplicative scheme turns out to be neither strongly nor weakly consistent, even for decoherent partitions, as the following example demonstrates.

\begin{example}\label{example:multiplicative consistency problem}
Consider the histories theory $\H$ where $\O=\{a_1,a_2,b_1,b_2\}$, $\EA=P\O$ and the measure is derived from the strongly positive decoherence functional
$$
D = \begin{pmatrix}
1 & 0 & -1 & 1 \\
0 & 1 & -1 & 1 \\
-1 & -1 & 2 & -2 \\
1 & 1 & -2 & 4
\end{pmatrix}
$$
The eigenvalues of $D$ are $(0,1,1,6)$, thus the decoherence functional obeys strong positivity. Under the coarse graining $\Lambda=\{A,B\}$ where $A=\{a_1,a_2\}$, $B=\{b_1,b_2\}$ this becomes
$$
D_\Lambda = \begin{pmatrix}
1 & 0 \\
0 & 2
\end{pmatrix}
$$
Thus $\Lambda$ is a decoherent partition. The only null set in this system is $Z=\{a_1,a_2,b_1\}$. It is easy to see that:
$${\cal{M}}(\Lambda,\EA_\Lambda,\mu_\Lambda)= \{A^*,B^*\},$$
whereas
$$\M = \{b_2^*\}.$$
We can see that $A^*(A)=1$ while $b_2^*(A)=0$, which means that the multiplicative scheme is not weakly consistent.
\flushright{$\square$}
\end{example}

\begin{figure}[t]
\begin{center}
\includegraphics[width=0.5\textwidth,angle=-90]{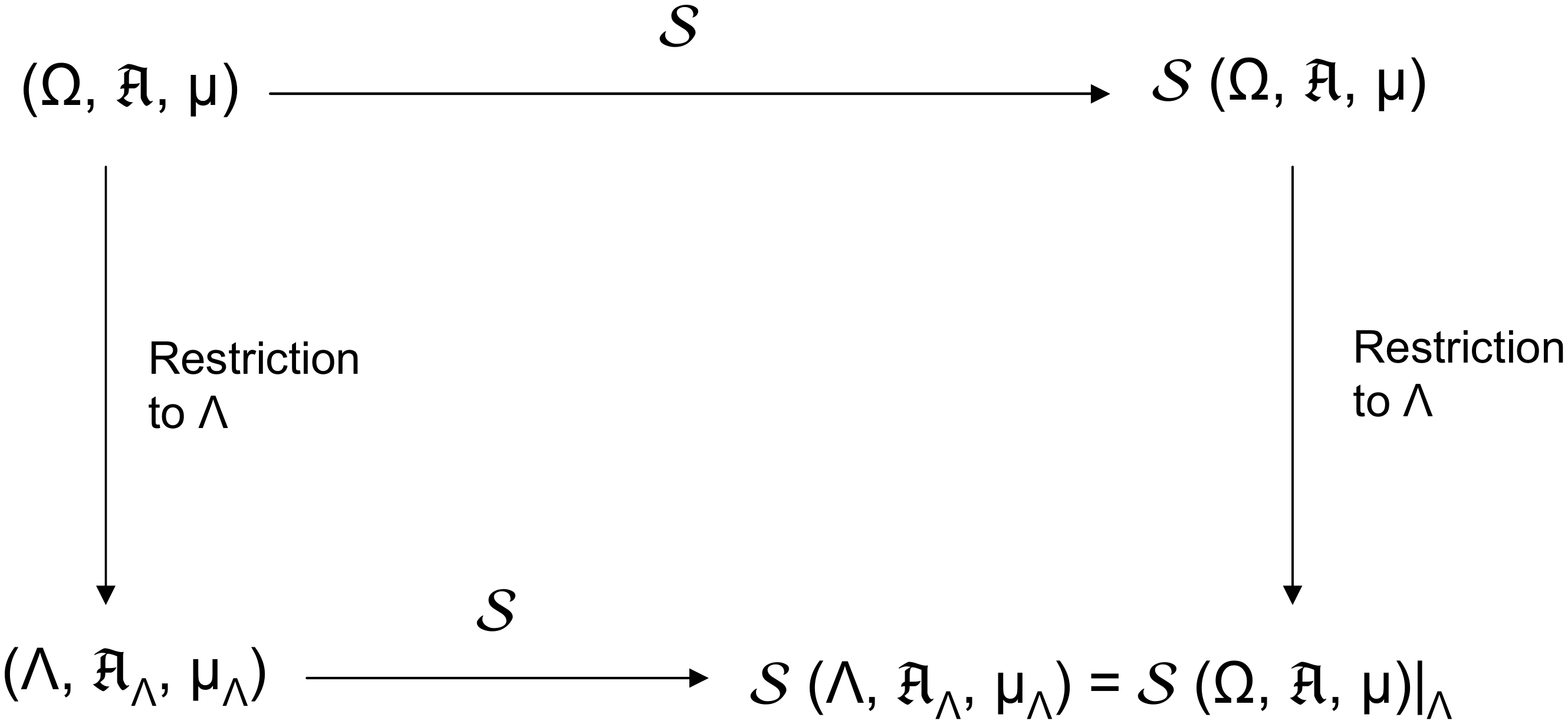}
\caption{\small{Total Consistency}} \label{fig:total consistency}
\end{center}
\end{figure}

Historically weak and strong consistency were introduced in an attempt to understand and compare the consistency of the ideal and multiplicative schemes. The following condition, illustrated in figure \ref{fig:total consistency} is stronger than both.

\begin{definition}
Let $\H$ be a histories theory and $\Lambda$ a finite coarse graining of the sample space, corresponding to the restricted histories theory $\HL$. The \textbf{restriction} $\S|_\Lambda$ of a co-event scheme $\S$ to $\HL$ is given by the restriction of the elements of $\S$:
$$\S|_\Lambda = \{\p|_\la | \p\in\S\},$$
where
\bea
\p|_\la:\la &\rightarrow& \Z2 \nonumber \\
\p|_\la(A) &=& \p(A), \nonumber
\eea
for all $A\in\EA_\la$. $\S$ is \textbf{totally consistent on the partition} if
$$\S|_\Lambda = {\cal{S}}\HL.$$
We say that ${\cal{S}}$ is \textbf{totally consistent} if it is totally consistent on every histories theory and finite coarse graining.
\end{definition}

Total consistency allows us to link together the notions of weak and strong emergent classicality, assuming we know what we mean by a `classical partition':

\begin{lemma}\label{lemma:total consistency and emergent classicality}
Assume that for every histories theory $\H$ with a finite sample space we can define the class of `classical partitions'. Further let ${\cal{S}}$ be a co-event scheme that obeys weak classicality and is totally consistent on all `classical partitions' then ${\cal{S}}$ obeys strong emergent classicality.
\end{lemma}

Note that this holds independently of what exactly constitutes a `classical partition', however for this lemma to me meaningful we need to define the term more rigorously.

\subsection{Classical Partitions}\label{sec:classical partitions}

The multiplicative scheme's failure to achieve even weak consistency is a serious hurdle for the theory. However all is not lost, as there are indications that this hurdle is surmountable. The key question, as mentioned above, is what exactly we mean by a `classical partition'. At the very least we would wish to ensure some form of `censorship' principle; since we do not observe anhomomorphic behaviour our theory must not predict any observable anhomomorphisms. Indeed, we would expect partitions corresponding to observable alternatives to be a proper subset of the set of all decoherent partitions, and would hope that on these `observable partitions' the multiplicative scheme does indeed reduce to the classical scheme.

The difficulty lies in characterising such partitions mathematically, in terms of the properties of the histories theory and the partition. Indeed, such a restriction of the set of partitions considered as classical might go some way in alleviating the `multiple consistent sets' criticism leveled at this approach and by the consistent histories interpretation in particular (see section \ref{sec:the consistent histories interpretation}).

The question can be reversed, we could ask on what types of partition the multiplicative scheme behaves consistently. If we could answer this we might then be in a position to determine whether or not this class of partition included all examples of `measurable partitions'. However, we are not yet in such a position, though we do have some interim results.

\subsubsection{Preclusive Separability}

Looking back at example \ref{example:multiplicative consistency problem}, we might ask what feature of the histories theory or the partition obstructs weak consistency. Of course the only aspect of the measure to affect the co-events is the structure of the null sets, and in particular this system possesses only one null set, $Z$. This null set contains one of the elements of the partition, $A$, and so prevents any subset of $A$ from corresponding to a preclusive co-event, so that $\p(A)=0$ for all $\p\in\M$. However though $Z$ contains $A$ it contains only parts of $B$, and thus does not feature in the subalgebra generated by the partition, which includes no null sets thus allowing the duals of its atoms, $A^*$ and $B^*$, to be primitive and preclusive. It is this `disappearing act' played by $Z$ that causes the discrepancy between the fine grained and coarse grained pictures.

Essentially the problem structures relate to null sets that intersect more than one element of the partition without completely containing every element they intersect. We could of course explicitly rule out this behaviour, but such a condition would mean that given any partition every null set would have to be a disjoint union of the elements of that partition. Since this would hold for every partition, the condition would be far too strong. The \textit{strong preclusive separability condition}, introduced by Rafael Sorkin \cite{Sorkin:private}, is a first attempt at a more realistic constraint.

\begin{definition}\label{def:strong preclusive seperability}
Let $\H$ be a histories theory. A finite partition of $\O$ into events $A_i$ obeys \textbf{strong preclusive separability} if the following holds: Let $B$ be an element of the event algebra. Then $B$ is precluded (null) if and only if its intersection with each $A_i$ is precluded.
\end{definition}

Since being first introduced, this constraint has been replaced by the weaker:

\begin{definition}\label{def:preclusive separability}
Let $\H$ be a histories theory. A finite partition of $\O$ into events $A_i$ obeys \textbf{preclusive separability} if the following holds: Let $B$ be an element of the event algebra. Then $B$ is precluded (null) if and only if its intersection with each $A_i$ is contained in a precluded subset $N_i$ of $A_i$.
\end{definition}

Note that these definitions make sense for infinite as well as finite sample spaces. It is easy to see that strong preclusive separability implies preclusive separability, so any property we can demonstrate to hold for preclusively separable partitions will automatically hold for strongly preclusively separable partitions. Notice that (strongly) preclusively separable partitions are not necessarily decoherent. However the condition is powerful enough to ensure that the co-events given by ${\cal{M}}$ are classical.

\begin{lemma}\label{lemma:preclusive separability implies clasical co-events}
Let $\H$ be a histories theory with a finite sample space and a strongly positive measure. Further, let $\Lambda=\{A_i\}$ be a preclusively separable partition of the sample space. Then we have
\bea
{\cal{M}}\HL &=& \{A_i^*|_\la | \mu(A_i)\neq 0 \}, \nonumber \\
&=& {\cal{C}}\HL. \nonumber
\eea
\end{lemma}
\begin{proof}
Let $A_i\in\Lambda$. Then if $A_i^*$ is not preclusive (over $\H$) there exists a null $Z\supset A_i$. Preclusive separability means that there exists a null $Z_i\in\EA$ contained in $A_i$ and containing $Z\cap A_i$. But $Z\supset A_i\Rightarrow Z\cap A_i = A_i$, so we have:
$$A_i = Z\cap A_i \subset Z_i \subset A_i.$$
Therefore $Z_i=A_i$, so that $A_i^*$ is always preclusive over $\H$, and thus over $\HL$, when $A_i$ is not itself null. Since there can be no finer grained element than $A_i$ in $\EA_\Lambda$, $A_i^*|_\la$ is preclusive and primitive in $\mu_\la$ whenever $A_i$ is not null.

Further, if $B\in \EA_\Lambda$ then it is a disjoint union of elements of the partition $\Lambda$, $B=\bigsqcup_{A_i\subset B} A_i$. If $\mu_\Lambda(A_i)=0 ~\forall A_i\subset B$ then strong positivity means that $\mu_\Lambda(B)=0$, so that $B^*$ is not preclusive. Conversely, if $B^*|_\la$ is preclusive in $\EA_\Lambda$ this means it must contain some non-null $A_i$, but then $A_i^*|_\la$ dominates $B^*|_\la$ which then can not be primitive. Therefore the $A_i^*|_\la$ where $A_i$ is non-null are all of the primitive preclusive co-events.
\end{proof}

This leads on to our central theorem regarding the consistency \& compatibility of the multiplicative scheme

\begin{theorem}\label{thm:Rafael's thm}
Let $\H$ be a histories theory with a finite sample space and strongly positive measure. Let $\Lambda=\{A_i\}$ be a preclusively separable partition, then
\begin{enumerate}
\item If $B^*\in\M$ then the restriction $B^*|_\Lambda$ of $B^*$ to $\EA_\Lambda$ is an element of ${\cal{M}}\HL$, further $B^*|_\Lambda=A_i^*|_\Lambda$ for some $i$
\item $\forall A_i\in\Lambda$ with $\mu(A_i)\neq 0$ $\exists~B^*\in\M$ satisfying the above.
\end{enumerate}
\end{theorem}
\begin{proof}
Let $B^*$ be an element of $\M$, we claim that $B\subset A_j$ for some $j$. To see this, assume that $B\not\subset A_i$ for any $i$, this is equivalent to assuming $B\neq B\cap A_i$ for any $i$. Now if any of the $(B\cap A_i)^*$ are preclusive, they will dominate $B^*$ which can not then be primitive.

Hence if our assumptions hold $(B\cap A_i)^*$ is not preclusive for any $i$, so $\forall i$ $\exists$ a measure zero set $Z_i$ such that $Z_i\supset B\cap A_i$. But then preclusive separability implies the existence of null sets $Z_i'\supset A_i$ such that $Z_i'\supset A_i\cap Z_i$, so $A_i\supset Z_i'\supset A_i\cap B$ and thus $\bigsqcup Z_i' \supset B$. But then strong positivity means that $\mu(Z_i')=0\Rightarrow \mu(\bigsqcup Z_i')=0$, so $B^*$ can not be preclusive, contradicting our assumption.

Thus $B\subset A_j$ for some $j$. But then $B^*(A_i) = \delta_{ij}$, and it is easy to see that restricting to the subalgebra generated by the $A_i$ we have $B^*|_\Lambda=A_j^*|_\Lambda$, proving the first part of the theorem.

To prove the second part, notice that the proof of lemma \ref{lemma:preclusive separability implies clasical co-events} implies that in a preclusively separable partition $\mu(A_i)\neq 0$ implies that $A_i^*$ is preclusive and thus $A_i$ is non-negligible, so by lemma \ref{lemma:mult existence primitivity} there exists some $B^*\in\M$ such that $B^*(A_i)=1$, meaning that $B\subset A_i$. It is then easy to see that $B^*|_\Lambda = A_i^*|_\Lambda$.
\end{proof}

Together with lemma \ref{lemma:preclusive separability implies clasical co-events} this immediately gives us:

\begin{corollary}\label{corollary:mult scheme consistency and compatibility}
The multiplicative scheme is totally consistent on preclusively separable partitions.
\flushright{$\square$}
\end{corollary}

Then using lemmas \ref{lemma:multiplicative emergent classicality} \& \ref{lemma:total consistency and emergent classicality} we have:

\begin{corollary}\label{corollary:mult scheme strong emergent classicality}
Let the class of `classical partitions' be a subset of the class of preclusively separable partitions. Then the multiplicative scheme obeys strong emergent classicality.
\end{corollary}

These are powerful results, however it will not be useful unless preclusively separable partitions can be considered `classical' in some way, enabling us to use this result to argue that the multiplicative scheme does not lead to macroscopic anhomomorphisms. In terms of our criteria (section \ref{sec:the criteria}) it ensures consistency, compatibility and strong emergent classicality if we are happy to assume that `classical partitions' are preclusively separable. Fortunately, strongly preclusively separable partitions include partitions according to final time projectors in histories theories derived from Hilbert space theories.

To see this, consider such a histories theory $\H$ derived from a Hilbert space theory $({\cal{H}},H,\ket{\psi},T)$. Recall that a (fine grained) history in $\O$ corresponds to a string of one dimensional projection operators associated with every $t\in T$, including a final time projector corresponding to the upper bound of $T$ (definition \ref{def:history}). Then let $\Lambda=\{A_i\}$ be partition of the sample space corresponding to the final time projectors $P_i$, so that $P_i$ is the final time projector for every (fine grained) history in the partition element $A_i$. Now not only is the system decoherent, $D(A_i,A_j)=0$ for $i\neq j$, but there is no interference whatsoever between the elements of the partition, $D(a_i,a_j)=0$ for $i\neq j$, where $a_i\in A_i$ and $a_j\in A_j$ are fine grained histories in $\O$. This is because $i\neq j$ implies $P_iP_j=\textbf{0}$ (then use equations \ref{eq:class operator}, \ref{eq:probability amplitude} and \ref{eq:decoherence functional on amplitudes}); intuitively histories `ending at different locations' do not interfere. This means the measure of any event $B$ can be decomposed into the sum of the measures of its intersections with the elements of the partition
$$\mu(B) = \sum_i \mu(B\cap A_i),$$
thus $B$ can only be null if $B\cap A_i$ is null for all $i$. This condition has been called \textit{super-decoherence}.

Does super-decoherence hold in practice? Typically an experimental set up as used in the investigations of quantum mechanics involves a histories theory which can be derived from a Hilbert space theory in which final time projectors represent the observable alternatives, in which case super-decoherence will certainly hold, as we have just seen. In particular note that this means that \emph{for single time `Copenhagen' measurements the rules of inference regarding the outcomes will be classical}. It has been suggested \cite{Sorkin:private} that it will also hold for any partition based on `records'. However the condition does appear rather strong; we would like to be able to associate classicality with weaker conditions on the partition.

In support of this argument, preclusive separability is itself stronger than is necessary to prove theorem \ref{thm:Rafael's thm}, as the following example demonstrates.

\begin{example}
Let $\H$ be a histories theory with a finite sample space and $\Lambda = \{A_1,A_2\}$ a partition thereof. Assume that the system admits only one null set, $Z$, which intersects both $A_1$ and $A_2$ but contains neither. Then $\g^*$ will be preclusive for all $\g\in\O$ such that $\g\not\in Z$. But then no element of $Z$ can participate in the dual of any primitive preclusive co-event $\p$, for if $\p^*\subset Z$ then $\p$ is not preclusive, and if $\p^*$ intersects but is not contained in $Z$ then it must contain some $\g\not\in Z$, and $\g^*$ will dominate $\p$ which then can not be primitive. Now by assumption $\exists~\g_1\in A_1,\g_2\in A_2$ such that $\g_i\not\in Z$. Then $\g_i^*\in\M$, $\g_i^*|_\la = A_i^*$. Further every co-event is of this sort, so $\M|_\la = \ML$. However the system is not weakly preclusively separable, as by assumption $Z$ is the only null set.
\flushright{$\square$}
\end{example}

It is generally accepted that a `finer' condition than weak preclusive separability is desirable \cite{Dowker:private,Sorkin:private,Wallden:private}. However what this condition should or even could be remains to be determined.

\section{Infinite Theories}\label{sec:infinite sample spaces}

So far we have only dealt with finite sample spaces, and indeed many of our co-event constructions would not be well defined if the sample space were infinite. Principally this is because we do not have a satisfactory understanding of quantum measure theory for infinite sample spaces, as we touched upon in section \ref{sec:quantum measure theory}. In particular, it is not clear what the event algebra should be when the sample space is no longer finite. While $P\O$ seems too large there is as yet no satisfactory `quantum' or higher order analogue to the concept of the $\sigma$-algebra of measurable sets in `classical' measure theory. Because of this `uncertainty' it is difficult to even begin extending our co-event theories to encompass the infinite case. We will nevertheless make some general and some specific comments on the issue.

The `rules' of the various co-event schemes can be generalised with no changes made, however the various concepts of primitivity are generally only well defined in the finite case, the exception being the multiplicative scheme that remains well defined when the sample space ceases to be finite. Conversely the polynomial and ideal based schemes would not be well defined, and it is not clear how they should be generalised to the infinite case.

Of more concern is the difficulty we have with the classical scheme. As mentioned in section \ref{sec:quantum measure theory}, even when the measure is classical an infinite sample space generally means the singleton sets are null, precluding any classical reality. This may be acceptable in co-event theory, for using the `ontological coarse graining' interpretation of the multiplicative scheme (section \ref{sec:the interpretation of the interpretation}) we could claim that the singleton sets are simply theoretical constructs and that reality itself is somewhat `coarse-grained' or `fuzzy' even when the measure is classical. Alternatively we might exclude the singleton (or even all finite) sets from the event algebra, in which case we may yet be able to associate reality with a homomorphism.

\subsection{Multiplicative Infinity}\label{sec:infinite multiplicative scheme}

The multiplicative scheme itself is really the only one of our schemes that remains well defined when the sample space becomes infinite; this is, in fact, one of its `selling points'. However some subtleties are introduced, in particular the notion of primitivity becomes confused; aiming as it does to ensure that the primitive co-events are classical in the event of a classical measure it is difficult to assess the success of any proposed condition when the status of the classical scheme itself is in flux. For now we will retain the concept of primitivity as it stands, and glance through the results of this chapter to review their status given an infinite sample space.

\subsubsection{Filters \& Duals in the Infinite Case}

A quick look at the proof of lemma \ref{lemma:mult coevents are filters} shows that it also holds in the infinite case, so that multiplicative co-events can still be associated with filters.

\begin{lemma}\label{lemma:mult coevents are filters infinite}
Let $\H$ be a histories theory. If $\p\in\M$ then the set of events $\p^{-1}(1)$,
\begin{equation*}
\p^{-1}(1) = \{ A \in \EA | \p(A) = 1\}\, ,
\end{equation*}
is a filter.
\end{lemma}
\begin{proof}
See proof of lemma \ref{lemma:mult coevents are filters}
\end{proof}

However, the concept of the dual is no longer well-defined in general, for in the infinite case a filter need not contain a `least element'. If $A\in\EA$ then the set $\{B|A^*(B)=1\}$ forms a principal filter, and the intersection of all of its elements gives back $A$. However if $\p$ corresponds to a \textit{free filter} then $\bigcap_{\p(B)=1} B = \emptyset$, so we can not identify $\p$ with any element of the event algebra; in other words we can not define $\p^*$ in general. Thus the map $*:\ \  \EA \rightarrow \EA^*$ is well-defined, however it is no longer surjective in general as $\EA^*$ can be `larger' than $\EA$.

The `co-finite filter' provides us with an example of a free filter. If $\O$ is infinite then we can define the filter $F\subset\EA$ to be the set of co-finite elements of the event algebra, in other words $F=\{A\in\EA | \O+A~\text{finite}\}$. It is straightforward to check that $F$ is indeed a filter. Firstly if $A,B\in F$ then $\O+A$, $\O+B$ are finite, so their intersection $(\O+A)(\O+B)=\O+A+B+AB$ is finite. But every element of $\EA$ is additively self-inverse, so $RHS=(\O+A)+(\O+B)+(\O+AB)$. Since the first two summands are finite by assumption $\O+AB$ must also be finite, so that $AB\in F$. Secondly if $A\in F$ and $A\subset B$ then $(\O+B)\subset (\O+A)$, and thus is finite so that $B\in F$, and $F$ is a filter. There are history theories for which $F$ is a preclusive co-event, for example any infinite theory with no null sets. However no history theories have been identified in which $F$, or any other free filter, is primitive, and the existence of such theories remains an open question.

It has been suggested that multiplicative co-events be restricted from filters to principal filters \cite{Dowker:private}, in part for simplicity and in part because this would be in line with the interpretation of multiplicative co-events as `ontological coarse grainings' of the sample space (see section \ref{sec:the interpretation of the interpretation}).

\subsubsection{Weak Emergent Classicality \& Existence in the Infinite Case}

In the infinite case weak emergent classicality becomes confused by the lack of clarity surrounding the classical scheme, it is not possible to establish weak emergent classicality while we are unsure of what classicality itself means. We can establish some basic results, for example if $\EA=P\O$ and there are no measure zero sets then $\p\in\M$ will always be a homomorphism. However such results are not useful given the nullity of all singleton sets under probability theory measures. Thus we do not  at this point have an analogue for lemma \ref{lemma:multiplicative emergent classicality}. The concept of existence suffers from similar problems, in part because we are unsure of our concept of primitivity in the infinite case. Even if we hold to the finite concept of maximal (principal) filters, the existence of primitive co-events for arbitrary measures is yet to be established.

\section{Conclusion}

The multiplicative scheme is the most successful co-event scheme to date, and at the time of writing has become the working model for co-event theory. Looking back at our criteria (section \ref{sec:the criteria}), the linear, quadratic \& higher order polynomial schemes (chapters \ref{chapter:linear scheme} \& \ref{chapter:polynomial}) all fail to be consistent due to the non-existence of unital co-events for certain measures (theorem \ref{thm:general poynomial scheme unitality problem} \& section \ref{sec:many slit}). More fatally, these examples can be used to construct gedankenexperiments in which these schemes fail to be compatible with quantum mechanics (appendix \ref{appendix:many slit}). On the other hand, when the sample space is finite both the multiplicative and ideal-based schemes can be shown to satisfy existence (section \ref{sec:mult existence} and \cite{Sorkin:private}), compatibility \& weak emergent classicality (section \ref{sec:mult prim and emergent classicality} and \cite{Sorkin:private}). Although the multiplicative scheme fails to be even weakly consistent (example \ref{example:multiplicative consistency problem})  we can argue that decoherence itself is too weak a condition for the identification of `classical partitions', and we see that the multiplicative scheme is totally consistent on preclusively separable partitions (corollary \ref{corollary:mult scheme consistency and compatibility}). Therefore if we are happy for `classical partitions' to be necessarily preclusively separable we see that the multiplicative scheme obeys strong emergent classicality (corollary \ref{corollary:mult scheme strong emergent classicality}). Note that for all histories theories derived from Hilbert space theories partitions based on final time projectors will be preclusively separable; in particular this means the multiplicative scheme yields `classical rules of inference' regarding the outcomes of single time `Copenhagen' measurements. Though the ideal-based scheme is weakly consistent it is not strongly (or totally) consistent \cite{Sorkin:private}, thus the preclusive separability results favor multiplicative scheme ahead of the ideal-based approach.

Further, unlike any of the other schemes the multiplicative scheme has its own interpretation based upon the concept of an `ontological coarse graining' (section \ref{sec:the interpretation of the interpretation}). This new interpretation seems clearer and more cogent than the more general interpretations of co-event theory, thus lending further credibility to the multiplicative scheme.

Finally, again unlike any of the other schemes and in particular in sharp contrast with the ideal-based scheme the multiplicative scheme seems open to generalisation to history theories with infinite sample spaces (section \ref{sec:infinite multiplicative scheme}). Although this generalisation, and indeed the generalisation of quantum measure theory itself to infinite sample spaces, is far from complete, such a generalisation appears possible and in fact is hampered more by our lack of understanding of infinite event algebras and classical co-events than by the nature of the multiplicative scheme itself.

%The Multiplicative Scheme
%\singlespacing
\chapter[Dynamics \& Predictions]{Dynamics \& Predictions the Co-Event Interpretation}\label{chapter:dynamics}
%\doublespacing

In a quantum measure theory $\H$ the dynamics is encoded in the measure $\mu$, and we expect that any predictions the theory makes should be derived from $\mu$. However since every $A\in\EA$ need not be observable, $\mu$ will not in general be equivalent to the theory's predictive content; indeed the nature of this predictive content remains an open area of research \cite{Dowker:private}. However, restricting to histories theories derived from Hilbert space theories (as we will throughout this chapter) we can identify (or define) the predictions of $\H$ with the usual `Copenhagen predictions' of the underlying $({\cal{H}},H,\ket{\psi_{t_0}},T)$. For simplicity, we will assume that the Hilbert space theory will identify a set of (Copenhagen) observable events, that each observable event is a member of an observable partition (a partition of $\O$ consisting entirely of observable events), and that the measure is classical on all such partitions. Using the results of sections \ref{sec:the axiomatic approach} \& \ref{sec:quantum measure theory}, we note that this means that if $A$ is an observable event, $\mu(A)$ can be interpreted as a probability $\P_\mu(A)$ ($=\mu(A)$), and that this probability agrees with the Copenhagen predicted probability for $A$. Thus the predictive content of our histories theory lies in these `probability statements' $\P_\mu(A)$.

Now as far as co-events go our predictive power is (so far) limited to statements concerning preclusion. Given an event algebra and a measure, our co-event approach identifies the primitive preclusive co-events (assuming a particular scheme) as the potential realities of our theory; however we cannot in general reconstruct the measure from the set of potentially real co-events. Further, were we to ignore the measure and use only the potentially real co-events themselves (without knowledge of which co-event is actually real), we would be unable to make any predictive statements beyond ruling out as impossible those events that are precluded (mapped to zero) by all the potentially real co-events, and identifying as certain all those events that are mapped to unity. This may not be quite as restrictive as it appears, for by \emph{conditioning} on known truth valuations of events (disregarding the co-events that disagree with these valuations) we may be able to narrow down the set of potentially real co-events to the point that such predictions are relatively powerful \cite{Sorkin:private}.

However there is reason to believe (depending on the `correct' definition of classicality) that conditioning on previous observations will not increase our power to predict other observable events; though conditioning on non-observable events may increase our ability to predict observable events and conditioning on observable events may increase our ability to predict non-observable events.

In any case, it is clear that such conditioning will not in general allow us to recover the \emph{entirety} of either the measure or its predictive content, whatever definition of classicality we adopt, whereas we would like to be able to express the full dynamical \& predictive content of a quantum measure theory in terms of the potentially real co-events rather than the histories.  We would thus be able to `start with' a set of potentially real co-events and make our predictions using only these co-events and structures defined in terms of them. In this chapter will explore attempts to achieve this; we are motivated in our endeavor by three arguments.

Firstly, we have in this thesis followed the second (and where applicable the third) interpretation of co-events as described in section \ref{sec:the interpretation of the interpretation}, treating the primitive co-events themselves as the potential realities, of the histories approach. The author feels that we should be able to express all meaningful statements concerning a theory in terms of the theory's potential realities; even if \emph{in practise} it is simpler to use the measure $\mu$ defined on the event algebra, at least \emph{in principle} this structure should `emerge' from a `deeper' expression of the predictive content of the theory in terms of the co-events themselves.

Secondly, there are few if any exactly null sets in the `real world', their presence in many gedankenexperimental frameworks (such as the double slit experiment) depends upon idealised assumptions (such as point slits) that do not hold in practice. Regardless of our interpretation, co-events as they now stand cannot distinguish between events of large or small measure, so our use of null sets to approximate `almost null' sets in idealised gedankenexperiments requires justification. As mentioned above, at present the range of predictive statements we can make with the set of potentially real co-events is extremely narrow; were we able to fully express the predictive content of quantum measure theory in terms of the potentially real co-events we may be able to better grapple with issues such as `almost null sets' that are phrased in terms of probability rather than preclusion.

Thirdly, were we able to express in terms of co-events the predictive content of those histories theories derived from Hilbert space theories, we might then be in a position to extrapolate our findings and better understand the predictive content of a general quantum measure theory.

In this chapter we will explore recent attempts to rephrase in terms of co-events the dynamical \& predictive content of histories theories. We first consider attempts to move wholesale from the `histories theory' structure to one of `co-event theories' in which the dynamics is defined directly in terms of co-events (as anticipated in \cite{Sorkin:2007}). We then turn our attention to the important concept of `approximate preclusion', and discuss the details of its implementation. Throughout, we will follow \cite{ApproximatePreclusion,ApproximatePreclusionProceedings} closely, and unless specifically noted otherwise will assume that all sample spaces are finite, and that all histories theories are derived from Hilbert space theories (we will continue to make the assumptions described above regarding observable events).

\section{Dynamical Co-Events}\label{sec:dynamical coevents}

\subsection{Placing a Measure on the Space of Co-Events}

One approach, anticipated by Sorkin in \cite{Sorkin:2007}, is to rephrase the dynamics (the quantum measure $\mu$) in terms of the co-events themselves; for example by placing a measure on the space of co-events. This would be a natural development, given that the co-events are the basic potential realities of the theory, and could lead to the rephrasing of the entirety of quantum mechanics in terms of co-events, bypassing our current need to construct dynamics upon the event algebra. We could then explain the rarity of low probability events (such as `almost null' sets) by using the dynamics on our co-events to suppress (assign a low probability to) those co-events that value them to $1$.

Perhaps the most obvious way to do this would involve literally `moving' the dynamics from the event algebra onto the co-events, which can thus realise their  status as the central objects of the theory. So far we have always begun with a histories theory $\H$ from whence we have derived our allowed co-events ${\cal{S}}\H$. Since we are asserting that ${\cal{S}}\H$ rather than $\O$ is the `true' sample space of `potential realities' it would seem more natural to place the co-events at the center of our structure, moving from a histories theory $\H$ to a full \textit{co-event theory} $(S,PS,\P_S)$, where $S$ is a set of (allowed) co-events, for example $S={\cal{S}}\H$, and $\P_S$ is a probability measure on $S$. We have assumed a probability measure, for if we were to use a more general higher order measure we would not have gained anything from our adoption of co-events in place of histories, encountering the same interpretational difficulties that led us to co-events. Indeed, following that route might even lead us to adopt co-co-events, co-co-co-events and so on. Conversely, since we are using the `one co-event is real' interpretation (see section \ref{sec:the interpretation of the interpretation}), the use of a probability measure echoes the use of probability in classical stochastic theories in which we believe that `one history is real'.

In a sense this approach is born of the Bayesian approach to probability \cite{Ramsey:1926,Finetti:1930,Cox:2001}, which looks on probability as representing a degree of information concerning a system and thus ascribes meaning to individual probability statements which can therefore be treated as `logical' predicates in their own right \cite{Cox:2001}. Following this philosophy we seek to express our incomplete information about the actual reality in terms of probability statements concerning the potential realities, much as we are accustomed to do in the classical theory. We can, therefore, think of $\P_S(\{\p\})$ as the probability that the co-event $\p$ is the actually real co-event; consequently $\P_S(\{\p\})=0$ could be interpreted to mean that $\p$ is not a potential reality. Indeed, we could even enforce our choice of $S$ as the set of potential realities by extending the domain of $\P_S$ to all of $\CE$, and constraining $\P_S$ to be zero on all co-events outside $S$.

Nevertheless our observations are in terms of events in $\EA$, so we must relate this new structure back to our Copenhagen probabilistic predictions. As stated above we are focusing only on histories theories $\H$ that are derived from a Hilbert space theory; we have assumed that in such theories we can identify the set of `observable events', that each observable event is an element of an `observable partition', and that $\mu$ is classical on all such partitions. We have noted that the results of sections \ref{sec:the axiomatic approach} \& \ref{sec:quantum measure theory} then imply that $A$ is an observable event, $\mu(A)$ can be interpreted as a probability, and that this probability agrees with the Copenhagen predicted probability for $A$. Then at the very least we want, for all observable $A$,

\beq\label{eq:binary probs on coevents}
\mu(A)=\left\{\begin{array}{c}1\\0\end{array}\right. \Rightarrow\P_S(\{\p\in S|\p(A)=1\})=\left\{\begin{array}{c}1\\0.\end{array}\right.
\eeq

As we will discuss below, non-binary (ie not equal to $0$ or $1$) probabilities themselves require interpretation, and can arguably be defined in terms of repeated trials and binary probabilities. However we will steer clear of this issue for now, and instead of using  equation \ref{eq:binary probs on coevents} to define probabilities through repeated trials we will instead use the stronger:
\beq\label{eq:probs on coevents weak}
\P_S(\{\p\in S|\p(A)=1 \})=\mu(A),
\eeq
for all observable events $A$. However it is not clear how `observable events' should be defined in our quantum measure theory (see for example section \ref{sec:classical partitions}). We sidestep this question by simply strengthening the condition once more to require:
\beq\label{eq:probs on coevents}
\P_S(\{\p\in S|\p(A)=1 \})=\mu(A) ~\forall A\in\EA.
\eeq
One difficulty with this condition is that in general $\mu(A)$ can exceed $1$. However, even when we are able to constrain $\mu$ to take values in $[0,1]$, equation \ref{eq:probs on coevents} is a problematic condition for multiplicative co-events, as the following theorem shows.
\begin{theorem}\label{thm:dynamical coevents are quadratic}
Let $\H$ be a histories theory with a finite sample space, let $S\subset\CE$ be a set of multiplicative co-events on $\EA$, so that
$$\p(AB)=\p(A)\p(B)~\forall A,B\in\EA,~\forall\p\in S.$$
Further, let $\P_S$ be a probability measure on $S$ such that
$$\P_S(\{\p\in S|\p(A)=1 \})=\mu(A) ~\forall A\in\EA.$$
Then for any $\p\in S$, $\P_S(\{\p\})=0$ unless for all $A,B,C\in\EA$
\bea
\p(A+ B+ C) &=& \p(A+ B)+\p(B+ C)+\p(C+ A) \nonumber \\
&& +\p(A)+\p(B)+\p(C). \nonumber
\eea

\end{theorem}

As noted above, the use of a probability measure on the space of co-events echoes its use on histories in classical stochastic theories. Then following our usual interpretation that dynamically precluded `events' do not occur, we conclude that the actually real co-event must be quadratic. Since the quadratic scheme has already been ruled out in sections \ref{sec:quadratic failure} \& \ref{sec:sixteen slit} this puts into question our idea of `moving' the dynamics onto the co-events by constructing `co-event theories' in place of histories theories, unless we are willing to abandon the multiplicative scheme.

Of course this approach is not entirely ruled out, to arrive at equation \ref{eq:probs on coevents} we have made several simplifying assumptions so there may still be some scope for further investigation. For example, if we accept $\M$ as the space of potential realities of a histories theory $\H$, we might attempt to define all other elements of the theory in terms of the co-events $\p\in\M$. In particular, we could define events as maps from $\M$ to $\Z2$ using the `dual map' $A[\p]=\p(A)$. Since we may not be able to distinguish between all elements of $\EA$ in this manner, we might perhaps choose to take equivalence classes of `distinguishable events' ($A\sim B$ if $A[\p]=B[\p]$ for all $\p\in\M$), and to attempt to define a measure on this set as a stepping stone to a `full co-event theory'. Such a structure is yet to be explored. Alternatively, we might explore the restrictions on $\mu$ that would enable equation \ref{eq:probs on coevents}, or we might turn to another scheme.

\subsection{Proof of Theorem \ref{thm:dynamical coevents are quadratic}}

In this section we will prove theorem \ref{thm:dynamical coevents are quadratic}. We will need the following technical lemmas.

\begin{lemma}\label{lemma:quantum sum rule implies quadratic rule}
Let $\H$ be a histories theory with a finite sample space and let $\p$ be a (not necessarily multiplicative) co-event. Then
\bea
\p(A\sqcup B\sqcup C) &=& \p(A\sqcup B)+\p(B\sqcup C)+\p(C\sqcup A) \nonumber \\
&& +\p(A)+\p(B)+\p(C),\label{eq:proof sum rule equivalent to quadratic rule 1}
\eea $\forall~disjoint~A,B,C\in\EA$ if and only if \bea
\p(A'+ B'+ C') &=& \p(A'+ B')+\p(B'+ C')+\p(C'+ A') \nonumber \\
&& +\p(A')+\p(B')+\p(C'),\label{eq:proof sum rule equivalent to quadratic rule 2} \eea $\forall A',B',C'\in\EA$.
\end{lemma}
\begin{proof}
First note that equation \ref{eq:proof sum rule equivalent to quadratic rule 1} is simply the restriction of equation \ref{eq:proof sum rule equivalent to quadratic rule 2} to disjoint sets, so that we immediately see that equation \ref{eq:proof sum rule equivalent to quadratic rule 1} $\Leftarrow$ equation \ref{eq:proof sum rule equivalent to quadratic rule 2}.

To prove the converse we consider sets $A_1,A_2,A_3\in\EA$ and break down their union $\bigcup_{i=1}^3A_i$ into disjoint components by defining:
\bea
A_iA_jA_k &=& A_i\cap A_j\cap A_k, \nonumber \\
\overline{A_iA_j}&=&(A_i\cap A_j)\setminus (A_1A_2A_3), \nonumber \\
\overline{A_i} &=& A_i\setminus (\overline{A_iA_j}\sqcup\overline{A_iA_k}\sqcup A_1A_2A_3). \nonumber \\
\eea
Then
\beq\nonumber
\bigcup_iA_i = (\bigsqcup_i\overline{A_i})\sqcup(\bigsqcup_{i<j}\overline{A_iA_j})\sqcup A_1A_2A_3.
\eeq
Assuming equation \ref{eq:proof sum rule equivalent to quadratic rule 1}, we can use this notation to decompose the terms in equation \ref{eq:proof sum rule equivalent to quadratic rule 2}. The left hand side becomes
\bea
\p(A_1+ A_2 + A_3) &=& \p((\bigsqcup_i\overline{A_i})\sqcup A_1A_2A_3), \nonumber \\
&=& \sum_{i<j}\p(\overline{A_i}\sqcup\overline{A_j}) + \sum_i \p(\overline{A_i}\sqcup A_1A_2A_3). \nonumber
\eea
Similarly turning our attention to the right hand side, for $i,j,k$ distinct we find that
\bea
\p(A_i+ A_j) &=& \p(\overline{A_i}\sqcup\overline{A_iA_k}\sqcup\overline{A_j}\sqcup\overline{A_jA_k}), \nonumber \\
&=& \p(\overline{A_i}\sqcup \overline{A_j})+\p(\overline{A_i}\sqcup \overline{A_iA_k})+ \p(\overline{A_i}\sqcup \overline{A_jA_k}) \nonumber \\
&& + \p(\overline{A_j}\sqcup \overline{A_iA_k})+ \p(\overline{A_j}\sqcup \overline{A_jA_k}) + \p(\overline{A_iA_k}\sqcup \overline{A_jA_k}), \nonumber
\eea
and
\bea
\p(A_i) &=& \p(\overline{A_i}\sqcup\overline{A_iA_j}\sqcup\overline{A_iA_k}\sqcup A_1A_2A_3), \nonumber \\
&=&\p(\overline{A_i}\sqcup\overline{A_iA_j})+\p(\overline{A_i}\sqcup\overline{A_iA_k})+\p(\overline{A_i}\sqcup A_1A_2A_3) \nonumber \\
&& + \p(\overline{A_iA_j}\sqcup\overline{A_iA_k})+ \p(\overline{A_iA_j}\sqcup A_1A_2A_3)+ \p(\overline{A_iA_k}\sqcup A_1A_2A_3). \nonumber
\eea
Comparing these it is easy to see that that the left and right hand sides of equation \ref{eq:proof sum rule equivalent to quadratic rule 2} are equal, hence the result.
\end{proof}

We now introduce some notation. First, given a histories theory $\H$ and a co-event $\p\in\CE$ we define
\bea
Q_{ABC}(\p) &=& \p(A\sqcup B\sqcup C) + \p(A\sqcup B)+\p(B\sqcup C)+\p(C\sqcup A) \nonumber \\
&& +\p(A)+\p(B)+\p(C),
\eea
where $A,B,C\in\EA$ are disjoint events. Then lemma \ref{lemma:quantum sum rule implies quadratic rule} tells us that $\p\in\CE$ is quadratic if and only if $Q_{ABC}(\p)=0$ for all disjoint $A,B,C\in\EA$.

Now though co-events take values in $\Z2$, probability measures take values in $\R$. It will be useful to algebraically combine the images of (the values returned by) co-events and probability measures, which we can achieve by thinking of $\Z2$ as a subset of $\R$; thus given a co-event $\p\in\CE$ we can construct the related map:
\bea
\tilde{\p}:\EA&\rightarrow&\R \nonumber \\
\tilde{\p}(A) &=& \left\{\ba{cc} 0 & \p(A)=0 \\ 1 & \p(A) = 1. \ea\right. \nonumber
\eea
Using this map we can define a real valued analogue of the function $Q_{ABC}$:
\bea
R_{ABC}:\CE &\rightarrow& \R \nonumber \\
R_{ABC}(\p) &=& \tilde{\p}(A\sqcup B\sqcup C)- \tilde{\p}(A\sqcup B)-\tilde{\p}(B\sqcup C)-\tilde{\p}(C\sqcup A)  \nonumber \\
&& +\tilde{\p}(A)+\tilde{\p}(B)+\tilde{\p}(C) \nonumber,
\eea
where $A,B,C\in\EA$ are disjoint events. We can show that if $\p$ obeys the multiplicative rule then $R_{ABC}$ is either $0$ or $1$, regardless of our choice of $A,B,C$.
\begin{lemma}\label{lemma:dynamical multiplicative coevents R is 0 or 1}
Let $\H$ be a histories theory with a finite sample space, and let $\p\in\CE$ be multiplicative. Then for all disjoint $A_1,A_2,A_3\in\EA$ we have:
$$R_{A_1A_2A_3}(\p)\in\{0,1\}.$$
\end{lemma}
\begin{proof}
Since $\p$ is multiplicative, for any $A\in\EA$ we know that $\p(A)=1$ if and only if $\p^*\subset A$. Then fixing our three events $A_1,A_2,A_3$ we have four cases:
\begin{enumerate}
  \item $\p^* \not\subset A_1\sqcup A_2\sqcup A_3$; then $\p( A_1\sqcup A_2\sqcup A_3)=\p( A_i\sqcup A_j)=\p(A_k)=0$ for all $i,j,k$, so that $R_{A_1A_2A_3}(\p)=0$.

  \item $\p^* \subset A_1\sqcup A_2\sqcup A_3$, $\p^*\not\subset A_i\sqcup A_j$ for any $i,j$; then $\p( A_1\sqcup A_2\sqcup A_3)=1$ but $\p( A_i\sqcup A_j)=\p(A_k)=0$ for all $i,j,k$, so that $R_{A_1A_2A_3}(\p)=1$.

  \item $\p^* \subset A_i\sqcup A_j$ for some $i,j$, but $\p^*\not\subset A_k$ for any $k$. Without loss of generality we can assume that $i,j=1,2$; then $\p( A_1\sqcup A_2\sqcup A_3)=1$, $\p(A_1\sqcup A_2) = 1$, and $\p(A_k)=0$ for all $k$. Further, $\p^*\not\in A_2\sqcup A_3$ otherwise $\p^*\in A_2$, similarly $\p^*\not\in A_3\sqcup A_1$, so $\p(A_2\sqcup A_3) = \p(A_3\sqcup A_1) = 0$. Therefore $R_{A_1A_2A_3}(\p)=1-1=0$.

  \item $\p^*\subset A_i$ for some $i$; without loss of generality assume that $i=1$. Then $\p^*\not\subset A_2,A_3$, since these sets are disjoint to $A_1$. Then $\p(A_k)=\delta_{1k}$, $\p(A_1\sqcup A_2)=\p(A_1\sqcup A_3)=1$, $\p(A_2\sqcup A_3)=0$, and finally $\p(A_1\sqcup A_2 \sqcup A_3)=1$. Thus $R_{A_1A_2A_3}(\p)=1-1-1+1=0$.
\end{enumerate}
Hence the result.
\end{proof}

Finally, we can relate $R_{ABC}$ and $Q_{ABC}$.

\begin{lemma}\label{lemma:dynamical coevents R and Q}
Let $\H$ be a histories theory with a finite sample space and let $\p\in\CE$ be a multiplicative co-event. Then if $A,B,C\in\EA$ are disjoint
$$R_{ABC}(\p)=0 \Rightarrow Q_{ABC}(\p)=0.$$
\end{lemma}
\begin{proof}
First, denote by $X~Mod~2$ the $\Z2$ element corresponding to the integer $X~mod~2$, where $X$ is an integer in $\R$. Then $\p(A)=\tilde{\p}(A)~Mod~2$ for any $A\in\EA$, so noting that $+$ and $-$ are equivalent in $\Z2$, for any disjoint $A,B,C\in\EA$ we have:
\bea
R_{ABC}(\p)~Mod~2 &=& \tilde{\p}(A\sqcup B\sqcup C)~Mod~2 \nonumber \\
&& - \tilde{\p}(A\sqcup B)~Mod~2 -\tilde{\p}(B\sqcup C)~Mod~2-\tilde{\p}(C\sqcup A)~Mod~2  \nonumber \\
&& +\tilde{\p}(A)~Mod~2+\tilde{\p}(B)~Mod~2+\tilde{\p}(C)~Mod~2, \nonumber \\
&=& \p(A\sqcup B\sqcup C) + \p(A\sqcup B)+\p(B\sqcup C)+\p(C\sqcup A) \nonumber \\
&& +\p(A)+\p(B)+\p(C), \nonumber \\
&=& Q_{ABC}(\p). \nonumber
\eea
Hence the result.
\end{proof}

We are now in a position to prove the theorem.

\begin{proof}{\textit{of theorem \ref{thm:dynamical coevents are quadratic}}}
\\By assumption our probability measure $\P_S$ obeys equation \ref{eq:probs on coevents}, then the sum rule obeyed by the quantum measure (equation \ref{eq:quantum measure sum rule}) gives us:
\bea
\P_S(\{\p|\p(A\sqcup B\sqcup C)=1\}) &=& \P_S(\{\p|\p(A\sqcup B)=1\})+\P_S(\{\p|\p(B\sqcup C)=1\}) \nonumber \\
&& +\P_S(\{\p|\p(C\sqcup A)=1\})-\P_S(\{\p|\p(A)=1\})\nonumber \\
&&-\P_S(\{\p|\p(B)=1\})-\P_S(\{\p|\p(C)=1\}), \label{eq:coevent prob sum rule}
\eea
for all disjoint $A,B,C\in\EA$. Now because $\P_S$ is a probability measure we can use the Kolmogorov sum rule to decompose its valuation on any set into a sum of its valuations on the elements of that set. Thus if we define $p_\p=\P_S(\{\p\})$ we get:
\bea
\P_S(\{\p|\p(A)=1\}) &=& \sum_{\p(A)=1} p_\p, \nonumber \\
&=& \sum_{\p\in S} p_\p \tilde{\p}(A), \label{eq:coevent prob decomposition}
\eea
for all $A\in\EA$. Putting this back into equation \ref{eq:coevent prob sum rule} we get:
\bea
\sum_{\p\in S} p_\p \tilde{\p}(A\sqcup B\sqcup C) &=& \sum_{\p\in S} p_\p \tilde{\p}(A\sqcup B)+\sum_{\p\in S} p_\p \tilde{\p}(B\sqcup C)+\sum_{\p\in S} p_\p \tilde{\p}(C\sqcup A) \nonumber \\
&& -\sum_{\p\in S} p_\p \tilde{\p}(A)-\sum_{\p\in S} p_\p \tilde{\p}(B)-\sum_{\p\in S} p_\p \tilde{\p}(C),
\eea
for all disjoint $A,B,C\in\EA$, which we can rewrite as:
\beq\label{eq:decomposed coevent prob sum rule}
0 = \sum_{\p\in S}p_\p R_{ABC}(\p)~~~~\forall~\text{disjoint}~A,B,C\in\EA.
\eeq
Now fix $A,B,C$ and let $Z_{ABC}\subset S$ be the set of co-events $\p$ such that $R_{ABC}=0$. Further denote by $\overline{Z}_{ABC}$ the set of co-events such that $R_{ABC}\neq 0$, so that $S=Z_{ABC}\sqcup\overline{Z}_{ABC}$. Now by assumption every $\p\in S$ is multiplicative, so by lemma \ref{lemma:dynamical multiplicative coevents R is 0 or 1} $\p\in\overline{Z}_{ABC}\Rightarrow R_{ABC}(\p)=1$. Then equation \ref{eq:decomposed coevent prob sum rule} gives us:
\bea
0 &=& \sum_{\p\in Z_{ABC}}p_\p R_{ABC}(\p) + \sum_{\p\in \overline{Z}_{ABC}}p_\p R_{ABC}(\p), \nonumber \\
&=& \sum_{\p\in \overline{Z}_{ABC}}p_\p.
\eea
Noting that $p_\p\geq 0$, this means $p_\p=0$ for all $\p\in\overline{Z}_{ABC}$. Since $A,B,C$ are arbitrary disjoint sets, we conclude that $p_\p=0$ unless $R_{ABC}(\p)=0$ for all disjoint $A,B,C$. But by lemma \ref{lemma:dynamical coevents R and Q} $R_{ABC}(\p)=0$ implies that $Q_{ABC}(\p)=0$, which by lemma \ref{lemma:quantum sum rule implies quadratic rule} implies that $\p$ is quadratic. This completes our proof.
\end{proof}

\section{Approximate Preclusion}\label{sec:approximate preclusion}

Rather than tackling the whole of the dynamics, an alternate approach is to focus directly on the predictive content of a histories theory. Thus instead of expressing $\mu$ in terms of the theory's allowed co-events, we would only be concerned with the probability statements that are the (experimentally falsifiable) predictions of our theory. As before, we seek to be able to `start' with a set of potentially real co-events, and to make our experimental predictions in terms of these co-events and structures defined in terms of them.

Our strategy will be to examine more closely the meaning (or interpretation) of probability \emph{as applied to (experimentally) falsifiable predictions}; thus we look to understand a dynamical statement $\mu(A)=q$ or $\P(B)=p$ through its (experimentally) falsifiable implications. We will begin by focusing on classical theories, where we can draw from a long history of thought upon the matter to pick a suitable interpretation of probability; we then seek to generalise this interpretation in terms of the potentially real co-events in such a manner that can be extended to quantum histories theories.

This strategy requires a somewhat rigorous approach to the interpretation of probability, for we are not simply attempting to justify pre-existing statistical practice, but to construct a predictive framework for a new ontology. In a world of co-events, what do we mean by $\mu(A)=q$, or even $\mathbb{P}(A)=p$? Since co-events have been connected to the measure through the preclusion of null sets, a natural approach is to seek an interpretation of probability based on this notion. We begin with a simple example that will guide us through this process.

\subsection{The Classical Coin}\label{sec:classical coin}

Consider a coin described by classical stochastic dynamics. We assume that if we throw the coin we will either have a `heads' or a `tails' result, we will further assume that all such throws are in some sense `equivalent' so that the probability of `heads' is always $p$ and the probability of `tails' always $1-p$. But what exactly do we mean by this?

One approach to its interpretation is to emphasise the role of probability in representing our knowledge or expectations about a system, so in this case we would for example say that $p=1/2$ if we had no information or view on the outcome of a coin toss \cite{Ramsey:1926,Finetti:1930,Cox:2001}. An alternate approach focuses on the practical testing of such assertions, which takes place through a set of repeated, independent trials \cite{Friedman:1999,Kolmogorov:1933}. In the latter framework a probability assertion $\mathbb{P}(heads)=1/2$ is a statement, not about a single trial, but about a theoretical ensemble of trials, or a prediction regarding a hypothesised set of future trials. Although the former approach is valid and useful, for example in the theory of decision making \cite{Anand:1993}, in light of our difficulties in constructing a probability measure on the space of co-events (section \ref{sec:dynamical coevents}) we will focus on the latter approach, which is more in accordance with the predictive and experimental nature of our field, and as we shall see lends itself naturally to the concept of preclusion. We begin by discussing how we might test an assertion $\P(heads)=p$ in practice.

We test a theory, or a statement within a theory, by testing its experimentally falsifiable predictions. But when we believe there is one realised outcome (the actually real co-event, or the real history if we using the naive interpretation) we face the problem that unless $p\in\{0,1\}$, the statement $\P(heads)=p$ can not be falsified by either outcome `heads' or `tails'. We can however make falsifiable statements about sequences of trials.

The frequentist approach, as articulated by Friedman \cite{Friedman:1999}, interprets a probability statement $\P(heads)=p$ by identifying it with the asymptotic relative frequency of the outcome `heads' in an infinite sequence of repeated trials; under the requirement that the sequence of trials conforms to a `randomness' criterion \cite{Friedman:1999,VonMises:1957}. Indeed, if we had a theoretical ensemble of infinitely many trials, the proportion of the trials resulting in heads would be exactly\footnote{More precisely it would be $p$ with probability $1$.} $p$; alternatively the assertion that `the proportion of heads is not $p$' is precluded by the measure on the ensemble, and herein lies our link with co-events.

However asymptotic properties of a sequence cannot be determined using a finite number of elements of the sequence, and an infinite ensemble of trials is not realisable in terms of an experiment, whereas we seek to phrase probability in terms of experimentally falsifiable predictions. Further, the co-event is meant to be `real', so it is not meaningful to say it precludes an imagined event in a theoretical ensemble that is not part of the `actual' event algebra on which the co-event is defined.

We are thus pushed into defining probabilities based upon finite trials, and will focus on one particular technique of testing probability assertions through a sequence of repeated trials; the standard statistical technique of hypothesis testing \cite{Tucker:1962}. We present a simple version of hypothesis testing that will suffice for the purposes of this chapter.

We begin with our assertion, or \emph{hypothesis}, $\mathbb{P}(heads)=p$ and repeat our trial $n$ times, resulting in a history $\g$ consisting of $n$ ordered `heads' or `tails' outcomes (we will ignore the issue of the `randomness' of this sequence for now). We will denote by $H_\g$ (or $H(\g)$) the number of these trials that result in a `heads' outcome, so that the proportion of heads is $H_\g/n$; to compare this `statistic' \cite{Tucker:1962} with the implications of our hypothesis we assume the product measure (which we shall also denote by $\P$) on the sequence of trials. Using $N_H$ to denote the event that the number of heads in the realised history is $H$, we define the \emph{cumulative probability} of $N_H$ to be:
\beq\label{eq:cumulative probability}
\CP(N_H) = \sum_{m=1}^{H}\P(N_m).
\eeq
Thus $\CP(N_k)$ is the probability that the number of heads in the realised history is less than or equal to $k$. For the purposes of this paper we will say that we \emph{reject the hypothesis at the $\e$ level} if $\CP(N_{H(\g)}) < \e$, where $0<\e\leq 1$ (typically $\e<<1$). If $\g$ is such that $\CP(N_{H(\g)}) \geq \e$ we \emph{fail to reject the hypothesis at the $\e$ level}\footnote{This is essentially a `one-tailed test'; while both one- and two-tailed tests would be equally valid in this context (though they would correspond to different values of $\e$), we will find the one-tailed test simpler to analyse.}. Note crucially that we have implicitly \textit{chosen} to organise the potential outcomes according to the number (or equivalently the proportion) of heads outcomes, based on our hypothesis; the importance of this choice will become clear below.

This technique (or a more sophisticated version thereof \cite{Tucker:1962}) can be used in the testing of scientific theories; $\e$ is then chosen to represent the `degree of certainty' we wish to test our theory to. However, assuming $p\not\in\{0,1\}$ no history (no sequence of outcomes) is actually precluded by our hypothesis; strictly speaking the hypothesis has no falsifiable implications for the realised outcomes, thus no realised history can falsify the hypothesis.

To justify our hypothesis testing technique, and to place it on a more rigorous footing, we turn to \emph{Cournot's Principle} \cite{Cournot:1843,Shafer:2005}; which has long been used to connect ``the mathematical formalism of probability to the empirical world'' \cite{Galvan:2008}.

\subsection{Cournot's Principle}\label{sec:cournot's principle}

Crudely speaking, in one way or another Cournot's Principle `rules out' events of small probability. This concept was certainly known to Bernoulli, who asserted that high probability can be treated as a `moral certainty' \cite{Bernoulli:1713}. Cournot himself made the connection with physics, arguing that events of small probability may be mathematically possible but `physically impossible' \cite{Cournot:1843}. Among many other references, the principle is used or alluded to by Levy \cite{Levy:1925,Levy:1937}, Markov \cite{Shafer:2005}, Borel \cite{Borel:1909} \& Kolmogorov \cite{Kolmogorov:1933,KolmogorovEnglish:1933}, under the name `Principle B'. More recently it has been applied by Goldstein et al to statistical mechanics \cite{Goldstein:2001} and Bohmian mechanics \cite{Durr:1992}, and by Galvan to quantum mechanics \cite{Galvan:2008}. For a more detailed discussion of the history of Cournot's principle see \cite{Shafer:2005}.

However it was the French mathematician Maurice Frechet who coined the term ``principe de Cournot'', which has come into English as `Cournot's Principle', or the `Cournot Principle' \cite{Shafer:2005}. Frechet distinguished between Strong and Weak forms of Cournot's Principle \cite{Frechet:1949}; Shafer describes these two formulations as follows \cite{Shafer:2005}:
\begin{quote}
The strong form refers to an event of small or zero probability that we single out in advance of a single trial: it says the event will not happen on that trial. The weak form says that an event with very small probability will happen very rarely in repeated trials. Some authors, including Levy, Borel, and Kolmogorov, adopted the strong principle. Others, including Chuprov and Frechet himself, preferred the weak principle.
\end{quote}
Kolmogorov's statement of Strong Cournot (his `Principle B) as concerns the probability $\P(E)$ of an event $E$ occurring in an experiment $C$ is \cite{Kolmogorov:1933,KolmogorovEnglish:1933} (as translated by Sahfer \& Vovk \cite{Shafer:2005vv}):
\begin{quote}
If $\P(E)$ is very small, one can be practically certain that when $C$ is carried out only once, the event $E$ will not occur at all.
\end{quote}
These statements contain some ambiguities, for example, in Kolmogorov's `Principle B', what exactly do we mean by `practically certain'? To apply Cournot's Principle to co-events, we will have to be more rigorous. We begin by giving a more precise specification of the Strong Cournot Principle:
\begin{description}
  \item[Literal Strong Cournot] Events of probability less than $\e<<1$ do not occur, for some $\e> 0$.
\end{description}
Where there is no danger of confusion with other variations of the Strong Cournot Principle, we will refer to Literal Strong Cournot simply as the `Strong Cournot Principle', or `Strong Cournot'. This literal specification of the Strong Cournot Principle is in some ways a `maximal' interpretation of the Cournot Principle, and as such may appear to be `neat' or `simple'; however it is problematic in the case of classical stochastic theories. Firstly what do we mean by a small probability, are we to take $\e$ as a constant of nature? Secondly there is nothing to prevent the probability of every single history in the repeated trial being much smaller than the probability of the events we wish to preclude, in which case a literal application of Strong Cournot would rule out all single histories, and thus any reality. In fact an example of such a situation is provided by our coin, given a sufficiently large number of trials.

Since Literal Strong Cournot is not appropriate for classical theories, we instead present a `minimal' version of Weak Cournot that allows us to perform the hypothesis testing described above, but little else.
\begin{description}
  \item[Operational Weak Cournot] In a repeated trial, an event of probability less than $\e<<1$, identified ahead of the repeated trial, will not occur, for some $\e> 0$. Only a single event can be considered for a given repeated trial.
\end{description}
Thus we can only make one prediction for each repeated trial. Where there is no danger of confusion with other variations of the Weak Cournot Principle, we will refer to Operational Weak Cournot simply as the `Weak Cournot Principle', or `Weak Cournot'. Notice that our reference to repeated trials is essentially semantic; a single trial can be considered as a special, `$n=1$', case of a repeated trial, and a general `$n$ times' repeated trial can be considered as a single trial in which sequences of length $n$ are thought of as single outcomes. We insist on the term `repeated trial' to emphasise the link between Operational Weak Cournot and hypothesis testing\footnote{Indeed, we regard Operational Weak Cournot as a variation on Shafer's `weak principle' precisely because of the shared emphasis on repeated trials. We note however, that in its insistence that certain `low probability' events \emph{will not} occur (rather than \emph{occurring rarely}, Operational Weak Cournot bears some resemblance to Shafer's strong principle'.}.

Operational Weak Cournot has the advantage that it fits in well with our actual methods of falsifying theories, such as the statistical hypothesis testing discussed in section \ref{sec:classical coin}. When we make the hypothesis $\P(heads)=p$, there may be certain outcomes that would convince us that this hypothesis has been falsified; for example we may reconsider the assertion $\P(heads)=1/2$ were we to toss our coin a million times only to find a heads outcome for every toss. We can turn this around by saying that \textit{those outcomes that would falsify a theory are precluded by it}. Of course there remains some ambiguity in the choice of $\e$, however this ambiguity is inherent and has not been introduced by our adoption of Weak Cournot; whatever interpretation of probability we adopt we would have to choose the $\e$ we use to falsify our theories (of course this $\e$ need not be unique). In a sense this identification takes the arbitrariness of our choice of $\e$ out of our theory and into the `meta' level on which we compare and reject theories. The `meta' level is always present and by using it to give us $\e$ we have avoided the addition of `new' ambiguity.

Applying this to our coin (section \ref{sec:classical coin}), we start with our hypothesis $\P(heads)=p$, then in the context of a `possible' experiment (or $n$-fold repeated trial) we \emph{choose} to consider the proportion of heads that would be realised in such as experiment, and thus single out the event\footnote{This is an event in the event algebra related to the sample space of all possible sequences of outcomes in our repeated trial.} $L$ that the realised history $\g$ will obey $\CP(H(\g))<\e$. Since the probability of this event is less than $\e$, using Operational Weak Cournot we are justified in asserting that it will not occur (we are assuming the hypothesis). The non-occurrence of $L$ then becomes a falsifiable prediction of our hypothesis; and once we perform the experiment, the occurrence of the event $L$ would falsify the hypothesis.

However, given the realised history $\g$ (once the experiment has been carried out), if $n$ is sufficiently large we will always be able to find an event that has occurred (ie it contains the realised history) yet is assigned probability $<\e$ by the hypothesis; for example, as mentioned above the event that the realised history $\g$ will occur will itself be of probability $<\e$ for sufficiently large $n$. Because of this, to rule out the problems faced by Literal Strong Cournot we have phrased Operational Weak Cournot so as to allow conclusions to be drawn concerning only the single\footnote{Each such prediction is an element of the event algebra related to the sample space of all possible sequences of outcomes in our repeated trial. Thus if we wish to make multiple predictions in the context of a single experiment, we can combine them using the logical (Boolean) operations in the event algebra to yield a single event.} falsifiable prediction (implied by our hypothesis) singled out before the experiment and being tested by it.

In this way Operational Weak Cournot provides us with a practical method of relating probability statements to experimental measurements, though we have had to adopt a `minimalist' and perhaps `restrictive' interpretation of probability, which might for example restrict our ability to treat probability statements as logical predicates \cite{Cox:2001}. However in the classical theory this has no impact on either our ontology or the `meaning' we attribute to dynamics, for we typically assume a deeper deterministic theory operating at a more fundamental level. This will not hold in the quantum case, which therefore will require us to take our chosen interpretation of probability more seriously.

\subsection{Approximate Co-Events}

In one way or another Cournot's principle rules out events of small probability. Because this is a straightforward generalisation of the concept of preclusion, which rules out null sets, we can easily express it in terms of co-events by introducing the concept of \textit{approximate preclusion}. For the sake of clarity we will henceforth refer to preclusion itself as \textit{exact preclusion}, thus an exactly preclusive co-event (which we will sometimes shorten to an \emph{exact co-event}) obeys
$$\mu(A) = 0 \Rightarrow \p(A)=0,$$
whilst an \textit{approximately preclusive co-event} is given by:
\begin{definition}\label{def:approximate preclusion}
Let $\H$ be a histories theory. Given $\e>0$ a co-event $\pe$ is \textbf{approximately preclusive} at the $\e$ level if
$$\mu(A)<\e \Rightarrow \pe(A)=0.$$
We say that $\pe$ is an \textbf{approximately preclusive co-event}, or simply an \textbf{approximate co-event}.
\end{definition}
Notice that this definition holds for a general histories theory, not simply in the classical case, and thus might be used to generalise Cournot's Principle.

Though the introduction of approximate co-events is a fundamental change in the the theory, its implications for the `internal structure' of the `allowed' co-events is less than radical. To be precise, we have effectively altered the precluded events, including `almost null' as well as null sets. We will refer to events of measure less then $\e$ as `approximately precluded' or `$\e$-null' sets, with `$\e$-negligible' sets defined similarly. However other than this we have made no changes, and have not altered the algebraic structure of the co-events, and so can define many of the same concepts and prove many of the same theorems that we could for exactly preclusive co-events. In particular we can define multiplicativity, domination and primitivity based upon approximate preclusion in exactly the same way we defined these concepts for exact preclusion, leading to a notion of `approximate co-event schemes' and in particular an approximate multiplicative scheme, which we shall henceforth assume. Further, we can prove `approximate co-event versions' of many of the results we have established for the exact multiplicative scheme, for example the following lemma, which shall make use of below, is analogous to lemma \ref{lemma:mult existence primitivity}:

\begin{lemma}\label{lemma:approximate mult existence primitivity}
Let $\H$ be a history theory with a finite sample space and let $A\in\EA$ be non-negligible (ie $A$ is not a subset of a null set). Then there exists a primitive approximately preclusive multiplicative co-event $\pe$ such that $\pe(A)=1$.
\end{lemma}
\begin{proof}
See proof of lemma \ref{lemma:mult existence primitivity}.
\end{proof}

In a classical theory, how we interpret $\p_\e$ depends on the version of Cournot's Principle we are following. Weak Cournot means that the $\p_\e$ are essentially theoretical tools, used to phrase (experimentally falsifiable) predictions in terms of preclusion. In this case $\e$ will be taken from our `meta' level choice of $\e$ used to falsify a given theory, and may not be the same for every system. On the other hand Strong Cournot means that the $\p_\e$ are the potential realities, in which case we may consider $\e$ as a constant of nature. Note that our basic ontology remains unchanged, the `actually real' co-event remains our description of reality and its internal structure (or logic) is still given by the multiplicative rule\footnote{We assume that the multiplicative scheme holds for all histories theories, in a classical theory it `happens' to coincide with the classical scheme.}. What we have done is to alter the set of potential realities \textit{given} a measure, thus essentially we have altered the role \& meaning of the measure.

Moving from classical stochastic theories to quantum mechanics and the multiplicative scheme, we are no longer treating a single history as real\footnote{More precisely we are no longer expecting our primitive multiplicative co-events to be classical} so our objections to Strong Cournot may no longer be relevant. This leads us to question whether it may be possible to achieve Strong Cournot in the context of quantum measure theory, and to take the $\pe$ literally. One advantage of such an approach lies in our practise of treating some events of small probability as null. At the beginning of this chapter we raised concerns regarding the preclusion of `approximately null' sets, arguing that true null sets are rare; in fact due to experimental inaccuracies the events we are characterising as null will in general be only of small probability. Approximate preclusion is tailor made to address such concerns, in particular if we are able to adopt Strong Cournot we can preclude such sets directly.

If Strong Cournot, as expressed by approximate co-event, is to be applicable to a general quantum measure theory, it must certainly make sense in the context of classical theories. In the next section we therefore use approximate preclusion to explore the application of Strong Cournot to hypothesis testing in a classical theory; returning to the example of our simple coin (section \ref{sec:classical coin}). In what follows we assume that all co-events are multiplicative.

\subsection{Can we Achieve Strong Cournot?}\label{sec:strong cournot}

\subsubsection{Approximate Preclusion for a Coin}

As before we assume that if we throw the coin we will either have a `heads' (`$h$') or a `tails' (`$t$') result, we will further assume that all such throws are in some sense `equivalent' so that the probability of `heads' is always $p$ and the probability of tails always $1-p$. Thus in the histories formalism, a single throw corresponds to the sample space: \beq\label{eq:coin sample space}
\O=\{h,t\}.
\eeq
In the language of our hypothesis testing technique (section \ref{sec:classical coin}), our hypothesis is that $\P(\{h\})=p$, so that the classical measure $\P$ is:
\bea
\P(\{h\})&=&p,\nonumber \\
\P(\{t\})&=&1-p,\nonumber \\
\P(\{h,t\})&=&1.\label{eq:coin dynamics}
\eea
An `experiment' consisting of $n$ trials (which we assume to be independent) has the sample space
\beq
\O = \{h,t\}^n,
\eeq
of histories (ordered sequences of outcomes) $\gamma=a_1,\ldots , a_n$, where $a_i\in\{h,t\}$. The corresponding event algebra is $P\O$ and since we have assumed that our trials are independent we use the product measure, which by abuse of notation we shall also denote by $\P$.

Now we have previously denoted the number of `heads' outcomes in a history $\g$ by $H(\g)$, so that the proportion of `heads' is $H(\g)/n$. Then we have:
\beq\label{eq:coin single history probability}
\P(\{\g\})=p^{H(\g)}(1-p)^{n-H(\g)}.
\eeq
We are less interested in the individual histories than in the proportion of heads, which, as discussed above, should reflect the underlying probability $p$ of a heads outcome in a single trial. We will denote by $N_H$\ the event that the number of heads realised in the sequence is $H$. Then we have:
\beq\label{eq:coin probability of H heads}
\P(N_H) = \left( \begin{array}{c} n \\ H \\ \end{array} \right) p^{H}(1-p)^{n-H}.
\eeq
Further, we will label by $L_H$ and $G_H$ the events that the number of heads in the realised sequence is less than or equal to or greater than $H$ respectively. Then we get:
\bea
\P(L_H)&=&\sum_{m\leq H}\P(m), \nonumber \\
&=& \CP(N_H). \label{eq:coin proabaility of H or less heads}
\eea
Thus if we performed such an experiment and realised a proportion of heads corresponding to a small $\P(L_H)<\e$ we would reject our hypothesis $\P(h)=p$ (at the $\e$ level). Turning this around, given the assumption $\P(h)=p$ we wish to preclude all events $L_H$ (and $N_H$) with $\P(L_H)<\e$ for some small $\e$.

We can make this more precise; since the measure $\P$ is classical and non-zero everywhere, and since $L_m$ is a proper subset of $L_{m+1}$, we can see that $\P(L_m)$ is monotonic in $m$. Thus defining $H_\e$ as the greatest $H$ such that $\P(L_H)<\e$, we have $\P(L_{H_\e+m})\geq\e ~\forall ~m>0$; in particular $\P(L_{H_\e})<\e\leq\P(L_{H_\e+1})$. Thus, following the argument of section \ref{sec:classical coin} we would like to say that $L_{H_\e}$ is precluded while $L_{H_\e+1}$ is not. Since $\P(L_H)>0$ in all cases, we turn to approximate preclusion.

Adopting Strong Cournot, we fix $\e$ and assume that events of measure less than $\e$ never occur. Then we can immediately rule out $L_{H_\e}$, indeed $\p_\e(L_{H_\e})=0$ for all approximately preclusive co-events $\p_\e$. However, this does not hold for $L_{H_\e+1}$, indeed since $\P(L_{H_\e+1})>\e$ by lemma \ref{lemma:approximate mult existence primitivity} there exists a primitive approximate co-event $\p_{H_\e}$ mapping it to unity. In other words:
\bea
\p_{H_\e}(L_{H_\e})&=&0, \nonumber \\
\p_{H_\e}(L_{H_\e+1})&=&1.
\eea
Thus every potential reality precludes $L_{H_\e}$, but not every potential reality precludes $L_{H_\e+1}$. This is enough to allow us to use the hypothesis testing technique outlined in section \ref{sec:classical coin}, thus enabling us to deal with (experimentally falsifiable) predictions in terms of our approximate co-events. Since approximate preclusion is defined in terms of a general histories theory, this seems to raise the possibility of a predictive approximate co-event framework for a general quantum measure theory. However, though widening the scope of preclusion by moving from exact to approximate co-events has introduced useful features to our theory, we must make sure that it has not also introduced problems.

\subsubsection{The Failure of Strong Cournot}

The above construction seems promising, it seems that our shift from exact to approximate preclusion has succeeded in introducing new \& useful features to co-event theory; however we must check that it has not inadvertently introduced new problems as well. In dealing with exactly preclusive co-events we can show that (under certain conditions) classical measures always imply classical outcomes meaning that the co-events will necessarily behave classically (see section \ref{sec:classical partitions}); do the approximate co-events always `behave'? Such concerns lead to three objections to the construction we outlined above:

\begin{description}

\item[I The Status of Single Histories]~

    The first question to be raised regards the status of single histories. In a system obeying non-classical dynamics we may be happy with single histories not being realised, however in this system with its classical dynamics we could still find all single histories to be ruled out for a sufficiently large number of trials. This raises interpretational issues, for example if we take $p=1/2$, $\e=10^{-3}$ and $n=10$ we find all single histories ruled out. However if $n=9$ all single histories are allowed. We have several problems here, firstly we may not be comfortable with the non-classical behaviour of the $n=10$ system. Secondly there is the question of the value of $\e$, which leads to classical outcomes in the $n=9$ case but not the $n=10$ case. Finally it seems odd that one additional trial will `disallow' classical behaviour, particularly given that a single trial behaves classically in and of itself.

    Note that we don't have a causality problem here; although in the $n=10$ system no single history is realised this will not become apparent in the first trial. The two outcomes of the first trial can be thought of as a coarse grained partition of the $n=10$ sample space, and as such each outcome has probability greater than $\e$. This will also hold in general, any single history in the $n=10-m$ system (where $1\leq m\leq 9$) corresponds to a coarse grained event in the $n=10$ system that has probability greater than $\e$ and so is not precluded.

    Further note that this problem of systems that behave classically becoming non-classical in a repeated trial can occur in the histories framework without the use of co-events. We could take the view that no system is truly classical, but rather that at the level of observable events we have an emergent classicality based on environmental decoherence. Adopting this view, observed `classical' behaviour is due to approximate decoherence \cite{Sorkin:private,Dowker:private}, and repeated trials of an approximately decohering system (such as a `quantum coin' with interference $\varepsilon$ between heads and tails) may lead to non-classical outcomes, notably the reappearance of quantum coherence \cite{Sorkin:private}.

\item[II The Problems of Assigning Ontology to the Approximate Co-Events]~

    The second question to be raised concerns the range of possible co-events `allowed' by approximate preclusion. We may be able to construct allowed co-events $\phi_\e$ such that some observable questions have no definite answer. In other words our framework allows potential realities that if realised would imply observable anhomomorphisms.

    For example, $\{L_{H_\e},G_{H_\e}\}$ is a partition of $\O$, both  elements of which are observable (since all events are observable in this classical system). Then using lemma \ref{lemma:approximate mult existence primitivity} there will be at least one primitive approximate co-event $\p_{G_\e}$ mapping $G_{H_\e}$ to $1$, and $\p_{G_\e}$ will map $L_{H_\e}$ to $0$ (as do all approximate co-events) because $\P(L_{H_\e})<\e$. However, although we can `cherry-pick' the co-event $\p_{G_\e}$ to treat the partition $\{G_{H_\e},~L_{H_\e}\}$ classically, we do not know if $\p_{H_\e}$ will map $G_\e$ to one and so treat this partition classically. More generally we do not have any guarantee that all the allowed co-events will treat this partition classically; in fact we can find primitive approximate co-events that map both elements of the observable partition $\{G_{H_\e},L_{H_\e}\}$ to zero.

    We can find subsets of histories $T\subset L_{H_\e}$ and $P\subset G_{H_\e}$ such that $\P(T),\P(P)<\e$ but $\P(T\sqcup P)>\e$.  Now $\P(T\sqcup P)>\e$ means that by lemma \ref{lemma:approximate mult existence primitivity} some subset $C$ of $T\sqcup P$ will be the dual of a primitive approximate co-event $C^*$. Further both $T$ and $P$ are approximately precluded so $C$ cannot be a subset of either; therefore $C$ has non-empty intersection with both $L_{H_\e}$ and $G_{H_\e}$, which means:
    \bea
    C^*(L_{H_\e})&=&0 \nonumber \\
    C^*(G_{H_\e})&=&0.
    \eea
    For an explicit example consider the case $p=1/2,~n=10^3$ and $\e=10^{-3}$; then $H_\e=450$ since $\P(L_{450})<\e$ whereas $\P(L_{451})\geq\e$. Now $L_{451}=L_{450}\sqcup N_{451}$, so there is some (not necessarily unique) subset $S\subset N_{451}$ such that $\P(S\sqcup L_{450})\geq\e$ whereas $\P((S\setminus\{\g\})\sqcup L_{450})<\e$ for any $\g\in S$. We can think of constructing $S$ by adding fine grained histories from $N_{451}$ to $L_{450}$ one by one until the measure is greater than $\e$; thus we can not reduce the set $S\sqcup L_{450}$ without its measure falling below $\e$. In fact, because $p=1/2$ (and the measure is classical) every single history $\g$ contributes the same amount, $p^n=2^{-10^3}$, to the probability of any event containing $\g$, so that $\P(S)=p^n |S|$. Further, since the measure is classical we know that $\e\leq\P(S)+\P(L_{450})<\e+p^n$. Therefore: \beq
    p^{-n}(\e - \P(L_{450})) \leq |S| < p^{-n}(\e - \P(L_{450})) + 1,
    \eeq
    so that $S$ could be any subset of $N_{451}$ of cardinality:
    \bea
    |S|&=&Int(\frac{\e-\P(L_{450})}{p^n})\nonumber \\
    &\approx& 1.4\times10^{297},
    \eea
    where $Int(x)$ denotes the least integer which is greater than or equal to $x$. Then the set $S\sqcup L_{450}$ has measure greater than or equal to $\e$, and so is not a subset of any $\e$-null set (since the measure is classical). Further, we cannot reduce our set without the remainder being $\e$-negligible. Therefore $S\sqcup L_{450}$ is the base of an approximate co-event $(S\sqcup L_{450})^*$ that maps both $G_{H_\e}$ and $L_{H_\e}$ to zero.

\item[III The Inconsistency of Multiple Partitions]~

     Finally, in the examples we have looked at so far, we considered a single partition $\O=G_{H_\e}\sqcup L_{H_\e}$ which was treated non-classically by some allowed co-events. Intuitively, both elements of this partition correspond to `meaningful' propositions; that the proportion of heads in the observed history is less than or equal to ($L_{H_\e}$), or greater than ($G_{H_\e}$), $H_\e/n$. However, we did not attach any particular `meaning' to our `problem co-event' $C^*$, indeed as a proposition the related event $C$ is `pathological' in that it would be difficult to express it as an `English sentence' as we were able to do for $L_{H_\e}$ and $G_{H_\e}$ above. This might lead one to speculate that we could perhaps focus on partitions and co-events corresponding to `good', `meaningful' or `useful' propositions, whilst ignoring partitions and co-events corresponding to `bad' or `pathological' propositions; indeed we were able to find the `good' approximate co-event $\p_{G_\e}$ that was classical on the `good' partition $\O=G_{H_\e}\sqcup L_{H_\e}$.

    However the situation is not so simple; in the above we focused on a single partition, whereas there are many `good' partitions, each with its associated `good' co-events. Unfortunately the co-events that treat one partition classically will not in general be classical on another partition, as the following example shows.

    When the number of trials of our coin is even, $n=2m$, we can partition the sample space $\O$ into even and odd coarse grainings as follows. Given a single history $\g=a_1,a_2,\ldots,a_{2m}$ (where $a_i\in\{h,t\}$) we can form the \textit{even} and \textit{odd} histories $E(\g)=a_2,a_4,a_6,\ldots,a_{2m}$ and $O(\g)=a_1,a_3,a_5,\ldots,a_{2m-1}$. Likewise we can form the even and odd coarse grainings $\O_E=\{E(\g)|\g\in\O\}$ and $\O_O=\{O(\g)|\g\in\O\}$, which inherit their associated measures from $\O$, and it is easy to see that $\O=\O_E\sqcup\O_O$. Building on this, we can treat both $\O_E$ and $\O_O$ in the same way that we previously treated $\O$ by looking at the number of heads in the even and odd trials, $H^E$ and $H^O$ respectively. Given an $\e$ we can go on to define $H^E_\e,H^O_\e$, and the subsets $G_{H^E_\e},G_{H^O_\e}$ and $L_{H^E_\e},L_{H^O_\e}$ of the two distributions. Then, following the analysis above, we can define the co-events $\p^E_{G^E_\e},~\p^O_{G^O_\e}$  corresponding to the questions `is the number of heads in the even distribution greater than or equal to $H^E_\e$?' and `is the number of heads in the odd distribution greater than or equal to $H^O_\e$?'. Now as before we have
    \bea
    \p^E_{G^E_\e}(L_{H^E_\e})&=&0 \nonumber \\
    \p^E_{G^E_\e}(G_{H^E_\e})&=&1, \label{eq:approx coevent multiple partitions classical}
    \eea
    however now we also have
    \bea
    \p^E_{G^E_\e}(L_{H^O_\e})&=&0 \nonumber \\
    \p^E_{G^E_\e}(G_{H^O_\e})&=&0, \label{eq:approx coevent multiple partitions nonclassical}
    \eea
    with $\p^O_{G^O_\e}$ showing similar behaviour.

    For an explicit example consider the case $p=1/2$, $m=10^3$ and $\e=10^{-3}$; thus the even and odd distributions are similar to the example considered above. Then $H^E_\e=H^O_\e=450$, and as before every single history $\g$ contributes the same amount, $p^{2m}=2^{-2*10^3}$, to the probability of any event containing $\g$; thus the measure of any event $S$ is given by $\P(S)=p^{2m} |S|$. From this we can see that an event $S$ is a subset of an approximately precluded set $T$ if and only if $S$ itself is approximately precluded, thus an approximate co-event $\p_\e$ is preclusive if and only if its dual $\p_\e^*$ has measure greater than or equal to $\e$; then if $\p_\e$ is preclusive
    \beq\nonumber
    \e \leq \P(\p_\e^*) = p^{2m}|\p_\e^*|,
    \eeq
    so that
    \beq\nonumber
    |\p_\e^*| \geq \e p^{-2m} = 10^{-3} \times 2^{2\times 10^3}.
    \eeq
    Thus $\p_\e$ is primitive if and only if $|\p_\e^*|=Int(\e p^{-2m})$, where $Int(x)$ denotes the least integer that is greater than or equal to $x$. Now let $\g_E=a_1,a_2,\ldots,a_{2m}$ be the history defined by the outcomes
    \beq\nonumber
    a_i = \left\{\ba{cc} h & i~even \\ t & i~odd. \ea\right.
    \eeq
    Then $E(\g_E) = hhhh\ldots$ and $O(\g_E) = tttt\ldots$, so $\g_E\in L_{H^O_\e}\cap G_{H^E_\e}$. Now it is easy to see that $\P(G_{H^E_\e})>\e$, so that $|G_{H^E_\e}|>\e p^{-2m}$ and thus we can find an event $C\subset G_{H^E_\e}$ of cardinality $Int(\e p^{-2m})$ that contains the history $\g_E$. But then setting $\p^E_{G^E_\e}=C^*$, we have found a primitive approximately preclusive co-event satisfying equations \ref{eq:approx coevent multiple partitions classical} \& \ref{eq:approx coevent multiple partitions nonclassical}. A similar construction can be made for $\p^O_{G^O_\e}$.

    This is in fact a coarse graining problem (as are most of the problems in the multiplicative scheme) reminiscent of the interpretational problems of the consistent histories approach. Essentially we can recover the probabilities in the measure (or decoherence functional) but we are also recovering the interpretational problems in the sense that different coarse grained partitions have become incompatible. In consistent histories each question `makes sense' in one decoherent partition, but may not have a classical answer in other decoherent partitions. In our case every approximate co-event (which corresponds in a natural way to a question via its dual) will treat at least one partition classically (for example the partition consisting of its dual and the complement thereof) but may not treat other partitions classically, even though these partitions are dynamically classical (in that they decohere). However this feature is more of an issue for approximate preclusion than for consistent histories, since, as our coin example shows, approximate preclusion may have difficulties even when the underlying fine grained histories obey classical dynamics, a problem not shared with consistent histories.
\end{description}

\subsection{Quantum Operational Weak Cournot}

Although Strong Cournot may be more philosophically satisfying, in particular with regard to co-event theory, the arguments of section \ref{sec:strong cournot} above conclusively demonstrate the failure of its application to multiplicative co-events due to violations of observable classicality. As in classical stochastic theories, this leads us to fall back on Weak Cournot. Though perhaps less philosophically satisfying than the strong variety, it may be the `best we can do', at least at the present time.

Our application of Operational Weak Cournot to quantum histories theories is similar in some ways to its classical application (section \ref{sec:cournot's principle}). We again evoke the `meta-level' process by which we falsify theories to give us $\e$, which consequently is no longer considered as a constant of nature and may be different for different systems. Our restriction of approximate preclusion to events singled out ahead of a repeated trial avoids the problems of multiple partitions and single histories encountered above. Essentially, we depart from Strong Cournot by introducing a split between the ontology and the predictive content of the theory; whereas in Strong Cournot both are described by a primitive approximate co-event we now propose:
\newline\newline
\textbf{Ontology}: The potential realities are primitive exactly preclusive (multiplicative) co-events. In the example of the classical coin above, since the measure is classical, these primitive co-events correspond to single fine grained histories.
\\\textbf{Predictions}: In an experiment consisting of repeated trials, an experimentally observable event $A$ singled out in advance, of measure less than $\e$, will not occur. We can of course phrase this in terms of approximate co-events, replacing the phrase `of measure less than $\e$' by `precluded by all approximate co-events $\p_\e$'. However this is tautological, and the use of approximate co-events is now rather vacuous since the ontology is given by primitive exactly preclusive co-events.
\newline\newline
This formalism aims to allow us to use the hypothesis testing technique described in section \ref{sec:classical coin} to test `Copenhagen predictions' (discussed in the preamble to this chapter) through experiments consisting of repeated trials, for both classical and non-classical systems. There are, however, several objections we can make.

Firstly, the restriction of our predictive ability to such experiments may obstruct the application of Weak Cournot beyond the Copenhagen framework, and the need to single out our predictions in advance of an experiment raises questions regarding observer independence; a hallmark of the histories approach. Thus even if Weak Cournot is successful in justifying the Copenhagen predictive mechanism from the perspective of quantum measure theory it is doubtful that it will assist us to develop a wider conception of the predictive content of a general quantum measure theory as hoped for in the preamble to this chapter. Simply by choosing to focus solely on (experimentally falsifiable) predictions we have constructed a formalism in which we cannot attach any (`physical') meaning to non-predictive dynamical statements. While such an `instrumentalist' view of dynamics may be appealing to some \cite{Dowker:private}, the authors feel that the dynamics should express something about the structure of reality, even in the absence of observation.

Had Literal Strong Cournot been applicable it would have avoided such issues, since Cournot's Principle would have been applied at the ontological level and thus its implications would have been observer independent. More importantly this split between the ontology and the predictive mechanism separates us not only from Strong Cournot but also from the application of Weak Cournot in classical physics. In a classical theory we typically assume an underlying deterministic dynamics; the dynamical statements of this `true theory' would then have observer independent meaning, allowing probability to be considered as an instrumentalist phenomena without any implication for the interpretation of this `true dynamics'.

Further, in a classical stochastic theory our statement of Weak Cournot is made in terms of the event algebra and the measure, which are both defined in terms of the histories; the potential realities of the theory. In quantum measure theory our potential realities are primitive exactly preclusive multiplicative co-events, however our formulation of Weak Cournot is still expressed in terms of events and the measure defined on them, or alternatively in terms of approximate co-events. Thus contrary to the goals we set out in the preamble to this chapter we have not been able to express the predictive content of quantum measure theory in terms of its potential realities.

Finally, by adopting the above formulation of Weak Cournot we have not made headway in our attempt to deal with `almost null' sets. Given our practice of treating `almost null' sets as `exactly null' by assuming idealised conditions our inability to address this issue is a cause for concern.

\section{Conclusion}

Our aim in this chapter has been to rephrase the dynamical \& predictive content of a quantum histories theory in terms of co-events, or at the very least to demonstrate that this is possible in principle even if in practice it may be simpler to work with event algebras. Further, we were concerned that due to experimental inaccuracies the events that we are characterising as null will in general be only of small probability.

Our first, and potentially most appealing proposal was to shift the dynamics wholesale onto the co-events, `completing' the co-event program by moving from `histories theories' to `co-event theories' in which we could use co-events and the dynamics defined on them to deal directly with physical systems. This would have allowed us to deal with events of small probability by assigning small probabilities to the co-events that found those events to be true. Unfortunately this approach has stalled.

Our second proposal was approximate preclusion. Following Strong Cournot we aimed to explain the emergence of probability from preclusion by altering the status of the measure with regard to the allowed co-events. This would have allowed us to deal with events of small probability by directly precluding them. Again, this approach has not met with success.

Finally we have been driven to adopt Weak Cournot, which can perhaps be used to justify the (existing) Copenhagen experimental framework from the vantage point of a quantum histories theory. However we do not manage to express our predictions in terms of the `potentially real' co-events themselves, and make no headway in extending our understanding of a histories theory's predictive content beyond the narrow Copenhagen framework. Further, we are left in a quandary regarding the application of preclusion in practice. As we have previously pointed out the events that we are assuming to be null, for example in the double slit experiment (see appendix \ref{appendix:many slit}), are only found to have zero measure following the assumption of an idealised system that we know cannot hold in practice. Thus for our results to have application to the real world an alternate solution to this problem must be found. %Dynamics
%\singlespacing
\chapter*{Conclusion}
%\doublespacing
\addcontentsline{toc}{chapter}{Conclusion}
\pagestyle{myheadings}
\markboth{CONCLUSION \hfill}{CONCLUSION \hfill}

In our opening comments we set out to establish a `realist, observer independent interpretation of quantum measure theory'. We now step back to review what we have achieved, and our progress toward this goal.

After formally introducing quantum measure theory as a reformulation of the histories approach in the image of classical stochastic mechanics we began by adopting the standard interpretation of the classical theory as the `naive interpretation' of quantum measure theory. We then constructed a gedankenexperiment (the PKS system) which the naive interpretation was unable to describe. The need to respond to the failure of the naive interpretation in describing quantum dynamics motivated the remainder of our work.

Our first response was to hold to the naive interpretation and alter the dynamics, albeit in a non-measurable fashion\footnote{More precisely, the alterations of the dynamics by stochastic collapse models can be measurable in principle, though the parameters are typically set so as not to contradict current experimental results.}. Since we were confident in our interpretation of classical stochastic theories we sought to resolve the interpretational difficulties of quantum measure theory by reducing the dynamics to a classical stochastic theory. Our means of achieving this was stochastic collapse theory, though other approaches including hidden variable theories follow the same philosophy. We judged our effort to be unsuccessful, though this in no way prejudices other developments in this direction, which remains a critical area of research.

Our second response was to accept the quantum dynamics unaltered and to abandon the naive interpretation. We proposed a new interpretation, based on a `shift in thinking' that would replace events with co-events as the central objects of the theory. The failure of the naive interpretation could then be viewed as the failure of the classical scheme. This led to the question of what condition homomorphism should be replaced by, and we presented the major co-event schemes that have been considered as candidates.

After demonstrating the failure of several co-event schemes, including the linear scheme which historically was the first to be suggested, we settled on the multiplicative scheme which, at the time of writing, is the working model of co-event theory. We finished by turning back to the dynamics, asking whether it could be phrased in terms of co-events and the multiplicative scheme in particular, an endeavor that has so far met with frustration.

The multiplicative scheme achieves all of the `criteria' we set out in section \ref{sec:the interpretation of the interpretation}, with two caveats which we shall discuss below. This is no mean feat, and establishes the co-event approach, and the multiplicative scheme in particular, on a solid footing as a viable interpretation of quantum measure theory. Looking ahead, a next step for co-event theory may be its application to the search for quantum gravity. As an example, a key obstruction to the development of a quantum theory of causal sets has been the need for a `quantum causality condition' to replace `Bell causality' \cite{Craig:2006}. Since causality is linked to the ontology, and we have not had a working picture of the ontology of quantum measure theory until now, this field has stalled. The application of the multiplicative scheme may well prove critical in moving causal set theory forward, and in fact this was one of the key motivations behind the development co-event theory itself \cite{Dowker:private}.

However the multiplicative scheme is still very much a work in progress, leading to the two caveats we mentioned above. The scheme is at present only defined in the finite case, and the question of what could (or should) constitute a classical partition is not at present resolved. In both of these areas there is case for optimism, we have not encountered any obstructions to the generalisation of the multiplicative scheme to the infinite case, rather we have indications both that such a generalisation is possible and of what it may look like. We have been held up only by the lack of the full development of quantum measure theory itself in the infinite case, and the author is confident that a generalisation of the multiplicative scheme to the infinite case is within reach. As to classical partitions, the preclusive separability results are themselves powerful, and we know that the conditions of theorem \ref{thm:Rafael's thm} can be weakened considerably. This remains an open area for future research, and the author is again confident of positive results in this field.

Our failure to express the dynamics in terms of the co-events is of more concern. Though the ontology of the multiplicative scheme is not challenged, and we are able to achieve an interpretation of probability compatible with it, we do this at the cost of adopting a narrow instrumentalist view of probability, and further are unable to express the dynamics in terms of the theory's potential realities. As a consequence we are so far unable to address the rarity of exact preclusion in non-idealised systems, which is an obstruction to the applicability of our results to the real world. 

\appendix
\part{Appendix}
\pagestyle{headings}
%\singlespacing
\chapter{Proof of Lemma \ref{lemma:decoh}}
%\doublespacing

%\section{Proof of Lemma \ref{lemma:decoh}}

\begin{proof}\;  \textit{{Of Lemma}} \ref{lemma:decoh}

Recall the definition of $D_{qc}$:
\begin{align*}
D_{qc}(Cyl(\Phi^n,\alpha^n)\,;\,& Cyl(\m{\Phi}{}^n,\m{\alpha}{}^n)) =\\
{}&D_q(Cyl(\Phi^{n})\,; Cyl(\m{\Phi}{}^{n}))\; \frac{X^{d(\Phi^n,\alpha^n)+
d(\m{\Phi}{}^n,\m{\alpha}{}^n)}}{(1+X^2)^{2n}} \;\delta(\alpha^n,\m{\alpha}{}^n)\;.
\end{align*}
When the sum is taken over all $\alpha^n$ and $\m{\alpha}{}^n$, field
configurations on the first $2n$ vertices,
 it results in
 \begin{align}
D_{qc}(Cyl(\Phi^n)\times \Omega_c\,;\,& Cyl(\m{\Phi}{}^n)\times \Omega_c) =
\nonumber \\
{}& \frac{1}{(1+X^2)^{2n}}\;  D_q(Cyl(\Phi^{n})\,; Cyl(\m{\Phi}{}^{n}))
\sum_{\alpha^n}  X^{d(\Phi^n,\alpha^n)+
d(\m{\Phi}{}^n,\alpha^n)}\;.
\end{align}

Let $d(\Phi^n, \m{\Phi}{}^n) = m$, which is the number of links on
which the values of the two fields differ. For the duration
of this proof only, we relabel the links on which the two
fields differ $l_1,l_2,\dots l_m$ and the rest,
on which the fields agree, are labelled $l_{m+1},\dots l_{2n}$.
Consider the exponent ${d(\Phi^n,\alpha^n)+
d(\m{\Phi}{}^n,\alpha^n)}$. The first $m$ links contribute $m$ to
the exponent whatever $\alpha^n$ is,
because for each link, $\alpha^n$ will agree with
exactly one of $\Phi^n$ and $\m{\Phi}{}^n$. Therefore
\begin{equation}
{d(\Phi^n,\alpha^n)+
d(\m{\Phi}{}^n,\alpha^n)} =
m + 2\tilde{d}(\alpha^n, \Phi^n)\,,
\end{equation}
where $\tilde{d}$ is the number of the last $2n -m$ links
on which $\alpha^n$ and $\Phi^n$ differ.

The sum over $\alpha^n$
can be expressed as a multiple
sum over the values of the $\alpha$ variable on each
link in turn. We first do the sum over the values on the $m$ links
on which $\Phi^n$ and $\m{\Phi}{}^n$ differ. The summand does not depend
on the values on those links and so that gives a factor of $2^m$
 \begin{equation}
\sum_{\alpha^n} X^{d(\Phi^n,\alpha^n)+
d(\m{\Phi}{}^n,\alpha^n)}
=  2^m X^m \sum_{\alpha^n_{m+1}} \dots \sum_{\alpha^n_{2n}}
X^{2\tilde{d}(\alpha^n, \Phi^n)}\;.
\end{equation}
The remaining sum is over all $\alpha$ configurations on the last
$2n-m$ links. There is one such configuration that agrees with
$\Phi^n$ on all $2n-m$ links, $\binom{2n-m}{1}$ configurations
that differ from $\Phi^n$ on one link,
$\binom{2n-m}{2}$  that differ from $\Phi^n$ on two links, {\textit {etc}}.
The remaining sum therefore gives $(1+X^2)^{2n-m}$ and we have
 \begin{equation}
\sum_{\alpha^n} X^{d(\Phi^n,\alpha^n)+
d(\m{\Phi}{}^n,\alpha^n)}
=  2^m X^m (1+X^2)^{2n-m}\,,
\end{equation}
and hence the result.

\end{proof}

\begin{claim}

\begin{equation}
\ket{\,\Phi^n, E^n}_{qe} =
\frac{X^{d(\Phi^n, E^n)}}{(1+X^2)^n}\; \ket{\Phi^n}_q
\ket{E^n_1}_{e_1} \ket{E^n_2}_{e_2}\dots
\ket{E^n_{2n}}_{e_{2n}}\otimes_{a=2n+1}^\infty \ket{0}_{e_a} \,,
\end{equation}
where $\ket{\Phi^n}_q$ is given by (3.1).

This is the claim in lemma 3.
\end{claim}
\begin{proof}

We use induction. It is trivially true for $n=0$.

We assume it is true for $n$. Let $\Phi^{n+1}|_n = \Phi^n$ and
$E^{n+1}|_n = E^n$. Then
\begin{align}
\ket{\,\Phi^{n+1}, E^{n+1}}_{qe} ={}&\;
Q_{2n+2}(E^{n+1}_{2n+2})\, P_{2n+2}(\Phi^{n+1}_{2n+2})
\, Q_{2n+1}(E^{n+1}_{2n+1})\, P_{2n+1}(\Phi^{n+1}_{2n+1})\nonumber\\
{}&\ \ \ \ U_{2n+2}\, U_{2n+1}\, R_{n+1}\ket{\Phi^n,E^n}_{qe}\,.
\end{align}

The $P$ projectors commute with the $Q$ projectors. The $P_{a}$ projectors
also commute with the partial measurement operators $U_{a}$ as can be seen from
the definition of $U$ (\ref{eq:defU}). So we have
\begin{align}
\ket{\,\Phi^{n+1}, E^{n+1}}_{qe} = \,&\,\frac{X^{d(\Phi^n, E^n)}}{(1+X^2)^n}\;
Q_{2n+2}(E^{n+1}_{2n+2})\, Q_{2n+1}(E^{n+1}_{2n+1})\,  U_{2n+2}\, U_{2n+1}
\nonumber\\
{}&\ \ \ \ \ \ \ \left[\,P_{2n+2}(\Phi^{n+1}_{2n+2})\,P_{2n+1}(\Phi^{n+1}_{2n+1})\,R_{n+1}
 \ket{\Phi^n}_q \,\right]\nonumber \\
{}&\ \ \ \ \ \ \ \ \ \ \ \ \ \ket{E^n_1}_{e_1} \dots
\ket{E^n_{2n}}_{e_{2n}} \ket{0}_{e_{2n+1}}\ket{0}_{e_{2n+2}}
\otimes_{a=2n+3}^\infty \ket{0}_{e_a} \,.
\end{align}

The factor in square brackets is $\ket{\Phi^{n+1}}_q \in H_q$ and is
unchanged by the $U$'s because it is an eigenstate of the field $\Phi$
on the links $l_{2n+1}$ and $l_{2n+1}$. The same factor is also
unchanged by the $Q$'s
which only act on the
environment states.
$U_{2n+1}$ turns $\ket{0}_{e_{2n+1}}$ into a linear combination
of $\ket{0}_{e_{2n+1}}$ and $\ket{1}_{e_{2n+1}}$, enhancing the
term which is correlated to the value $\Phi^{n+1}_{2n+1}$.
Similarly for $U_{2n+2}$. Finally $Q_{2n+1}(E^{n+1}_{2n+1})$
projects onto the state $\ket{E^{n+1}_{2n+1}}_{e_{2n+1}}$ and
similarly for $Q_{2n+2}(E^{n+1}_{2n+2})$ with the result
\begin{equation}
\ket{\,\Phi^{n+1}, E^{n+1}}_{qe} = \frac{X^{d(\Phi^n, E^n)}}{(1+X^2)^n}\;
\frac{ X^{ 2 - \delta(\Phi^{n+1}_{2n+2}, E^{n+1}_{2n+2})-
\delta(\Phi^{n+1}_{2n+1}, E^{n+1}_{2n+1}) }}{(1+X^2)} \,
\ket{\Phi^{n+1}}_q \ket{E^{n+1}}_e \,.
\end{equation}
The $\delta$'s in the exponent of $X$
are Kronecker deltas and combining the factors of
$X$ gives the result.

\end{proof}

%\singlespacing
\chapter{The Existence of Quadratic Co-Events}\label{appendix:quadratic coevent existence}
%\doublespacing

\begin{theorem}\label{thm:quadratic coevent existence}
Let $\H$ be a histories theory with finite $\O$, and let $D$ be the associated decoherence functional. Then there exist
non-zero
\begin{eqnarray}
\p:\EA &\longrightarrow& \Z2 \nonumber
\\D_{\Z2}:\EA\times \EA &\longrightarrow& \Z2 \nonumber
\end{eqnarray}
such that:
\begin{eqnarray}
D(A,A)=0 &\Rightarrow& D_{\Z2}(A,A) = \p(A) = 0 \nonumber
\\\p(A) &=& D_{\Z2}(A,A) \nonumber
\\\p(A+ B + C) &=&
\p(A + B) + \p(A + C) + \p(B + C) \nonumber \\
&& + \p(A) + \p(B) + \p(C) \nonumber
\end{eqnarray}
\end{theorem}

Recall that the existence of quadratic co-events is an immediate corollary (corollary \ref{corollary:quadratic existence}). We build up our proof in several steps, beginning with decoherence functionals with integer entries (section \ref{sec:quadratic existence integer decoherence functional}) before generalising to the general case via decoherence functionals with rational entries (section \ref{sec:quadratic existence general decoherence functional}).

\section{Integer Decoherence Functional}\label{sec:quadratic existence integer decoherence functional}

Start with a histories theory $(\O,\EA,\mu_{\ZZ})$ with a finite sample space $\O = \{e_{1},\ldots,e_{n}\}$ and a measure derived from an integer valued decoherence functional $D_{\ZZ}$:

$$D_{\ZZ}:\EA \times \EA \longrightarrow \ZZ$$

We can regard $D_{\ZZ}$ as a symmetric matrix over the formal complex vector space of sums of the elements of $\Omega$, with $ij$ entry $D_{\ZZ}(e_{i},e_{j})$\footnote{Throughout we will abuse notation by writing $D(e_{i},e_{j})$ in place of $D(\{e_{i}\},\{e_{j}\})$} . Now if $N\in\EA$ is a null set with elements $N=\{e_{i_{1}},\ldots ,e_{i_{m}}\}$ then
\beq\label{eq:quadratic construction integer measure null action}
\mu_{\ZZ}(N) = D_{\ZZ}(N,N) = 0 \Leftrightarrow \sum_{j,k=1}^{m}D_{\ZZ}(e_{i_{j}},e_{i_{k}}) = 0
\eeq
We now wish to construct a $\mathbb{Z}_{2}$ valued form $\widetilde{D_{\Z2}}$ from $D_{\ZZ}$ to fit the requirements of the theorem. The most obvious way to do this is to set:
\bea
\widetilde{D_{\Z2}}:\EA \times \EA
&\longrightarrow& \Z2 \nonumber
\\\widetilde{D_{\Z2}}(e_{i},e_{j}) &=& (D_{\ZZ}(e_{i},e_{j})/L) \ mod \ 2
\label{eq:quadratic construction tildeDZ2 def}
\eea
Where $L$ is the greatest common divisor of the matrix elements $D_{\ZZ}(e_{i},e_{j})$. Dividing by $L$ ensures that at least one of the $D_{\ZZ}$ matrix elements will be odd, hence at least one of the $\widetilde{D_{\Z2}}$ matrix elements will be non-zero. Then we can mimic the construction of $\mu_{\ZZ}$ from $D_{\ZZ}$ we to set $\widetilde{\p}(A) = \widetilde{D_{\Z2}}(A,A)$. It is easy to see that $\mu_{\ZZ}(A) = 0\Rightarrow \widetilde{\p}(A) = 0$, however we have no guarantee that $\widetilde{\p}$ is non-zero. We know that at least one matrix element $\widetilde{D_{\Z2}}(e_i,e_j)$ is non-zero, but this term may be off-diagonal whereas it is easy to see that $\widetilde{\p}$ depends only on the diagonal terms, on which it is linear:
\bea
\widetilde{\p}(A) &=& \widetilde{D_{\Z2}}(A,A) \nonumber \\&=& \sum_{j,k}
\widetilde{D_{\Z2}}(e_{i_{j}},e_{i_{k}}) \nonumber
\\ &=& \sum_{j} \widetilde{D_{\Z2}}(e_{i_{j}},e_{i_{j}}) +
\overbrace{2*\sum_{j<k}\widetilde{D_{\Z2}}(e_{i_{j}},e_{i_{k}})}^{= 0}
\nonumber
\\ &=& \sum_{j}\widetilde{D_{\Z2}}(e_{i_{j}},e_{i_{j}})\nonumber
\\ &=& \sum_{j}(D_{\ZZ}(e_{i_{j}},e_{i_{j}}) \ mod \ 2)\nonumber
\\ &=& (\sum_{j} D_{\ZZ}(e_{i_{j}},e_{i_{j}})) \ mod \ 2\nonumber
\\ &=& \sum_j \mu_{\ZZ}(\{e_{i_j}\})  \ mod \ 2
\eea
To avoid this problem we must adjust the definition of our $\Z2$ valued form, which we do at the level of the original $\ZZ$ valued form $D_{\ZZ}$:
$$
\\\widehat{D_{\ZZ}}(e_{i},e_{j}) = \left\{\begin{array}{cc}2D_{\ZZ}(e_{i},e_{j}) & i > j
\\D_{\ZZ}(e_{i},e_{j}) & i = j
\\0 & i < j\end{array}\right.
$$
Note that we still have
\bea
\widehat{D_{\ZZ}}(A,A) &=& D_{\ZZ}(A,A) \nonumber \\
&=& \mu_{\ZZ}(A) \label{eq:quadratic construction adjusted dec functional compatible with measure}
\eea
$\forall A\in\EA$. As in equation \ref{eq:quadratic construction tildeDZ2 def} above we can use $\widehat{D_{\ZZ}}$ to define a $\Z2$ valued form:
\bea
D_{\Z2}:\EA \times \EA &\longrightarrow&
\mathbb{Z}_{2} \nonumber
\\D_{\Z2}(e_{i},e_{j}) &=& (\widehat{D_{\ZZ}}(e_{i},e_{j})/\widehat{L}) \ mod \ 2
\nonumber
\eea
where as before $\widehat{L}$ is the greatest common divisor of the matrix elements of $\widehat{D_{\ZZ}}$, and linearity ensures that we need only define $D_{\Z2}$ on the elements of the sample space. Note however that $D_{\Z2}$ is not symmetric in general. Following the constructions of $\mu_{\ZZ}$ and $\widetilde{\p}$, we can now define:
\bea
\p:\EA &\longrightarrow& \mathbb{Z}_{2} \nonumber \\
\p(A) &=& D_{\Z2}(A,A) \nonumber
\eea
Then using equation \ref{eq:quadratic construction adjusted dec functional compatible with measure} we can check that $\p$ is preclusive:
\begin{eqnarray}
\p(A) &=& \sum_{jk} D_{\Z2}(e_{i_{j}},e_{i_{k}}) \nonumber \\
&=& \sum_{jk} [\widehat{D_{\ZZ}}(e_{i_{j}},e_{i_{k}})/\widehat{L}] \ mod \ 2 \nonumber \\
&=& [\mu_{\ZZ}(A)/\widehat{L}] \ mod \ 2 \nonumber \\
\end{eqnarray}
Which is zero whenever $\mu_{\ZZ}(A)=0$. Unlike $\widetilde{\p}$, the contribution of the off-diagonal terms means that $\p$ is not linear, however it does obey the quadratic sum rule, similar to that followed by the quantum measure:
\begin{lemma}
$\p$ is not in general linear, however:
$$\p(A+ B + C) = \p(A + B) + \p(A + C) + \p(B + C) + \p(A) + \p(B) + \p(C)$$
\end{lemma}
\begin{proof}
First we define the \textit{indexing set} of $A\subset \Omega$ to be the unique $I_{A}\subset \mathbb{Z}$ such that $A = \{e_{i}|i\in I_{A}\}$. Now $i\in I_{A\sqcup B} \Rightarrow i\in A$ xor $i\in B$, so that $\sum_{i\in I_{A\sqcup B}} = \sum_{i\in I_{A}} + \sum_{i\in I_{B}}$. But then, if we are summing over $\mathbb{Z}_{2}$:
\bea
\sum_{i\in I_{A} + I_{B}} &=& \sum_{i\in I_{A\backslash A \cap B}} +
\sum_{i\in I_{B\backslash A \cap B}} + \overbrace{2*\sum_{i\in I_{A
\cap B}}}^{=0} \nonumber
\\&=& \sum_{i\in I_{A+ B}} \nonumber
\eea
This is the key result, dependant upon our use of $\Z2$, and we can now follow the derivation of the sum rule for the quantum measure. Noting that
$$D_{\Z2}(A,B) = \sum_{i\in I_{A}, j\in I_{B}} D_{\Z2}(e_{i_{j}},e_{i_{k}})$$
We see that
\begin{eqnarray}
D_{\Z2}(A+ B,C) &=& \sum_{i\in I_{A+ B}, j\in I_C} D_{\Z2}(e_{i_{j}},e_{i_{k}})\nonumber \\
&=& (\sum_{i\in I_{A}, j\in I_C} + \sum_{i\in I_{B}, j\in I_C}) D_{\Z2}(e_{i_{j}},e_{i_{k}})\nonumber \\
&=& D_{\Z2}(A,C) + D_{\Z2}(B,C)\nonumber
\end{eqnarray}
Then
\begin{eqnarray}
\p(A+ B) &=& D_{\Z2}(A+ B,A+ B)\nonumber \\
&=& D_{\Z2}(A,A) + D_{\Z2}(B,B) + D_{\Z2}(A,B) + D_{\Z2}(B,A)\nonumber \\
&=& \p(A) + \p(B) + E_{\Z2}(A,B)\nonumber
\end{eqnarray}
Where we have set $E_{\Z2}(A,B) = D_{\Z2}(A,B)+D_{\Z2}(B,A)$ for simplicity. $E_{\Z2}$ is symmetric, and follows the same linear sum rule as $D_{\Z2}$ since it is itself linear in $D_{\Z2}$. Since $E_{\Z2}(A,B)$ is not zero in general (because $D_{\Z2}$ is not symmetric in general), $\p$ will not in general be linear, however:
\begin{eqnarray}
\p(A+ B + C) &=& \p(\{A+ B\} + C)\nonumber \\
&=& \p(A+ B) + \p(C) + C(A+ B,C)\nonumber \\
&=& \p(A) + \p(B) +C(A,B) + \p(C) + E_{\Z2}(A,C) + E_{\Z2}(B,C)\nonumber \\
&=& \p(A) + \p(B) +  \p(C) + E_{\Z2}(A,B)  + E_{\Z2}(A,C) + E_{\Z2}(B,C) \nonumber \\
&& +2*(\p(A)+\p(B)+\p(C))\nonumber \\
&=& \p(A)+\p(B)+\p(C) \nonumber\\
&& + [\p(A)+\p(B)+E_{\Z2}(A,B)] \nonumber\\
&& + [\p(B)+\p(C)+E_{\Z2}(B,C)] \nonumber\\
&& + [\p(A)+\p(C)+E_{\Z2}(A,C)]\nonumber \\
&=& \p(A)+\p(B)+\p(C) + \p(A+ B) + \p(B+ C) + \p(A+C)\nonumber
\end{eqnarray}
\end{proof}

\section{General Decoherence Functional}\label{sec:quadratic existence general decoherence functional}

We now turn to the more general case, of a histories theory $\H$ with a finite sample space $\O = \{e_{1},\ldots,e_{n}\}$ where the measure is derived from the decoherence functional $D$, whose image is not longer constrained to be a subset of $\ZZ$. To prove the theorem it suffices to find an integer valued decoherence functional $D_{\ZZ}$ over $\EA$ whose null sets include the null sets associated with $D$. For then the maps $\p$ and $D_{\Z2}$ we found in section \ref{sec:quadratic existence integer decoherence functional} would meet the conditions of the theorem, and in particular $\p$ would be preclusive over $\H$. In other words we want to find an integer valued $D_{\ZZ}$ such that:
$$D_{\ZZ}(A,A) = 0 \Rightarrow D(A,A) = 0$$
Alternatively, denoting the set of null sets associated with $D$ by $N(D)$, we want a non-zero $D_{\ZZ}$ such that:
$$N(D)\subset N(D_{\ZZ})$$
Now considering $D$ as a complex matrix, we can decompose it into its real and imaginary components:
$$D = D^{Re} + D^{Im}$$
Then the hermiticity of $D$ means that $D^{Im}$ is skew-symmetric, so that:
$$D^{Im}(A,A) = 0 \ \forall \ A\in\EA$$
Hence $D$ and $D^{Re}$ have the same null sets. The next step is a jump from the reals to the rationals \cite{Mannan:private}:
\begin{lemma}
Given a (non-zero) real symmetric matrix $D^{Re}$ there exists a (non-zero) rational symmetric matrix $D_{\QQ}$ such that $Z(D_{\QQ})\supset Z(D^{Re})$.
\end{lemma}
\begin{proof}
Let $d_{ij}$ denote the matrix element $D^{Re}(e_i,e_j)$, with $i,j$ ranging from $1$ to $n$. Then each null set corresponds to a linear constraint upon the $d_{ij}$:
\beq\label{eq:quadratic construction null set constraint}
\sum_{i,j}\alpha_{ijl}d_{ij}=0
\eeq
Where $\alpha_{ijl}=\alpha_{jil}\in\Z2$, and $l\in\{1,\ldots ,m\}$ labels the null sets. We can linearly order the matrix elements $d_{ij}$, writing $c_{k}=d_{ij}$ where $k=n(i-1)+j$. Then our constraints in equation \ref{eq:quadratic construction null set constraint} can be written:
\beq
\sum_{k}\alpha_{kl}c_{k}=0
\eeq
This linear system can be `row-reduced' - assume there are $p$ linearly independent constraints, then the system is solvable (ie $\exists$ a non-zero solution) iff $p\leq n$. But since $D^{Re}$ exists, we know that the system is solvable, hence $p\leq n$.

Row-reducing will yield $n-p$ `independent' $c_{k}'s$, $c_{k_{1}},\ldots ,c_{k_{n-p}}$, and linear equations defining the remaining $c_{k}'s$ in terms of the independent ones. The coefficients of these equations will be rational combinations of the $\alpha_{kl}$, hence requiring the independent $c_{k_{1}},\ldots ,c_{k_{n-p}}$ to take integer values will result in all the $c_{k}'s$ being rational. Further, setting the vector $(c_{k_{1}},\ldots ,c_{k_{n-p}})$ to equal $(1,0,\ldots ,0),(0,1,\ldots ,0),\ldots ,(0,0,\ldots ,1)$ will yield a set of rational matrices that form a basis for all matrices $M$ that satisfy $Z(M)\supset Z(D^{Re})$.
\end{proof}

Finally, since multiplying by a scalar will no affect the sets of measure zero, given any rational decoherence functional $D_{\QQ}$ we can define an integer decoherence functional $D_{\ZZ}$ by
\beq
D_{\ZZ}(A,B) = M*D_{\QQ}(A,B)
\eeq
Where $M$ is the least common multiple of the denominators of the matrix elements of $D_{\QQ}$. Clearly $N(D_{\ZZ}) = N(D_{\QQ})$, hence using the result of section \ref{sec:quadratic existence integer decoherence functional} we have proved the theorem.

%\singlespacing
\chapter{The Many Slit System}\label{appendix:many slit}
%\doublespacing

Throughout this thesis we have made reference to many slit systems subject to certain key constraints, assuming always that the relevant gedankenexperimental realisation was possible. In this appendix we will justify our assumption, beginning with the simple double slit experiment before building on this to construct the general case.

Throughout, unless explicitly stated otherwise, we will assume idealised conditions, including pin-point slits, detector and source, as well as an idealised particle with a fully determined momentum. For definiteness we can assume we are using photons. Then if $\g$ is the distance from the source, at time $t$ the amplitude along a single path (ignoring superpositions for now) is given by
\beq\nonumber
\Psi(r,t) = \frac{A}{r}e^{i(\frac{rp}{\hbar}-\omega t)}
\eeq
where $\omega,A$ are constants. We will make the far field approximation $1/r=const$, which is accurate when $r$ is much larger than the wavelength $\lambda =h/p$. Thus absorbing $1/r$ into $A$ the amplitude is approximated by a plane wave
\beq\nonumber
\Psi(r,t) \approx Ae^{\frac{2\pi i r}{\lambda}}e^{-\omega t}
\eeq
Note that, with reference to chapter \ref{chapter:dynamics}, our use of this approximation actually means that the dark fringes we will calculate correspond to approximate rather than exact preclusion, even in the simple double slit gedankenexperiment. However as this may be true for most or even all `real-world' cases of idealised exact preclusion. We will not explore this issue here, but will simply proceed with our approximation, the reader is referred to chapter \ref{chapter:dynamics} for a discussion of recent attempts to deal with the notions of `almost null' events and approximate preclusion (section \ref{sec:approximate preclusion}).

\section{The Double Slit Gedankenexperiment}

We refer to the double slit system in section \ref{sec:double slit}, where our set up is subject to two constraints. Firstly, we require the initial state to be symmetric in the states corresponding to the two slits. Secondly we constrain the detector to lie in a dark fringe. In wavefunction terms this can be restated:
\begin{enumerate}
  \item The amplitude must be equivalent at the two slits
  \item The detector must lie in a dark fringe
\end{enumerate}
This system is realised by the Young Double Slit experiment \cite{Young:1802,Young:1804}, one of the canonical and foundational experiments in the development of quantum mechanics which, originally realised with photons \cite{Young:1802,Young:1804,Taylor:1909} has now also been realised with electrons \cite{Jonsson:1961,Merli:1976,Tonomura:1989} and other forms of matter (see for example \cite{Carnal:1991}). With an eye on the generalisation to the many slit case we will tweak the system slightly.

\begin{figure}
\begin{center}
\includegraphics[width=0.7\textwidth,angle=-90]{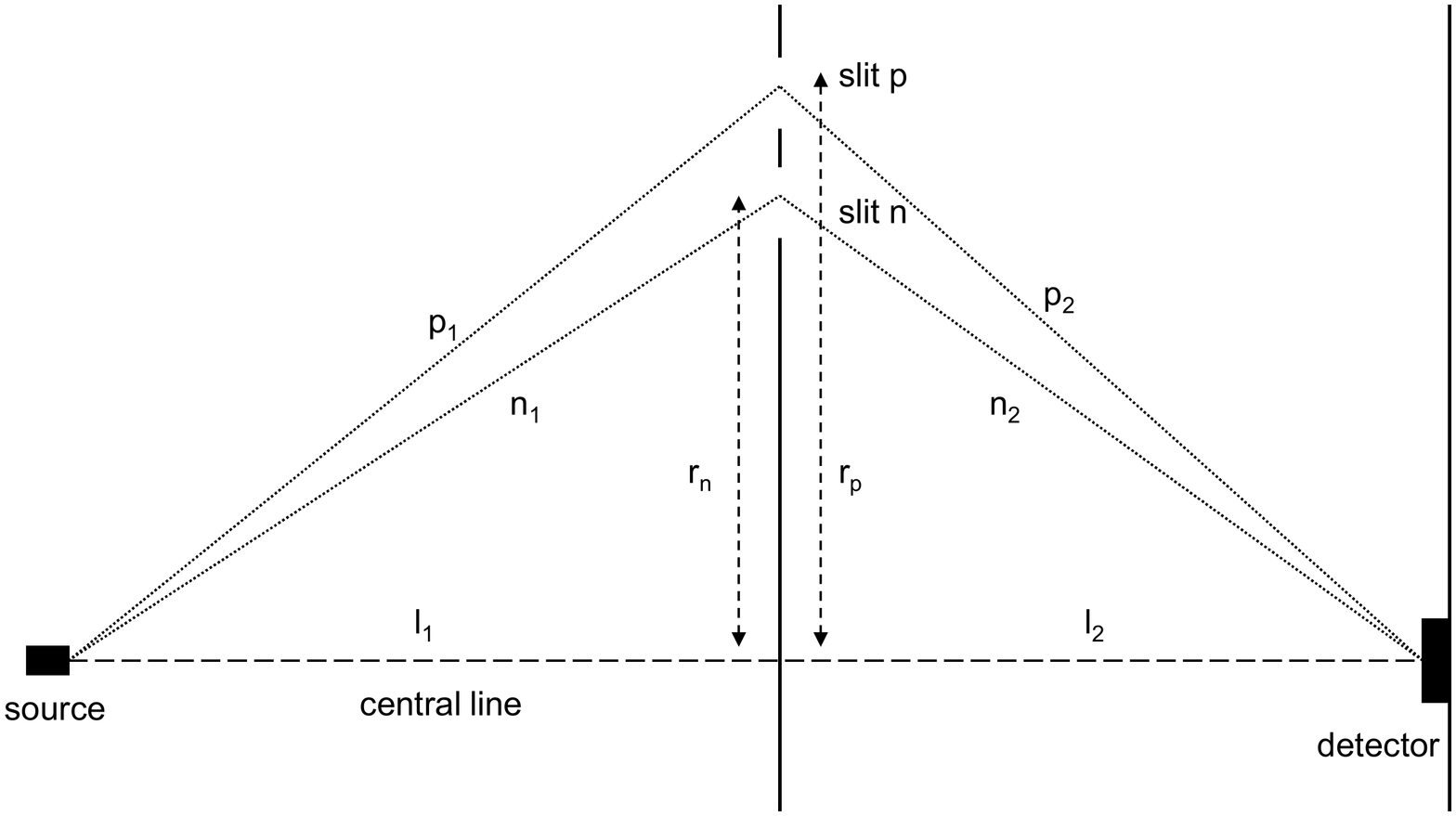}
\caption{\small{The Double Slit Experiment}} \label{fig:DoubleSlit}
\end{center}
\end{figure}

Our apparatus is shown in figure \ref{fig:DoubleSlit}. We label the amplitude at slits $p$ and $n$ $\Psi(p)$ and $\Psi(n)$ respectively. Our first constraint then becomes

\bea
\Psi(p_a,t) &=& \Psi(n_a,t) \nonumber \\
\Rightarrow p_a - n_a &=& m\lambda \label{eq:double slit experiment first constraint}
\eea
for $m\in\ZZ$. In the standard double slit experiment this is achieved by setting $r_p=-r_n$, however with an eye on our generalisation to the many slit case we require both $r_p$ and $r_n$ to be positive. Now noting that we have defined $\Psi$ \textit{along} a single path, our second constraint becomes
\bea
\Psi(p_a+p_b,t) &=& -\Psi(n_a+n_b,t) \nonumber \\
\Rightarrow p_b-n_b &=& \frac{l\lambda}{2} \label{eq:double slit experiment second constraint}
\eea
where $l\in\ZZ$. We now have two constraints, equations \ref{eq:double slit experiment first constraint} \& \ref{eq:double slit experiment second constraint}, and four degrees of freedom $l_1,l_2,r_p,r_n$. Thus our system will have a multiplicity of solutions.

\section{The Many Slit Gedankenexperiment}

We make reference to the triple, quadruple and sixteen slit systems in sections \ref{sec:triple slit}, \ref{sec:four slit} \& \ref{sec:sixteen slit} respectively, as well as using a general many slit system in section \ref{sec:many slit}. Since the general case encompasses the specific cases, we need only deal with the many slit system outlined in section \ref{sec:many slit}.

In constructing an $n$-slit gedankenexperiment to realise the system outlined in section \ref{sec:many slit}, we begin with the $n$ real numbers $c_i$ given to us by the proof of theorem \ref{thm:general poynomial scheme unitality problem}. Following section \ref{sec:many slit} we wish to ensure that the projection of the initial state onto the state corresponding to the $i$th slit is $N_\psi \sqrt{|c_i|}$, and that the projection of the state corresponding to the detector on the state corresponding to the $i$th slit is $N_\psi sign(c_i)\sqrt{|c_i|}$ where $N_\psi$ is the normalisation factor introduced in section \ref{sec:many slit}. As discussed in section \ref{sec:many slit} this will ensure the desired pattern of null sets outlined in lemma \ref{lemma:general polynomial problem structure}. Because theorem \ref{thm:general poynomial scheme unitality problem} has an existence proof, the actual numbers $c_i$ are never constructed. Thus the simplest means to proceed is to ensure that we can achieve the required projections given \textit{any} set of (suitably normalised) real numbers $c_i$. Since an overall normalisation will not affect our null set structure we can simply assume such a normalisation without explicitly constraining the $c_i$, so in practise we can set $N_\psi=1$ without altering the null set structure.

In terms of wavefunctions we can summarise this as two constraints. Given real numbers $\{c_i\}_{i=1}^n$ we can set $a_i=\sqrt{|c_i|}$ and $s_i=sign(c_i)$ so that $c_i=s_i a_i^2$. Then enumerating our slits $A_i$ we seek to be should be able to arrange the apparatus such that
\begin{enumerate}
\item The amplitude at slit $A_i$ should equal $a_i$
\item The contribution to the amplitude at the detector from the path passing through slit $A_i$ should be $s_i a_i$
\end{enumerate}

Now we turn back to the double slit experimental construction shown in figure \ref{fig:DoubleSlit}, which we outlined above. Crucially, we note that the intermediate screen is two dimensional and that as regards the positions of our slits $p$ \& $n$, we have only constrained them to lie at certain distances, $r_p$ \& $r_n$, from the central `straight line path' connecting the source and the detector (see figure \ref{fig:DoubleSlit}). Further, note that due to the symmetry of the system we could rotate either or both of the slits around the central line while preserving both the amplitude arriving at each slit from the source and the amplitude arriving at the detector from each slit. This is because, as we saw above, these amplitudes depend only on the lengths of the paths concerned, which would not be affected by such a rotation. Thus we could alter our apparatus to include many slits, $\{p_i\}_{i=1}^{m_p}$ and $\{n_i\}_{i=1}^{m_n}$, arranged on the circles of radius $r_p$ and $r_n$ around the central line.

As noted above, we are only concerned with the relative ratios of our amplitudes, which will be time independent, so we can use the amplitude at the slit $p_1$ as a unit. Thus we use the notation $\psi_s(x)$ to label the amplitude at slit $x$ divided by the amplitude at slit $p_1$, and the notation $\psi_d(x)$ to label the contribution of slit $x$ to the amplitude at the detector, divided by the amplitude at slit $p_1$\footnote{Because of the time dependence there will be moments when the amplitude at $p_1$ is zero, however at this point all amplitudes will be zero, at all other moments the ratios will be constant. Thus we can `smooth over' such points using continuity}. Thus we have $\psi_s(p_i)=\psi_s(n_i)=\psi_d(p_i)=1$ and $\psi_d(n_i)=-1$ for all $i$. This would provide us with the requisite $a_i$ for the double, triple and quadruple slit experiments (sections \ref{sec:double slit},\ref{sec:triple slit} and \ref{sec:four slit}), but we will need to make a further innovation to achieve the sixteen slit system and the general case of arbitrary $a_i$, $s_i$. We have two options:

\begin{enumerate}

\item
    To begin with we will assume that the $a_i$ are all rational. Then let $l$ be the least common multiple of the denominators of the $a_i$, so that $l a_i\in\ZZ$. Now we employ our final trick, constructing our $i$th `slit' $A_i$ not from a single slit $p_i$ or $n_i$, but from a multiplicity of slits. Noting that we can construct our apparatus to possess an arbitrary number of $p_i$ or $n_i$ slits, when $s_i=1$ we group together $l a_i$ of the $p$ slits and label this group $A_i$. Conversely, when $s_i=-1$ we group together $l a_i$ of the $n$ slits and label this group $A_i$. The state corresponding to $A_i$ will then be a linear superposition of the states corresponding to its constituent slits, and thus we will have
    \bea
    \psi_s(A_i) &=& a_i  \\
    \psi_d(A_i) &=& s_i a_i
    \eea
    as required. Note that in essence we have achieved the required system as a coarse graining of another many slit system, with each of our `slits' $A_i$ corresponding in reality to a group of slits.

    However, we have no reason to believe that the $a_i$ will all be rational. Nevertheless, we can approximate irrational $a_i$ with rational numbers to an arbitrary degree of accuracy. Although this will mean that our `null sets' will now in general be `approximately' rather than `exactly' of zero measure, our entire construction is subject to approximation, as noted above. In particular, we have assumed the far field approximation, and can require that the accuracy of our rational approximation of $a_i$ is far higher than the accuracy of our existing assumptions.

\item
    We can vary the size of the slits. In reality point slits are unattainable, and a more accurate assumption would be that each slit $A_i$ is circular with radius $R_i<<\lambda$, the wavelength. Varying $R_i$ will affect both $\psi_s(A_i)$ and $\psi_d(A_i)$, for example setting $R_i=2R_1$ is equivalent to grouping together four slits in the approach above. Thus given $S_i,a_i$, if $s_i=1$ we set $A_i$ to be a `$p$ slit' with its center at a distance of $r_p$ from the central line, similarly if $s_i=-1$ we set $A_i$ to be an `$n$ slit' with its center at a distance of $r_n$ from the central line. We can then adjust the radius of the slit:
    \beq
    R_i = R_1 \sqrt{a_i}
    \eeq
    to get
    \bea
    \psi_s(A_i) &=& a_i  \\
    \psi_d(A_i) &=& s_i a_i
    \eea
    as required. As with the first approach, this mechanism has its inaccuracies. In particular, varying the size of the slits may lead to approximate rather than exact preclusion at the detector. However, idealised point slits were never realisable in practise, and as mentioned earlier in this appendix we refer the reader to chapter \ref{chapter:dynamics} for a discussion of recent attempts to deal with the notions of `almost null' events and approximate preclusion (section \ref{sec:approximate preclusion}).

\end{enumerate}

\bibliography{Bib}
\bibliographystyle{plain}

\end{document}